\documentstyle[12pt,meqnarr,epsf]{book1}

\font\icalsymb calsymb at 8.0pt

\def\imercury{\hbox{\icalsymb \char 1}}

\makeatletter

\def\numberline#1{\hbox to\@tempdima{\hfil #1\kern 0.4em}}

\def\refname{\Large\bf \hspace*{-40mm} Bibliography}
\def\thebibliography#1{\vskip10mm\centerline{\refname}

\list
 {\arabic{enumi}.}{\settowidth\labelwidth{[#1]}
 \leftmargin\labelwidth
 \advance\leftmargin\labelsep
 \usecounter{enumi}}
 \def\newblock{\hskip .11em plus .33em minus .07em}
 \sloppy\clubpenalty4000\widowpenalty4000
 \sfcode`\.=1000\itemsep -1pt\relax}

\def\section{\@startsection{section}{1}{\z@}{-3.5ex plus-1ex minus
    -.2ex}{7mm}{\reset@font\Large\bf\raggedright}}

\def\ps@logunov{\let\@mkboth\@gobbletwo
 \def\@oddhead{}%
 \def\@oddfoot{{}\hfil \rm\thepage}%
 \def\@evenhead{}%
 \def\@evenfoot{\rm\thepage \hfil {}}%
 \def\sectionmark##1{}\def\subsectionmark##1{}}
 \ps@logunov
\makeatother
\def\thesubsection{\thesection\arabic{subsection}.}
\def\thefigure{\arabic{figure}}

\begin{document}
\def\dalam{\raise 0.43ex\hbox{\fbox{}}}

\newcommand{\smb}[3]{\stackrel{\scriptstyle #2}
{ #1}\,\!\!\!^{#3}}

\newcommand{\be}{\begin{equation}}
\newcommand{\ee}{\end{equation}}
\newcommand{\ba}{\begin{eqnarray}}
\newcommand{\ea}{\end{eqnarray}}
\newcommand{\pa}{\partial}
\newcommand{\f}{\frac}
\newcommand{\st}{\stackrel}
\newcommand{\tk}{\tilde\kappa}
\newcommand{\ep}{\epsilon}
\newcommand{\ds}{\displaystyle}

\newlength{\ots}
\settowidth{\ots}{\copyright\ }
\def\pr{\hspace*{\ots}}
\widowpenalty=10000

\newpage
\addtolength{\leftmargin}{0.6cm}
\setcounter{page}{5}

\section*{Preface}\addcontentsline{toc}{section}{Preface}

This monograph sums up studies performed in developing the
relativistic theory of gravity (RTG) and presented in
refs.~\cite{1,
1_1, 1_2, 1_3,
1_4,
1_5, 1_6,  1_7,  1_8, 1_9,  1_10, 1_11, 15,19}.
Detailed references to earlier works, that to a certain extent served
as
scaffolding in the construction of RTG, are given in the
monograph~\cite{1_3},
written together with prof.~M.~A.~Mes\-t\-vi\-ri\-sh\-vi\-li and
published
in
1989.
Therein, also, critical comments are presented concerning general
relativity
theory (GRT), which still remain in force. In order to facilitate
reading
in section~14 we provide elements of tensor analysis and Riemannian
geometry. As a rule, we make use of the set of units in which
$G=c=\hbar=1$.
However, in the final expressions we restore the dependence on the
constants
$G,c,\hbar$. Throughout the book, Greek letters assume values
0,1,2,3, while
Latin letters assume ---1,2,3.

The creation of this monograph advanced together with the completion
of
studies of individual issues, so it inevitably contains
recurrences, especially concerning such issues that are important for
understanding the essence of both RTG and GRT.

The hypothesis underlying RTG asserts that the gravitational field,
like
all other physical fields, develops in Min\-kow\-ski space, while the
source
of this field is the conserved energy-momentum tensor of matter,
including
the gravitational field itself. This approach permits constructing,
in a
unique manner, the theory of the gravitational field as a gauge
theory.
Here, there arises an effective Riemannian space, which literally has a
field nature. In GRT the space is considered to be Riemannian owing to the
presence of matter, so gravity is considered a consequence of
space--time
exhibiting curvature. The RTG gravitational field has spins 2 and 0
and
represents a physical field in the Faraday--Maxwell spirit. The
complete
set of RTG equations follows directly from the least action
principle.
Since all physical fields develop in Minkowski space, all fundamental
principles of physics --- the integral conservation laws of
energy--momentum
and of angular momentum --- are strictly obeyed in RTG. In the theory
the
Mach principle is realized: an inertial system is determined by the
distribution of matter. Unlike GRT, acceleration has an absolute
sense.
Inertial and gravitational forces are separated, and they differ in their
nature. The theory, unlike GRT, provides a unique explanation for all
gravitational effects in the Solar system. {\bf GRT does not comply
with
the equivalence principle, does not explain the equality of the
inert
and active gravitational masses, and gives no unique prediction for
gravitational effects. It does not contain the usual conservation
laws of
energy--momentum and of angular momentum of matter. }

It should be especially noted that the known post-New\-to\-ni\-an
approximation
do satisfy the equivalence principle, do provide a unique description
of
gravitational effects in the Solar system, and also establish the
equality
between the inertial and active gravitational masses. However, it
does not
follow uniquely from the GRT equations, since its derivation relies
on
additional assumptions, that do not follow from the theory, i.e. a
departure
occurs beyond the limits of GRT, which is based on the gravitational
field
being represented as a physical field, although this is not so in
GRT.
Therefore, this approximation cannot be considered a unique
consequence of
the GRT equations. It has rather been guessed, then derived from the
theory,
while,  according to RTG, the post--Newtonian approximation
follows
uniquely from equations of the theory. Thus, the post--Newtonian
approximation, previously applied for the description of
gravitational
effects follows directly from our theory. RTG introduces essential
changes
into the character of the development of the Universe and into the
collapse
of large masses.

Analysis of the development of a homogeneous and isotropic Universe
within
RTG leads to the conclusion that the Universe is infinite, and that
it is
``flat''. Its development proceeds cyclically from a certain maximum
density
down to a minimum and so on. Thus, no pointlike Big Bang occurred in
the past. There existed a state of high density and high temperature
at each
point in space.

According to RTG, the so-called cosmological ``expansion'' of the
Universe, observed by the red shift, is explained by changes in
the gravitational field, but not by relative motion ---  galaxies
escaping from each other, which actually does not take place.
Matter in the Universe is in a state of rest relative to an
inertial coordinate system. The peculiar velocities of galaxies
relative to an inertial system arose owing to a certain structure
of the inhomogeneity of the distribution of matter during the
period, when the Universe became transparent. This means that in
the past the distance between galaxies was never zero. The theory
predicts the existence in the Universe of a large hidden mass of
``dark matter". According to RGT, ``black holes" cannot exist: a
collapsing star cannot disappear beyond its gravitational radius.
Objects with large masses can exist, and they are characterized
not only by mass, but also by a distribution of matter density.
Since, in accordance with GRT, objects with masses exceeding three
solar masses transform, at the conclusive stage of their
evolution, into ``black holes", an object found to have a large
mass is usually attributed to ``black holes". Since RTG
predictions concerning the behaviour of large masses differ
essentially from GRT predictions, observational data of greater
detail are required for testing the conclusions of theory. Thus,
for example, in RTG spherically symmetric accretion of matter onto
a body of large mass, that is at its conclusive stage of evolution
(when the nuclear resources are exhausted), will be accompanied by
a significant release of energy owing to the fall of matter onto
the body's surface, while in GRT the energy release in the case of
spherically symmetric accretion of matter onto a ``black hole" is
extremely small, since the falling matter takes the energy with it
into the ``black hole". Observational data on such objects could
answer the question whether ``black holes" exist in Nature, or
not. Field concepts of gravity necessarily require introduction of
the graviton rest mass, which can be determined from observational
data: the Hubble ``constant" and the deceleration parameter~$q$.
According to the theory, the parameter $q$ can only be positive,
at present, i.e. deceleration of ``expansion" of the Universe
takes place, instead of acceleration. For this reason, the latest
observational data on acceleration of the ``expansion" must be
checked carefully, since the conclusions of theory concerning
``deceleration" follow from the general physical principles
mentioned above.

I sincerely wish to thank my teacher acad.~N.N.~Bogolubov who, during
the
hard years of searches and struggle, provided spiritual support as
well as
valuable advice that stimulated the research.

I am grateful to {\bf Providence} for my wife Anna Nikolaevna, who
for over
forty years was my support.

I am profoundly grateful to prof.~M.A.~Mestvirishvili for many years
of
joint work on the construction of relativistic theory of gravity.

I am grateful to acad.~A.M.~Baldin, acad.~V.S.~Vladimirov,
acad.~V.G.~Kadyshevsky, acad.~A.N.~Tavkhelidze for valuable
discussions.

I take advantage of the occasion to express my deep gratitude to
professors S.S.~Gershtein, V.I.~Denisov, Yu.M.~Loskutov, to
associate professor A.A.~Vlasov and Candidate of phy\-sics and
mathematics Yu.V.~Chugreev for common work and for numerous
discussions of the problems at issue. I am also grateful to
professors V.A.~Petrov, N.E.Tyurin, A.A.~Tyapkin and
O.A.~Khrustalev for useful discussions.

I express profound gratitude to acad.~A.M.~Baldin, corresponding
member of RAS S.S.~Gershtein and
prof.~M.A.~Mes\-t\-vi\-ri\-sh\-vi\-li, who read the entire
manuscript and made a number of valuable advices and comments.

\vspace*{5mm}
\noindent
 \hspace*{8.4cm} {\it A.A.~Logunov} \\
\hspace*{8.4cm} {\it April 2000}

\thispagestyle{empty}
\newpage

\pagestyle{logunov}

\section*{Introduction}\addcontentsline{toc}{section}{Introduction}

Since construction of the relativistic theory of gravity (RTG) is
based on special relativity theory (SRT), we shall deal with the
latter
in greater detail and in doing so we shall examine both the approach
of
Henri Poincar\'e and that of Albert Einstein. Such an analysis will
permit
a more profound comprehension of the difference between these
approaches
and will make it possible to formulate the essence of relativity
theory.

In analyzing the Lorentz transformations, H. Poincar\'e sho\-wed
that these transformations, together with all spatial rotations,
form a {\bf group} that does not alter the equations of
electrodynamics. Richard Feynman wrote the following about this:
{\it ``Precisely Poincar\'e proposed investigating what could be
done with the equations without altering their form.  It was
precisely his idea to pay attention to the symmetry properties of
the laws of physics"}$\,$\footnote{R.Feynman. The character of
physical laws. M.:Mir, 1968, p.97.}. H.Poincar\'e did not restrict
himself to studying electrodynamics; he discovered the equations
of relativistic mechanics and extended the Lorentz transformations
to all the forces of Nature. Discovery of the group, termed by
H.Poincar\'e the Lorentz group, made it possible for him to
introduce four-dimensional space-time with an {\bf invariant}
subsequently termed the interval

\vspace*{-0.1cm}
$$
d\sigma^2=(dX^0)^2-(dX^1)^2-(dX^2)^2-(dX^3)^2\;.\eqno{(\alpha)}
$$
Precisely from the above it is absolutely clear that time and spatial
length are {\bf relative}.

Later, a further development in this direction was made by Herman
Min\-kow\-ski, who introduced the concepts of timelike and
spacelike intervals. Following H.Poincar\'e and H.Min\-kow\-ski
exactly, the essence of relativity theory may be formulated thus:
{\bf all physical phenomena proceed in space--time, the geometry
of which is pseudo-Euclidean and is determined by the interval
$(\alpha)$.} Here it is important to emphasize, that {\bf the
geometry of space-time reflects those general dynamic properties,
that represent just what makes it universal.}  In four-dimensional
space (Minkowski space) one can adopt a quite arbitrary reference
frame

\vspace*{-0.1cm}
\[
 X^\nu=f^\nu(x^\mu)\;,
 \]

\vspace*{-0.3cm}
\noindent
realizing a mutually unambiguous correspondence with a Jacobian
differing
from zero. Determining the differentials
\[
 dX^\nu=\frac{\partial f^\nu}{\partial x^\mu}dx^\mu\;,
 \]
and substituting these expressions into $(\alpha)$ we find
$$
\!\!\!\!\!\!\!\!\!\!\!\!\!\!\!\!\!\!\!\!\!\!\!\!\!\!\!\!
\!\!\!\!\!\!\!\!\!\!
d\sigma^2=\gamma_{\mu\nu}(x)dx^\mu dx^\nu\;,
 \;\;\mbox {where}\;\;
$$
$$
\ds \gamma_{\mu\nu}(x)=\epsilon_\sigma\frac{\partial
f^\sigma}{\partial x^\mu}
\frac{\partial f^\sigma}{\partial x^\nu}\;,\,\,
\epsilon_\sigma =(1,-1,-1,-1)\;. \nonumber
\eqno{(\beta)}
$$

It is quite evident that the transition undergone to an arbitrary
reference
system did not lead us beyond the limits of pseudo-Euclidean
geometry.  But
hence it follows that non-inertial reference systems can also be
applied in
SRT. The forces of inertia arising in transition to an accelerated
reference system are expressed in terms of the Christoffel symbols of
Minkowski space. The representation of SRT stemming from the work of
H.Poincar\'e and H.Minkowski was more general and turned out
to
be
extremely
necessary for the construction of SRT, since it permitted
introduction of
the metric tensor $\gamma_{\mu\nu}(x)$ of Minkowski space in
arbitrary
coordinates and thus made it possible to introduce in a covariant
manner
the gravitational field, upon separation of the forces of inertia
from
gravity.

From the point of view of history it must be noted that in his
earlier
works$\,$\footnote{H.Poincar\'e. The principle of relativity.
M.:Atomizdat, 1973, pp.19, 33.}, ``The measurement of time" and ``The
present and future of
mathematical physics", H.Poincar\'e discussed in detail issues
of
the
constancy
of the velocity of light, of the simultaneity of events at different
points
of space determined by the synchronization of clocks with the aid of
a light
signal.   Later, on the basis of the relativity principle, which he
formulated in 1904 for all physical phenomena, as well as on the work
published by H.Lorentz the same year, {\bf H.Poincar\'e
discovered a
transformation group in 1905 and termed it the Lorentz group. This
permitted
him to give the following essentially accurate formulation of the
relativity
theory: the equations of physical processes must be invariant
relative to the
Lorentz group.} Precisely such a formulation was given by A.Einstein
in 1948:
{\it ``With the aid of the Lorentz transformations the special
principle of
relativity can be formulated as follows: The laws of Nature are
invariant
relative to the Lorentz transformation (i.e. a law of Nature should
not
change if it is referred to a new inertial reference frame with the
aid of
the Lorentz transformation for $x,y,z,t)"$}$\,$\footnote{Einstein A.
Collection of scientific works, Moscow: Nauka, 1966, vol.2, art.133,
p.660.}.

The existence of a group of coordinate-time transformations
signifies that there exists an infinite set of equivalent
(inertial) reference frames related by the Lorentz
transformations. From the invariance of equations it follows, in a
trivial manner, that physical equations in the reference frames
$x$ and $x'$, related by the Lorentz transformations, are
identical. But this means that any phenomenon described both in
$x$ and $x'$ reference systems under identical conditions will
yield identical results, i.e. the relativity principle is
satisfied in a trivial manner. Certain, even prominent, physicists
understood this with difficulty not even long ago, while others
have not even been able to. There is nothing strange in this fact,
since any study requires certain professionalism. What is
surprising is the following: they attempt to explain their
incomprehension, or the difficulty they encountered in
understanding, by\break H.Poincar\'e allegedly ``not having taken
the decisive step", ``not having gone to the end". But these
judgements, instead of the level of the outstanding results
achieved by H.Poincar\'e in relativity theory, characterize their
own level of comprehension of the problem.

Precisely for this reason W.Pauli wrote the
following in 1955 in connection with the 50-th anniversary of
relativity
theory: ``{\it Both Einstein and Poincar\'e relied on the preparatory
works
performed by H.A.Lorentz, who was very close to the final result, but
was
not able to take the last decisive step. In the results, obtained by
Einstein and Poincar\'e independently of each other, being identical
I
see
the profound meaning of the harmony in the mathematical method and
analysis
performed with the aid of thought experiments and based on the entire
set
of data of physical experiments"}$\,$\footnote{W.Pauli. Essays in
physics.
M.:Nauka, 1975, p.189.}.

Detailed investigation by H.Poincar\'e of the Lorentz group
invariants resulted in his discovery of the pseudo-Euclidean
geometry of space-time. Precisely on such a basis, he established
the four-dimensionality of physical quantities: force, velocity,
momentum, current. H.Poincar\'e's first short work appeared in
the reports of the French Academy of sciences before A.Einstein's
work was even submitted for publication. That work contained an
accurate and rigorous solution of the problem of electrodynamics
of moving bodies, and at the same time it extended the Lorentz
transformations to all natural forces, of whatever origin they
might be. Very often many historians, and, by the way, physicists,
also, discuss priority issues. A very good judgement concerning
this issue is due to academicians V.L.Ginzburg and
Ya.B.Zel'dovich, who in 1967 wrote:``{\it Thus, no matter what a
person has done himself, he cannot claim priority, if it later
becomes known that the same result was obtained earlier by
others}"$\,$\footnote{V.L.Ginzburg, Ya.B.Zel'dovich. Familiar and
unfamiliar Zel'dovich. M.:Nauka, 1993, p.88.}.

A.Einstein proceeded toward relativity theory from an analysis of
the concepts of simultaneity and of synchronization for clocks at
different points in space on the basis of the principle of
constancy of the velocity of light. <<{\it Each ray of light
travels in a reference frame at ``rest" with a certain velocity
$V$, independently of whether this ray of light is emitted by a
body at rest or by a moving body.}>> But this point cannot be
considered a principle, since it implies a certain choice of
reference frame, while a physical principle should clearly not
depend on the method of choosing the reference frame. In essence,
A.Einstein accurately followed the early works of H.Poincar\'e.
However, within such an approach it is impossible to arrive at
non-inertial reference frames, since in such reference frames it
is impossible to take advantage of clock synchronization, so the
notion of simultaneity loses sense, and, moreover, the velocity of
light cannot be considered constant.

In a reference frame undergoing acceleration the proper time
$d\tau$, where $$ d\sigma^2=d\tau^2-s_{ik}dx^idx^k,\,\,
d\tau=\frac{\gamma_{0\alpha}dx^\alpha}{\sqrt{\gamma_{00}}},\,\,
s_{ik}=-\gamma_{ik}+\frac{\gamma_{0i}\gamma_{0k}}{\gamma_{00}}\;
$$ is not a complete differential, so the synchronization of
clocks at different points in space depends on the synchronization
path. This means that such a concept cannot be applied for
reference frames undergoing acceleration. It must be stressed that
the coordinates in expression $(\beta)$ have no metric meaning, on
their own. Physically measurable quantities must be constructed
with the aid of coordinates and the metric coefficients
$\gamma_{\mu\nu}$. But all this remained misunderstood for a long
time in SRT, since it was usual to adopt A.Einstein's approach,
instead of the one of H.Poincar\'e and H.Minkowski. Thus, the
starting points introduced by A.Einstein were of an exclusively
limited and partial nature, even though they could create an
illusion of simplicity. It was precisely for this reason that even
in 1913 A.Einstein wrote: ``{\it In usual relativity theory only
linear orthogonal transformations are
permitted"}$\,$\footnote{Einstein A. Collection of scientific
works. Moscow: Nauka, 1965, vol.1, art.21, p.232.}. Or somewhat
later, in the same year, he writes: ``{\it In the original
relativity theory the independence of physical equations of the
specific choice of reference system is based on postulating the
fundamental invariant $ds^2=\sum dx^2_i$, while now the issue
consists in constructing a theory (general relativity theory is
implied -- A.L.), in which the role of the fundamental invariant
is performed by a linear element of the general form}
\[
ds^2=\sum_{i,k}g_{ik}dx^i dx^k\,\,
\mbox{''}\,
\footnote{Einstein A. Collection of
scientific works, Moscow: Nauka, 1965, vol.1, art.22,p.269.}.
\]
A.Einstein wrote something similar in 1930: ``{\it In special
relativity theory only such coordinate changes (transformations)
are allowed that provide for the quantity $ds^2$ (a fundamental
invariant)
in the new coordinates having the form of the sum of square
differentials
of the new coordinates. Such transformations are called Lorentz
transformations"}$\,$\footnote{Einstein A. Collection of scientific
works,
Moscow: Nauka, 1966, vol.2, art.95, p.281.}.

Hence it is seen that the approach adopted by A.Einstein did not lead
him
to the notion of space-time exhibiting a pseudo-Euclidean geometry. A
comparison of the approaches of H.Poincar\'e and A.Einstein to
the
construction of SRT clearly reveals H.Poincar\'e's approach to
be more
profound and general, since precisely H.Poincar\'e had defined
the
pseudo-Eu\-cli\-de\-an structure of space-time. A.Einstein's approach
essentially
restricted the boundaries of SRT, but, since the exposition of SRT in
the literature usually followed A.Einstein, SRT was quite a long time
considered valid only in inertial reference systems. Minkowski space
was
then treated like a useful geometric interpretation or like a
mathematical
formulation of the principles of SRT within the approach of Einstein.
Let
us now pass over to gravity. In 1905 H.Poincar\'e wrote: {\it
``... that
forces
of whatever origin, for example, the forces of gravity, behave in the
case of
uniform motion (or, if you wish, under Lorentz transformations)
precisely
like electromagnetic forces''}$\,$\footnote{H.Poincar\'e.
Special
relativity principle. M.:Atomizdat, 1973, p.152.}. This is precisely
the path we shall
follow.

A.Einstein, having noticed the equality of inertial and gravitational
masses, was convinced that the forces of inertia and of gravity are
related,
since their action is independent of a body's mass. In 1913 he
arrived at
the conclusion that, if in expression $(\alpha)$ ``{\it... we
introduce new
coordinates $x_1, x_2, x_3, x_4$, with the aid of some arbitrary
substitution, then the motion of a point relative to the new
reference frame
will proceed in accordance with the equation}
\[
\delta\{\int ds\}=0\;,
\]
{\it and}
\[
ds^2 =\sum_{\mu,\nu}g_{\mu\nu}dx^\mu dx^\nu\;."
\]
and he further pointed out: <<{\it The motion of a material point
in the new reference system is determined by the quantities
$g_{\mu\nu}$, which in accordance with the preceding paragraphs
should be understood as the components of the gravitational field,
as soon as we decide to consider this new system to be ``at
rest">>}$\,$\footnote{Einstein A. Collection of scientific works,
Moscow: Nauka, 1965, vol.1, art.23, p.286.}. {\bf Identifying in
such a manner the metric field, obtained from $(\alpha)$ with the
aid of coordinate transformations, and the gravitational field is
without physical grounds, since transformations of coordinates do
not lead us beyond the framework of pseudo-Euclidean geometry.}
From our point of view, it is not permitted to consider such a
metric field as the gravitational field, since this contradicts
the very essence of the concept of a field as a physical reality.
Therefore, it is impossible to agree with the following reasoning
of A.Einstein: <<{\it The gravitational field ``exists" with
respect to the system $K'$ in the same sense as any other physical
quantity that can be defined in a certain reference system, even though
it does not exist in system $K$. There is nothing
strange, here, and it may be readily demonstrated by the following
example taken from classical mechanics. Nobody doubts the
``reality" of kinetic energy, since otherwise it would be
necessary to renounce energy in general. It is clear, however,
that the kinetic energy of bodies depends on the state of motion
of the reference system: by an appropriate choice of the latter it
is evidently possible to provide for the kinetic energy of uniform
motion of a certain body to assume, at a certain moment of time, a
given positive or zero value set beforehand. In the special case,
when all the masses have equal in value and equally oriented
velocities, it is possible by an appropriate choice of the
reference system to make the total kinetic energy equal to zero.
In my opinion the analogy is complete>>}$\,$\footnote{Einstein A.
Collection of scientific works, Moscow: Nauka, 1965, vol.1,
art.46, p.620.}.

As we see, Einstein renounced the concept of a classical field,
such as the Faraday--Maxwell field possessing density of
energy-momentum, in relation to the gravitational field. Precisely
this path led him up to the construction of GRT, to gravitational
energy not being localizable, to introduction of the pseudotensor
of the gravitational field. If the gravitational field is
considered as a physical field, then it, like all other physical
fields, is characterized by the energy-momentum tensor
$t^{\mu\nu}$. If in some reference frame, for instance, $K'$,
there exists a gravitational field, this means that certain
components (or all of them) of the tensor $t^{\mu\nu}$ differ from
zero. The tensor $t^{\mu\nu}$ cannot be reduced to zero by a
coordinate transformation, i..e, if a gravitational field exists,
then it represents a physical reality, and it cannot be
annihilated by a choice of reference system. It is not correct to
compare such a gravitational field with kinetic energy, since the
latter is not characterized by a covariant quantity. It must be
noted that such a comparison is not admissible, also, in GRT,
since the gravitational field in this theory is characterized by
the Riemann curvature tensor. If it differs from zero, then the
gravitational field exists, and it cannot be annihilated by a
choice of reference system, even locally.

Accelerated reference systems have played an important
he\-u\-ri\-stic role in A.Einstein's creative work, although they
have nothing to do with the essence of GRT. By identifying
accelerated reference systems to the gravitational field,
A.Einstein came to perceive the metric space-time tensor as the
principal characteristic of the gravitational field. But the
metric tensor reflects both the natural properties of geometry and
the choice of reference system. In this way the possibility arises
of explaining the force of gravity kinematically, by reducing it
to the force of inertia. But in this case it is necessary to
renounce the gravitational field as a physical field. ``{\it
Gravitational fields} (as A.Einstein wrote in 1918) {\it may be
set without introducing tensions and energy
density."}$\,$\footnote{Einstein A. Collection of scientific
works, Moscow: Nauka, 1965, vol.1, art.47, p.627.}. But that is a
serious loss, and one cannot consent to it. However, as we shall
further see, this loss can be avoided in constructing RTG.

Surprisingly, even in 1933 A.Einstein wrote: <<{\it In special
Relativity
theory --- as shown by H.Minkowski --- this metric was
quasi-Euclidean,
i.e. the square ``length" $ds$ of a linear element represented a
certain
quadratic function of the coordinate differentials. If, on the other
hand,
new coordinates are introduced with the aid of a linear
transformation,
then $ds^2$ remains a homogeneous function of the coordinate
differentials,
but the coefficients of this function $(g_{\mu\nu})$ will no longer
be
constant, but certain functions of the coordinates. From a
mathematical
point of view this means that the physical (four-dimensional) space
possesses a Riemannian metric}>>$\,$\footnote{Einstein A. Collection of
scientific works, Moscow: Nauka, 1966, vol.2, art.110, p.405.}.

This is certainly wrong, since a pseudo-Euclidean metric cannot be
transformed into a Riemannian metric by transformation of the
coordinates. But the main point, here, consists in something else,
namely, in that in this way, thanks to his profound intuition,
A.Einstein arrived at the necessity of introducing precisely
Riemannian space, since he considered the metric tensor
$g_{\mu\nu}$ of this space to describe gravity. This was
essentially how the tensor nature of gravity was revealed. The
unity of the Riemannian metric and gravity is the main principle
underlying general relativity theory. V.A.Fock wrote about this
principle: ``{\it ... precisely this principle represents the
essence of Einstein's theory of gravity"}$\,$\footnote{V.A.Fock.
Theory of space, time and gravity. M.:Gostekhizdat, 1961, p.308.}.
From a general point of view, however, the answer to the following
question still remains unclear: why is it necessary to relate
gravity precisely to Riemannian space, and not to any other.

The introduction of Riemannian space permitted using the scalar
curvature $R$ as the Lagrangian function and, with the aid of the
least
action principle, to obtain the Hilbert--Einstein equation. Thus,
the
construction of Einstein's general relativity theory was completed. Here,
as
particularly stressed by J.L.Synge: ``{\it In Einstein's theory the
presence
or absence of a gravitational field depends on whether the Riemann
tensor
differs from or equals to zero. This is an absolute property, which is
in no
way related to the world line of any
observer"}$\,$\footnote{J.L.Synge.
Relativity: the general theory. M.:Foreign literature publishers, 1963,
p.9.}.

In GRT, however,
difficulties arose with the conservation laws of energy-momentum and
angular
momentum. D.Hilbert wrote in this connection: ``{\it ... I claim that
within
general relativity theory, i.e. in the case of general invariance of
the
Hamiltonian function, there definitely exist no energy equations ...
corresponding to the energy equations in orthogonal-invariant
theories, I
could even point to this circumstance as a characteristic feature of
general
relativity theory"}$\,$\footnote{V.P.Vizgin. Relativistic theory of
gravity. M.:Nauka, 1981, p.319.}. All the above is explained by the
absence in Riemannian
space of the ten-parameter group of motion of space-time, so it is
essentially impossible to introduce energy-momentum and angular
momentum
conservation laws, similar to those that hold valid in any other
physical
theory.

Another feature peculiar to GRT, as compared to known theories,
consists
in the presence of second-order derivatives in the Lagrangian
function
$R$. About fifty years ago Nathan Rosen demonstrated that if, together
with
the Riemannian metric $g_{\mu\nu}$ one introduces the metric
$\gamma_{\mu\nu}$
of Minkowski space, then it becomes possible to construct the scalar
density of the Lagrangian of the gravitational field, which will not
contain
derivatives of orders higher than one. Thus, for example, he
constructed
such a density of the Lagrangian which led to the Hilbert-Einstein
equations. Thus came into being the bimetric formalism. However,
such
an approach immediately complicated the problem of constructing a
theory of
gravity, since, when using the tensors $g_{\mu\nu}$ and
$\gamma_{\mu\nu}$,
one can write out a large number of scalar densities, and it is
absolutely
not clear which scalar density must be chosen as the Lagrangian
density for
constructing the theory of gravity. Although the GRT mathematical
apparatus
does permit introducing, instead of ordinary derivatives, covariant
derivatives of Minkowski space, the metric $\gamma_{\mu\nu}$ not
being
present in the Hilbert-Einstein equations renders its utilization in
GRT
devoid of any physical meaning, because the solutions for the metric
$g_{\mu\nu}$ are independent of the choice of $\gamma_{\mu\nu}$. It
must
be noted that substitution of covariant derivatives for ordinary
derivatives
in Minkowski space leaves the Hilbert-Einstein equations intact. This
is
explained by the fact that, if in Minkowski space one substitutes
covariant
derivatives for ordinary ones in the Riemann curvature tensor, it
will not
change. Such a substitution in the Riemann tensor is nothing, but an
identical transformation. Precisely for this reason such a freedom in
writing out the Riemann tensor cannot be taken as an advantage  within
the
framework of GRT, since the metric tensor of Minkowski space does not
enter
into the Hilbert-Einstein equations.

In constructing RTG, this freedom in writing the Riemann tensor
turns out to be extremely necessary. But in this case the metric
of Minkowski space enters into the equations of the gravitational
field, and the field itself is considered as a physical field in
Minkowski space. In GRT we only deal with the metric of Riemannian
space as the main characteristic of gravity, in which both the
features of geometry itself and the choice of reference frame are
reflected. When the gravitational interaction is switched off,
i.e. when the Riemann curvature tensor equals zero, we arrive at
Minkowski space. It is precisely for this reason that in GRT the
problem arises of satisfying the equivalence principle, since it
is impossible to determine in which reference frame (inertial or
accelerated) we happened to be when the gravitational field was
switched off.

The relativistic theory of gravity, presented in this work is
constructed as a field theory of the gravitational field within the
framework of special relativity theory. The starting point is the
hypothesis
that the energy-momentum tensor --- which is a universal
characteristic of
matter --- serves as the source of gravity. The gravitational field
is
considered to be a universal physical field with spins 2 and 0, owing
to
the action of which the effective Riemannian space arises. This permits
to find the gauge group and to construct unambiguously the Lagrangian
density
of the gravitational field. The set of equations of this theory is
generally
covariant and form-invariant with respect to the Lorentz group. Here,
it is
necessary in the theory to introduce the graviton mass. The graviton
mass
essentially influences the evolution of the Universe and alters the
character of the gravitational collapse.

The goal of this work
is a further development of the ideas by H.Poincar\'e, H.Minkowski,
A.Einstein,
D.Hilbert, N.Rosen, V.A.Fock, S.Gupta, V.Thirring, R.Feynman,
S.Weinberg
in the domain of theory of relativity and gravity.

\thispagestyle{empty}
\newpage
\section{The geometry of space-time}

In Chapter~II, ``Space and time", of his book ``Recent ideas",
H.Poincar\'e
wrote: ``{\it The principle of physical relativity may serve for
defining of space. It can be said to provide us with a novel instrument for measurement.
I
shall explain. How can a solid body serve for measuring or, to be
more
correct, for constructing space? The point is the following: in
transferring
a solid body from one place to another we, thus, note that it can be
first
applied to one figure, then to another, and we conventionally agree
to
consider these figures equal to each other. Geometry originated from
this
convention. Geometry is nothing but a science of mutual
interrelationships
between such transformations or, speaking in the mathematical
language, a
science of the structure of the group formed by these
transformations, i.e.
of the group of motions of solid bodies.

Now, consider another group, the group of transformations not
altering our differential equations. We arrive at a new way for
defining the equality between two figures. We no longer say: two
figures are equal, if one and the same solid body can be applied
to both one and the other figures. We shall say: two figures are
equal, when one and the same mechanical system, so distant from
its neighbours that it may be considered isolated, being first
thus situated so its material points reproduce the first figure,
and then so they reproduce the second figure, behaves in the
second case precisely like in the first. Do these two approaches
differ in essence? No.

A solid body represents a mechanical system, just like any other. The
only
difference between the previous and the new definitions of space
consists
in that the latter is broader, since it allows substitution of any
mechanical system for the solid body. Moreover, our new convention
not only
defines space, but time, also. It provides us with an explanation of
what are
two equal time intervals or of what is represented by a time interval
twice
as long as another"}$\,$\footnote{H.Poincar\'e. On science.
M.:Nauka, 1938, p.427.}.

Precisely in this way, by discovering the
group of transformations not altering the Maxwell--Lorentz equations,
H.Poin\-ca\-r\'e introduced the notion of four-dimensional
space-time
exhibiting
pseudo-Euclidean geometry. This concept of geometry was later
developed
by H.Minkowski.

We have chosen the pseudo-Euclidean geometry of space-time as the
basis of
the relativistic theory of gravity presently under development,
since it is the fundamental Minkowski space for all physical fields,
including the gravitational field. Minkowski space cannot be
considered
to exist a priori, since it reflects the properties of matter and,
hence,
cannot be separated from it. Although formally, precisely owing to
the
structure of space being independent of the form of matter, it is
sometimes
dealt with abstractly, separately from matter. In Galilean
coordinates of
an inertial reference system in Minkowski space, the interval that
characterizes the structure of geometry and that is an invariant by
construction, has the form
\[
d\sigma^2=(dx^0)^2-(dx^1)^2-(dx^2)^2-(dx^3)^2\;.
\]
Here $dx^\nu$ represent differentials of the coordinates.
In spite of the fact that the interval $d\sigma$, as a geometric
characteristic of space-time, is independent of the choice of
reference
system, which is due to its very construction, one can still
encounter in
modern text-books on theoretical physics (see, for instance,
Ref.~\cite{2})
``proofs" of the interval being the same in all inertial reference
systems
although it is an invariant and is independent of the choice of
reference
system.

Even~ such~ an~ outstanding~ physicist~ as~ L.I.Mandelstam wrote in his
book~\cite{3}: ``{\it ... special relativity theory cannot answer the
question,
how a clock behaves when moving with acceleration and why it slows
down,
because it does not deal with reference systems moving with
acceleration".}
The incorrect assertions in \cite{4,4_2,4_1,4_3} can be explained by
Minkowski
space
being considered by many people to be only some formal geometrical
interpretation of SRT within A.Einstein's approach, instead of a
revelation
of the geometry of space-time. The issues of such limited concepts as
the
constancy of the speed of light, the synchronization of clocks, the
speed of light being independent of the motion of its source became
the
most discussed topics. All this narrowed the scope of SRT and
retarded the
understanding of its essence. {\bf And its essence actually consists
only in
that the geometry of space-time, in which all physical processes
occur, is
pseudo-Euclidean.}

In an arbitrary reference system the interval assumes the form
\[
d\sigma^2=\gamma_{\mu\nu}(x)dx^\mu dx^\nu\;,
\]
$\gamma_{\mu\nu}(x)$ is the metric tensor of Minkowski space. We note
that
one cannot, in principle, speak of the synchronization of clocks or
of the
constancy of the speed of light in an non-inertial reference system
\cite{5}. Most likely, precisely the lacking clarity on the essence of SRT
led
A.Einstein to concluding: ``{\it that within the framework of special
relativity theory there is no place for a satisfactory theory of
gravity"}$\,$\footnote{Einstein A. Collection of scientific works.
Moscow:
Nauka, 1967, vol.4, art.76, p.282.}.
Free motion of a test body in an arbitrary reference system takes
place along
a geodesic line of Minkowski space:
 \[
\frac{DU^\nu}{d\sigma}=\frac{d U^\nu}{d\sigma}
+\gamma^\nu_{\alpha\beta}U^\alpha U^\beta=0\;,\]
where  $U^\nu=$\Large$\frac{dx^\nu}{d\sigma}\,$\normalsize
, $\gamma^\nu_{\alpha\beta}(x)$ are Christoffel symbols defined by the
expression
\[
\gamma^\nu_{\alpha\beta}(x)=\frac{1}{2}\gamma^{\nu\sigma}
(\partial_\alpha\gamma_{\beta\sigma}+
\partial_\beta\gamma_{\alpha\sigma}
-\partial_\sigma\gamma_{\alpha\beta})\;.
\]

In 1921, in the article ``Geometry and experiment", A. Einstein
wrote:
{\it ``The issue of whether this continuum has an Euclidean, Riemannian
or
any other structure is a physical issue, which can only be settled by
experiment, and not an issue of convention concerning a choice of
simple
expedience...}"$\,$\footnote{Einstein A. Collection of scientific
works.
Moscow: Nauka, 1965, vol.2, art.61, p.87.}. This is, naturally,
correct. But there immediately arises
a question: what experiment?  There may exist quite many experimental
facts.
Thus, for example, it is possible, in principle, by studying the
motion of
light and of test bodies, to establish unambiguously the geometry of
space-time. Must a physical theory be based on it?  At first sight,
the
answer to this question could be positive. And the issue would seem
settled.
Precisely such was the path that A.Einstein took in constructing GRT.
Test
bodies and light move along geodesic lines of Riemannian space-time. So
he based the theory on Riemannian space. However, the situation is much
more
complex. All types of matter satisfy conservation laws of
energy-momentum
and of angular momentum. Precisely these laws, that originated from a
generalization of numerous experimental data, characterize the
general
dynamic properties of all forms of matter by introducing universal
characteristics permitting quantitative description of the
transformation
of some forms of matter into others. And all this also represents
experimental facts, which have become fundamental physical
principles. What
should be done with them? If one follows A.Einstein and retains
Riemannian
geometry as the basis, then they must be discarded. That price
would be
too high. It is more natural to retain them for all physical fields,
including the gravitational field. But, in this case, theory must,
then, be
based on Minkowski space, i.e. on the pseudo-Euclidean geometry of
space-time. We have adopted precisely this approach, following
H.Poincar\'e.
The fundamental principles of physics, that reflect the numerous
available
experimental facts, indicate what geometry of the space-time it is
actually
necessary to use as the basis of gravity theory. Thus, the issue of
the
structure of the space-time geometry is actually a physical issue,
that
should be resolved by experiment, and, from our point of view, the
structure
of the geometry of space-time is not determined by specific
experimental
data on the motion of test bodies and of light, but by fundamental
physical
principles based on the entire set of existing experimental facts. It
is
precisely here that our initial premises for constructing the theory
of
gravity differ completely from the ideas applied by A.Einstein as the
basis
of GRT. But they are fully consistent with the ideas of H.Poincar\'e.

We have chosen the pseudo-Euclidean geometry of space-time as the
basis of
the relativistic theory of gravity, but that certainly does not mean
that
the effective space will also be pseudo-Euclidean. The influence of
the
gravitational field may be expected to lead to a change in the
effective
space. We shall deal with this issue in detail in the next section.
The
metric of Minkowski space permits introducing the concepts of
standard
length and time intervals, when no gravitational field is present.

\thispagestyle{empty}
\newpage
\section{The energy-momentum tensor of matter as the source of the
gravitational field}

Owing to the existence in Minkowski space of the Poincar\'e
ten-parameter
group of motion, there exist for any closed physical system ten
integrals
of motion, i.e. the conservation laws of energy-momentum and angular
momentum hold valid. Any physical field in Minkowski space is
characterized
by the density of the energy-momentum tensor $t^{\mu\nu}$, which is a
general universal characteristic of all forms of matter that
satisfies both
local and integral conservation laws. In an arbitrary reference
system the
local conservation law is written in the form
\[
D_\mu t^{\mu\nu}=\partial_\mu t^{\mu\nu}
+\gamma^\nu_{\alpha\beta}t^{\alpha\beta}=0\;.
\]
Here $t^{\mu\nu}$ is the total conserved density of the
energy-momentum
tensor for all the fields of matter;  $D_\mu$ represents the
covariant
derivative in Minkowski space. Here and further we shall always deal
with
the densities of scalar and tensor quantities defined in accordance
with
the rule
 \[
 \tilde\phi=\sqrt{-\gamma}\phi\;,\,\,
\tilde\phi^{\mu\nu}=\sqrt{-\gamma}\phi^{\mu\nu},\,\,
\gamma=\det (\gamma_{\mu\nu})\;.
\]
The introduction of densities is due to an invariant volume element
in
Minkowski space being determined by the expression
\[
\sqrt{-\gamma}d^4x\;,
\]
while an invariant volume element in Riemannian space is given by the
expression
\[
\sqrt{-g}d^4x\;, g=\det (g_{\mu\nu})\;.
\]
Therefore, the principle of least action assumes the form
\[
\delta S=\delta\int Ld^4x=0\;,
\]
where $L$ is the scalar density of the Lagrangian of matter. In
deriving
Euler's equations with the aid of the principle of least action we
shall
automatically have to deal precisely with the variation of the
Lagrangian
density. According to D.Hilbert, the density of the energy-momentum
tensor $t^{\mu\nu}$ is expressed via the scalar density of the
Lagrangian
$L$ as follows:
\be
t^{\mu\nu}=-2\frac{\delta L}{\delta\gamma_{\mu\nu}}\;, \label{1}
\ee
where
\[
\frac{\delta L}{\delta\gamma_{\mu\nu}}=
\frac{\partial L}{\partial\gamma_{\mu\nu}}-
\partial_\sigma\left (
\frac{\partial L}{\partial\gamma_{\mu\nu,\sigma}}\right )\;,\,\,
\gamma_{\mu\nu,\sigma}=\frac{\partial\gamma_{\mu\nu}}{\partial
x^\sigma}\;.
\]

{\bf Owing to gravity being universal, it would be natural to assume
the
conserved density of the energy-momentum tensor of all fields of
matter,
$t^{\mu\nu}$, to be the source of the gravitational field.} Further,
we
shall take advantage of the analogy with electrodynamics, in which
the
conserved density of the charged vector current serves as the source
of
the electromagnetic field, while the field itself is described by the
density of the vector potential
$\tilde A^\nu$:
\[
\tilde A^\nu=(\tilde \phi, \tilde A)\;.
\]

In the absence of gravity, Maxwell's equations of electrodynamics
will have the following form in arbitrary coordinates: \ba
&\gamma&^{\alpha\beta}D_\alpha D_\beta \tilde A^\nu +\mu^2\tilde
A^\nu=4\pi j^\nu, \nonumber\\*
&D&_\nu \tilde A^\nu=0\;,\nonumber
\ea
Here, for generalization we have introduced the parameter $\mu$,
which, in
the system of units
$\hbar=c=1$ is the photon rest mass.

Since we have decided to consider the conserved density of the
energy-momentum $t^{\mu\nu}$ to  be the source of the gravitational
field,
it is natural to consider the gravitational field a tensor field and
to
describe it by the density of the symmetric tensor
$\tilde\phi^{\mu\nu}$:
\[
\tilde\phi^{\mu\nu}=\sqrt{-\gamma}\phi^{\mu\nu},
\]
and in complete analogy with Maxwell's electrodynamics the equations
for
the gravitational field can be written in the form
\ba
&\gamma&^{\alpha\beta}D_\alpha D_\beta
\tilde\phi^{\mu\nu}+m^2\tilde\phi^{\mu\nu}=\lambda t^{\mu\nu}\;,
\label{2}
\\*
&D&_\mu\tilde\phi^{\mu\nu}=0\;. \label{3}
\ea
Here $\lambda$ is a certain constant which, in accordance with the
principle of correspondence to Newton's law of gravity, should be equal
to
$16\pi$. Equation (\ref{3}) excludes spins~1~and~$0'$, only retaining
those
polarizational properties of the field, that correspond to
spins~2~and~0.

The density of the energy-momentum tensor of matter $t^{\mu\nu}$
consists
of the density of the energy-momentum tensor of the gravitational
field,
$t^{\mu\nu}_{g}$, and of the energy-momentum tensor of matter,
$t_M^{\mu\nu}$. We understand matter to comprise all the fields of
matter,
with the exception of the gravitational field,
\[
t^{\mu\nu}=t^{\mu\nu}_g+t^{\mu\nu}_{M}\;.
\]
The interaction between the gravitational field and matter is
taken into account in the density of the energy-momentum tensor of
matter, $t^{\mu\nu}_M$.

Back in 1913 A.Einstein wrote \cite{6}: ``{\it ... the tensor of
the
gravitational field $\vartheta_{\mu\nu}$ is the source of a field
together
with the tensor of material systems $\Theta_{\mu\nu}$. The energy of
the
gravitational occupying a special position as compared with all other
forms
of energy would result in inadmissible consequences"} ~\cite{6}. We
have
adopted
precisely this idea of A.Einstein as the basis for constructing the
relativistic theory of gravity (RTG). In constructing general
relativity
theory (GRT) A.Einstein was not successful, since instead of the
energy-momentum tensor of the gravitational field there arose in GRT
the
pseudotensor of the gravitational field. All this happened because
A.Einstein did not consider the gravitational field a physical field
(such
as the Faraday--Maxwell field) in Minkowski space. Precisely for this
reason the equations of GRT do not contain the metric of Minkowski
space.
From equations (\ref{2}) it follows that they will also  be
non-linear for
the gravitational field proper, since the density of the tensor
$t^{\mu\nu}_g$ is the source of the gravitational field.

Equations (\ref{2}) and (\ref{3}), which we formally declared the
equations
of gravity by analogy with electrodynamics, must be derived from the
principle of least action, since only in this case we will have an
explicit
expression for the density of the energy-momentum tensor of the
gravitational field and of the fields of matter. But, to this end it
is
necessary to construct the density of the Lagrangian of matter and of
the
gravitational field. Here it is extremely important to realize this
construction on the basis of general principles. Only in this case
one can 
speak of the theory of gravity. The initial scalar density of the
Lagrangian
of matter may be written in the form
$$
L=L_g (\gamma_{\mu\nu}, \tilde\phi^{\mu\nu})
+L_M (\gamma_{\mu\nu}, \tilde\phi^{\mu\nu}, \phi_A)\;,
$$
here $L_g$ is the density of the Lagrangian of the gravitational
field;
$L_M$ is the density of the Lagrangian of the fields of matter;
$\phi_A$ represents the fields of matter.

The equations for the gravitational field and the fields of matter
have,
in accordance with the principle of least action, the form
\ba
\frac{\delta L}{\delta\tilde\phi^{\mu\nu}}&=&0\;,\label{4}\\*
\frac{\delta L_M}{\delta\phi_A}&=&0\;. \label{5}
\ea
Equations (\ref{4}) differ from equations (\ref{2}), first of all, in
that
the variational derivative of the density of the Lagrangian there
is the
derivative with respect to the field $\tilde\phi^{\mu\nu}$, while the
variational derivative in equations (\ref{2}) is, in agreement with
definition (\ref{1}), taken from the density of the Lagrangian
over the
metric $\gamma_{\mu\nu}$. For equations (\ref{4}) to reduce to
equations
(\ref{2}) for any form of matter it is necessary to assume the tensor
density $\tilde\phi^{\mu\nu}$ to be always present in the density of
the
Lagrangian together with the tensor density $\tilde \gamma^{\mu\nu}$
via
some common density $\tilde g^{\mu\nu}$ in the form
\be
\tilde g^{\mu\nu}=\tilde\gamma^{\mu\nu}+\tilde\phi^{\mu\nu},\,\,
\tilde g^{\mu\nu}=\sqrt{-g}g^{\mu\nu}. \label{6}
\ee

{\bf Thus arises the effective Riemannian space with the metric
$g^{\mu\nu}(x)$. Since the gravitational field
$\tilde\phi^{\mu\nu}(x)$,
like all other physical fields in Minkowski space, can be described
within a
sole coordinate map, it is evident from expression (2.6) that the
quantity $\tilde g^{\mu\nu}(x)$ can also be fully defined in a sole
coordinate map.} For description of the effective Riemannian space due to the
influence of the gravitational field, no atlas of maps is required,
which
is usually necessary for describing Riemannian space of the general
form.
This means that our effective Riemannian space has a simple topology. In
GRT topology is not simple. Precisely for this reason, GRT cannot, in
principle, be constructed on the basis of ideas considering gravity a
physical gravitational field in Minkowski space.

If condition (2.6) is taken into account, the density of the
Lagrangian
$L$ assumes the form
\[
L=L_g(\gamma_{\mu\nu}, \tilde g^{\mu\nu})+
L_M (\gamma_{\mu\nu}, \tilde g^{\mu\nu}, \phi_A)\;.
\]

It must be stressed that condition (\ref{6}) permits substituting the
variational derivative with respect to $\tilde g^{\mu\nu}$ for the
variational derivative with respect to $\tilde\phi^{\mu\nu}$, and to
express the variational derivative with respect to $\gamma_{\mu\nu}$
through the variational derivative with respect to $\tilde
g^{\mu\nu}$
and the variational derivative with respect to $\gamma_{\mu\nu}$
entering
explicitly into the Lagrangian $L$. Indeed,
\ba
\frac{\delta L}{\delta \tilde\phi^{\mu\nu}}=
\frac{\delta L}{\delta \tilde g^{\mu\nu}}=&0&\;,\label{7}\\*
\frac{\delta L}{\delta \gamma_{\mu\nu}}=
\frac{\delta^\star L}{\delta \gamma_{\mu\nu}}+
\frac{\delta L}{\delta \tilde g^{\alpha\beta}}&\cdot&
\frac{\partial\tilde g^{\alpha\beta}}{\partial \gamma_{\mu\nu}}\;.
\label{8}
\ea
The derivation of the latter formula is presented in detail in
Appendix
(A.17). The asterisk in formula (\ref{8}) indicates the variational
derivative of the density of the Lagrangian with respect to the
metric
$\gamma_{\mu\nu}$ which is explicitly present in $L$. In agreement
with
(\ref{1}), formula (\ref{8}) can be written in the form
\[
t^{\mu\nu}=-2\frac{\delta L}{\delta\tilde g^{\alpha\beta}}\cdot
\frac{\partial\tilde g^{\alpha\beta}}{\partial\gamma_{\mu\nu}}
-2\frac{\delta^\star L}{\delta\gamma_{\mu\nu}}\;.
\]
Taking equation (\ref{7}) into account in the above expression we
obtain
\be
t^{\mu\nu}=-2\frac{\delta^\star L}{\delta\gamma_{\mu\nu}}\;.
\label{9}
\ee
Comparing equations (\ref{9}) and (\ref{2}) we obtain the condition
\be
-2\frac{\delta^\star L}{\delta\gamma_{\mu\nu}}
=\frac{1}{16\pi}
[\gamma^{\alpha\beta}D_\alpha D_\beta\tilde\phi^{\mu\nu}
+m^2\tilde\phi^{\mu\nu}]\;, \label{10}
\ee
which, in case it is fulfilled, makes it possible to derive the
equations
of the gravitational field, (\ref{2}) and (\ref{3}), directly from
the
principle of least action. Since the fields of matter are not present
in the right-hand side of (\ref{10}), this means that the variation
in
density of the Lagrangian of matter, $L_M$, with respect to the
explicitly
present metric $\gamma_{\mu\nu}$ must be zero. For no additional
restrictions on the motion of matter determined by equations
(\ref{5})
to arise, it hence follows directly that the tensor $\gamma_{\mu\nu}$
does
not explicitly enter into the expression for the density of the
Lagrangian
of matter $L_M$. Expression (\ref{10}) then assumes the form
\be
-2\frac{\delta^\star L_g}{\delta\gamma_{\mu\nu}}
=\frac{1}{16\pi}
[\gamma^{\alpha\beta}D_\alpha D_\beta\tilde\phi^{\mu\nu}
+m^2\tilde\phi^{\mu\nu}]\;. \label{11}
\ee
Thus, everything reduces to finding the density of the Lagrangian of
the gravitational field proper, $L_g$, which would satisfy condition
(\ref{11}).

At the same time, from the previous arguments we arrive at the
important
conclusion that the density of the Lagrangian of matter, $L$, has the
form
\be
L=L_g(\gamma_{\mu\nu}, \tilde g^{\mu\nu})+
L_M (\tilde g^{\mu\nu}, \phi_A)\;. \label{12}
\ee

{\bf Thus, from the requirement that the density of the
energy-momentum
tensor of matter be the source of the gravitational field it follows
in
a natural way that the motion of matter should take place in
effective
Riemannian space.} This assertion has the character of a theorem. Hence
it
becomes clear, why Riemannian space arose, instead of some other.
Precisely
this circumstance provides us with the possibility of formulating, in
section~3, the gauge group, and then to construct the density of the
Lagrangian (\ref{55*}) satisfying condition (\ref{11}), in accordance
with
(B.20).

An interesting picture arises consisting in that the motion of matter
in Minkowski space with the metric $\gamma_{\mu\nu}$ under the influence
of the gravitational field $\phi^{\mu\nu}$ is identical to the motion of
matter in effective Riemannian space with the metric $g_{\mu\nu}$,
determined by expression (\ref{6}). We term such interaction of the
gravitational field with matter the
g~e~o~m~e~t~i~z~a~t~i~o~n
p~r~i~n~c~i~p~l~e.
The geometrization principle is a consequence of the initial
assumption
that a universal characteristic of matter --- the density of the
energy-momentum tensor --- serves as the source of the gravitational
field.
Such a density structure of the Lagrangian of matter indicates that a
unique
possibility is realized for the gravitational field to be attached
inside
the Lagrangian density of matter directly to the density of the
tensor
$\tilde\gamma^{\mu\nu}$.

{\bf The effective Riemannian space is literally
of a field origin, owing to the presence of the gravitational field.}
Thus,
the reason that the effective space is Riemannian, and not any other,
lies
in the hypothesis that a universal conserved quantity --- the density
of
the energy-momentum tensor --- is the source of gravity. We shall
explain
this fundamental property of gravitational forces by comparing them
with the
electromagnetic forces.

In the case of a homogeneous magnetic field, a
charged particle in Minkowski space is known to undergo, due to the
Lorentz
force, motion along a circle in the plane perpendicular to the
magnetic
field. However, this motion is far from identical even for charged
particles, if their charge-to-mass ratio differ. Moreover, there
exist
neutral particles, and their trajectories in a magnetic field are
just
straight lines. Therefore, owing to the non-universal character of
electromagnetic forces their action cannot be reduced to the geometry
of
space-time. Gra\-vi\-ty is another issue. It is universal, any test
bodies
move along identical trajectories given identical initial
conditions. In
this case, owing to the hypothesis claiming the energy-momentum
tensor of
matter to be the source of the gravitational field, it turns out to
be
possible to describe these trajectories by geodesic lines in the
effective
Riemannian space-time due to the presence of the gravitational field in
Minkowski space. In those regions of space, where a whatever small
gravitational field is present, we have metric properties of space
which
up to a high precision approach the actually observed properties of
pseudo-Euclidean space. On the other hand, when the gravitational
fields
are strong, the metric properties of the effective space become
Riemannian.
But in this case, also, the pseudo-Euclidean geometry does not vanish
without trace --- it is observable and manifests itself in that the
motion
of bodies in effective Riemannian space is not free by inertia, but
proceeds
with acceleration with respect to pseudo-Euclidean space in Galilean
coordinates. Precisely for this reason, acceleration in RTG, unlike
GRT,
has an absolute sense. Consequently, ``Einstein's lift" cannot serve
as an
inertial reference frame. This is manifested in that a charge at rest
in
``Einstein's lift" will emit electromagnetic waves. This physical
phenomenon
should also testify in favour of the existence of Minkowski space. As
we
shall further see, the metric of Minkowski space can be defined from
studies
of the distribution of matter and of the motion of test bodies and
light in
effective Riemannian space. We shall raise this issue again in
section~7.

The equation of motion of matter does not include the metric tensor
$\gamma_{\mu\nu}$ of Minkowski space. Minkowski space will only
influence
the motion of matter by means of the metric tensor $g_{\mu\nu}$ of
Riemannian space, derived, as we shall further see, from the equations
of
gravity, which contain the metric tensor $\gamma_{\mu\nu}$ of
Minkowski
space. Since the effective Riemannian metric arises on the basis of the
physical field given in Minkowski space, hence it follows that
effective
Riemannian space has a simple topology and is presented in a single map.
If, for instance, matter is concentrated in a region of the
island-type,
then in Galilean coordinates of an inertial reference system the
gra\-vi\-ta\-ti\-o\-nal field $\tilde\phi^{\mu\nu}$ cannot decrease slower than
$1/r$, but this circumstance imposes a strong restriction on the
asymptotic
behaviour of the metric $g_{\mu\nu}$ of effective Riemannian geometry
\be
g_{\mu\nu}=\eta_{\mu\nu}+0\left(\frac{1}{r}\right),\;
\mbox{here}\; \eta_{\mu\nu}=(1,-1,-1,-1)\;. \label{13}
\ee

If, on the other hand, one simply takes as the starting point the
Riemannian metric, without assuming it to have originated from the
action of a physical field, then such restrictions do not arise,
since the asymptotics of the metric $g_{\mu\nu}$ even depends on
the choice of three-dimensional space coordinates.  Physical
quantity, however, in principle, cannot depend on the choice of
the three-dimensional space coordinates. RTG imposes no
restrictions on the choice of reference system. The reference
system may be arbitrary, if only it realizes a one-to-one
correspondence between all the points of the inertial reference
system in Minkowski space and provides for the following
inequalities, necessary for introducing the concepts of time and
spatial length, to be satisfied:
\[
\gamma_{00}>0,\, dl^2 = s_{ik} dx^i\, dx^k>0;\, i,k = 1,2,3,
\]
where
\[
s_{ik} = - \gamma_{ik} + \frac{\gamma_{0i} \gamma_{0k}}
{\gamma_{00}}.
\]

In our theory of gravity the geometrical characteristics of Riemannian
space arise as field quantities in Minkowski space, and for this
reason
their transformational properties become tensor properties, even if
this
was previously not so, from the conventional point of view. Thus, for
instance, the Christoffel symbols, given as field quantities in
Galilean
coordinates of Minkowski space become tensors of the third rank. In a
similar manner, ordinary derivatives of tensor quantities in
Cartesian
coordinates of Minkowski space are also tensors.

{\baselineskip=17pt The question may arise: why is no division of
the metric, like (\ref{6}), performed in GRT by introduction of
the concept of gravitational field in Minkowski space? The
Hilbert-Einstein equations only contain the quantity $g_{\mu\nu}$,
so, consequently, it is impossible to say unambiguously with the
help of which metric $\gamma_{\mu\nu}$ of Minkowski space we
should define, in accordance with (\ref{6}), the gravitational
field. But the difficulty consists not only in the above, but,
also, in that the solutions of Hilbert-Einstein equations are
generally found not in one map, but in a whole atlas of maps. Such
solutions for $g_{\mu\nu}$ describe Riemannian space with a
complex topology, while the Riemannian spaces, obtained by
representation of the gravitational field in Minkowski space, are
described in a sole map and have a simple topology. It is
precisely for these reasons that field representations are not
compatible with GRT, since they are extremely rigorous. But this
means that no field formulation of GRT in Minkowski space can
exist, in principle, no matter how much someone and who might want
this to happen. The apparatus of Riemannian geometry is inclined
towards the possibility of introducing covariant derivatives in
Minkowski space, which we took advantage of in constructing RTG.
But to implement this, it was necessary to introduce the metric of
Minkowski space into the gravitational equations, and it, thus,
turned out to be possible to realize the functional relationship
of the metric of Riemannian space, $g_{\mu\nu}$, with the metric
of Minkowski space, $\gamma_{\mu\nu}$. But this will be dealt with
in detail in subsequent sections.
\par}

\thispagestyle{empty}
\newpage
\section{The gauge group of transformations}

Since the density of the Lagrangian of matter has the form
\begin{equation}
L_M(\tilde g^{\mu\nu},\, \phi_A), \label{14}
\end{equation}
it is easy to find the group of transformations, under which the
density
of the Lagrangian of matter is only changed by the divergence. To
this
end we shall take advantage of the action
\begin{equation}
S_M = \int \, L_M(\tilde g^{\mu\nu}, \, \phi_A)\, d^4x \label{15}
\end{equation}
being invariant under infinitesimal transformations of
coordinates,
\begin{equation}
x^{\prime\,\alpha} = x^\alpha + \xi^\alpha(x), \label{16}
\end{equation}
where $\xi^\alpha$ is the four-vector of an infinitesimal
displacement. The field functions $\tilde g^{\mu\nu}$, $\phi_A$ vary
as
follows under these transformations of  coordinates:
\be
\begin{array}{l}
\tilde g^{\prime\, \mu\nu}(x^{\prime}) = \tilde g^{\mu\nu}(x) +
\delta_\xi \tilde g^{\mu\nu} (x) + \xi^\alpha(x) D_\alpha
\tilde g^{\mu\nu}(x), \\[1,5mm]
\phi^\prime_A(x^\prime) = \phi_A(x) + \delta_\xi \phi_A(x) +
\xi^\alpha (x) D_\alpha\phi_A(x), \label{17}
\end{array}
\ee
where the expressions
\be
\begin{array}{l}
\delta_\xi \tilde g^{\mu\nu} (x) = \tilde g^{\mu\alpha}D_\alpha
\xi^\nu(x) + \tilde g^{\nu\alpha}D_\alpha
\xi^\mu(x) - D_\alpha(\xi^\alpha \tilde g^{\mu\nu}),
\\[1,5mm]
\delta_\xi \phi_A(x) = -\xi^\alpha(x) D_\alpha \phi_A(x) +
F_{A; \beta}^{B; \alpha}\, \phi_B(x) D_\alpha \xi^\beta (x)
\label{18*}
\end{array}
\ee
are Lie variations.

The operators $\delta_\xi$ satisfy the conditions of Lie algebras,
i.e.
the commutation relation
\begin{equation}
[\delta_{\xi_1}, \; \delta_{\xi_2}](\cdot) = \delta_{\xi_3}\, (\cdot)
\label{19}
\end{equation}
and the Jacobi identity
\begin{equation}
[\delta_{\xi_1},\, [\delta_{\xi_2},\, \delta_{\xi_3}]] +
[\delta_{\xi_3},\, [\delta_{\xi_1}, \, \delta_{\xi_2}]] +
[\delta_{\xi_2}, \, [\delta_{\xi_3}, \, \delta_{\xi_1}]] = 0,
\label{20}
\end{equation}
where
$$
\xi^\nu_3 = \xi^\mu_1 D_\mu \xi^\nu_2 - \xi^\mu_2 D_\mu \xi^\nu_1 =
\xi^\mu_1 \partial_\mu \xi^\nu_2 - \xi^\mu_2 \partial_\mu \xi^\nu_1.
$$
For (\ref{19}) to hold valid the following conditions must be
satisfied:
\begin{equation}
F^{B; \,\mu}_{A;\, \nu} \;F^{C;\, \alpha}_{B; \,\beta}
- F^{B;\, \alpha}_{A;\, \beta}\, F^{C;\, \mu}_{B; \,\nu} =
f^{\mu \alpha;\, \tau}_{\nu \beta;\, \sigma} \; F^{C;\, \sigma}_{A;\,
 \tau},
\label{21}
\end{equation}
where the structure constants $f$ are
\begin{equation}
f^{\mu \alpha;\, \tau}_{\nu \beta; \, \sigma} =
\delta^\mu_\beta \delta^\alpha_\sigma \delta^\tau_\nu -
\delta^\alpha_\nu \delta^\mu_\sigma \delta^\tau_\beta. \label{22}
\end{equation}
It is readily verified that they satisfy the Jacobi equality
\begin{equation}
f^{\alpha \nu; \, \sigma}_{\beta \mu;\, \tau} \;
f^{\tau \rho;\, \omega}_{\sigma \varepsilon;\, \delta} +
f^{\nu \rho; \, \sigma}_{\mu \varepsilon; \, \tau} \;
f^{\tau \alpha; \, \omega}_{\sigma \beta; \, \delta} +
f^{\rho \alpha; \, \sigma}_{\varepsilon \beta; \, \tau}\;
f^{\tau \nu; \, \omega}_{\sigma \mu;\, \delta} = 0 \label{23}
\end{equation}
and have the property of antisymmetry,
\begin{displaymath}
f^{\alpha \nu;\, \rho}_{\beta \mu; \, \sigma} = - f^{\nu \alpha; \,
\rho}_
{\mu \beta; \, \sigma}.
\end{displaymath}

The variation of action under the coordinate
transformation
(\ref{16}) equals zero:
\begin{equation}
\delta_c S_³ = \int\limits_{\Omega^\prime} \; L^\prime_³(x^\prime)
\, d^4x^\prime - \int\limits_\Omega \; L_M(x)\, d^4x = 0. \label{24}
\end{equation}
The first integral in (\ref{24}) can be written in the form
\begin{displaymath}
\int\limits_{\Omega^\prime} \; L^\prime_M(x^\prime)\, d^4x^\prime =
\int\limits_\Omega\; J\, L^\prime_M(x^\prime)\, d^4x,
\end{displaymath}
where
\begin{displaymath}
J = \det\Biggl(\frac{\partial x^{\prime \alpha}}{\partial\,
x^\beta}\biggr)\,.
\end{displaymath}
In the first order of $\xi^\alpha$ the determinant $J$ equals
\begin{equation}
J = 1 + \partial_\alpha \xi^\alpha(x). \label{25}
\end{equation}
Taking into account the expansion
\begin{displaymath}
L^\prime_M(x^\prime) = L^\prime_M(x) + \xi^\alpha(x) \frac{\partial
L_M}{\partial x^\alpha}\, ,
\end{displaymath}
as well as (\ref{25}), one can represent the expression for the
variation
in the form
\begin{displaymath}
\delta_c S_M = \int\limits_\Omega\, [\delta L_M(x) +
\partial_\alpha(\xi^\alpha L_M(x))]\, d^4x = 0.
\end{displaymath}

Owing to the integration volume $\Omega$ being arbitrary, we have the
identity
\begin{equation}
\delta\, L_M(x) = -\partial_\alpha(\xi^\alpha (x) L_M(x)), \label{26}
\end{equation}
where the Lie variation  $\delta L_M$ is
\begin{eqnarray}
\delta L_M(x) &=& \frac{\partial L_M}{\partial \tilde g^{\mu \nu}}\,
\delta \tilde g^{\mu \nu} + \frac{\partial
L_M}{\partial(\partial_\alpha
\tilde g^{\mu \nu)}} \, \delta(\partial_\alpha \tilde g^{\mu
\nu}) +\nonumber \\*
&+&\frac{\partial L_M}{\partial \phi_A} \, \delta \phi_A +
\frac{\partial L_M}{\partial(\partial_\alpha \phi_A)} \,
\delta(\partial_\alpha \phi_A). \label{27}
\end{eqnarray}
Hence, for instance, it follows that if the scalar density depends
only
on $\tilde g^{\mu\nu}$ and its derivatives, it will vary under
transformation (\ref{18*}) only by the divergence
\begin{aequation}{26}{a}
\delta L(\tilde g^{\mu \nu}(x))
= -\partial_\alpha(\xi^\alpha(x)\,
L(\tilde g^{\mu \nu}(x))), \label{26a}
\end{aequation}
where the Lie variation $\delta L$ is
$$
\delta L(\tilde g^{\mu \nu}(x)) = \frac{\partial L}{\partial \tilde
g^{\mu
\nu}}
\, \delta \tilde g^{\mu \nu} + \frac{\partial
L}{\partial(\partial_\alpha
\tilde g^{\mu \nu})} \, \delta(\partial_\alpha \tilde g^{\mu \nu})+
$$
\begin{aequation}{27}{a}
\!\!\!\!\!\!\!\!\!\!\!\!\!\!\!\!\!\!\!\!\!\!\!\!\!\!\!\!\!\!\!\!\!\!\
!
\!\!\!\!
\!\!\!\!\!\!\!\!\!\!\!\!\!\!\!
+\frac{\partial L}{\partial
 (\partial_\alpha\partial_\beta\tilde g^{\mu\nu})}
\delta (\partial_\alpha\partial_\beta \tilde g^{\mu\nu}). \label{27a}
\end{aequation}

The Lie variations (\ref{18*}) were established within the context of
the
coordinate transformations (\ref{16}). But one may also adopt another
standpoint, in accordance with which transformations (\ref{18*}) can
be
considered gauge transformations. In this case an arbitrary
infinitesimal
four-vector $\xi^\alpha (x)$ will already be a gauge vector, but no
longer the displacement vector of the coordinates. To stress the
difference
between the gauge group and the group of coordinate transformations,
we
shall further use the notation $\varepsilon^\alpha (x)$ for the group
parameter and call the transformation of field functions
\begin{eqnarray}
&&\tilde g^{\mu \nu}(x) \rightarrow \tilde g^{\mu \nu}(x) +
\delta \tilde g^{\mu \nu}(x), \nonumber\\*[-0.3cm] \label{28}
\\*[-0.3cm]
&&\phi_A(x) \rightarrow \phi_A(x) + \delta \phi_A(x) \nonumber
\end{eqnarray}
with the variations
\ba
&&\delta_\varepsilon \tilde g^{\mu \nu}(x) =
\tilde g^{\mu \alpha} D_\alpha \varepsilon^{\nu}(x) +
\tilde g^{\nu \alpha} D_\alpha \varepsilon^{\mu}(x) -
D_\alpha(\varepsilon^\alpha \tilde g^{\mu \nu}),\nonumber
\\*[-0.1cm] \label{29} \\*[-0.2cm]
&&\delta_\varepsilon \phi_A(x)=-\varepsilon^\alpha(x)
D_\alpha
\phi_A(x) +
F^{B;\, \alpha}_{A;\, \beta} \, \phi_B(x)\, D_\alpha
\,\varepsilon^\beta(x) 
\nonumber
\ea
{\bf gauge transformations}.

In full compliance with formulae (\ref{19}) and (\ref{20}),
the operators
satisfy the same Lie algebra, i.e. the commutation relation
\begin{equation}
[\delta_{\varepsilon_1},\; \delta_{\varepsilon_2}](\cdot) =
\delta_{\varepsilon_3}(\cdot) \label{30}
\end{equation}
and the Jacobi identity
\begin{equation}
[\delta_{\varepsilon_1}, \, [\delta_{\varepsilon_2}, \,
\delta_{\varepsilon_3}]] +
[\delta_{\varepsilon_3}, \, [\delta_{\varepsilon_1}, \,
\delta_{\varepsilon_2}]] + [\delta_{\varepsilon_2},\, [\delta_
{\varepsilon_3}, \, \delta_{\varepsilon_1}]] = 0.  \label{31}
\end{equation}
Like in the preceding case, we have
$$
\varepsilon^\nu_3 = \varepsilon^\mu_1 D_\mu \varepsilon^\nu_2 -
\varepsilon^\mu_2 D_\mu \varepsilon^\nu_1 =
\varepsilon^\mu_1 \partial_\mu \varepsilon^\nu_2 - \varepsilon^\mu_2
\partial_\mu \varepsilon^\nu_1.
$$

The gauge group arose from the geometrized structure of the scalar
density
of the Lagrangian of matter, $L_M(\tilde g^{\mu \nu},\, \phi_A)$,
which
owing to identity (\ref{26}) only changes by the divergence under
gauge
transformations (3.16). Thus, the geometrization principle, which
determined the universal character of the interaction of matter and
of
the gravitational field, has provided us with the possibility of
formulating
the non-commutative infinite-dimensional gauge group (3.16).

The essential difference between the gauge and coordinate
transformations
will manifest itself at the decisive point of the theory in the
course of
construction of the scalar density of the Lagrangian of the
gravitational
field proper. The difference arises owing to the metric tensor
~$\gamma_{\mu \nu}$ not changing under gauge transformation, and,
consequently, owing to (\ref{6}) we have
$$
\delta_\varepsilon \tilde g^{\mu \nu}(x) = \delta_\varepsilon
\tilde \phi^{\mu \nu} (x).
$$

From (3.16) the transformation for the field  follows
$$
\delta_\varepsilon \tilde \phi^{\mu\nu}(x) = \tilde g^{\mu \alpha}\,
D_\alpha \, \varepsilon^\nu(x) + \tilde g^{\nu \alpha}\,
D_\alpha \, \varepsilon^\mu(x) - D_\alpha(\varepsilon^\alpha\,
\tilde g^{\mu \nu}),
$$
but this transformation for the field differs essentially from its
transformation in the case of displacement of the coordinates:
$$
\delta_\xi\, \tilde \phi^{\mu \nu}(x) = \tilde \phi^{\mu \alpha}\,
D_\alpha \xi^\nu(x) + \tilde \phi^{\nu \alpha}\, D_\alpha \xi^\mu (x)
-
D_\alpha(\xi^\alpha\, \tilde \phi^{\mu \nu}).
$$
Under the gauge transformations (3.16) the equations of motion
for
matter do not change, because under any such transformations the
density
of the Lagrangian of matter is altered only by the divergence.

\thispagestyle{empty}
\newpage
\section{Density of the Lagrangian and the equations of motion for
the
gravitational field proper}

It is known to be impossible, only using the sole tensor $g_{\mu
\nu}$, to
construct the scalar density of the Lagrangian of the gravitational
field
proper with respect to arbitrary coordinate transformations in the
form of
a quadratic form of derivatives of order not exceeding the first.
Therefore,
such a density of the Lagrangian will certainly contain the metric
$\gamma_{\mu \nu}$ together with the metric $g_{\mu \nu}$. But, since
the
metric $\gamma_{\mu \nu}$ is not altered under the gauge
transformation
(3.16), there, consequently, must be imposed strong restrictions
on
the structure of the density of the Lagrangian of the gravitational
field
proper for it to change only by the divergence under this
transformation.
It is precisely here that there arises an essential difference
between
gauge and coordinate transformations.

While coordinate transformations impose nearly no restrictions on the
structure of the scalar density of the Lagrangian of the
gravitational
field proper, gauge transformations will permit us to find the
density of
the Lagrangian. A straightforward general method for constructing the
Lagrangian is presented in the monograph~\cite{1_3}.

Here we shall choose a more simple method for constructing the
Lagrangian.
On the basis of (\ref{26a}) we conclude that the most simple scalar
densities $\sqrt{- g}$ and $\tilde R = \sqrt{- g} R,$ where $R$~is
the
scalar curvature of effective Riemannian space, vary as follows under
the
gauge transformation (3.16):
\be
\sqrt{- g} \rightarrow \sqrt{- g} - D_\nu (\varepsilon^\nu \sqrt{-
g}),
\label{32}
\ee
\be
\tilde R \rightarrow \tilde R - D_\nu(\varepsilon^\nu \tilde R).
\label{33}
\ee
The scalar density $\tilde R$ is expressed via the Christoffel
symbols
\be
\Gamma^\lambda_{\mu \nu} = \frac{1}{2}\, g^{\lambda \sigma}
(\partial_\mu\, g_{\sigma \nu} + \partial_\nu\, g_{\sigma \mu} -
\partial_\sigma\, g_{\mu \nu}) \label{34}
\ee
as follows:
\begin{equation}
\tilde R = - \tilde g^{\mu \nu} (\Gamma^\lambda_{\mu \nu}\,
\Gamma^\sigma_{\lambda \sigma} - \Gamma^\lambda_{\mu \sigma}\,
\Gamma^\sigma_{\nu \lambda}) - \partial_\nu
(\tilde g^{\mu \nu}\, \Gamma^\sigma_{\mu \sigma} -
\tilde g^{\mu \sigma} \, \Gamma^\nu_{\mu \sigma}). \label{35}
\end{equation}
Since the Christoffel symbols are not tensor quantities, no summand
in (\ref{35}) is a scalar density. However, if one introduces the
tensor
quantities
~$G_{\mu \nu}^\lambda$
\begin{equation}
G^\lambda_{\mu \nu} = \frac{1}{2}\, g^{\lambda \sigma}
(D_\mu g_{\sigma \nu} + D_\nu g_{\sigma \mu} - D_\sigma g_{\mu \nu}),
\label{36}
\end{equation}
then the scalar density can be identically written in the form
\begin{equation}
\tilde R = -\tilde g^{\mu \nu} (G_{\mu \nu}^\lambda\, G_{\lambda
\sigma}^\sigma - G^\lambda_{\mu \sigma} \, G^\sigma_{\nu \lambda}) -
D_\nu(\tilde g^{\mu \nu}\, G^\sigma_{\mu \sigma} -
\tilde g^{\mu \sigma}\, G^\nu_{\mu \sigma}). \label{37}
\end{equation}

Note that under arbitrary coordinate transformations each group of
terms
in (\ref{37}) individually exhibits the same behaviour as scalar
density.
We see that the apparatus of Riemannian geometry is inclined toward the
introduction of covariant, instead of ordinary, derivatives in
Minkowski
space, but the metric tensor $\gamma_{\mu\nu}$, used for determining
the
covariant derivatives, is in no way fixed here.

With account of (\ref{32}) and (\ref{33}), the expression
\begin{equation}
\lambda_1(\tilde R + D_\nu\, Q^\nu) + \lambda_2 \sqrt{- g} \label{38}
\end{equation}
varies only by the divergence under arbitrary gauge transformations.
Choosing the vector density $Q^\nu$ to be
$$
Q^\nu = \tilde g^{\mu \nu}\, G^\sigma_{\mu \sigma} - \tilde g^{\mu
\sigma} \,
G^\nu_{\mu \sigma},
$$
we exclude from the preceding expression terms containing derivatives
of
orders higher, than the first, and obtain the following density of
the
Lagrangian:
\begin{equation}
 - \lambda_1 \tilde g^{\mu \nu} (G^\lambda_{\mu \nu}\,
G^\sigma_{\lambda \sigma} -
G^\lambda_{\mu \sigma} \,G^\sigma_{\nu \lambda}) + \lambda_2 \sqrt{-
g}.
\label{39}
\end{equation}

Thus, we see that the requirement for the density of the Lagrangian
of the
gravitational field proper to vary under the gauge transformation
(3.16)
only by the divergence, unambiguously determines the structure of the
Lagrangian's density (\ref{39}). But, if one restricts oneself only
to
considering this density, then the equations of the gravitational
field will
be gauge invariant, while the metric of Minkowski space,
$\gamma_{\mu\nu}$,
will not be present in the set of equations determined by the density
of the
Lagrangian~(\ref{39}). Since within such an approach the metric of
Minkowski
space disappears, the possibility of representing the gravitational
field
as a physical field of the Faraday--Maxwell type in Minkowski space disappears also.

In the case of the density of the Lagrangian (\ref{39}), introduction
of
the metric $\gamma_{\mu \nu}$ with the aid of equations~(\ref{3})
will not
save the situation, since physical quantities --- the interval and
the
curvature tensor of Riemannian space, as well as the tensor
~$t^{\mu\nu}_g$
of the gravitational field --- will depend on the choice of gauge,
which
is inadmissible from a physical point of view. Thus, for example,
\ba
&&\delta_\epsilon R_{\mu\nu}=-R_{\mu\sigma} D_\nu\epsilon^\sigma
-R_{\nu\sigma} D_\mu \epsilon^\sigma -\epsilon^\sigma D_\sigma
R_{\mu\nu}\;,
\nonumber \\*
&&\delta_\epsilon R_{\mu\nu\alpha\beta}=
R_{\sigma\nu\alpha\beta}
D_\mu \epsilon^\sigma-
R_{\mu\sigma\alpha\beta}
D_\nu \epsilon^\sigma-
\nonumber \\*
&&-R_{\mu\nu\sigma\beta}D_\alpha\epsilon^\sigma -
R_{\mu\nu\alpha\sigma} D_\beta\epsilon^\sigma -\epsilon^\sigma
D_\sigma
R_{\mu\nu\alpha\beta}.\nonumber
\ea

To retain the concept of a field in Minkowski space and to exclude
the above
ambiguity it is necessary to add, in the density of the Lagrangian of
the
gravitational field, a term violating the gauge group. It is
precisely here
that there arises an essentially new way, which for a long time
evaded being
revealed. At first sight, it may seem that a significant
arbitrariness should arise here, since the group can be violated in extremely
diverse
ways. However, it turns out not to be so, because our physical
requirement,
concerning the polarization properties of the gravitational field
which is
a field of spins 2 and 0, imposed by equations~(\ref{3}), results in
the
term violating the group (3.16) being necessarily chosen so as to
make
equations (\ref{3}) a consequence of the set of equations of the
gravitational field and of fields of matter, since only in this case
we
have no over-determined set of differential equations arising. To
this end
we introduce into the scalar density of the Lagrangian of the
gravitational
field a term of the form
\begin{equation}
\gamma_{\mu \nu} \tilde g^{\mu \nu}, \label{40}
\end{equation}
which, given conditions (\ref{3}), also varies under transformations
(\ref{29}) by the divergence, but only on the class of vectors
satisfying
the condition
\begin{equation}
g^{\mu \nu} D_\mu D_\nu \varepsilon^\sigma(x) = 0. \label{41}
\end{equation}

In electrodynamics a nearly analogous situation occurs with the
photon rest
mass differing from zero. With account of (\ref{39})-(\ref{40}) the
general
scalar density of the Lagrangian has the form:
\ba
&&L_g = -\lambda_1 \tilde g^{\mu \nu} (G^\lambda_{\mu \nu}\, G^\sigma
_{\lambda \sigma} - G^\lambda_{\mu \sigma}\, G^\sigma_{\nu \lambda})+
\nonumber
\\
&&+ \lambda_2 \sqrt{-g} + \lambda_3\, \gamma_{\mu \nu}\, \tilde
g^{\mu
\nu} +
\lambda_4 \sqrt{-\gamma}. \label{42}
\ea
We have introduced the last constant term in (\ref{42}) in order to
use it
for reducing to zero the density of the Lagrangian in absence of the
gravitational field. The narrowing of the class of gauge vectors due
to
introduction of the term (\ref{40}) automatically results in
equations
(\ref{3}) being a consequence of the equations of the gravitational
field.
We shall further verify this directly.

In accordance with the principle of least action, the equations for
the
gravitational field proper are of the form
\begin{equation}
\frac{\delta L_g}{\delta \tilde g^{\mu \nu}} =
\lambda_1 R_{\mu \nu} + \frac{1}{2}\, \lambda_2\, g_{\mu \nu} +
\lambda_3\, \gamma_{\mu \nu} = 0, \label{43}
\end{equation}
here
\[
\frac{\delta L_g}{\delta\tilde g^{\mu\nu}}
=\frac{\partial L_g}{\partial\tilde g^{\mu\nu}}
-\partial_\sigma \left (\frac{\partial L}{\partial(\partial_\sigma
\tilde g^{\mu\nu})}\right ),
\]
where we write the Ricci tensor $R_{\mu \nu}$ in the form
\begin{equation}
R_{\mu \nu} = D_\lambda\, G^\lambda_{\mu \nu} - D_\mu\,
G^\lambda_{\nu \lambda}
+G^\sigma_{\mu \nu}\, G^\lambda_{\sigma \lambda} -
G^\sigma_{\mu \lambda}\, G^\lambda_{\nu \sigma}. \label{44}
\end{equation}
Since in absence of the gravitational field equations (\ref{43}) must
be
satisfied identically, hence follows
\begin{equation}
\lambda_2 = -\, 2\, \lambda_3. \label{45}
\end{equation}

Let us now find the density of the energy-momentum for the
gravitational
field in Minkowski space
\ba
&t&^{\mu \nu}_g = - 2 \frac{\delta L_g}{\delta\, \gamma_{\mu \nu}}
2 \sqrt{-\gamma}(\gamma^{\mu \alpha} \gamma^{\nu \beta} -
\frac{1}{2} \gamma^{\mu \nu} \gamma^{\alpha \beta})\;
\frac{\delta L_g}{\delta \tilde g^{\alpha \beta}} +
\nonumber\\*
&+& \lambda_1\, J^{\mu \nu} - 2\lambda_3\, \tilde g^{\mu \nu} -
\lambda_4\, \tilde \gamma^{\mu \nu}, \label{46}
\ea
where
\begin{equation}
J^{\mu \nu} = D_\alpha D_\beta (\gamma^{\alpha \mu}
\tilde g^{\beta \nu} + \gamma^{\alpha \nu}\tilde g^{\beta \mu} -
\gamma^{\alpha \beta}\, \tilde g^{\mu \nu} - \gamma^{\mu \nu}\,
\tilde g^{\alpha \beta}). \label{47}
\end{equation}
(see Appendix (B.19)). If the dynamic equations (\ref{43}) are taken
into
account in expression (4.15), then we obtain equations for the
gravitational field proper in the form
\begin{equation}
\lambda_1\, J^{\mu \nu} - 2\, \lambda_3 \tilde g^{\mu \nu} -
\lambda_4\, \tilde \gamma^{\mu \nu} = t_g^{\mu \nu}. \label{48}
\end{equation}
For this equation to be satisfied identically in the absence of the
gravitational field, it is necessary to set
\begin{equation}
\lambda_4 = -2\, \lambda_3. \label{49}
\end{equation}
Since the equality
\begin{equation}
D_\mu\, t_g^{\mu \nu} = 0, \label{50}
\end{equation}
always holds valid for the gravitational field proper, from equation
(\ref{48}) it follows that
\begin{equation}
D_\mu\, \tilde g^{\mu \nu} = 0. \label{51}
\end{equation}

Thus, equations (\ref{3}) determining the polarization states of the
field
follow directly from equations (\ref{48}). With account of equations
(\ref{51}), one can write the field equations (\ref{48}) in the form
\begin{equation}
\gamma^{\alpha \beta}\, D_\alpha\, D_\beta\, \tilde \phi^{\mu \nu} -
\frac{\lambda_4}{\lambda_1}\, \tilde \phi^{\mu \nu} =
-\frac{1}{\lambda_1}\; t_g^{\mu \nu}. \label{52}
\end{equation}
In Galilean coordinates this equation has the simple form
\begin{equation}
\dalam \, \tilde \phi^{\mu \nu} - \frac{\lambda_4}{\lambda_1}\,
 \tilde \phi^{\mu \nu} =
- \frac{1}{\lambda_1}\, t_g^{\mu \nu}. \label{53}
\end{equation}
It is natural to consider the numerical factor
$- \frac{\lambda_4}{\lambda_1} = m^2$ to represent the square
graviton mass
and to set the value of $- 1/\lambda_1$ equal to $16 \pi$, in
accordance
with the equivalence principle. Thus, all the unknown constants
present in
the density of the Lagrangian have been defined:
\begin{equation}
\lambda_1 = - \frac{1}{16 \pi}, \quad \lambda_2 = \lambda_4 =
-2\, \lambda_3 = \frac{m^2}{16 \,\pi}. \label{54}
\end{equation}

The constructed scalar density of the Lagrangian of the gravitational
field
proper will have the form
\ba
&L_g& = \frac{1}{16\, \pi} \, \tilde g^{\mu \nu}(G^\lambda_{\mu
\nu}\,
G^\sigma_{\lambda \sigma}
- G^\lambda_{\mu \sigma}\, G^\sigma_
{\nu \lambda}) -\nonumber \\
&-& \frac{m^2}{16\, \pi} \biggl(
\frac{1}{2}\, \gamma_{\mu \nu} \tilde g^{\mu \nu} -
\sqrt{- g} - \sqrt{- \gamma}\biggr). \label{55*}
\ea
The corresponding to it dynamic equations for the gravitational field
proper
can be written down in the form
\begin{equation}
J^{\mu \nu} - m^2\, \tilde \phi^{\mu \nu} =
- 16 \pi\, t_g^{\mu \nu}, \label{56}
\end{equation}
or
\begin{equation}
R^{\mu \nu} - \frac{m^2}{2}(g^{\mu \nu} -
g^{\mu \alpha} g^{\nu \beta} \gamma_{\alpha \beta}) = 0. \label{57}
\end{equation}
These equations impose significant limits on the class of gauge
transformations, retaining only the trivial ones satisfying the
Killing
conditions in Minkowski space. Such transformations are a consequence
of
Lorentz invariance and are present in any theory.

The density of the Lagrangian constructed above leads to equations
(\ref{57}) from which it follows that equations (\ref{51}) are their
consequence, and, therefore, outside matter we shall have ten
equations for
ten unknown field functions. The unknown field functions $\phi^{0
\alpha}$
are readily expressed with the aid of equations (\ref{51}) via the
field
functions $\phi^{i k}$, where the indices $i$ and $k$ run through the
values 1, 2, 3.

Thus, the structure of the mass term violating the gauge
group in the density of the Lagrangian of the gravitational field
proper is
unambiguously determined by the polarization properties of the
gravitational
field. {\bf The field approach to gravity, that declares the
energy-momentum
tensor of all matter to be the source of the field, necessarily
requires
introduction of the graviton rest mass in the theory.}

\thispagestyle{empty}
\newpage
\section{Equations of motion for the gravitational field and for
matter}

The total density of the Lagrangian of matter and of the
gravitational field
is
\begin{equation}
L = L_g + L_M(\tilde g^{\mu \nu}, \, \phi_A), \label{58}
\end{equation}
where $L_g$ is determined by expression (\ref{55*}).

On the basis of (\ref{58}) we shall obtain, with the aid of the least
action principle, the complete set of equations for matter and for
the
gravitational field:
\ba
\frac{\delta L}{\delta \tilde g^{\mu \nu}} &=& 0,\label{59}
\\*[-0.005cm]
\frac{\delta L_M}{\delta \phi_A} &=& 0. \label{60}
\ea
Since in the case of an arbitrary infinitesimal variation of the
coordinates
the variation of the action, $\delta_c S_M$, is zero,
$$
\delta_c S_M = \delta_c \,\int \,L_M(\tilde g^{\mu \nu}, \, \phi_A)\,
d^4 x = 0,
$$
it is hence possible to obtain an identity (see Appendix (C.16)) in
the form
\begin{equation}
g_{\mu \nu} \nabla_\lambda T^{\lambda \nu} = -D_\nu
\biggl( \frac{\delta L_M}{\delta \phi_A} \, F^{B;\, \nu}_{A;\, \mu}\,
\phi_B(x)\biggr) - \frac{\delta L_M}{\delta \phi_A} \, D_\mu\,
\phi_A(x).
\label{61}
\end{equation}
Here $T^{\lambda \nu} = -2 \frac{\delta L_M}{\delta g_{\lambda \nu}}$
is the
density of the tensor of matter in Riemannian space; $\nabla_\lambda$ is
the covariant derivative in this space with the metric $g_{\lambda
\nu}$.
From identity (\ref{61}) it follows that, if the equations of motion
of
matter (\ref{60}) are satisfied, then the following equation occurs:
\begin{equation}
\nabla_\lambda\, T^{\lambda \nu} = 0. \label{62}
\end{equation}

When the number of equations (\ref{60}) for matter equals four, the
equivalent equations (\ref{62}) may be used, instead. Since we shall
further only deal with such equations for matter, we shall always
make use
of the equations for matter in the form (\ref{62}). Thus, the
complete set
of equations for matter and for the gravitational field will have the
form
\ba
\frac{\delta L}{\delta \tilde g^{\mu \nu}} =& 0&, \label{63} \\*
\nabla_\lambda\, T^{\lambda \nu} & =& 0. \label{64}
\ea
Matter will be described by velocity $\vec v$, the density of matter
$\rho$,
and pressure $p$. The gravitational field will be determined by ten
components of the tensor $\phi^{\mu \nu}$.

Thus, we have 15 unknowns. For determining them it is necessary to
add to
the 14 equations (\ref{63}),(\ref{64}) the equation of state for
matter.
If the relations (see Appendices B$^*$.18,
B$^*$.19)
\ba
\frac{\delta L_g}{\delta \tilde g^{\mu \nu}}& =&
- \frac{1}{16\, \pi}\, R_{\mu \nu} + \frac{m^2}{32\, \pi}\, (g_{\mu
\nu} -
\gamma_{\mu \nu}), \label{65} \\*
\frac{\delta L_M}{\delta \tilde g^{\mu \nu}} &=&
\frac{1}{2 \sqrt{- g}}\biggl(T_{\mu \nu} - \frac{1}{2} \, g_{\mu \nu}
\, T
\biggr), \label{66}
\ea
are taken into account, then the set of equations (\ref{63}),
(\ref{64})
may be represented as
\ba
&&\left (
R^{\mu \nu} - \frac{1}{2} \, g^{\mu \nu} R\right
) + \frac{m^2}{2}
\Biggl [ g^{\mu \nu} + (g^{\mu \alpha} g^{\nu \beta} -
\Biggr.\nonumber
\\
&&-\frac{1}{2}\, g^{\mu \nu} g^{\alpha \beta})\, \gamma_{\alpha
\beta}  \Biggl.\Biggr ] =
\frac{8\, \pi}{\sqrt{- g}} \, T^{\mu \nu},   \label{67}
\ea
\be
\nabla_\lambda\, T^{\lambda \nu} = 0. \label{68}
\ee
Owing to the Bianchi identity
$$
\nabla_\mu (R^{\mu \nu} - \frac{1}{2} \, g^{\mu \nu}\, R) = 0
$$
from equations (\ref{67}) we have
\begin{equation}
m^2 \sqrt{- g} (g^{\mu \alpha} g^{\nu \beta} - \frac{1}{2}\, g^{\mu
\nu}
g^{\alpha \beta}) \nabla_\mu\, \gamma_{\alpha \beta} =
16\, \pi\, \nabla_\mu\, T^{\mu \nu}. \label{69}
\end{equation}
Taking into account expression
\begin{equation}
\nabla_\mu \, \gamma_{\alpha \beta} = - G^\sigma_{\mu \alpha}\,
\gamma_{\sigma
\beta} - G^\sigma_{\mu \beta}\, \gamma_{\sigma \alpha}, \label{70}
\end{equation}
where $G^\sigma_{\mu \alpha}$ is defined by formula (\ref{36}), we
find
\ba
&&(g^{\mu \alpha} g^{\nu \beta} - \frac{1}{2}\, g^{\mu \nu} g^{\alpha
\beta})
\, \nabla_\mu \, \gamma_{\alpha \beta} = \nonumber
\\
&&=\gamma_{\mu\lambda} g^{\mu\nu}
(D_\sigma \,g^{\sigma \lambda} + G^\beta_{\alpha \beta}\, g^{\alpha
\lambda}),
\label{71}
\ea
but since (see formulae (\ref{b20}))
\begin{equation}
\sqrt{- g}(D_\sigma\, g^{\sigma \lambda} + G^\beta_{\alpha \beta}\,
g^{\alpha \lambda}) = D_\sigma\, \tilde g^{\lambda \sigma},
\label{72}
\end{equation}
expression (\ref{71}) assumes the form
\begin{equation}
\sqrt{- g}(g^{\mu \alpha} g^{\nu \beta} - \frac{1}{2}\,
g^{\mu \nu} g^{\alpha \beta})\, \nabla_\mu\,
\gamma_{\alpha \beta} = \gamma_{\mu \lambda} \, g^{\mu \nu}\,
D_\sigma\, \tilde g^{\lambda \sigma}. \label{73}
\end{equation}
With the aid of (\ref{73}) expression (\ref{69}) can be represented
in the
form
$$
m^2 \gamma_{\mu \lambda}\, g^{\mu \nu}D_\sigma \tilde g^{\lambda
\sigma} =
16 \pi\, \nabla_\mu\, T^{\mu \nu}.
$$
This expression can be rewritten in the form
\begin{equation}
m^2 D_\sigma \tilde g^{\lambda \sigma} = 16\, \pi \, \gamma^{\lambda
\nu}
\nabla_\mu\, T^\mu_\nu. \label{74}
\end{equation}
With the aid of this relation, equation (\ref{68}) can be replaced by
the equation
\begin{equation}
D_\sigma \tilde g^{\nu \sigma} = 0. \label{75}
\end{equation}
Therefore, the set of equations (\ref{67}), (\ref{68}) is reduced to
the set of gravitational equations in the form
\ba
&&\left(\!R^{\mu \nu} - \frac{1}{2}\, g^{\mu \nu} R\!\right) +
\frac{m^2}{2}
\left [g^{\mu \nu} + (g^{\mu \alpha} g^{\nu \beta} -
\right.\nonumber \\
&&\left. - \frac{1}{2}\, g^{\mu \nu}\, g^{\alpha \beta})
\gamma_{\alpha \beta}\right ]
=\frac{8\, \pi}{\sqrt{-g}}\, T^{\mu \nu}, \label{76}\\
&&D_\mu\, \tilde g^{\mu \nu} = 0. \label{77}
\ea

These equations are universally covariant with respect to arbitrary
transformations of coordinates and form-invariant only with respect
to such
transformations of coordinates that leave the Minkowski metric
$\gamma_{\mu\nu}(x)$ form-invariant. Hence, for instance, it follows
that
in any inertial (Galilean) reference system phenomena are described
by
identical equations. Equations involving the graviton mass had arisen
previously; however, owing to
misunderstanding of the fundamental fact 
that special relativity theory is also valid in non-inertial
reference
systems, they were not considered seriously, since they were not
universally
covariant. Usually, following A.Einstein, the metric
$\eta_{\alpha\beta}=(1,-1,-1,-1)$ was considered to be a tensor only
with
respect to the Lorentz transformations. But, actually, the metric of
Minkowski space, $\gamma_{\mu\nu}(x)$, is a tensor with respect to
arbitrary
transformations of coordinates. The set of equations
(\ref{76}) and (\ref{77}) is hyperbolic. In the case of static
problems,
it is elliptic. By adding the equation of state to the set of
equations
(\ref{76}) and (\ref{77}) we obtain a complete set of equations for
determining the unknown physical quantities
$g_{\mu\nu},\;\; \vec v,\;\; \rho, p$ for one or another formulations
of
the problem.

A concrete inertial Galilean reference system is singled out by
formulation
of the actual physical problem (by the initial and boundary
conditions).
The descriptions of a given formulated physical problem in different
inertial (Galilean) reference systems are different, naturally, but
this
does not contradict the relativity principle. If one introduces the
tensor
$$
N^{\mu \nu} = R^{\mu \nu} - \frac{m^2}{2}
\lbrack g^{\mu \nu} - g^{\mu \alpha} g^{\nu \beta} \gamma_{\alpha
\beta}
\rbrack,
\> \> N = N^{\mu \nu} g_{\mu \nu},
$$
then the set of equations (\ref{76}) and (\ref{77}) can be written in
the
form
\begin{aequation}{76}{a}
N^{\mu \nu} - \frac{1}{2}\, g^{\mu \nu} N = \frac{8\, \pi}{\sqrt{-
g}}\,
T^{\mu \nu}, \label{76a}
\end{aequation}
\begin{aequation}{77}{a}
D_\mu\, \tilde g^{\mu \nu} = 0. \label{77a}
\end{aequation}
It may also be represented in the form
\begin{equation}
N^{\mu \nu} = \frac{8\, \pi}{\sqrt{- g}} (T^{\mu \nu} - \frac{1}{2}\,
g^{\mu \nu}\, T), \label{78}
\end{equation}
\begin{equation}
D_\mu \tilde g^{\mu \nu} = 0, \label{79}
\end{equation}
or
\begin{aequation}{78}{a}
N_{\mu \nu} = \frac{8\, \pi}{\sqrt{- g}} (T_{\mu \nu} - \frac{1}{2}\,
g_{\mu \nu}\, T), \label{78a}
\end{aequation}
\begin{aequation}{79}{a}
D_\mu \tilde g^{\mu \nu} = 0. \label{79a}
\end{aequation}
It must be especially stressed that both sets of equations (\ref{78})
and (\ref{79}) contain the metric tensor of Minkowski space.

Transformations of coordinates, which leave the metric of Minkowski
space
form-invariant, relate physically equivalent reference systems. The
most
simple of these are inertial reference systems. For this reason,
possible
gauge transformations satisfying the Killing conditions
$$
D_\mu \varepsilon_\nu + D_\nu \varepsilon_\mu = 0,
$$
do not remove us from the class of physically equivalent reference
systems.

Let us deal with this issue in more detail. To this end we write the
equation of RTG, (\ref{78}) and (\ref{79}), in the expanded form:
\ba
&R&\!^{\mu\nu}(x)-\frac{m^2}{2}
[ g^{\mu\nu} (x) - g^{\mu\alpha} g^{\nu\beta}
\gamma_{\alpha\beta} (x)]=\nonumber \\
&=& 8\pi
\left [T^{\mu\nu}(x)-\frac{1}{2}g^{\mu\nu}T (x)\right ],\label{523}
\ea
\be
D_\mu \tilde g^{\mu\nu}=0.\label{524}
\ee
Consider that, given appropriate conditions of the problem, these
equations
have the solution $g^{\mu\nu} (x)$ in Galilean coordinates $x$ in an
inertial reference system, when the distribution of matter is
$T^{\mu\nu} (x)$. In another inertial reference system, in Galilean
coordinates $x'$ satisfying the condition
\ba
x'^\nu=x^\nu+\epsilon^\nu(x),\nonumber
\\* [-0.3cm] \label{525}\\* [-0.3cm]
D^\mu\epsilon^\nu+D^\nu\epsilon^\mu=0.\nonumber
\ea
We obtain with the aid of tensor transformations the following:
\ba
&&R\,'^{\mu\nu}(x')-\frac{m^2}{2}
[g'^{\mu\nu}(x')-g'^{\mu\alpha}g'^{\nu\beta}
\gamma_{\alpha\beta}(x')]=\nonumber
\\
&&= 8\pi \left [T'^{\mu\nu}(x')-\frac{1}{2}
g'^{\mu\nu}T'(x')\right ]. \label{526}
\ea

Since equations (\ref{523}) are form-invariant with respect to the
Lorentz
transformations, we can return to the initial variables~$x$:
\ba
&&R\,'^{\mu\nu} (x) -\frac{m^2}{2}
[g'^{\mu\nu}(x) - g'^{\mu\alpha}
g'^{\nu\beta}\gamma_{\alpha\beta}(x)]=
\nonumber  \\
&&=8\pi
\left [T'^{\mu\nu}(x)-\frac{1}{2} g'^{\mu\nu}
T'(x)\right ] \label{527}
\ea
Hence it is clear that the solution $g'^{\mu\nu}(x)$ does not
correspond
to the distribution of matter $T^{\mu\nu}(x)$, but to another
distribution
$T'^{\mu\nu}(x)$. The quantity $g'^{\mu\nu}(x)$  in equations
(\ref{527})
is
\be
g'^{\mu\nu}(x)=g^{\mu\nu}(x)+\delta_\epsilon g^{\mu\nu},\label{528}
\ee
where
\be
\delta_\epsilon g^{\mu\nu}=g^{\mu\lambda}
D_\lambda\epsilon^\nu +g^{\nu\lambda}
D_\lambda\epsilon^\mu -\epsilon^\lambda
D_\lambda g^{\mu\nu}. \label{529}
\ee
In the case of transformations (\ref{525}) we have
\ba
&&R'^{\mu\nu}(x)=R^{\mu\nu}(x)+\delta_\epsilon R^{\mu\nu},\nonumber
\\
&&T'^{\mu\nu}(x)=T^{\mu\nu}(x)+\delta_\epsilon
T^{\mu\nu},\label{530}\\*
&&T'(x)=T(x)+\delta_\epsilon T. \nonumber
\ea
Here
\ba
&&\delta_\epsilon R^{\mu\nu}=R^{\mu\lambda} D_\lambda
\epsilon^\nu +R^{\nu\lambda}D_\lambda\epsilon^\mu
-\ep^\lambda D_\lambda R^{\mu\nu},\nonumber \\*
&&\delta_\ep T^{\mu\nu}=T^{\mu\lambda} D_\lambda
\ep^\nu
+T^{\nu\lambda}D_\lambda\ep^\mu -
\ep^\lambda D_\lambda T^{\mu\nu}, \label{531}\\*
&&\delta_\ep T=-\ep^\lambda D_\lambda T=-\ep^\lambda
\partial_\lambda T.\nonumber
\ea

We obtained expression (\ref{527}) with the aid of the coordinate
transformations (\ref{525}), but identical equations are also
obtained
in the case of the gauge transformation (\ref{529}) with the vectors
$\ep^\lambda$ satisfying condition (\ref{525}). Thus, gauge
transformations
result in the metric field $g'^{\mu\nu}(x)$ in the case of matter
exhibiting the distribution $T'^{\mu\nu}(x)$. Although we considered
the
transition from one inertial reference system in Galilean coordinates
to
another, our formulae (5.25), (5.31) are of a general nature, they
are
valid,
also, for a noninertial reference system in Minkowski space. The same
situation occurs in electrodynamics.

In GRT the situation is completely different, since owing to the
Hilbert--Einstein equations being form-invariant with respect to
arbitrary
transformations of coordinates, there exist, for one and the same
distribution of matter, $T^{\mu\nu}(x)$, any amount of metrics
$g_{\mu\nu}(x), g'_{\mu\nu}(x)...$ satisfying the equations. It is
precisely for this reason that in GRT there arises an ambiguity in
the
description of gravitational phenomena.

If one imagines that it is possible to perform experimental measurements of
the
characteristics of Riemannian space and of the motion of matter with
whatever
high precision, then it would be possible, on the basis of equations
(\ref{78a}) and (\ref{79a}), to determine the metric of Minkowski
space and
to find Galilean (inertial) reference systems. Thus, Minkowski space
is
observable, in principle.

The existence of Minkowski space is reflected in the conservation
laws, and,
therefore, testing their validity in physical phenomena serves at the
same
time for testing the structure of space-time.

It must be especially noted that both sets of equations
(\ref{76}) and
(\ref{77}) contain the metric tensor of Minkowski space. The presence
of
the cosmological term in the equations of GRT is known not to be
obligatory,
and this issue is still being discussed. The presence of the
cosmological
term in the equations of gravity is obligatory in RTG. However, the
cosmological term in equations (\ref{76}) arises in combination with
the
term related to the metric $\gamma_{\mu\nu}$ of Minkowski space, and
with
the same constant factor equal to a half of the square graviton mass.
The
presence in equations (\ref{76}) of the term connected with the
metric
$\gamma_{\mu\nu}$ significantly alters the character of the collapse
and
development of the Universe. In accordance with equations (\ref{76}),
in
absence of matter and of the gravitational field, the metric of the
space
becomes the Minkowski metric, and it coincides exactly with the one
previously chosen in formulating the physical problem. If the metric
of
Minkowski space were absent in the equations of the gravitational
field,
then it would be absolutely unclear, in which reference system of
Minkowski
space we would happen to be in the absence of matter and of the
gravitational field.

The graviton mass is essential for this theory, since only its
introduction
permits construction of the theory of gravity in Minkowski space. The
graviton mass violates the gauge group or, in other words, it removes
the
degeneracy. Therefore, one cannot exclude the possibility of the
graviton
mass tending toward zero in the final results, when gravitational
effects
are studied. However, the theory with a graviton mass and the theory
involving violation of the gauge group \cite{7} (with the graviton
mass
subsequently tending toward zero) are essentially different theories.
Thus,
for example, while in the first theory the Universe is homogeneous
and
isotropic, no such Universe can exist in the second one.

Let us now touch upon the equivalence principle. Any physical theory
must
comply with the equivalence principle. Gravitational interactions
alter
the equations of motion of matter. The requirement imposed by the
equivalence principle reduces to the requirement that, when the
gravitational interaction is switched off, i.e. when the Riemann
curvature
tensor turns to zero, these equations of motion become ordinary
equations
of motion of SRT in the chosen reference system.

In formulating the physical problem in RTG we choose a certain
reference
system with the metric tensor of Minkowski space,
$\gamma_{\mu\nu}(x)$.
In RTG, the equation of motion of matter in effective Riemannian space
with
the metric tensor $g_{\mu\nu}(x)$, determined by the equations of the
gravitational field (\ref{76}) and (\ref{77}), has the form
$$
\nabla_\mu T^{\mu\nu}(x)=0.\eqno{(\sigma)}
$$
For example, we shall take dustlike matter with the energy-momentum
tensor $T^{\mu\nu}$ equal to
\[
T^{\mu\nu}(x)=\rho U^\mu U^\nu,\, \;\; U^\nu=\frac{dx^\nu}{ds},
\]
$ds$ is the interval in Riemannian space.

On the basis of equations $(\sigma)$, using the expression for
$T^{\mu\nu}$,
we find the equation for the geodesic line in Riemannian space,
\[
\frac{dU^\nu}{ds}+\Gamma^\nu_{\alpha\beta}(x)U^\alpha U^\beta=0.
\]
When the gravitational interaction is switched off, i.e. when the
Riemann
curvature tensor turns to zero, from the equations of the
gravitational
field (\ref{76}) and (\ref{77}) it follows that the Riemannian metric
$g_{\mu\nu}(x)$ transforms into the previously chosen metric of
Minkowski
space, $\gamma_{\mu\nu}(x)$. In this case the equation of motion of
matter
$(\sigma)$ assumes the form
$$
D_\mu t^{\mu\nu}(x)=0.\eqno{(\lambda)}
$$
Here the energy-momentum tensor $t^{\mu\nu}(x)$ is
\[
t^{\mu\nu}(x)=\rho u^\mu u^\nu,\, u^\nu=\frac{dx^\nu}{d\sigma},
\]
$d\sigma$ is the interval in Minkowski space.

On the basis of $(\lambda)$, using the expression for $t^{\mu\nu}$,
we find
the equations for the geodesic line in Minkowski space,
\[
\frac{du^\nu}{d\sigma}+\gamma^\nu_{\alpha\beta}u^\alpha u^\beta=0,
\]
i.e., we have arrived at the ordinary equations for the free
motion of particles in SRT in the previously chosen reference
system with the metric tensor $\gamma_{\mu\nu}(x)$. Thus, the
equation of motion of matter in a gravitational field in the
chosen reference system automatically transforms, when the
gravitational interaction is switched off, i.e., when the Riemann
curvature tensor turns to zero, into the equation of motion of
matter in Minkowski space in the same reference system with the
metric tensor $\gamma_{\mu\nu}(x)$, i.e., the equivalence
principle is obeyed. This assertion in RTG is of a general
character, since when the Riemann tensor turns to zero the density
of the Lagrangian of matter in the gravitational field,
$L_M(\tilde g^{\mu\nu}, \Phi_A)$, transforms into the ordinary SRT
density of the Lagrangian, $L_M(\tilde \gamma^{\mu\nu}, \Phi_A)$,
in the chosen reference system.

In GRT the equation of motion of matter also has the form $(\sigma)$.
But,
since the Hilbert--Einstein equations do not contain the metric
tensor of
Minkowski space, then when the gravitational interaction is switched
off,
i.e., the Riemann curvature tensor turns to zero, it is impossible to
say
in which reference system (inertial or accelerated) of Minkowski
space we
happen to be, and therefore it is impossible to determine which
equation
of motion of matter in Minkowski space we will obtain when the
gravitational
interaction is switched off. Precisely for this reason, the
equivalence
principle cannot be complied with in GRT within the framework of this
theory. Usually such correspondence in GRT is achieved precisely
within the
field approach, when a weak gravitational field is considered to be a
physical field in Minkowski space in Galilean coordinates. Thus, GRT
is made
to include what it essentially does not contain, since, as
A.~Einstein
wrote, ``{\it gravitational fields can be given without introducing
tensions
and energy density}"$\,$\footnote{Einstein A. Collection of
scientific
works, Moscow: Nauka, 1965, vol.1, art.47, p.627.}, so no existence
of any physical field in GRT can even
be spoken of.

In conclusion, we note that RTG revives all those concepts (inertial
reference system, the law of inertia, acceleration relative to space,
the
conservation laws of energy-momentum and of angular momentum), which
occurred
in classical Newtonian mechanics and in special relativity theory,
and which
had to be renounced by A.~Einstein in constructing GRT.

In~1955 A.Einstein wrote: <<{\it A significant achievement of
general relativity theory consists in that it rids physics of the
necessity of introducing the ``inertial system" (or ``iner\-ti\-al
sys\-tems")}>>$\,$\footnote{Einstein A. Collection of scientific
works, Moscow: Nauka, 1965, vol.2, art.146, p.854.}. From our
point of view, the fields of inertia and of gravity cannot be
identified with each other even locally, since they are of totally
different natures. While the former can be removed by a choice of
the reference system, no choice whatever of the reference system
can remove the fields of gravity. Regrettably, this circumstance
is not understood by many persons, since they do not apprehend
that ``{\it in Einstein's theory}", as especially stressed by
J.L.Synge, ``{\it the existence or absence of a gravitational field
depends on whether the Riemann curvature tensor differs from or
equals zero}"$\,$\footnote{J.L.Synge. Relativity: the general
theory. M.:Foreign literature publishers, 1963, p.9.}.

\thispagestyle{empty}
\newpage
\section{The causality principle in RTG}

RTG was constructed within the framework of SRT, like the theories of
other
physical fields. According to SRT, any motion of a pointlike test
body
(including the graviton) always takes place within the causality
light cone
of Minkowski space. Consequently, non-inertial reference systems,
realized
by test bodies, must also be inside the causality cone of
pseudo-Euclidean
space-time. This fact determines the entire class of possible
non-inertial
reference systems. Local equality between the three-dimensional force
of
inertia and gravity in the case of action on a material pointlike
body will
occur, if the light cone of the effective Riemannian space does not go
beyond
the limits of the causality light cone of Minkowski space. Only in
this case
can the three-dimensional force of the gravitational field acting on
the
test body be locally compensated by transition to the admissible
non-inertial reference system, connected with this body.

If the light cone of the effective Riemannian space were to reach beyond
the
causality light cone of Minkowski space, this would mean that for
such a
``gravitational field" no admissible non-inertial reference system
exists,
within which this ``force field" could be compensated in the case of
action
on a material pointlike body. In other words,~~ local~~ compensation~~ of~~
the
3-force of gravity by the force of inertia is possible only when the
gravitational field, acting as a physical field on particles, does
not lead
their world lines outside the causality cone of pseudo-Euclidean
space-time.
This condition should be considered the causality principle
permitting
selection of solutions of the set of equations (\ref{76}) and
(\ref{77})
having physical sense and corresponding to the gravitational fields.

The causality principle is not satisfied automatically. There is
nothing
unusual in this fact, since both in electrodynamics, and in other
physical
theories, as well, the causality condition for matter in the form
$d\sigma^2=\gamma_{\mu\nu}dx^\mu dx^\nu\geq 0$ is always added (but
not
always noted) to the main equations, which actually provides for it
being
impossible for any form of matter to undergo motion with velocities
exceeding the speed of light. In our case it is necessary to take
into
account that the gravitational interaction enters into the
coefficients of
the second-order derivatives in the field equations, i.e. there
arises an
effective geometry of space-time. This feature is only peculiar to
the
gravitational field. The interaction of all other known physical
fields
usually does not involve the second-order derivatives of the field
equations, and therefore does not alter the initial pseudo-Euclidean
geometry of space-time.

We shall now present an analytical formulation of the ca\-u\-sa\-li\-ty
principle in
RTG. Since in RTG the motion of matter under the action of the
gravitational
field in pseudo-Euclidean space-time is equivalent to the motion of
matter
in the corresponding effective Riemannian space-time, we must for events
(world
lines of particles and of light) related by causality, on the one
hand, have
the condition
\begin{equation}
d\, s^2 = g_{\mu \nu}\, dx^\mu\, dx^\nu \> \ge \> 0, \label{80}
\end{equation}
and, on the other hand, the following inequality must hold valid for
such
events:
\begin{equation}
d \sigma^2 = \gamma_{\mu \nu}\, dx^\mu\, dx^\nu \> \ge \> 0.
\label{81}
\end{equation}
The following condition is valid for the chosen reference system
realized
by physical bodies:
\begin{equation}
\gamma_{00} \> > \> 0. \label{82}
\end{equation}
We single out in expression (\ref{81}) the time- and spacelike parts:
\begin{equation}
d\, \sigma^2 = \biggl(\sqrt{\gamma_{00}} \, dt +
\frac{\gamma_{0\, i}\, dx^i}{\sqrt{\gamma_{00}}} \biggr)^2 -
s_{i\, k}\, dx^i \, dx^k, \label{83}
\end{equation}
here the Latin indices $i,\, k$ run through the values 1, 2, 3;
\begin{equation}
s_{i\, k} = -\, \gamma_{i\, k} + \frac{\gamma_{0\, i} \gamma_{0\, k}}
{\gamma_{00}}, \label{84}
\end{equation}
$s_{i\, k}$ is the metric tensor of three-dimensional space in
four-dimensional pseudo-Euclidean space-time. The square spatial
distance
is determined by the expression
\begin{equation}
d\, l^2 = s_{i\, k}\, dx^i\, dx^k. \label{85}
\end{equation}

Now we represent the velocity $v^i = \frac{dx^i}{dt}$ as  $v^i =
ve^i$,
where $v$ is the absolute value of the velocity and $e^i$ is an
arbitrary
unit vector in three-dimensional space,
\begin{equation}
s_{i\, k} e^i\, e^k = 1. \label{86}
\end{equation}
In absence of the gravitational field the velocity of light in the
chosen
reference system is readily determined from expression (\ref{83}) by
setting it equal to zero:
$$
\biggl( \sqrt{\gamma_{00}}\, dt + \frac{\gamma_{0i} \, dx^i}
{\sqrt{\gamma_{00}}} \biggr)^2 = s_{i\, k}\, dx^i\, dx^k.
$$
Hence, we find
\begin{equation}
v = \sqrt{\gamma_{00}}/{\biggl( 1 - \frac{\gamma_{0i}\, e^i}{\sqrt
{\gamma_{00}}}\,\biggr)} . \label{87}
\end{equation}
Thus, an arbitrary four-dimensional isotropic vector in Min\-ko\-wski
space,
$u^\nu$, is
\begin{equation}
u^\nu = (1, \, v\, e^i). \label{88}
\end{equation}

For both conditions (\ref{80}), (\ref{81}) to be satisfied
simultaneously, it is necessary and sufficient that for any isotropic
vector
\begin{equation}
\gamma_{\mu \nu}\, u^\mu\, u^\nu = 0 \label{89}
\end{equation}
the causality condition
\begin{equation}
g_{\mu \nu}\, u^\mu \, u^\nu \> \le \> 0, \label{90}
\end{equation}
hold valid, which precisely indicates that the light cone of the
effective
Riemannian space does not go beyond the causality light cone of
pseudo-Euclidean space-time. The causality condition may be written
in the
following form:
\begin{aequation}{89}{a}
g_{\mu \nu}\, v^\mu\, v^\nu = 0, \label{89a}
\end{aequation}
\begin{aequation}{90}{a}
\gamma_{\mu \nu} v^\mu\, v^\nu \> \ge \> 0. \label{90a}
\end{aequation}

In GRT, physical meaning is also attributed to such solutions of the
Hilbert-Einstein equations, which satisfy the inequality
$$
g < 0,
$$
as well as the requirement known as the energodominance condition,
which
is formulated as follows. For any timelike vector $K_\nu$ the
inequality
$$
T^{\mu \nu}\, K_\mu\, K_\nu \> \ge \> 0,
$$
must be valid, and the quantity $T^{\mu \nu}\, K_\nu$ must form, for
the
given vector $K_\nu$, a non-spacelike vector.

In our theory, such solutions of equations (\ref{78a}) and
(\ref{79a}) have
physical meaning, which, besides these requirements, must also
satisfy the
causality condition (\ref{89a}) and (\ref{90a}). The latter can be
written,
on the basis of equation (\ref{78a}), in the following form:
\ba
&&R_{\mu \nu}\, K^\mu K^\nu \; \le \; \frac{8\, \pi}{\sqrt{- g}}
(T_{\mu \nu} - \frac{1}{2}\, g_{\mu \nu}\, T)\,
K^\mu K^\nu + \nonumber \\
&&+\frac{m^2}{2}\, g_{\mu \nu}\, K^\mu K^\nu.  \label{91}
\ea
If the density of the energy-momentum tensor is taken in the form:
\vspace*{1mm}
\[
T_{\mu\nu}=\sqrt{-g}[(\rho+p) U_\mu U_\nu-p g_{\mu\nu}],
\vspace*{2mm}
\]
then on the  basis of (\ref{78a}) it is possible to establish between
the
interval of Minkowski space, $d\sigma$, and the interval of the
effective
Riemannian space, $ds$, the following relationship:
\[
\frac{m^2}{2}d\sigma^2=ds^2
[4\pi (\rho+3p)+\frac{m^2}{2}
-R_{\mu\nu}U^\mu U^\nu],
\]
here $\displaystyle U^\mu=\frac{dx^\mu}{ds}$.

Owing to the causality principle the inequality
\[
R_{\mu\nu}U^\mu U^\nu <4\pi (\rho +3p)+\frac{m^2}{2},
\]
which is a special case of inequality (\ref{91}), or
\begin{aequation}{91}{a}
{\sqrt{-g}}R_{\mu \nu}\,v^{\mu}v^{\nu}\,\le \; 8\, \pi T_{\mu \nu}
v^{\mu}v^{\nu} \label{91a}
\end{aequation}
must hold valid.

Let us now consider the motion of a test body under the influence of
gravity
in GRT and RTG. In 1918 A.~Einstein gave the following formulation of
the
equivalence principle: ``{\it Inertia and gravity are identical; hence
and
from the results of special relativity theory it inevitably follows
that the
symmetric  $\ll$fundamental tensor$\gg\, g_{\mu\nu}$ determines the
metric
properties of space, of the motion of bodies due to inertia in it,
and, also,
the influence of gravity}"$\,$\footnote{Einstein A. Collection of
scientific works, Moscow: Nauka, 1965, vol.1, art.45, p.613.}.
Identifying in GRT the gravitational field and
the metric tensor $g_{\mu\nu}$ of Riemannian space permits, by an
appropriate
choice of the reference system, to equate to zero all the components
of the
Christoffel symbol at all points of an arbitrary non-selfintersecting
line.
Precisely for this reason, motion along a geodesic line in GRT is
considered
free. But, in this case, the choice of reference system cannot remove
the
gravitational field in GRT, also, because the motion of two close
material
pointlike bodies will not be free due to the existence of the
curvature
tensor, which can never be equated to zero by a choice of reference
system
owing to its tensor properties.

The gravitational field in RTG is a physical field in the spirit of
Faraday--Maxwell, so the gravitational force is described by a
four-vector, and,
consequently, the forces of inertia can  be made to compensate the
three-dimensional part of the force of gravity by a choice of
reference
system only if conditions (\ref{89}) and (\ref{90}) are satisfied.
Now,
the motion of a material pointlike body in the gravitational field
can
never be free. This is especially evident, if the equation of the
geodesic
line is written in the form \cite{8}
\[
\frac{DU^\nu}{d\sigma}=-G^\rho_{\alpha\beta}
U^\alpha U^\beta (\delta^\nu_\rho -U^\nu U_\rho).
\]
Here
$$
d\sigma^2=\gamma_{\mu\nu}dx^\mu dx^\nu,\,
U^{\nu}=\frac{dx^\nu}{d\sigma}.
$$

Free motion in Minkowski space is described by the equation:
\[
\frac{DU^\nu}{d\sigma}=
\frac{dU^\nu}{d\sigma}+\gamma^\nu_{\mu\lambda}
U^\mu U^\lambda =0,
\]
$\gamma^\nu_{\mu\lambda}$ are the Christoffel symbols of
Minkowski
space. We see that motion along a geodesic line of Riemannian space is
the
motion of a test body under the action of the force $F^\nu$:
\[
F^\nu = -G^\rho_{\alpha\beta}U^{\alpha}U^\beta
(\delta^\nu_\rho - U^\nu U_\rho),
\]
and this force is a four-vector. If the test body were charged, it
would
emit electromagnetic waves, since it moves with acceleration.

In SRT there exists an essential difference between the forces of
inertia and physical forces. The forces of inertia can always be
made equal to zero by a simple choice of reference system, while
essentially no choice of reference system can turn physical forces
into zero, since they are of a vector nature in Minkowski space.
Since in RTG all forces, including gravitational forces, are of a
vector nature, this means that they cannot be equated to zero by a
choice of reference system. A choice of reference system can only
make  the force of inertia compensate a three-dimensional force,
acting on a material pointlike body, the force being of any
nature, including gravitational. In GRT, as noted by
J.L.Synge~\cite{9}, ``{\it... the concept of the force of
gravity does not exist, since gravitational properties are
organically present in the structure of space-time and are
manifested in the curvature of space-time, i.e. in that the
Riemann tensor $R_{ijkm}$ differs from zero.}" Precisely in this
connection, J.L.Synge wrote: ``{\it According to the famous
legend, Newton was inspired to create his theory of gravity, when
he once observed an apple falling from the branch of a tree, and
those who study Newtonian physics even now are ready to claim that
the acceleration (980$~\mbox{cm}/\mbox{s}^2$) of a falling apple
is due to the gravitational field. In accordance with relativity
theory} (GRT is intended --- A.L.), {\it this point of view is
completely erroneous. We shall undertake a thorough investigation
of this problem and verify that the gravitational field (i.e., the
Riemann tensor) actually plays an extremely insignificant role in
the phenomenon of a falling apple, while the acceleration
980$~\mbox{cm}/\mbox{s}^2$ is really due to the curvature of the
world line of the tree's branch.}"

According to RTG, the gravitational field is a physical field, and
therefore, unlike the case of GRT, it fully retains the concept of
the force
of gravity. Precisely owing to the force of gravity does the free
fall of
bodies occur, i.e., everything proceeds like in Newtonian physics.
Moreover,
all gravitational effects in the Solar system (the displacement of
the
perihelion of Mercury, the deflection of light by the Sun, the time
delay
of a radiosignal, the precession of a gyroscope) are caused precisely
by
the action of the force of gravity, but not by the curvature tensor
of
space-time, which in the Solar system is quite small.

The local identity between inertia and gravity was seen by Einstein
as the
main reason for the inertial and gravitational masses to be equal to
each
other. However, in our opinion, as it can be seen from equations
(\ref{2}),
the reason for this equality lies in that the source of the
gravitational
field is the conversed total density of the tensor of matter and of
the
gravitational field. Precisely for this reason, the inertial and
gravitational masses being equal to each other does not require the
forces
of gravity and of inertia to be locally identical.

\thispagestyle{empty}
\newpage
\section{Mach's principle}

In formulating the laws of mechanics Newton introduced the notion
of absolute space, which always remains the same and is
motionless. He defined the acceleration of a body precisely with
respect to this space. This acceleration had an absolute
character. The introduction of such an abstraction as absolute
space turned out to be extremely fruitful. Hence, for instance,
arose the concepts of inertial reference systems in the entire
space, the relativity principle for mechanical processes, and the
idea came into being of states of motion, that are physically
singled out. In this connection Einstein wrote the following in
1923: ``{\it Reference systems that are in such states of motion
are characterized by the laws of Nature formulated in these
coordinates assuming the most simple form."} And further: {\it
<<...according to classical mechanics there exists ``relativity of
velocity'', but not ``relativity of
acceleration"}>>$\,$\footnote{Einstein A. Collection of scientific
works, Moscow: Nauka, 1965, vol.2, art.70, p.122.}.

Thus in
theory was the notion established of inertial reference systems, in
which
material pointlike bodies, not subject to the action of forces, do
not
experience acceleration and remain at rest or in their state of
uniform
motion along a straight line. However, Newton's absolute space or
inertial
reference system were actually introduced a priori, without any
relation
to the character of the distribution of matter in the Universe.

Mach displayed much courage in seriously criticizing the main
points of Newton's mechanics. He later wrote that he succeeded in
publishing his ideas with difficulty. Although Mach did not
construct any physical theory free of the disadvantages he himself
pointed out, he greatly influenced the development of physical
theory. He drew the attention of scientists to the analysis of the
main physical concepts.

We shall quote some statements made by Mach~\cite{10}, which in the
literature have been termed the ``Mach  principle".  {\it``No one can
say
anything about absolute space and absolute motion, this is only
something
that can be imagined and is not observable in experiments"}. And
further:
{\it ``Instead of referring a moving body to space (to some reference
system), we shall directly consider its relation to
$\>$~\hbox{b o d i e s} $\>$ of the world, only by which it is
possible to
\hbox{d e f i n e} $\>$ a reference system. ...even in the most
simple
case, when we apparently consider the interaction between only
\hbox{t w o} $\>$ masses, it is \hbox{i m p o s s i b l e} $\>$ to
become
distracted from the rest of the world. ... If a body revolves with
respect
to the sky of motionless stars, then there arise centrifugal forces,
while
if it revolves round \hbox{a n o t h e r} $\>$  body, instead of the
sky
of motionless stares, no centrifugal forces will arise. I have
nothing
against calling the first revolution \hbox{a b s o l u t e,} $\>$ if
only
one does not forget that this signifies nothing but revolution
\hbox{r e l a t i v e} $\>$ to the sky of motionless stars."}

Therefore Mach wrote: {\it ``...there is no necessity for relating the
Law
of inertia to some special absolute space. The most natural approach
of a
true naturalist is the following: first to consider the law of
inertia as quite an approximate law, then to establish its
relationship
in space to the motionless sky of stars, ...and then one should
expect
corrections or some development of our knowledge on the basis of
further
experiments. Not long ago Lange published a critical article, in
which he
exposes how it would be possible, in accordance with his principles,
to
introduce a \hbox{n e w} $\>$ reference system, if the ordinary rough
reference to the motionless starry sky were to become no longer
suitable
owing to more precise astronomical observations. There exists no
difference
between the opinion of Lange and my own relative to the
\hbox{t h e o r e t i c a l} $\>$ formal value of Lange's
conclusions,
namely, that at present the motionless starry sky is the only
\hbox{p r a c t i c a l l y} $\>$ suitable reference system, and,
also,
relative to the method of defining a new reference system by
gradually
introducing corrections."}~\cite{10}.  Further, Mach quotes
S.~Neumann:{\it ``Since
all
motions must be referred to the reference system alpha (the reference
system of inertia), it evidently represents an indirect relationship
between
all the processes taking place in the Universe, and, consequently, it
contains, so to say, a universal law which is just as mysterious as
it is
complex".} In this connection Mach notes: {\it ``I think anyone will
agree
with this"}~\cite{10}.

From Mach's statements it is obvious that, since the issue concerns
the
law of inertia, in accordance with which, following Newton, {\it
``...each
individual body, being left to itself, retains its state of rest or
uniform
motion along a straight line..."}, there naturally arises the
question
of inertial reference systems and of their relations to the
distribution
of matter. Mach and his contemporaries quite clearly understood that
such
a relation should exist in Nature. Precisely this meaning will
further be
attributed to the concept of  ``Mach's principle".

Mach wrote: {\it ``Although I think that at the beginning
astronomical observations will necessitate only very insignificant
corrections, I anyhow do think it possible that the law of inertia
in the simple form given it by Newton plays for us, human beings,
only a limited and transient role."}~\cite{10}. As we shall
further see, Mach did not turn out to be right, here. Mach did not
give a mathematical formulation of his idea, and therefore very
often diverse authors attribute to Mach's principle diverse
meanings. We shall try, here, to retain the meaning, attributed to
it by Mach himself.

Poincar\'e, and later Einstein, generalized the relativity principle
to
all
physical phenomena. Poincar\'e's formulation \cite{11} goes as
follows:
{\it ``...the relativity principle, according to which the laws
governing
physical phenomena should be identical for an observer at rest and
for an
observer undergoing uniform motion along a straight line, so we have
and
can have no method for determining whether we are undergoing similar
motion
or not."} Application of this principle to electromagnetic phenomena
led
Poincar\'e, and then Minkowski, to the discovery of the
pseudo-Euclidean
geometry of space-time and thus even more reinforced the hypothesis
of
inertial reference systems existing throughout the entire space. Such
reference systems are physically singled out, and therefore
acceleration
relative to them has an absolute sense.

In general relativity theory (GRT) no inertial reference systems
exist
in all space. Einstein wrote about this in 1929: {\it ``The starting
point
of theory is the assertion that there exists no singled out state of
motion,
i.e. not only velocity, but acceleration has no absolute
sense"}$\,$\footnote{Einstein A. Collection of scientific works,
Moscow:
Nauka, 1966, vol.2, art.92, p.264.}.

Mach's principle, in his own formulation, turned out not to have any
use.
It must, however, be noted that the ideas of inertial reference
systems
throughout the space have quite a weighty basis, since, for instance,
in
passing from a reference system bound to the Earth to a reference
system
bound to the Sun and, then, further to the Metagalaxy we approach,
with an
increasing precision, the inertial reference system. Therefore, there
are
no reasons for renouncing such an important concept as the concept of
an
inertial reference system. On the other hand, the existence of the
fundamental conservation laws of energy-momentum and of angular
momentum
also leads with necessity to the existence of inertial reference
systems
in the entire space. The pseudo-Euclidean geometry of space reflects
the
general dynamic properties of matter and at the same time introduces
inertial reference systems. Although the pseudo-Euclidean geometry of
space-time resulted from studies of matter, and therefore cannot be
separated from it, nevertheless, it is possible to speak of Minkowski
space
in the absence of matter. However, like earlier in Newtonian
mechanics, in
special relativity theory no answer exists to the question of how
inertial
reference systems are related to the distribution of matter in the
Universe.

The discovery of the pseudo-Euclidean geometry of space and time
permitted
considering not only inertial, but ac\-ce\-le\-ra\-ted reference systems,
also,
from a unique standpoint. A large difference was revealed between the
forces
of inertia and forces caused by physical fields. It consists in that
the
forces of inertia can always be equated to zero by choosing an
appropriate
reference system, while forces caused by physical fields cannot, in
principle, be made equal to zero by a choice of reference system,
since they
are of a vector nature in four-dimensional space-time. Since the
gravitational field in RTG is a physical field in the spirit of
Faraday-Maxwell, forces caused by such a field cannot be equated to
zero
by a choice of reference system.

Owing to the gravitational field having a rest mass, the main
equations
of RTG, (\ref{76}) and (\ref{77}), contain, together with the Riemannian
metric, the metric tensor of Minkowski space, also, but this means
that,
in principle, the metric of this space can be expressed via the
geometric
characteristics of the effective Riemannian space and, also, via
quantities
characterizing the distribution of matter in the Universe. This is
readily
done by passing in equations (\ref{76}) from contravariant to
covariant
quantities. In this way we obtain
\begin{equation}
\frac{m^2}{2}\, \gamma_{\mu\nu} (x) = \frac{8\,\pi}{\sqrt{-g}}
(T_{\mu\nu} - \frac{1}{2}\, g_{\mu\nu}\,T) - R_{\mu\nu} +
\frac{m^2}{2}\, g_{\mu\nu}\, . \label{92}
\end{equation}

Hence, we see that in the right-hand side of the equations there occur
only
geometric characteristics of the effective Riemannian space and
quantities
determining the distribution of matter in this space.

Experimental investigation of the motion of particles, and of light,
in
Riemannian space, in principle, allows to find the metric tensor of
Minkowski
space and, consequently, to construct an inertial reference system,
also.
Thus, RTG constructed within the framework of special relativity
theory
permits to establish the relation between an inertial reference
system and
the distribution matter. For this reason, motion relative to space is
motion
relative to matter in the Universe. The existence of an inertial
reference
system, determined by the the distribution of matter, makes
acceleration
absolute. We see that the special relativity principle is of general
significance, independent of the form of matter.

The requirements of this principle in the case of the gravitational
field
are expressed by the condition that equations (\ref{76}) and
(\ref{77}) be
form-invariant relative to the Lorentz group. Lorentz form-invariance
of
physical equations remains a most important physical principle in
constructing a theory, since precisely this principle provides the
possibility of introducing universal characteristics for all forms of
matter.

A.~Einstein wrote in his work of 1950: {\it ``...should one not
finally try
to retain the concept of an inertial system, renouncing all attempts
at
explaining the fundamental feature of gravitational phenomena, which
manifests itself in Newton's system as the equivalence of inert and
gravitating masses?"}$\,$\footnote{Einstein A. Collection of
scientific
works, Moscow: Nauka, 1966, vol.2, art.137, p.724.}. The concept of
an inertial system is retained in RTG,
and at the same time it is shown that the equivalence of inert and
gravitating masses is a direct consequence of the hypothesis that the
conserved density of the energy-momentum tensor of matter is the
source of
the gravitational field. Thus, the equality between inert and
gravitating
masses in no way contradicts the existence of an inertial reference
system.
Moreover, these conditions organically complement each other and
underlie
RTG.

Contrary to our conclusion, A.Einstein gave the following answer to
his
own question: ``{\it Who believes in the comprehensibility of Nature
should
answer --- no.}" The existence of inertial reference systems permits
resolving Mach's paradox, since only in this case can one speak of
acceleration relative to space. V.A.Fock wrote in this connection:
<<{\it As to Mach's paradox, it is known to be based on the
consideration
of a rotating liquid, having the shape of an ellipsoid, and of a
spherical
body that does not rotate. The paradox arises, here, only if the
concept
``rotation relative to space"  is considered to be senseless;
then, indeed,
both bodies (the rotating one and the one not rotating) are
apparently
equivalent, and it becomes incomprehensible why one of them is
spherical
and the other one is not. But the paradox vanishes as soon as we
acknowledge
the legitimacy of the concept of ``acceleration relative to
space"}>>$\,$\footnote{V.A.Fock. Theory of space, time and gravity.
M.:Gostekhizdat, 1961, p.499.}.

Mach's ideas profoundly influenced Einstein's views on gra\-vi\-ty during
the
construction of general relativity theory. Einstein wrote in one of
his
works: {\it ``Mach's principle: the G-field is fully determined by the
masses of bodies."} But this statement turns out to be not valid in
GRT,
since there exist solutions in the absence of matter, also. Attempts
at
eliminating this circumstance by introduction of the $\lambda$-term
did
not lead to the desired result. It turned out to be that equations
with the
$\lambda$-term also have solutions differing from zero in the absence
of
matter. We see that Einstein attached a totally different meaning to
the
concept of ``Mach's principle". But within such an interpretation,
also, no
place was found in GRT for Mach's principle.

Is there any place in RTG for Mach's principle as formulated by
Einstein?
Unlike GRT, in this theory spacelike surfaces are present throughout
the
entire space (global Cauchy surfaces), owing to the causality
principle.
And if no matter is present on one of such surfaces, then the
requirement of
energodominance imposed on the tensor of matter will result in matter
always
being absent \cite{12}. It will be shown in section 10 that a
gravitational
field cannot arise without matter.

Only solutions of the set of inhomogeneous gravitational equations
have a
physical sense, when matter exists in some part of space or
throughout the
entire space. This means that the gravitational field and the
effective
Riemannian space in the actual Universe could not arise without the
matter
that produced them. Solution of the equations for the metric of
effective
Riemannian space in the absence of matter can, for example, be
considered a limit case of the solution obtained for a homogeneous
and
isotropic distribution of matter in space, as the density of matter
tends
subsequently toward zero. We see that Mach's principle, even as
formulated
by Einstein, is realized in relativistic theory of gravity.

There exists, however, an essential difference between the
understanding of the G-field in our theory and in GRT. Einstein
understood the G-field to be the Riemannian metric, while in our
opinion the gravitational field is a physical field. Such a field
is present in the Riemannian metric together with the plane
metric, and therefore the metric does not vanish in the absence of
matter and of the gravitational field, but remains the metric of
Minkowski space.

In the literature there also exist other formulations of Mach's
principle,
differing in meaning from the ideas of both Mach and Einstein, but
since,
in our opinion, these formulations are not sufficiently clear, we
have not
dealt with them. Since gravitational forces in RTG are due to a
physical
field of the Faraday--Maxwell type, any common unique essence of the
forces
of inertia and of gravity is, in principle, out of the question.

Sometimes the essence of Mach's principle is seen to consist in that
the
forces of inertia are determined, allegedly in compliance with this
principle, by interaction with matter in the Universe. From a field
standpoint such a principle cannot exist in Nature. The point is
that,
although inertial reference systems, as we have seen above, are
related to
the distribution of matter in the Universe, forces of inertia do not
result
from the interaction with matter in the Universe, because any
influence of
matter can only be exerted via physical fields, but this means that
the
forces produced by these fields, owing to their vector nature, cannot
be
made equal to zero by a choice of reference system. Thus, forces of
inertia
are directly determined not by physical fields, but by a rigorously
defined
structure of geometry and by the choice of reference system.

The pseudo-Euclidean geometry of space--time, which reflects dynamic
properties common to all forms of matter, on the one hand confirmed
the
hypothetical existence of inertial reference systems, and on the
other hand
revealed that forces of inertia, arising under an appropriate choice
of
reference system, are expressed via the Christoffel symbols of
Minkowski
space. Therefore, they are independent of the nature of the body. All
this
became clear when it was shown that special relativity theory is
applicable
not only in inertial reference systems, but also in non-inertial
(accelerated) systems.

This made it possible to provide in Ref.~\cite{5} a more general
formulation of the
relativity principle: {\it ``Whatever physical reference system
(inertial or
non-inertial we choose, it is always possible to indicate an
infinite set
of such other reference systems, in which all physical phenomena
proceed
like in the initial reference system, so we have and can have no
experimental for determining precisely in which reference system of
this
infinite set we happen to be."} Mathematically this is expressed as
follows:
consider the interval in a certain reference system of Minkowski
space to
be
\[
d\sigma^2=\gamma_{\mu\nu}(x)dx^\mu dx^\nu,
\]
then there exists another reference system $x'$:
\[
x'^{\nu}=f^\nu (x),
\]
in which the interval assumes the form
\[
d\sigma^2=\gamma_{\mu\nu}(x')dx'^{\mu}dx'^{\nu},
\]
where the metric coefficients $\gamma_{\mu\nu}$ have the same
functional
form as in the initial reference system. In this case it is said that
the {\bf metric is form-invariant} relative to such transformations,
and
all {\bf physical equations are also form-invariant}, i.e. they have
the
same form both in the primed and in the not primed reference systems.
The
transformations of coordinates that leave the metric form-invariant
form a
group. In the case of Galilean coordinates in an inertial reference
system
these are the usual Lorentz transformations.

In RTG there exists an essential difference between the forces of
inertia
and the forces of gravity consisting in that as the distance from
bodies
increases, the gravitational field becomes weaker, while the forces
of
inertia may become indefinitely large, depending on the choice of
reference
system. And only in an inertial reference system are they equal to
zero.
Therefore, it is a mistake to consider forces of inertia inseparable
from
forces of gravity. In everyday life the difference between them is
nearly
obvious.

The construction of RTG has permitted to establish the relationship
between
an inertial reference system and the distribution of matter in the
Universe
and, thus, to understand more profoundly the nature of forces of
inertial
and their difference from material forces. In our theory forces of
inertia
are assigned the same role as the one they assume in any other field
theories.

\thispagestyle{empty}
\newpage
\section{Post-Newtonian approximation}

The post-Newtonian approximation is quite sufficient for studying
gravitational effects in the Solar system. In this section we shall
construct this approximation. Technically, our construction takes
advantage
of many results previously obtained by V.A.Fock \cite{13}, and it
turns
out to be possible to further simplify the method of deriving the
post-Newtonian approximation.

We shall write the main equations of theory in the form (see Appendix
D)
\be
\tilde\gamma^{\alpha\beta}D_\alpha
D_\beta\tilde\Phi^{\epsilon\lambda}+
m^2\sqrt{-\gamma}
\tilde\Phi^{\epsilon\lambda}=-16\pi g (T_M^{\epsilon\lambda}+
\tau_g^{\epsilon\lambda}), \label{93}
\ee
\be
D_\lambda\tilde\Phi^{\epsilon\lambda}=0\;. \label{94}
\ee
where $T_M^{\epsilon\lambda}$ is the energy-momentum tensor of
matter;
$\tau_g^{\epsilon\lambda}$ is the energy-momentum tensor of the
gravitational field.

The expression for the energy-momentum tensor of the gravitational
field
can be represented in the form
\ba
&-&16\pi g\tau_g^{\epsilon\lambda}=\frac{1}{2}(\tilde
g^{\epsilon\alpha}
\tilde g^{\lambda\beta} -\frac{1}{2}\tilde g^{\epsilon\lambda}
\tilde g^{\alpha\beta})
(\tilde g_{\nu\sigma}\tilde g_{\tau\mu}
-\frac{1}{2}\tilde g_{\tau\sigma}\tilde g_{\nu\mu})
D_\alpha \tilde\Phi^{\tau\sigma}\times\nonumber \\*
&\times&D_\beta\tilde\Phi^{\mu\nu}+
\tilde g^{\alpha\beta}\tilde g_{\tau\sigma}
D_\alpha\tilde\Phi^{\epsilon\tau}
D_\beta\tilde\Phi^{\lambda\sigma}-\tilde g^{\epsilon\beta}
\tilde g_{\tau\sigma}D_\alpha\tilde\Phi^{\lambda\sigma}
D_\beta\tilde\Phi^{\alpha\tau}-\nonumber \\*
&-&\tilde g^{\lambda\alpha}\tilde g_{\tau\sigma}
D_\alpha\tilde\Phi^{\beta\sigma}
D_\beta\tilde\Phi^{\epsilon\tau}
+\frac{1}{2}\tilde g^{\epsilon\lambda}\tilde g_{\tau\sigma}
D_\alpha\tilde\Phi^{\sigma\beta}
D_\beta\tilde\Phi^{\alpha\tau} +\nonumber \\*
&+&D_\alpha\tilde\Phi^{\epsilon\beta}
D_\beta\tilde\Phi^{\lambda\alpha}-\tilde\Phi^{\alpha\beta}D_\alpha
D_\beta \tilde\Phi^{\epsilon\lambda}-\nonumber \\*
&-&m^2(\sqrt{-g}\tilde g^{\epsilon\lambda}-\sqrt{-\gamma}
\tilde\Phi^{\epsilon\lambda}
+\tilde g^{\epsilon\alpha}
\tilde g^{\lambda\beta}\gamma_{\alpha\beta}-
\frac{1}{2}\tilde g^{\epsilon\lambda}
\tilde g^{\alpha\beta}\gamma_{\alpha\beta})\;. \label{95}
\ea
This expression is written in an arbitrary reference system in
Minkowski
space. We shall further perform all computations in the Galilean
coordinates
of the inertial reference system,
\be
\gamma_{\mu\nu}=(1,-1,-1,-1)\;. \label{96}
\ee

In constructing the series of perturbation theory it is natural to
apply as
a small parameter such a quantity $\epsilon$ that
\be
v\sim \epsilon,\,\, U\sim\epsilon^2,\,\, \Pi\sim\epsilon^2,\,\,
p\sim\epsilon^2\;. \label{97}
\ee
Here $U$ is the Newtonian potential of the gravitational field; $\Pi$
is
the specific internal energy of the body considered; $p$ is the
specific
pressure.\\
For the Solar system the parameter $\epsilon^2$ is of the order of
\be
\epsilon^2\sim 10^{-6}\;. \label{98}
\ee

We shall use the expansions of the components of the density of the
tensor:
\begin{equation}
\tilde g^{00}=1+\smb{\tilde\Phi}{(2)}{00}+
\smb{\tilde\Phi}{(4)}{00}+..., \label{99}
\end{equation}
\begin{equation}
\tilde g^{0i}=\smb{\tilde\Phi}{(3)}{0i}+
\smb{\tilde\Phi}{(5)}{0i}+..., \label{100}
\end{equation}
\begin{equation}
\tilde g^{ik}=\tilde\gamma^{ik}+
\smb{\tilde\Phi}{(2)}{ik}+
\smb{\tilde\Phi}{(4)}{ik}+... \label{101}
\end{equation}
We shall adopt the ideal fluid model of matter, the energy-momentum
tensor
of which has the form
\begin{equation}
T^{\epsilon\lambda}=[p+\rho (1+\Pi)]u^\epsilon u^\lambda -
pg^{\epsilon\lambda}, \label{102}
\end{equation}
where $\rho$ is the invariant density of an ideal fluid;\\
$p$ is the specific isotropic pressure;\\
$u^\lambda$ is the velocity four-vector.

We shall now write the expansion in the small parameter $\epsilon$
for the
energy-momentum tensor of matter:
\begin{equation}
T_M^{00}=\smb{T}{(0)}{00}+\smb{T}{(2)}{00}+... , \label{103}
\end{equation}
\begin{equation}
T_M^{0i}=\smb{T}{(1)}{0i}+\smb{T}{(3)}{0i}+... ,  \label{104}
\end{equation}
\begin{equation}
T_M^{ik}=\smb{T}{(2)}{ik}+\smb{T}{(4)}{ik}+...  \label{105}
\end{equation}
In the Newtonian approximation, i.e., when we neglect the forces of
gravity,
we have for the four-vector the following:
\begin{equation}
u^0=1+0(\epsilon^2),\,\,u^i=v^i(1+0(\epsilon^2))\;. \label{106}
\end{equation}
Substituting these expressions into (\ref{102}) we find
\begin{equation}
\smb{T}{(0)}{00}=\rho,\,\,
\smb{T}{(1)}{0i}=\rho v^i,\,\,
\smb{T}{(0)}{ik}=0\;. \label{107}
\end{equation}
In this approximation, on the basis of (5.7), we have
\begin{equation}
\partial_0\rho +\partial_i (\rho v^i)=0\;.  \label{108}
\end{equation}

Hence it can be seen that in the Newtonian approximation the total
inert
mass of a body is a conserved quantity:
\begin{equation}
M=\int\rho d^3x\;. \label{109}
\end{equation}
On the basis of equations (\ref{93}) we have in the Newtonian
approximation:
\begin{equation}
\Delta\smb{\tilde \Phi}{(2)}{00} =-16\pi\rho\;, \label{110}
\end{equation}
\begin{equation}
\Delta\smb{\tilde \Phi}{(3)}{0i} =-16\pi\rho v^i\;, \label{111}
\end{equation}
\begin{equation}
\Delta\smb{\tilde \Phi}{(2)}{ik} =0\;. \label{112}
\end{equation}

In an inertial reference system the mass of the graviton, owing to
its
smallness, plays an insignificant role for effects in the Solar
system, and
therefore in deriving equations (\ref{110})~--~(\ref{112}) we did not
take it
into account. But even in this case its influence is manifested in
that
equations (\ref{94}) have to exist together with the set of equations
(\ref{93}). Such equations in Galilean coordinates were also applied
in
V.A.Fock's theory of gravity, but unlike the case of RTG, they did
not
follow from the least action principle, so it was not clear why precisely
they
had to be applied, instead of some other equations. V.A.~Fock chose
them as
coordinate conditions and applied them in studying island systems. In
RTG
these equations arise from the least action principle, and for this
reason
they are universal. It is precisely owing to equations (8.2) that we
obtain
a complete set of equations for determining physical quantities. It
must be
noted that in the general case of a non-inertial reference system or
in the
case of strong gravitational fields the term with the graviton mass
$m$
can no longer be dropped. Thus, for example, even for a static body
in the
region close to the Schwarzschild sphere the influence of the
graviton mass
is very significant, so it can no longer be neglected.

The solution of equations (\ref{110})~--~(\ref{112}) has the form
\begin{equation}
\smb{\tilde \Phi}{(2)}{00} =4U,\,\,\;\;
U=\int\frac{\rho}{|x-x'|}d^3x', \label{113}
\end{equation}
\begin{equation}
\smb{\tilde \Phi}{(3)}{0i} =-4V^i,\,\,\;\;
V^i=-\int\frac{\rho v^i}{|x-x'|}d^3x'\;,  \label{114}
\end{equation}
\begin{equation}
\smb{\tilde \Phi}{(2)}{ik} =0\;. \label{115}
\end{equation}
On the  basis of equations (\ref{94}) we have
\begin{equation}
\partial_0 \smb{\tilde \Phi}{(2)}{00}+\partial_i \smb{\tilde
\Phi}{(3)}{0i}=0\;. \label{116}
\end{equation}
Substituting (\ref{113}) and (\ref{114}) we find
\begin{equation}
\partial_0 U  - \partial_i V^i=0\;. \label{117}
\end{equation}
Hence, it is evident that differentiation of the potential $U$ with
respect
to time increases the order of smallness in $\epsilon$. We shall take
advantage of this circumstance in calculating the energy-momentum
tensor of
the gravitational field, $\tau_g^{\epsilon\lambda}$. We note that
equation
(\ref{117}) is satisfied identically by virtue of equations
(\ref{108}).

From (\ref{114}) and (\ref{115}) it follows that of all the density
components of the tensor $\tilde\Phi^{\epsilon\lambda}$ only the
component
$\smb{\tilde\Phi}{(2)}{00}$, determined by expression (\ref{113}), is
seen
to remain in the second approximation. Precisely this circumstance
significantly simplifies the method of finding the post-Newtonian
approximation, when at each stage of the construction we make use of
the
densities of tensor quantities.

Making use of (\ref{113})~--~(\ref{115}) with a precision up to the
second
order inclusively we obtain
\ba
&&\sqrt{-g}g^{00}=1+4U,\,\,
\sqrt{-g}g^{11}= \nonumber \\
&&=\sqrt{-g}g^{22}=\sqrt{-g}g^{33}=-1\;.  \label{118}
\ea
Hence, we have
\begin{aequation}{118}{a}
-g=1+4U\;,\label{118a}
\end{aequation}
consequently,
\begin{equation}
g_{00}=1-2U,\,\,g_{11}=g_{22}=g_{33}=-(1+2U)\;. \label{119}
\end{equation}

We see from (\ref{118}) that in the Newtonian approximation, when it
suffices to consider only one component of the density of the tensor
of
matter, $T^{00}$, the gravitational field is described, as it was
expected, by only a sole component $\tilde\Phi^{00}$, while the
metric
tensor $g_{\mu\nu}$ has in this approximation, also, several
components,
in accordance with (\ref{119}). Working with the field components
$\tilde\Phi^{\mu\nu}$, instead of the metric tensor $g_{\mu\nu}$,
significantly simplifies the entire computational process of
constructing
the post-Newtonian approximation. Precisely for this reason,
introduction of
the density of the tensor of the gravitational field
$\tilde\Phi^{\mu\nu}$
is important not only from a general theoretical point of view, but
from a
practical standpoint, also. Thus, the metric tensor of the effective
Riemannian space is
\begin{equation}
g_{00}=1-2U,\,\,
g_{0i}=4\gamma_{ik}V^k,\,\,
g_{ik}=\gamma_{ik} (1+2U)\;. \label{120}
\end{equation}
From expression (\ref{113}) for $U$ it follows that the inert mass
(\ref{109}) is equal to the active gravitational mass. In RTG, as we
have
seen, this equality arose because the energy-momentum tensor is the
source
of the gravitational field.

We shall now proceed to construct the next approximation for the
component
of the metric tensor $g_{00}$. For this purpose we shall find the
contribution of the energy-momentum tensor of the gravitational
field.
Since in expression (\ref{95}) it is necessary under the derivative
sign
to take into account only $\smb{\tilde\Phi}{(2)}{00}$, the first term
in
(\ref{95}) will give a contribution equal to
\begin{equation}
2(\mbox{grad}\,U)^2, \label{121}
\end{equation}
while the second term contributes
\begin{equation}
-16(\mbox{grad}\,U)^2. \label{122}
\end{equation}
The contribution from all the remaining terms in this approximation
will be
zero. Discarded are also the terms with time derivatives of the
potential $U$,
since, owing to (\ref{117}), they are also all of a higher order of
smallness in $\epsilon$. From (\ref{121}) and (\ref{122}) we have
\begin{equation}
-16\pi g\tau_g^{00}=-14(\mbox{grad}\,U)^2\;.  \label{123}
\end{equation}
Making use of (\ref{123}), equation (\ref{93}) for component
$\tilde\Phi^{00}$ in this approximation assumes the form
\begin{equation}
\Delta \smb{\tilde \Phi}{(4)}{00}=16\pi gT^{00}+14(\mbox{grad}\,U)^2+
4\partial_0^2U\;. \label{124}
\end{equation}
Since on the basis of (\ref{120}) the interval equals the following
in the
second order in $\epsilon$:
\begin{equation}
ds=dt (1-U+\frac{1}{2}v_iv^i)\;, \label{125}
\end{equation}
we hence obtain
\begin{equation}
u^0=\frac{dt}{ds}=1+U-\frac{1}{2}v_iv^i\;. \label{126}
\end{equation}
Substituting this expression into (\ref{102}) we find
\begin{equation}
\smb{T}{(2)}{00}=\rho [2U+\Pi-v_iv^i]\;. \label{127}
\end{equation}

On the basis of (\ref{118a}) and (\ref{127}) we obtain from equations
(\ref{124}) the following:
\ba
&&\Delta \smb{\tilde \Phi}{(4)}{00} =-96\pi\rho U+
16\pi\rho v_iv^i+\nonumber \\
&&+14(\mbox{grad}\,U)^2-16\pi\rho\Pi+4\partial^2_0U\;.
\label{128}
\ea
Now we shall take advantage of the obvious identity
\begin{equation}
(\mbox{grad}\,U)^2=\frac{1}{2}\Delta U^2-U\Delta U\;. \label{129}
\end{equation}
But, since
\begin{equation}
\Delta U=-4\pi\rho\;, \label{130}
\end{equation}
then equation (\ref{128}), upon utilization of (\ref{129}) and
(\ref{130}),
assumes the form
\begin{equation}
\Delta (\smb{\tilde \Phi}{(4)}{00}-7U^2)=
16\pi\rho v_iv^i-40\pi\rho U-16\pi\rho \Pi+4\partial^2_0 U\;.
\label{131}
\end{equation}
Hence, we have
\begin{equation}
\smb{\tilde \Phi}{(4)}{00}=
7U^2+4\Phi_1+10\Phi_2+4\Phi_3-\frac{1}{\pi}\partial^2_0
\int\frac{U}{|x-x'|}d^3x'\;, \label{132}
\end{equation}
where
\be
\begin{array}{l}
\displaystyle\Phi_1=-\int\frac{\rho v_iv^i}{|x-x'|}d^3x', \;\;\;,
\displaystyle\Phi_2=\int\frac{\rho U}{|x-x'|}d^3x',\,\, \\[5mm]
\displaystyle\Phi_3=\int\frac{\rho \Pi}{|x-x'|}d^3x'.
\end{array}\label{133}
\ee

Thus, in the post-Newtonian approximation we find:
\ba
&\tilde g^{00}=1+4U+7U^2+4\Phi_1+10\Phi_2+ \nonumber
\\
&+4\Phi_3 -\frac{1}{\pi}\partial_0^2
\int\frac{U}{|x-x'|}d^3x'. \label{134}
\ea
We now have to find the determinant of $g$ in the post-Newtonian
approximation. To this end we represent $\tilde g^{ik}$ in the form:
\begin{equation}
\tilde g^{ik}=\tilde\gamma^{ik}+
\smb{\tilde \Phi}{(4)}{ik}. \label{135}
\end{equation}
It must be especially underlined that calculation of the determinant
of $g$
is most readily performed if one takes advantage for this purpose of
the
tensor density $\tilde g^{\mu\nu}$ and takes into account that
\begin{equation}
g=det (\tilde g^{\mu\nu})=det (g_{\mu\nu})\;. \label{136}
\end{equation}
From (\ref{134}) and (\ref{135}) we find
\ba
&&\sqrt{-g}=1+2U+\frac{3}{2}U^2+2\Phi_1+5\Phi_2+\nonumber \\
&&+2\Phi_3 -\frac{1}{2}\Phi-\frac{1}{2\pi}
\partial_0^2
\int\frac{U}{|x-x'|}d^3x'. \label{137}
\ea
Here
\begin{equation}
\Phi=
\smb{\tilde \Phi}{(4)}{11}+\smb{\tilde \Phi}{(4)}{22}+
\smb{\tilde \Phi}{(4)}{33}. \label{138}
\end{equation}
Since in the considered approximation $g_{00}g^{00}=1$,
from expressions (\ref{134}) and (\ref{137}) we obtain
\ba
&&g_{00}=1-2U+\frac{5}{2}U^2-2\Phi_1-5\Phi_2- \nonumber
\\
&&-2\Phi_3 -\frac{1}{2}\Phi+\frac{1}{2\pi}
\partial_0^2
\int\frac{U}{|x-x'|}d^3x'. \label{139}
\ea
For determining $g_{00}$ we need to calculate the quantity $\Phi$.
Since
$\Phi$, in accordance with (\ref{138}), was derived by summation, it
is
possible to make use of equation (\ref{93}) and by summation to
obtain
directly equations for function~$\Phi$.

From expression (\ref{95}) by summation we derive from the first term
the
following expression:
\begin{equation}
-16\pi g \tau_g^{ii}=-2(\mbox{grad}\,U)^2. \label{140}
\end{equation}
All the remaining terms present in expression (\ref{95}) give no
contribution in this approximation. With the aid of expression
(\ref{102})
for the energy-momentum tensor we find
\begin{equation}
-16\pi g\smb{\tilde T}{(2)}{ii}=
-16\pi\rho v_iv^i+48\pi p\;. \label{141}
\end{equation}
Taking into account (\ref{140}) and (\ref{141}), we write the
equation
for $\Phi$ as follows:
\begin{equation}
\Delta\Phi=16\pi\rho v_iv^i-48\pi p+2(\mbox{grad}\,U)^2. \label{142}
\end{equation}
Taking advantage of identity (\ref{129}) and of equation (\ref{130})
we
obtain
\begin{equation}
\Delta (\Phi-U^2)=16\pi\rho v_iv^i+8\pi\rho U-48\pi p\;. \label{143}
\end{equation}
Hence, we find
\be
\Phi=U^2+4\Phi_1-2\Phi_2+12\Phi_4,\,\,\,\label{144}
\ee
where
$$
\Phi_4=\int\frac{p}{|x-x'|}d^3x'.
$$

Substituting expression (\ref{144}) into (\ref{139}) we have
\ba
&&g_{00}=1-2U+2U^2-4\Phi_1-4\Phi_2-\nonumber \\
&&-2\Phi_3-6\Phi_4
+\frac{1}{2\pi}\partial^2_0\int\frac{U}{|x-x'|}d^3x'. \label{145}
\ea
Making use of the identity
\[
\frac{1}{2\pi}\int\frac{U}{|x-x'|}d^3x'=-\int\rho|x-x'|d^3x',
\]
we write expression (\ref{145}) as
\ba
&g_{00}=1-2U+2U^2-4\Phi_1-4\Phi_2-\nonumber \\
&-2\Phi_3-6\Phi_4
-\partial^2_0\int\rho|x-x'|d^3x'. \label{146}
\ea

The solutions (\ref{146}) and (\ref{120}) are calculated in an
inertial
reference system in Galilean coordinates. The effective Riemannian
metric that
arises is due to the presence of the gravitational field, while the
forces
of inertia are totally excluded. It is quite obvious that these
solutions
retain their functional form in the Galilean coordinates of any
inertial
reference system. Since all physical quantities are independent of
transformations of the time variable, then if the following
transformation
is applied:
\begin{equation}
x'^0=x^0+\eta^0(x),\,\, x'^i=x^i\;, \label{147}
\end{equation}
the metric coefficients will change as follows:
\begin{equation}
g'_{00}=g_{00}-2\partial_0\eta^0,\,\,
g'_{0i}=g_{0i}-\partial_i\eta^0,\,\,
g'_{ik}=g_{ik}\;. \label{148}
\end{equation}
It must be noted that transformation (\ref{147}) does not take us
beyond the
inertial reference system, since such a transformation is nothing
but
another choice of clock. All physically measurable quantities are
independent
of this choice.

Assuming function $\eta^0$ to be
\begin{equation}
\eta^0=-\frac{1}{2}\partial_0\int\rho|x-x'|d^3x', \label{149}
\end{equation}
and taking into account the identity
\begin{equation}
\array{l}
\displaystyle \partial_i\eta^0=\frac{1}{2}
(\gamma_{ik}V^k-N_i),\,\, \\[0.7mm]
\displaystyle N_i=\int\frac{\rho
v^k(x_k-x_k')(x_i-x_i')}{|x-x'|^3}d^3x'
\endarray \label{150}
\end{equation}
upon substitution into (\ref{148}) of expressions (\ref{120}) for
$g_{0i}$
and $g_{ik}$ and, also, of expression (\ref{146}) for $g_{00}$, and
taking
into account(\ref{149}) and (\ref{150}), we find the metric
coefficients of
effective Riemannian space in the so-called "canonical form":
\begin{eqnarray}
g_{00}&=&1-2U+2U^2-4\Phi_1-4\Phi_2-2\Phi_3-6\Phi_4, \nonumber \\*
g_{0i}&=&\frac{7}{2}\gamma_{ik}V^k+\frac{1}{2}N_i, \label{151}\\*
g_{ik}&=&\gamma_{ik}(1+2U)\;. \nonumber
\end{eqnarray}
These expressions coincide precisely with the formulae that are
obtained
on the basis of GRT. The difference only consists in that here they
follow
exactly from RTG, while for deriving them from GRT equations it is
necessary
to apply additional assumptions, that do not follow from theory, i.e.
it is
necessary to go beyond the limits of GRT. But we shall specially deal
with
this issue.

In the case of a static spherically symmetric body the post-Newtonian
approximation at a distance from the body assumes, in accordance with
(\ref{151}), the form
$$
\!\!\!\! g_{00}=1-\f{2MG}{r}
+2\left (\f{MG}{r}\right )^2,\;\;
g_{0i}=0,
$$
\vspace*{-0.7cm}
\begin{aequation}{151}{a}
\label{151a}  \\
\end{aequation}
\vspace*{-0.7cm}
$$
g_{ik}=\gamma_{ik}
\left (
1+\f{2MG}{r}
\right ),\;\;
M=\int \rho (x) d^3x.
$$

On the basis of expressions (\ref{151}) the post-Newtonian
No\-r\-d\-t\-wedt--Will
parameters in RTG assume the following values:
\[
\gamma=1,\,
\beta=1,\,
\alpha_1=\alpha_2=\alpha_3=\xi_1=\xi_2=\xi_3=\xi_4=\xi_W=0.
\]
We have calculated the metric coefficients (\ref{151}) in RTG in an
inertial
reference system. We shall now present the expressions for the
components of
the energy-momentum tensor of matter in the next approximation, as
compared
with (\ref{107}). Taking into account expression (\ref{126}) for
$u^0$ and,
also, that
\begin{equation}
u^i=\frac{dx^i}{ds}=v^i (1+U-\frac{1}{2}v_kv^k)\;,  \label{152}
\end{equation}
we find from formula (\ref{102})
\begin{equation}
\smb{T}{(3)}{0i}=\rho v^i(2U+\Pi-v_kv^k)+p v^i, \label{153}
\end{equation}
\begin{equation}
\smb{T}{(2)}{ik}=\rho v^iv^k-p \gamma^{ik}. \label{154}
\end{equation}

The component $\smb{T}{(2)}{00}$ is determined by expression
(\ref{127}).
On the basis of expression (\ref{151}), making use of the equations
for the
geodesic line, it is possible to calculate all effects in the Solar
system.
When gravitational effects in the Solar system are calculated in GRT
on the
basis of the post-Newtonian approximation, the results obtained are
correct,
and no ambiguity is present in the description of the effects. At the
same
time, if the exact solutions of GRT are applied, ambiguity arises in
the
description of the effects.

In conclusion we shall deal in somewhat greater detail with the
comparison of RTG and GRT in analyzing effects occurring in a weak
gravitational field. The set of equations (\ref{93}) and
(\ref{94}) together with the equation of state determines all the
physical quantities of one or another gravitational problem. All
the calculations performed above in the post-Newtonian
approximation were made in an inertial reference system. In GRT
there in principle exists no inertial reference system. In this
connection A.~Einstein wrote: {\it ``The starting point of theory
consists in the assertion that there exists no state of motion
physically singled out, i.e. not only velocity, but acceleration,
also, have no absolute meaning"}$\,$\footnote{Einstein A.
Collection of scientific works, Moscow: Nauka, 1966, vol.2,
art.92, p.264.}. But if no inertial reference system exists, to
which reference system must one consider calculations performed
within GRT to pertain?

In calculating gravitational effects V.~A.~Fock made use of the
harmonicity
conditions in Cartesian coordinates. He called them coordinate
conditions.
Thus, in a work published in 1939~\cite{13} he wrote: {\it ``In
solving
Einstein's equations we took advantage of a reference system, which
we have
termed harmonic, but which merits being called inertial."} Further in
the
same article he noted: {\it ``It seems to us that the possibility of
introducing in general relativity theory a definite inertial
reference
system in an unambiguous manner is noteworthy."} And, finally, in
Ref.~\cite{14} he wrote: {\it ``The relativity principle expressed by
the
Lorentz transformations is possible in inhomogeneous space, also,
while
a general relativity principle is not possible." }

All these statements of V.~A.~Fock were due to his aspiration to
clarify the essence of GRT, freeing it of general relativity
devoid of any physical meaning. However, V.~A.~Fock, here,
actually went beyond the limits of GRT. Precisely owing to this
fact he arrived at the striking conclusion on the va\-li\-di\-ty
of the relativity principle in inhomogeneous space, also. If one
remains within Riemannian space, and no other space exists in GRT,
then this assertion contradicts the correct conclusion made by
V.~A.~Fock ``{\it that in general relativity theory there,
generally speaking, exists no relativity.}" ~\cite{14}. But to
realize his goal it is necessary to introduce the concept of a
gravitational field in Minkowski space. Where did V.~A.~Fock go
beyond GRT? In applying the conditions of harmonicity he actually
considered Cartesian coordinates:
\be
\frac{\partial\tilde g^{\mu\nu}}{\partial x^\mu}=0\;,  \label{155}
\ee
where $x^\mu$ are Cartesian coordinates.
In Cartesian coordinates \\
$\gamma (x)=\det \gamma_{\mu\nu}=-1$.
Therefore, in accordance with the tensor law of transformations we
have
\be
\tilde g^{\mu\nu}(x)=
\frac{\partial x^\mu}{\partial y^\alpha}\cdot
\frac{\partial x^\nu}{\partial y^\beta}\cdot
\frac{\tilde g^{\alpha\beta}(y)}{\sqrt{-\gamma (y)}}\;. \label{156}
\ee

We shall write equations (\ref{155}) in the form
\be
\partial_\mu\tilde g^{\mu\nu}(x) =
\frac{\partial y^\tau}{\partial x^\mu}\cdot
\frac{\partial \tilde g^{\mu\nu}(x)}{\partial y^\tau}. \label{157}
\ee
For further calculations we present the formulae
\be
\frac{\partial}{\partial y^\tau}
\left (\frac{1}{\sqrt{-\gamma (y)}}\right )=
-\frac{1}{\sqrt{-\gamma}}\gamma^\lambda_{\tau\lambda},\,\,
\gamma^\nu_{\alpha\beta}=\frac{\partial^2 x^\sigma}{\partial
y^\alpha\partial y^\beta}\cdot
\frac{\partial y^\nu}{\partial x^\sigma}\;. \label{158}
\ee
Upon substitution of (\ref{156}) into (\ref{157}) and taking into
account (\ref{158}) we obtain
\ba
&&\partial_\mu\tilde g^{\mu\nu}(x)=
\frac{1}{\sqrt{-\gamma}}\frac{\partial x^\nu}{\partial y^\sigma}\cdot
\frac{\partial \tilde g^{\alpha\sigma}(y)}{\partial y^\alpha}+
\nonumber \\
&&+\frac{1}{\sqrt{-\gamma}}\tilde g^{\alpha\beta}(y)
\frac{\partial^2 x^\nu}{\partial y^\alpha\partial y^\beta}=0\;.
\label{159}
\ea
We shall write the multiplier of the second term in the form
\[
\frac{\partial^2x^\nu}{\partial y^\alpha\partial y^\beta}=
\frac{\partial x^\nu}{\partial y^\sigma}\cdot
\frac{\partial y^\sigma}{\partial x^\tau}\cdot
\frac{\partial^2x^\tau}{\partial y^\alpha\partial y^\beta}=
\frac{\partial x^\nu}{\partial y^\sigma}\cdot
\gamma^\sigma_{\alpha\beta}\;.
\]

Substituting this expression into the preceding one we find
\[
\partial_\mu\tilde g^{\mu\nu}(x)=
\frac{1}{\sqrt{-\gamma}}\cdot
\frac{\partial x^\nu}{\partial y^\sigma}
\left (
\frac{\partial \tilde g^{\alpha\sigma}(y)}{\partial y^\alpha}+
\gamma^\sigma_{\alpha\beta}(y)\tilde g^{\alpha\beta}(y)\right )=0\;,
\]
i.e. we have
\be
\partial_\mu\tilde g^{\mu\nu}(x)=
\frac{1}{\sqrt{-\gamma}}\cdot
\frac{\partial x^\nu}{\partial y^\sigma}D_\mu\tilde
g^{\mu\sigma}(y)=0\;.
\label{160}
\ee

Thus, we have established that the density of the tensor
$\tilde g^{\mu\sigma}(y)$ in arbitrary coordinates automatically
satisfies
the general covariant equation
\[
D_\lambda\tilde g^{\lambda\sigma}=0\;,
\]
if the initial condition of harmonicity (\ref{155}) is written in
Cartesian
coordinates. But this means that the harmonicity condition is not a
coordinate condition, but a field equation in Minkowski space. Thus,
application of the condition of harmonicity in Cartesian coordinates
is not
an innocent operation, but it implies going beyond the framework of
GRT by
introduction of Minkowski space.

The obtained equation coincides with equation (\ref{77}) of RTG. In
RTG it
follows from the least action principle. Performing transformation
from
coordinates $y$ to coordinates $z$ we obtain (see Appendix~(E.12))
$$
\dalam\; y^\lambda =-\gamma^\lambda_{\alpha\beta} (y)
g^{\alpha\beta}(y),
$$
where $\dalam$ denotes the operator
$$
\dalam=\frac{1}{\sqrt{-g(z)}}\cdot\frac{\pa}{\pa z^\nu}
\left (\tilde g^{\nu\sigma}\f{\pa}{\pa z^\sigma}\right ).
$$
Therefore, when V.~A.~Fock wrote down the harmonicity conditions in
the
form
$$
\dalam\;y^\lambda=0,
$$
he~~ actually~~ dealt~~ with~~ Cartesian~~ coordinates,~~ for~~ which
$\gamma^\lambda_{\alpha\beta}(y)=0$, i.e. with Minkowski space in
Galilean coordinates. In choosing harmonic coordinates in the form of
conditions (8.63) V.~A.~Fock actually made use of Minkowski space in
Galilean coordinates, while equations (8.63) played the part of field
equations, instead of coordinate conditions. But why it was necessary
to
add to the Hilbert--Einstein equations precisely equations (8.63) in
Galilean
coordinates, instead of some others, in order to obtain the complete
set of
gravity equations within V.A.Fock's approach, remained unclear.
Here,
V.A.Fock was most likely guided by physical intuition, and also by
the
mathematical simplification that arose in the course of calculations.

Did V.~A.~Fock attempt to consider the gravitational field in
Minkowski
space? No, he was far from this idea, and in this connection he
wrote~\cite{13}: {\it ``We recall it here only in connection with the
sometimes observed tendency (certainly not shared by us) to pack the
theory
of gravity into the framework of Euclidean space."} As we have seen,
application of the conditions of harmonicity in Cartesian coordinates
makes
us go beyond the framework of GRT. But this means that the set of
equations
of gravity, studied by V.~A.~Fock, differs from the set of equations
of
GRT, i.e. V.~A.~Fock's theory of gravity based on the conditions of
harmonicity in Cartesian coordinates and Einstein's GRT are different
theories. V.~A.~Fock's approach turns out to be closer to the ideas
of RTG.
Everything that V.~A.~Fock attempted to introduce in the theory of
gravity
(inertial reference systems, acceleration relative to space) is fully
inherent in RTG, but this is achieved by consideration of the
gravitational
field, like all other physical fields, in Minkowski space. Here, all
the
geometric characteristics of Riemannian space are now field quantities
in
Minkowski space.

In~~ analyzing~~ gravitational~~ effects~~ in~~ the~~ Solar~~
system V.A. Fock actually made use of Minkowski space, since he
referred all the calculated gravitational effects to an inertial
reference system. Precisely this circumstance permitted him to
obtain correct expressions for the effects. Thus, for example, he
wrote~\cite{14}: {\it ``How should one define a straight line: as
a light ray or as a straight line in that Euclidean space in which
the harmonic coordinates $x_1, x_2, x_3$ serve as Cartesian
coordinates? We think the second definition to be the only correct
one. We actually made use of it in saying that a light ray has the
shape of a hyperbola in the vicinity of the Sun"}, and further on,
{\it ``the argument that a straight line, like a ray of light, is
more directly observable, but it has no sense: in the definitions
it is not the direct observability that is decisive, but the
correspondence to Nature, even though this correspondence may be
established by indirect reasoning."}

In RTG gravitational effects are determined unambiguously, because in
accordance with equations (\ref{93}) and (\ref{94}) written in the
Galilean
coordinates of an inertial reference system, the motion of light or
of a
test body, when the gravitational field is switched off, indeed
proceeds
along a straight line, which is a geodesic line in Minkowski space.
It is
absolutely clear that in a non-inertial reference system the geodesic
line
in Minkowski space will no longer be a straight line. But this means
that
in RTG, in an non-inertial reference system, for revealing a
gravitational
effect motion in effective Riemannian space must be compared precisely
with
the geodesic motion of the accelerated reference system.

In calculating gravity effects in the Solar system, when the
influence of
the graviton mass can be neglected, only the RTG set of equations
(\ref{93})
and (\ref{94}) in Galilean coordinates coincides with the set of
equations
dealt with by V.~A.~Fock in harmonic (Cartesian) coordinates.
If one remains
within the framework of GRT, then in any other, for instance
non-inertial,
reference system they differ essentially. This takes place because
the
covariance of V.~A.~Fock's set of equations is not general, unlike
the set
of RTG equations. V.~A.~Fock obtained the complete set of
gravitational
equations (for island systems) by adding the harmonicity conditions
to the
Hilbert-Einstein equations. But why must precisely the harmonicity
conditions be added, instead of some other conditions, remained
unclear. In
accordance with RTG, the complete set of gravitational equations
(8.1)
and (8.2) arises from the least action principle. Hence it becomes
clear,
why conditions (8.2), which in Cartesian coordinates coincide with
the
harmonicity conditions, arise, instead of some other conditions. But
these
equations become universal, valid not only for island systems. But if
V.~A.~Fock had realized that in applying the harmonicity conditions
he
actually had to deal with Cartesian coordinates of Minkowski space,
he would
have readily obtained expression (\ref{160}). As we already noted
earlier,
the harmonicity conditions in Cartesian coordinates successfully
applied by
V.~A.~Fock took him beyond the framework of Einstein's GRT. This fact
was
noted in 1957 by L.~Infeld who wrote:{\it ``Thus, for Fock the choice
of the
harmonicity coordinate condition becomes a certain fundamental law of
Nature, which alters the very character of Einstein's general
relativity
theory and transforms it into a theory of the gravitational field,
valid
only in inertial reference systems''}$\,$\footnote{L.Infeld. Most
recent
problems in gravity. M.:Foreign literature publishers,
1961,
p.162.}.

If one remains within the framework of
GRT, then it is absolutely incomprehensible, from the point of view
of
physics, why it is necessary to choose the harmonicity conditions,
instead
of any other conditions. While in RTG, owing to the existence of the
graviton mass, these conditions arise as a consequence of the
validity of
the equations for matter [see (5.7) and (5.17)], i.e. they follow
from
the least action principle, and they therefore have universal
significance.
However, in GRT similar expressions for the post-Newtonian
approximation are,
nevertheless, obtained without application of the har\-mo\-ni\-ci\-ty
conditions in
Cartesian coordinates. Why is this so? The reason consists in that
Minkowski
space in Galilean coordinates is once again introduced and that the
gravitational field is actually considered as a physical field in
this
space.

The metric of Minkowski space in Galilean coordinates is taken as
the zero order approximation for the Riemannian metric. It is
complemented with various potentials with arbitrary post-Newtonian
parameters, each of which decreases like $0(\frac{1}{r})$. In this
way the arbitrariness contained in GRT is discarded. Substitution
of the Riemannian metric $g_{\mu\nu}$ in this form into the
Hilbert-Einstein equation permits one to determine the values of
the post-Newtonian parameters, and we again arrive at the same
post-Newtonian approximation. Precisely here gravity is considered
to be a physical field in Minkowski space, the behaviour of which
is described by the introduced gravitational potentials. Such a
requirement imposed on the character of the metric of Riemannian
space does not follow from GRT, since in the general case the
asymptotics of the metric is quite arbitrary and even depends on
the choice of the three-dimensional space coordinates. Therefore
it is impossible to impose physical conditions on the metric. But
if it is effective and its arising is due to the physical field,
then the physical conditions are imposed on the metric in a
natural manner.

In RTG the gravitational equations (\ref{76}) and (\ref{77}) are
generally
covariant, but not form-invariant with respect to arbitrary
transformations.
They are form-invariant relative to the Lorentz transformations. But
this
means that in Lorentz coordinates, in case the solution $G(x)$ exists
for
the tensor of matter $T_{\mu\nu}(x)$, there exists, in the new
Lorentz
coordinates $x'$, the solution $G'(x')$ for the tensor of matter
$T'_{\mu\nu}(x')$, and, consequently, in the coordinates $x$ the
solution
$G'(x)$ is possible only for the tensor of matter $T'_{\mu\nu}(x)$.

In RTG a unique correspondence is established between the
Riemannian metric and the Minkowski metric, which permits one to
compare motion under the influence of the gravitational field and
in its absence, when calculation is performed of the gravitational
effect. When the gravitational field is switched off in RTG the
Riemann tensor turns to zero, and at the same time transition
occurs from Riemannian metric to the Minkowski metric, previously
chosen in formulating the physical problem. This is precisely what
provides for the equivalence principle to be satisfied in RTG.

For calculation of the gravitational effect it is necessary to
compare
motion in Riemannian space with motion in absence of the gravitational
field.
This is precisely how the gravitational effect is determined. If in
GRT
one refers the set of solutions for $g_{\mu\nu}$ to a certain
inertial
reference system, then it is quite obvious that one will obtain a
whole
set of various values for the gravitational effect. Which one of them
should
be chosen? Since the Hilbert-Einstein equations do not contain the
metric of
Minkowski space, it is impossible to satisfy the equivalence
principle,
because it is impossible to determine, in which (inertial or
non-inertial)
reference system one happens to be, when the gravitational field is
switched
off.

To conclude this section we note that the post-Newtonian approximation
(\ref{151}) satisfies the causality principle (\ref{90}).

 \thispagestyle{empty}
\newpage
\section{On the equality of inert and gravitational masses}

Owing to the density of the energy-momentum tensor being the source
of the
gravitational field, the inert and gravitational masses were shown in
section~8 to be equal. In this section we shall show that the field
approach
to gravity permits obtaining in a trivial manner the metric of
effective
Riemannian space in the first approximation in the gravitational
constant $G$.
This is especially simple to establish on the basis of equations
(\ref{2}).
In the case of a spherically symmetric static body, equations
(\ref{2}) have
the following form in the Galilean coordinates of an inertial
reference
system:
\be
\Delta\tilde\Phi^{00} - m^2\tilde\Phi^{00}=-16\pi t^{00}, \label{161}
\ee
\be
\Delta\tilde\Phi^{0i} - m^2\tilde\Phi^{0i}=0,\;\
\Delta\tilde\Phi^{ik}-m^2\tilde\Phi^{ik}=0,\;\;
i,k=1,2,3.  \label{162}
\ee
For a static body the sole component $t^{00}$ differs from zero.

From equations (\ref{162}) we have
\be
\tilde\Phi^{0i}=0,\;\; \tilde\Phi^{ik}=0\;. \label{163}
\ee
Far away from the body, from equation (\ref{161}) we find
\be
\tilde\Phi^{00}\simeq \frac{4M}{r}e^{-mr},\;\;
M=\int t^{00}d^3x\;, \label{164}
\ee
$M$ is the inert mass of the body, that creates the gravitational
field. In
the Solar system the exponential factor can be neglected, owing to
the
quantity $mr$ being small.
\be
\tilde\Phi^{00}\simeq \frac{4M}{r}\;. \label{165}
\ee

We shall now find the components of the density of the metric tensor
of
effective Riemannian space, $\tilde g^{\mu\nu}$. On the basis of
(\ref{6}) we
have
\be
\tilde g^{\mu\nu}=\tilde\gamma^{\mu\nu}+\tilde\Phi^{\mu\nu},\;\;
\tilde g^{\mu\nu}=\sqrt{-g} g^{\mu\nu}. \label{166}
\ee
Hence, taking into account (\ref{163}) and (\ref{165}), we obtain the
following $\tilde g^{\mu\nu}$ components, that differ from zero:
\be
\tilde g^{00}=1+\frac{4M}{r},\;\;
\tilde g^{11}=\tilde g^{22}=\tilde g^{33}=-1. \label{167}
\ee
They satisfy equation (2.3) exactly. On the basis of (\ref{167})
we find
\be
g_{00}=\frac{\sqrt{-g}}{1+\frac{4M}{r}},\;\;
g_{11}=g_{22}=g_{33}=-\sqrt{-g}. \label{168}
\ee
\be
-g=\;-\;\tilde g^{00}\tilde g^{11}\tilde g^{22}\tilde g^{33}=\left
(1+\frac{4M}{r}\right )\;. \label{169}
\ee
Substituting the expressions for $g$ into formulae (\ref{168})
we obtain
\be
g_{00}\simeq \left
(1-\frac{2M}{r}\right ),\;\;
g_{11}=g_{22}=g_{33}=
-\left
(1+\frac{2M}{r}\right )\;. \label{170}
\ee

It must be especially underlined that at the place, where in
accordance
with Newton's law of gravity there should be an active gravitational
mass,
there appears the inert mass $M$. Thus, the equality of the inert and
active
gravitational masses is a direct consequence of the density of the
energy-momentum tensor being the source of the gravitational field.
So the
reason that the inert and gravitational masses are equal is not the
local
identity of the forces of inertia and of gravity (this actually does
not
occur in GRT), but the universality of the conserved source of the
gravitational field, of the energy-momentum tensor of matter.

The interval in effective Riemannian space has the form
\be
ds^2=\left
(1-\frac{2M}{r}\right )dt^2-
\left
(1+\frac{2M}{r}\right )
(dx^2+dy^2+dr^2)\;. \label{171}
\ee
Classical effects of gravity, such as the gravitational red shift of
spectral lines, the deviation of a light ray by the Sun, the time
delay of
a radiosignal, the precession of a gyroscope on the Earth's orbit,
are
fully described by this interval.

From expression (\ref{170}) it is evident that the forces of
gra\-vity are attractive, since the quantity $M$, being an inert
mass, is always positive. As to GRT, in accordance with this
theory it is not possible to prove the equality of inert and
active gravitational masses. A detailed analysis of this issue is
presented in joint works performed with prof. V.~I.~Denisov. This
circumstance is dealt with in detail in the monograph~\cite{1_3}.
The essence of the issue consists in that the expression for inert
mass, determined from the pseudotensor of the gravitational field,
depends on the choice of the three-dimensional coordinates, which
is physically inadmissible. Precisely by a simple choice of
three-dimensional space coordinates (which is always permitted)
one can show that in GRT inert mass is not equal to active
gravitational mass. Since the equality of physically measurable
quantities in GRT depends on the choice of the three-dimensional
coordinates, this means that not everything in it is alright here,
also. Sometimes the opinion is voiced that within the framework of
GRT it is possible to construct the energy-momentum tensor of the
gravitational field by substitution of covariant derivatives in
Minkowski space for the ordinary derivatives in the expression for
the pseudotensor. However, here, on the one hand, it is impossible
to say with definiteness which metric in Minkowski space must be
taken for such a substitution, and, on the other hand, in
Riemannian space no global Cartesian coordinates exist, and,
consequently, no Minkowski space, so such an approach does not
remove the essential difficulty of GRT: the absence of integral
conservation laws of energy-momentum and of angular momentum  for
matter and gravitational field taken together.

\thispagestyle{empty}
\newpage
\section{Evolution of the homogeneous and isotropic
Universe}


We write the equations of RTG in the form
\be
R_{\mu\nu}-
\frac{m^2}{2}
(g_{\mu\nu}-\gamma_{\mu\nu})=
\frac{8\pi}{\sqrt{-g}}
\left (T_{\mu\nu}-\frac{1}{2}g_{\mu\nu}T\right )\;, \label{172}
\ee
\be
D_\mu\tilde g^{\mu\nu}=0\;. \label{173}
\ee

For~~ convenience~~ we~~ have~~ chosen~~ the~~ set~~ of units
$G=\hbar=c=1$. In the
final
expressions we shall restore the dependence upon these constants. The
density of the energy-momentum tensor has the form
\be
T_{\mu\nu}=\sqrt{-g}
[(\rho+p)U_\mu U_\nu -g_{\mu\nu}p],\;\;
U^\nu=\frac{dx^\nu}{ds}\;. \label{174}
\ee
Here $\rho$ is the density of matter, $p$ is pressure, $ds$ is the
interval
in effective Riemannian space. For a homogeneous and isotropic model of
the
Universe the interval of effective Riemannian space $ds$ has the general
form
\be
ds^2=U(t)dt^2-V(t)
\left [\frac{dr^2}{1-kr^2} + r^2
\!(d\Theta^2+\sin^2\Theta d\Phi^2)\right ].
\!\!    \label{175}
\ee
Here $k$ assumes the values $1,-1,0;\;\;\; k=1$ corresponds to the
closed
Universe, $k=-1$ --- to the hyperbolic Universe, and $k=0$ --- to the ``flat"
Universe.

Since the set of RTG equations (\ref{172}) and (\ref{173}) together
with the
equation of state is complete, then in the case of appropriate
initial
conditions it can yield only a single solution describing the
development of
a homogeneous and isotropic model of the Universe. At the same time
the
equations of GRT for the same model yield three well-known scenarios
for
the development of the Universe. The scenario of the development of
the
Universe obtained on the basis of the RTG does not coincide with any
of the
scenarios based on GRT. We shall follow ~\cite{15}. 

All our analysis will be made in an inertial reference system in
spherical
coordinates $r,\Theta,\Phi$. An interval in Minkowski space, in this
case,
will have the form
\ba
&&d\sigma^2=dt^2-dx^2-dy^2-dz^2=\nonumber \\
&&=dt^2-dr^2-r^2
(d\Theta^2+\sin^2\Theta d\Phi^2)\;. \label{176}
\ea
The determinant $g$, composed of components $g_{\mu\nu}$, equals
\be
g=-UV^3(1-kr^2)^{-1}r^4\sin^2\Theta. \label{177}
\ee
The tensor density
\be
\tilde g^{\mu\nu}=\sqrt{-g} g^{\mu\nu} \label{178}
\ee
has, in accordance with (\ref{175}), the following components:
\ba
&&\tilde g^{00}=
\sqrt{\frac{r^4 V^3}{U(1-kr^2)}}\cdot\sin \Theta,\;\nonumber \\
&&\tilde g^{11}=-r^2\sqrt{UV(1-kr^2)}\cdot\sin\Theta\;,\nonumber
\\[-0.1cm] \label{179} \\[-0.2cm]
&&\tilde g^{22}=
-\sqrt{\frac{UV}{1-kr^2}}\cdot\sin \Theta,\;\; \nonumber \\
&&\tilde
g^{33}=-\sqrt{\frac{UV}{1-kr^2}}\cdot\frac{1}{\sin\Theta}.\nonumber
\ea

The Christoffel symbols of Minkowski space are
\ba
\gamma^1_{22}&=&-r,\;\;
\gamma^1_{33}=-r\sin^2\Theta,\;\;
\gamma^2_{12}=\gamma^3_{13}=\frac{1}{r}\;,\nonumber
\\*[-0.3cm] \label{180} \\*[-0.3cm]
\gamma^2_{33}&=&-\sin\Theta\cos\Theta,\;\;
\gamma^3_{23}=\cot \Theta\;.\nonumber
\ea
Equation (\ref{173}) has the form
\be
D_\mu\tilde g^{\mu\nu}=\partial_\mu \tilde g^{\mu\nu}+
\gamma^\nu_{\alpha\beta}\tilde g^{\alpha\beta}=0\;. \label{181}
\ee
Substituting (\ref{180}) into (\ref{181}) we obtain
\be
\frac{\partial}{\partial t}
\left (\frac{V^3}{U}\right )=0\;, \label{182}
\ee
\be
\frac{\partial}{\partial r}(r^2\sqrt{1-kr^2})=2r(1-kr^2)^{-1/2}.
\label{183}
\ee
From equation (\ref{182}) it follows
$$
V=aU^{1/3},
$$
here $a$ is an integration constant.

From equation (\ref{183}) we directly find
\be
k=0\;, \label{184}
\ee
i.e. {\sf the space metric is Euclidean}. It must be stressed that
this
conclusion for a homogeneous and isotropic Universe follows directly
from
equation (\ref{173}) for the gravitational field and does not depend
on
the density of matter. Thus, equation (\ref{173}) excludes the closed
and
hyperbolic models of the Universe. A homogeneous and isotropic
Universe can
only be ``flat"  according to RTG. In other words, the
well-known
problem of the Universe's flatness does not exist within the
framework of
RTG. With account of (\ref{182}) and (\ref{184}) the effective
Riemannian
metric (\ref{175}) assumes the form
\be
ds^2=U(t)dt^2-a U^{1/3}
[dr^2+r^2(d\Theta^2+\sin^2\Theta d\Phi^2)]\;. \label{185}
\ee

If one passes to the proper time $d\tau$
\be
d\tau=\sqrt{U} dt \label{186}
\ee
and introduces the notation
\be
R^2=U^{1/3}(t)\;, \label{187}
\ee
the interval (\ref{185}) assumes the form
\be
ds^2=d\tau^2-aR^2(\tau)[dx^2+dy^2+dz^2]\;. \label{188}
\ee
Here and further in this section $R$ is a scaling factor. We are
compelled
to make use of the notation adopted in the literature for this
quantity, in
spite of the fact that in the previous sections $R$ stood for the
scalar
curvature. For the given metric the Christoffel symbols assume the
form
\ba
&&\Gamma^1_{22}=-r,\;
\Gamma^1_{33}=-r\sin^2\Theta,\;
\Gamma^2_{12}=\Gamma^3_{13}=\frac{1}{r},\;\nonumber
\\*[-0.3cm] \label{189} \\*[-0.3cm]
&&\Gamma^2_{33}=-\sin \Theta\cos\Theta,\;
\Gamma^3_{23}=\cot \Theta\;,\nonumber\\
\nonumber\\*[-0.3cm]
&&\Gamma^0_{ii}=aR\frac{dR}{d\tau},\;\;
\Gamma^i_{0i}=\frac{1}{R}\frac{dR}{d\tau},\;\;
i=1,2,3\;. \label{190}
\ea
Making use of expression (\ref{44}) we have
\ba
&&R_{00}=-\frac{3}{R}\frac{d^2R}{d\tau^2},\;\;
R_{11}=2a\left (\frac{dR}{d\tau}\right )^2+aR
\frac{d^2R}{d\tau^2}\;,\nonumber
\\*[-0.3cm] \label{191} \\*[-0.3cm]
&&R_{22}=r^2R_{11},\;\; R_{33}=\sin^2\Theta\cdot R_{22},\;\;
R_{0i}=0\;,\nonumber\\
\nonumber\\*[-0.3cm]
&&R_{\mu\nu}g^{\mu\nu}=-\frac{6}{R}\cdot
\frac{d^2 R}{d\tau^2} -
\frac{6}{R^2}\cdot
\left (\frac{dR}{d\tau}\right )^{2}. \label{192}
\ea
Since $g_{0i}=0,\; R_{0i}=0$, then from equation (\ref{172}) it
follows
directly that
\be
T_{0i}=0\;. \label{193}
\ee
Hence on the basis of (\ref{174}) we have
\be
U_i=0\;. \label{194}
\ee
This means that matter is at rest in an inertial reference system.
Thus, the
so-called  ``expansion" of the Universe, observed by the red shift, is
due to
the change of the gravitational field in time. Therefore, there
exists no
expansion of the Universe, related to the motion of objects with
respect to
each other. The red shift is not due to the motion of galaxies, which
is
absent, but to the variation of the gravitational field in time.
Therefore,
the red shift does not indicate that the galaxies were at a time
close to
each other. At the same time, in accordance with GRT {\it ``All
versions of
the Friedman model have in common that at a certain moment of time in
the
past (ten--twenty thousand million years ago) the distance between
adjacent
galaxies should have been equal to zero"}$\,$\footnote{S.~Hawking.
From
the Big Bang to Black Holes. M.:Mir, 1990, p.46.}. We, here, have
quoted S.~Hawking,
and below we shall establish the reason for the difference between
the
conclusions concerning the development of the Universe in \mbox{RTG}
and
\mbox{GRT}. We shall now deal in a somewhat greater detail with the
nature
of the red shift.

From (\ref{188}) it follows that the speed of a light ray equals
$$ \f{dr}{d\tau}=\f{1}{\sqrt{a}R(\tau)}. $$ Let us put the
observation point at the origin of the reference system $(r=0)$.
Consider a light signal emitted from point $r$ during the time
interval between $\tau$ and $\tau+d\tau$, and let its arrival at
the point $r=0$ take place during the time interval between
$\tau_0$ and $\tau_0+d\tau_0$; then, for light emitted at moment
$\tau$ and arriving at the point $r=0$ at the moment $\tau_0$ we
have $$ \int\limits^{\tau_0}_{\tau} \f{d\tau}{R(\tau)}=\sqrt{a}r,
$$ similarly, for light emitted at a moment $\tau+d\tau$ and
arriving at the point $r=0$ at the moment $\tau_0+d\tau_0$ we find
$$ \int\limits^{\tau_0+d\tau_0}_{\tau+d\tau}
\f{d\tau}{R(\tau)}=\sqrt{a}r. $$ Equating these expressions we
obtain $$ \f{d\tau}{R(\tau)}=\f{d\tau_0}{R(\tau_0)}. $$ Or,
passing to the light frequency, we have $$
\omega=\f{R(\tau_0)}{R(\tau)}\omega_0. $$ Hence, it is obvious that
the light frequency $\omega$ at the point of emitting is not equal
to the frequency of the light $\omega_0$ at the point of
observation.

Introducing the {\sf red shift parameter} $z$
$$
z=\f{\omega-\omega_0}{\omega_0},
$$
we have
$$
z=\f{R(\tau_0)}{R(\tau)}-1.
$$
We see that the red shift is only related to variation of the scaling
factor
$R(\tau)$, in the case of such variation there exists no motion of
matter,
in accordance with (\ref{194}). Thus, the nature of the red shift is
not
related to the scattering of galaxies, which is absent, but to
variation of
the gravitational field with time, i.e. it is related to the fact
that
$R(\tau_0)>R(\tau)$.

It must be especially stressed that a given inertial reference system
is
singled out by Nature itself, i.e. in the considered theory the Mach
principle is satisfied automatically.

Substituting (\ref{191}) and (\ref{174}) into equation (\ref{172}),
with
account of (\ref{194}), we have
\ba
\frac{1}{R}\frac{d^2R}{d\tau^2}=&-&\frac{4\pi G}{3}
\left (\rho+\frac{3p}{c^2}\right )
- 2\omega \left (1-\frac{1}{R^6}\right )\;, \label{195} \\*
\left ( \frac{1}{R}\frac{dR}{d\tau}\right )^2& = &\frac{8\pi
G}{3}\rho-
\frac{\omega}{R^6} \left (1-\frac{3R^4}{a}+2R^6 \right )\;.
\label{196}
\ea
where
\be
\omega=\frac{1}{12}\left (
\frac{mc^2}{\hbar}\right )^2. \label{197}
\ee
From (\ref{195}) it is seen that for small values of the scaling
factor $R$
there arises an initial acceleration owing to the second term. This
is
precisely what ``incites" the ``expansion" of the Universe. The initial
acceleration appears at the moment when the density of matter stops
growing in the preceding cycle. From (\ref{196}) it follows that in
the
region  $R>>1$ the contemporary density of matter in the Universe
equals
\be
\rho (\tau)=\rho_c(\tau)+\frac{1}{16\pi G} \left (
\frac{mc^2}{\hbar}\right )^2, \label{198} \ee where $\rho_c(\tau)$
is the critical density determined by the Hubble ``constant"
\be
\rho_c=\frac{3H^2(\tau)}{8\pi G},\;\;
H(\tau) =\frac{1}{R}\cdot \frac{dR}{d\tau}\;. \label{199}
\ee
Hence, the necessity for the existence of ``dark" matter, which is in
accordance with modern observational data.

From equations (\ref{195}) and (\ref{196}) one can obtain the
expression for
the deceleration parameter of the Universe, $q(\tau)$:
\be
q=-\frac{\ddot R}{R}\frac{1}{H^2}=\frac{1}{2}+\frac{1}{4}
\left (
\frac{c}{H}\right )^2
\left (
\frac{mc}{\hbar}\right )^2. \label{200}
\ee
Thus, the parameter $q$ at present is positive, i.e. ``expansion" of
the
Universe has slowed down, instead of being accelerated. The relation
(\ref{200}) makes it possible, in principle, to determine the mass of
the
graviton from two observable quantities, $H$ and $q$. From the
causality
principle (\ref{89}), (\ref{90}) it follows that
\be
R^2(R^4-a)\leq 0\;. \label{201}
\ee
To satisfy the causality condition throughout the entire region of
variation
of $R(\tau)$ it is natural to set
\be
a=R^4_{\max}. \label{202}
\ee
From the condition that the left-hand side of equation (\ref{196}) is
not
negative it follows that the expansion should start from some minimum
value
$\displaystyle R_{\min}$, corresponding to the value
$\frac{dR}{d\tau}=0$. On the
other
hand, if $R>>1$, the expansion should stop at $R_{\max}$, when the
density
(\ref{198}) reaches its minimum value
\be
\rho_{\min}=\frac{1}{16\pi G}
\left (
\frac{mc^2}{\hbar}\right )^2
\left (
1-\f{1}{R^6_{\max}}
\right ).\label{203}
\ee
and the process starts of compression down to $R_{\min}$.

Thus, in RTG there exists no cosmological singularity, and the
presence of
the graviton mass results in the evolution of the Universe exhibiting
a
cyclical character. The time required for the Universe to expand from
the
maximum to its minimum density is mainly determined by the stage at
which
nonrelativistic matter is dominant and is
\be
\tau_{\max}\simeq \sqrt{\frac{2}{3}}\frac{\pi\hbar}{mc^2}\;.
\label{204} \ee From the covariant conservation law, that is a
consequence of equations (\ref{76}), (\ref{77}) $$ \nabla_\mu
\tilde T^{\mu\nu} + \Gamma^\nu_{\alpha\beta}\tilde
T^{\alpha\beta}=0 $$ it is possible to obtain the equation
\be
\frac{1}{R} \frac{dR}{d\tau} =-\frac{1}{3(\rho+\frac{p}{c^2})}
\frac{d\rho}{d\tau}\;. \label{205}
\ee
For the stage of development of the Universe dominated by radiation
$$
p=\frac{1}{3}\rho c^2
$$
from equation (\ref{205}) we obtain the following expression for the
radiation density $\rho_r$:
\be
\rho_r(\tau)=\frac{A}{R^4(\tau)}\;. \label{206}
\ee
Here $A$ is an integration constant. At the stage of development of
the
Universe, when nonrelativistic matter is dominant and pressure can be
neglected, from equation (\ref{205}) we find
\be
\rho_m (\tau) =\frac{B}{R^3 (\tau)}\;, \label{207}
\ee
$B$ is an integration constant.

Consider that at a certain moment of time $\tau_0$ the radiation
density
$\rho_r(\tau_0)$ becomes equal to the density of matter, $\rho_m
(\tau_0)$
\be
\rho_r (\tau_0)=\rho_m(\tau_0)\;, \label{208}
\ee
then
$$
A=BR(\tau_0)=BR_0.
$$
Since at later stages of the development of the Universe matter is
dominant,
we have from formula (\ref{207}) the following:
\be
B=\rho_{\min}\cdot R^3_{\max}. \label{209}
\ee

Thus,
\be
\rho \simeq \rho_r =
\frac{\rho_{\min}R_0\cdot R^3_{\max}}{R^4},\;\;
R\leq R_0, \label{210}
\ee
\be
\rho \simeq \rho_m =
\rho_{\min}\left (
\frac{R_{\max}}{R}\right )^3,\;\;
R\geq R_0. \label{211}
\ee
According to observational data (see, for example,~\cite{16}), the
present-day density of radiation (including the three sorts of
neutrinos,
which we for definiteness consider massless) and the critical density
of
matter, are
\be
\rho_r (\tau_c)=8\cdot 10^{-34}\mbox{g/cm}^3,\;\;
\rho_m (\tau_c)=10^{-29}\mbox{g/cm}^3. \label{212}
\ee
The ``hidden" mass must be attributed to the density of matter,
$\rho_m(\tau_c)$, for our choice of the graviton mass
$(m=10^{-66}\mbox{g})$
it is close to the critical density $\rho_c$ determined by the Hubble
``constant". We intend matter to actually be all forms of matter, with
the
exception of the gravitational field.

In accordance with formulae (\ref{210}), (\ref{211}) and (\ref{212})
we have
\be
\rho_r(\tau_c)=
\frac{\rho_{\min}R_0\cdot R^3_{\max}}{R^4(\tau_c)}=
8\cdot 10^{-34}\mbox{g/cm}^3, \label{213}
\ee
\be
\rho_m(\tau_c)=
\rho_{\min}\left (
\frac{R_{\max}}{R(\tau_c)}\right )^3
=10^{-29}\mbox{g/cm}^3. \label{214}
\ee
Hence, we find
\be
R_0=
\frac{\rho_r(\tau_c)}{\rho_m^{4/3}(\tau_c)}
R_{\max}\cdot \rho^{1/3}_{\min}=
3,7 \cdot 10^{5}\rho^{1/3}_{\min}\cdot
R_{\max}. \label{215}
\ee

Let us introduce the notation
\be
\sigma=\frac{4}{3}R_0\cdot R^3_{\max}. \label{216}
\ee
In accordance with (\ref{210}), (\ref{215}) and (\ref{216}) we obtain
\be
\rho(\tau)\simeq\rho_r(\tau)=
\frac{3}{4}\cdot
\frac{\sigma\cdot \rho_{\min}}{R^4(\tau)},\;\;
R\leq R_0. \label{217}
\ee
Assuming radiation to be dominant at the initial stage of expansion
in the
hot Universe model, from equation (\ref{196}) by
taking into account (\ref{202}), (\ref{203}) and (\ref{217}), we obtain the following:
\be
H^2=\left ( \frac{1}{R} \frac{dR}{d\tau}\right )^2= \omega \left [
\frac{3\sigma}{2R^4}-2+ \frac{3}{R^4_{\max}\cdot
R^2}-\frac{1}{R^6}\right ]\;. \label{218} \ee Equation (\ref{218})
makes it possible to determine the law of expansion of the
Universe at the initial stage. It is readily seen that the
right-hand side of equation (\ref{218}) turns to zero at
sufficiently small values of $R=R_{\min}$. The main role, here, is
due to the first term in brackets in equation (\ref{196}), that is
responsible for the graviton mass.

By introducing the variable $x=R^{-2}$ one can readily find
approximate
values for the roots of the equation
\be
\frac{3}{2}\sigma x^2-2+\frac{3}{R^4_{\max}} x-x^3=0\;, \label{219}
\ee
which are the turning points
\be
x_1=\frac{3}{2}\sigma +
0\left (\frac{1}{\sigma^2}\right ),\;\;
x_{2,3}=\pm \sqrt{\frac{4}{3}}\frac{1}{\sqrt{\sigma}}+0
\left (\frac{1}{\sigma^2}\right )\;. \label{220}
\ee
Hence, we find the turning point
\be
R_{\min} =\sqrt{\frac{2}{3\sigma}}\;.  \label{221}
\ee

Thus, owing to the graviton mass, there exists no cos\-mo\-lo\-gi\-cal
singularity
in RTG, and expansion of the Universe starts from the finite non-zero
value
$R=R_{\min}$. On the basis of (\ref{217}) we obtain
\be
\rho_{\max}=\frac{3}{4}\cdot
\frac{\sigma \rho_{\min}}{R^4_{\min}}=
\frac{27}{16}\sigma^3\cdot \rho_{\min}. \label{222}
\ee
In accordance with (\ref{220}) expression (\ref{218}) can be written
in the
form
\be
H^2=\omega (x_1-x) (x-x_2)(x-x_3)\;.  \label{223}
\ee
Within the range of variation of the scaling factor
\be
R_{\min} \leq R\leq R_0 \label{224}
\ee
the expression for $H^2$ is significantly simplified:
\be
H^2\simeq \omega x^2 (x_1-x)=
\frac{3\sigma\omega}{2R^6}(R^2-R^2_{\min})\;.  \label{225}
\ee
Within this approximation equation (\ref{218}) assumes the form
\be
\frac{1}{R^2}\left (
\frac{dR}{d\tau}\right )^2=
\frac{3\sigma\omega}{2R^6}
(R^2-R_{\min})\;. \label{226}
\ee
Upon integration we find
\be
\tau =\frac{R^2_{\min}}{\sqrt{6\sigma\omega}}
[Z\sqrt{Z^2-1}+\ln (Z+\sqrt{Z^2-1})]\;, \label{227}
\ee
where
$$
Z=R/R_{\min}.
$$
Utilizing expressions (\ref{221}) and (\ref{222}) we obtain
\be
\frac{R^2_{\min}}{\sqrt{6\sigma\omega}}=
\frac{1}{2\sqrt{2\omega}}\left (
\frac{\rho_{\min}}{\rho_{\max}}\right )^{1/2}. \label{228}
\ee
Substituting into this expression the value $\rho_{\min}$ from
(\ref{203})
we find
\be
\frac{R^2_{\min}}{\sqrt{6\sigma\omega}}=
\sqrt{\frac{3}{32\pi G\rho_{\max}}}\;. \label{229}
\ee
Taking into account (\ref{229}) in (\ref{227}) we obtain
\be
\tau =\sqrt{\frac{3}{32\pi G\rho_{\max}}}
[Z\sqrt{Z^2-1}+\ln (Z+\sqrt{Z^2-1})]\;. \label{230}
\ee
In the vicinity of $R\simeq R_{\min}$ from (\ref{230}) we find
\be
R(\tau)=R_{\min}
\left [ 1+\frac{4\pi G}{3}\rho_{\max} \cdot \tau^2\right ]\;.
\label{231}
\ee
In the region $R_{\min}<<R<R_0$ we obtain
\be
R(\tau)=R_{\min}\left (
\frac{32\pi G}{3}\rho_{\max}\right )^{1/4}\!\cdot\tau^{1/2}.
\label{232}
\ee
In this region the dependence on time of the density of matter
determined
by equation (\ref{217}), with account of (\ref{221}), (\ref{222}) and
(\ref{232}), has the form
\be
\rho (\tau) =\frac{3}{32\pi G \tau^2}\;, \label{233} \ee i.e.
coincides with the known equation that yields the Friedman model
in GRT for a ``flat" Universe. We shall now determine the time
corresponding to transition from the stage of expansion of the
Universe dominated by radiation to the stage dominated by
nonrelativistic matter. According to (\ref{232}) we have
\be
R^2_0=R^2_{\min}\left (
\frac{32\pi G}{3}\rho_{\max}\right )^{1/2}\!\cdot\tau_0. \label{234}
\ee
Hence, taking into account (\ref{215}), (\ref{216}) and (\ref{229})
we find
\ba
&&\tau_0=\frac{\rho_r^{3/2}(\tau_c)}{\rho^2_m(\tau_c)}
\sqrt{\frac{3}{32\pi G}}=\nonumber \\
&&=2,26\cdot 10^8
\sqrt{\frac{3}{32\pi G}}\simeq 1,5\cdot 10^{11}\mbox{s}\;.
\label{235}
\ea

Now consider development of the Universe, when pressure can be
neglected. At
this stage of evolution we write equation (\ref{196}) as
\be
\left (
\frac{dx}{d\tau}\right )^2=
\frac{2\omega x^2}{\alpha} (x-1)
[(\alpha -x^3)(x^2+x+1)-3x^2]\;. \label{236}
\ee
Here $x=R_{\max}/R,\; \alpha=2R^6_{\max}$. Taking into account that
\be
\alpha >> 3\;, \label{237}
\ee
we find
\be
\tau =\tau_0+\sqrt{\frac{\alpha}{2\omega}}
\int\limits^{x_0}_{x}
\frac{dy}{y\sqrt{(y^3-1)(x^3_1-y^3)}}\;. \label{238}
\ee
Here $x_0=R_{\max}/R_0,\;\; x_1=2^{1/3}\cdot R^2_{\max}$. Upon
integration
of (\ref{238}) we obtain
\be
\tau = \tau_0 +\frac{1}{3}
\sqrt{\frac{\alpha}{2x_1\omega}}[\arcsin f(x_0) -
\arcsin f (x)]. \label{239}
\ee
Here
\be
f(x) =\frac{(x^3_1+1)x^3-2x^3_1}{x^3(x^3_1-1)}\;. \label{240}
\ee
Note that
\be
f (x_0)\simeq 1-\frac{2}{x^3_0}\;, \label{241}
\ee
\be
\arcsin f (x_0)\simeq \arccos \frac{2}{x_0^{3/2}}
=\frac{\pi}{2}-\frac{2}{x_0^{3/2}}\;. \label{242}
\ee
Taking into account (\ref{242}) we find
\be
\tau=\tau_0 -\frac{2}{3\sqrt{2\omega}x^{3/2}_0}
+\frac{1}{3\sqrt{2\omega}}
\left [
\frac{\pi}{2} - \arcsin f (x)\right ]\;. \label{243}
\ee
Taking into account the equality
\be
\tau_0 =\frac{1}{2\sqrt{2\omega}x_0^{3/2}}\;, \label{244}
\ee
expression (\ref{243}) can be written in the form
\be
3\sqrt{2\omega} (\tau+\beta\tau_0)=\frac{\pi}{2}-\arcsin f(x)\;.
\label{245}
\ee
Hence, we have
\be
\cos\lambda (\tau+\beta\tau_0)=
\frac{(\alpha+1)x^3-2\alpha}{x^3 (\alpha-1)}\;. \label{246}
\ee
Here
\be
\lambda=3\sqrt{2\omega} =\sqrt{\frac{3}{2}}\left (
\frac{mc^2}{\hbar}\right ),\;\;
\beta=1/3\;. \label{247}
\ee
From expression (\ref{246}) we find
\be
\!  R(\tau)=\left [\frac{\alpha}{2}\right ]^{1/6}\!\cdot
\left [
\frac{(\alpha+1)-(\alpha-1)\cos\lambda (\tau+\beta\tau_0)}{2\alpha}
\right ]^{1/3}\!. \label{248}
\ee
Owing to the equality (\ref{211}) the following relation occurs:
\be
\frac{\rho_m(\tau)}{\rho_{\min}}=
\left [
\frac{R_{\max}}{R(\tau)}\right ]^3, \label{249}
\ee
taking into account (\ref{249}) we obtain
\be
\rho_m(\tau)=
\frac{2\alpha\rho_{\min}}
{(\alpha+1)-(\alpha-1)\cos\lambda(\tau+\beta\tau_0)}\;. \label{250}
\ee
Since $\alpha>>1$, from (\ref{250}) we have
\be
\rho_m(\tau)=\frac{\rho_{\min}}
{\sin^2\frac{\lambda(\tau+\beta\tau_0)}{2}}\;, \label{251}
\ee
in a similar way from formula (\ref{249}) we have
\be
R(\tau)=R_{\max}\sin^{2/3}\frac{\lambda (\tau+\beta\tau_0)}{2}.
\label{252}
\ee
In the region of values
$\frac{\lambda(\tau+\beta\tau_0)}{2}\ll 1$
we
have
\be
\rho_m(\tau)=\frac{1}{6\pi G(\tau+\beta\tau_0)^2}\;, \label{253}
\ee
\be
R(\tau)=R_{\max}\left [
\frac{\lambda(\tau+\beta\tau_0)}{2}\right ]^{2/3}. \label{254}
\ee
For $\tau>>\beta\tau_0$ formulae (\ref{253}) and (\ref{254}) yield
for
$\rho_m(\tau)$ and $R(\tau)$ time dependencies similar to those
obtained
within the  Friedman model in GRT for a "flat" Universe.

Making use of formulae (\ref{215}), (\ref{216}) and (\ref{222}) one
can
readily establish the following relation:
\be
R_{\max}=
\frac{\rho_m^{1/3}(\tau_c)}
{\rho_r^{1/4}(\tau_c)}
\left (
\frac{\rho_{\max}}
{4\rho^2_{\min}}
\right )^{1/12}
\!\!\simeq
3,6\cdot 10^{-2}
\left (
\frac{\rho_{\max}}
{\rho^2_{\min}}\right )^{1/12}\!\! . \label{255}
\ee
In a similar manner with the aid of expression (\ref{228}) and of
(\ref{221}) one can express, via $\rho_{\max}$, the second turning
point
$R_{\min}$:
\be
R_{\min}=\left (
\frac{\rho_{\min}}{2\rho_{\max}}\right )^{1/6}. \label{256}
\ee
From (\ref{255}) and (\ref{256}) it is clear that the existence of
the
graviton mass not only removes the cosmological singularity, but also
stops the expansion process of the Universe, which undergoes
transition to
the compression phase. Thus, the evolution of a homogeneous and
isotropic
Universe is determined by modern observational data (\ref{212}), by
the
maximum density of matter and the graviton mass. The scalar curvature
is
the largest at the beginning of the ``expansion", and on the basis of
(\ref{192}),
(\ref{195}) and
(\ref{256}) equals the following value:
$$
R_{\mu\nu} g^{\mu\nu}=-16\pi G\cdot \f{\rho_{\max}}{c^2},
$$
while its minimum value at the end of  ``expansion" is
$$
R_{\mu\nu} g^{\mu\nu}=\f{3}{2}\left (\f{mc}{\hbar}\right )^2.
$$
The initial acceleration, which serves as the initial ``push" that led
to
the ``expanding" Universe, is, in accordance with  (\ref{195}),
(\ref{203})
and
(10.85), the following:
$$
\f{d^2R}{d\tau^2}=\f{1}{3}
\left (
8\pi G \rho_{\max}
\right )^{5/6}
\left (
\f{mc^2}{2\hbar}
\right )^{1/3}.
$$
It arises at the moment when the density of matter stops growing
during the
preceding cycle. The maximum density of matter in the Universe in
this
model remains undefined. It is related to the integral of motion. The
latter
is readily established. We write equation (\ref{195}) in the form
\be
\frac{d^2R}{d\tau^2}=-4\pi G
\left(\rho+\frac{p}{c^2}\right )R+
\frac{8\pi G}{3}\rho R-2\omega \left( R-\frac{1}{R^5}\right )\;.
\label{257}
\ee
Determining from equation
\be
\frac{1}{R}\frac{dR}{d\tau}=-\frac{1}{3(\rho+\frac{p}{c^2})}
\frac{d\rho}{d\tau} \label{258}
\ee
the value of $\left (\rho+\frac{p}{c^2}\right )$ we find
\be
\rho+\frac{p}{c^2}=-\frac{1}{3}R\frac{d\rho}{dR}\;. \label{259}
\ee
Substituting this value into equation (\ref{257}) we obtain
\be
\frac{d^2R}{d\tau^2}=\frac{4\pi G}{3}\cdot\frac{d}{dR}
(\rho R^2)-\omega\frac{d}{dR}
\left (R^2+\frac{1}{2R^4}\right )\;. \label{260}
\ee

Introducing the notation
\be
V=-\frac{4\pi G}{3}\rho R^2+
\omega \left (
R^2+\frac{1}{2R^4}\right )\;, \label{261}
\ee
one can write equation (\ref{260}) in the form of the Newton equation
of
motion
\be
\frac{d^2R}{d\tau^2}=-\frac{dV}{dR}\;, \label{262}
\ee
where $V$ plays the role of the potential. Multiplying (\ref{262}) by
$\frac{dR}{d\tau}$ we obtain
\be
\frac{d}{d\tau} \left [
\frac{1}{2}\left (\frac{dR}{d\tau}\right )^2+V\right ]=0\;.
\label{263}
\ee
Hence, we have
\be
\frac{1}{2}\left (\frac{dR}{d\tau}\right )^2+V=E\;, \label{264}
\ee
where $E$ is an intergal of motion, the analog of energy in classical
mechanics. Comparing (\ref{264}) with (\ref{196}) and taking into
account
(\ref{202}) we obtain
\be
R^4_{\max}=\frac{1}{8E}\left (
\frac{mc^2}{\hbar}\right )^2. \label{265}
\ee
Substituting into (\ref{265}) expression (\ref{255}) we find
\be
E=7,4\cdot 10^{4}\left [
\frac{\left (\frac{mc^2}{\hbar}\right )^{10}}
{(16\pi G)^2\rho_{\max}} \right ]^{1/3}. \label{266}
\ee
This quantity is extremely small.

Thus, $\rho_{\max}$ is actually an integral of motion, determined by
the
initial conditions of the dynamic system. The analysis performed
reveals
that the model of a homogeneous and isotropic Universe develops, in
accordance with RTG, cyclically starting from a certain finite
maximum
density $\rho_{\max}$ down to the minimum density, and so on. The
Universe
can only be ``flat". Theory predicts the existence in the Universe of
a large
``hidden" mass of matter. The Universe is infinite and exists for an
indefinite time, during which an intense exchange of information took
place between its regions, which resulted in the Universe being
homogeneous
and isotropic, with a certain inhomogeneity structure. In the model
of a
homogeneous and isotropic Universe this inhomogeneity is not taken
into
account, for simplifying studies. The information obtained is
considered
a zeroth approximation, that usually serves as a background in
considering
the development of inhomogeneities caused by gravitational
instability.
``Expansion" in a homogeneous and isotropic Universe, as we are
convinced,
is due to variation of the gravitational field, and no motion of
matter
occurs, here. The existence of a certain inhomogeneity structure in
the
distribution of matter in space introduces a significant change,
especially
in the period after the recombination of hydrogen, when the Universe
becomes transparent and the pressure of radiation no longer hinders
the
collection of matter in various parts of the Universe.

This circumstance results in the motion of matter relative to the
inertial reference system. Thus arise the peculiar
ve\-lo\-ci\-ti\-es of galaxies with respect to the inertial
reference system. A reference system related to the relic
radiation can, with a great precision, be considered inertial.
Naturally, a re\-fe\-rence system related to the relic
gravitational radiation would to an extremely high degree be close
to an inertial system. What was the maximum density of matter,
$\rho_{\max}$, earlier in the Universe? An attractive possibility
is reflected in the hypothesis that $\rho_{\max}$ is determined by
the world constants. In this case, the Planck density is usually
considered to be $\rho_{\max}$. Here, however, there exists the
problem of overproduction of monopoles arising in Grand
Unification theories. To overcome this problem, one usually
applies the ``burning out" mechanism of monopoles during the
inflational expansion process due to the Higgs bosons. Our model
provides another, alternative, possibility. The quantity
$\rho_{\max}$ may even be significantly smaller, than the Planck
density. In this case the temperature of the early Universe may
turn out to be insufficient for the production of monopoles, and
the problem of their overproduction is removed in a trivial
manner. This, naturally, does not exclude the possibility of
inflational expansion of the Universe, if it turns out to be that
at a certain stage of its development the equation of state is
$p=-\rho$.

Thus, in accordance with RTG, no pointlike Big Bang occurred, and,
consequently, no situation took place, when the distance between
galaxies
were extremely small. Instead of the explosion, at each point of the
space
there occurred a state of matter of high density and temperature, and
it
further developed till the present moment, as described above. The
difference between the development of a homogeneous and isotropic
Universe
in \mbox{RTG} and \mbox{GRT} arose owing to the scaling factor
$R(\tau)$ in
RTG not turning into zero, while in GRT it becomes zero at a
certain
moment in the past.

In GRT with the cosmological term $\lambda$ the homogeneous and
isotropic
model of the Universe is also possible in the absence of matter. The
solution of the GRT equations for this case was found by de~Sitter.
This
solution corresponds to curved four-dimensional space-time. This
signifies
the existence of gravity without matter. What is the source of this
gravity?
Usually, it is vacuum energy identified with the cosmological
constant
$\lambda$ that is considered to be this source. In RTG, when matter
is absent
$(\rho=0)$, in accordance with equations (\ref{196}) and (\ref{202}),
the
right-hand side should turn to zero, which is possible only if
$R_{\max}=1$.
Hence it follows that $R\equiv 1$, and, consequently, the geometry of
space-time in the absence of matter will be pseudo-Euclidean. Thus, in
accordance with RTG, when matter is absent in the Universe, there
also exists
no gravitational field, and, consequently, vacuum possesses no
energy, as it
should be. According to RTG, the Universe cannot exist without
matter.

In conclusion let us determine the horizon of particles and the
horizon of
events. For a light ray, in accordance with the interval (\ref{188}),
we
have
\be
\frac{dr}{d\tau}=\frac{1}{\sqrt{a}R(\tau)}\;. \label{267}
\ee
The distance covered by the light by the moment $\tau$ is
\be
d_r(\tau)=\sqrt{a}R(\tau)\int\limits^{r(\tau)}_{0}
dr=R(\tau)\int\limits^{\tau}_{0}
\frac{d\sigma}{R(\sigma)}\;. \label{268}
\ee
If the gravitational field were absent, the distance covered by the
light
would be $c\tau$. As $R$ we should have substituted expression
(\ref{230})
for the interval $(0,\tau_0)$, and expression (\ref{252}) for the
time
interval $(\tau_0, \tau)$. We shall estimate $d_r(\tau)$
approximately by
the expression
\be
R(\tau) =R_{\max}\sin^{2/3}\frac{\pi \tau}{2\tau_{\max}}
\label{269} \ee Throughout the entire integration interval \ba
&&d_r(\tau)=\frac{2\tau_{\max}}{\pi}\left [
\sin\frac{\pi\tau}{2\tau_{\max}}\right ]^{2/3}\;\cdot
\int\limits^{\sqrt{y}}_{0}
\frac{dx}{x^{2/3}\sqrt{1-x^2}}=\nonumber \\
&&=\frac{6\tau_{\max}}{\pi}\sqrt{y} F\left
(\frac{1}{2},\frac{1}{6},\frac{7}{6},y\right )\;. \label{270} \ea
Here $y=\sin^2\frac{\pi\tau}{2\tau_{\max}}, F(a,b,c,y)$ is a
hypergeometric function.

We shall give the values for some quantities determining the
evolution of a
homogeneous and isotropic Universe. We set the graviton mass to
$m=10^{-66}$g, while the present-day Hubble ``constant" is
\be
H_c\simeq 74\frac{\mbox{\small{km}}}{\mbox{\small{s  Mpc       }}}\;.
\label{271}
\ee
Then for the present-day moment of time $\tau_c, q_c$ will be equal
to
\be
\tau_c\simeq 3\cdot 10^{17}\mbox{s},\;\;
q_c=0,59,\;\;
\rho_c=10^{-28}\frac{\mbox{\small{g}}}{\mbox{\small{cm}}^3}\;.
\label{272}
\ee
According to formula (\ref{204}) the half-period of cyclical
development is
\be
\tau_{\max}=9\pi\cdot 10^{17}\mbox{s}\;. \label{273}
\ee

It must be stressed that the parameters $\tau_c, q_c$ determining the
evolution of the Universe are practically independent of the maximum
density of matter $\rho_{\max}$. The maximum temperature (and, hence
the
maximum density), that could occur in the Universe, may be determined
by
such phenomena, that took place in these extreme conditions, and the
consequences of which may be observed today. A special role, here,
is
played by the gravitational field, which contains the most complete
information on the extreme conditions in the Universe. In the model,
considered above, of an isotropic and homogeneous Universe the known
problems of singularities, of causality, of flatness, that are
present
in GRT, do not arise.

Making use of (\ref{270}) and (\ref{272}) we find the size of the
observable part of the Universe at the moment $\tau_c$:
$$
d_r(\tau_c)\simeq 3c\tau_c=2,7\cdot 10^{28}\mbox{cm}\;.
$$
We see that the path covered by light in the gravitational field of
the
Universe during the time $\tau_c$ is three times larger than the
corresponding distance in absence of the gravitational field,
$c\tau_c$.
During the half-period of evolution, $\tau_{\max}$, the horizon of
particles
be
\be
d_r(\tau_{\max})=\frac{c\tau_{\max}}{\sqrt{\pi}}\cdot
\frac{\Gamma (1/6)}{\Gamma(2/3)}\;. \label{274}
\ee
The horizon of events is determined by the expression
\be
d_c=R(\tau)\int\limits^{\infty}_{\tau}\frac{d\sigma}{R(\sigma)}\;.
\label{275}
\ee
Since the integral (\ref{275}) turns to infinity, the horizon of
events in
our case does not exist. This means that information on events taking
place
in any region of the Universe at the moment of time $\tau$ will reach
us.
This information can be obtained with the aid of gravitational waves,
since
they are capable of passing through periods, when the density of
matter was
high.

We shall especially note that within the framework of RTG a
homogeneous and
isotropic Universe can exist only if the graviton mass differs from
zero.
Indeed, in accordance with (\ref{217}) and (\ref{222}), the constant
$A$ in
expression (\ref{206}) equals
$A=\rho_{\max}^{1/3}\left(\frac{\rho_{\min}}{2}\right )^{2/3}$.
Therefore,
if $\rho_{\max}$ is fixed, then the constant $A$ turns to zero, when
$m=0$.

\thispagestyle{empty}
\newpage
\section{The gravitational field of a spherically symmetric static
body}
\baselineskip=1.1\normalbaselineskip
{The issue of what takes place in the vicinity of the Schwar\-zschild
sphere,
when the graviton has a rest mass, was first dealt with in
relativistic
theory of gravity in ref.~\cite{17}, in which the following
conclusion was
made: in vacuum the metric coefficient of effective Riemannian space,
$g_{00}$,
on the Schwarzschild sphere differs from zero, while $g_{11}$ has a
pole.
These changes, that in the theory are due to the graviton mass,
result in
the ``rebounding" effect of incident particles and of light from the
singularity on the Schwarzschild sphere, and consequently, in the
absence
of ``black holes".

Further, in ref.~\cite{18} a detailed analysis of this problem in RTG
was
performed, which clarified a number of issues, but which at the same
time
revealed that the ``rebounding" takes place near the Schwarzschild
sphere.
In the present work we follow the article~\cite{19}, in which it was
shown
in a most simple and clear manner that at the point in vacuum, where
the
metric coefficient of effective Riemannian space $g_{11}$ has a pole,
the other
metric coefficient $g_{00}$ does not turn to zero. The re\-sul\-ting
singularity
cannot be removed by a choice of the reference system, so the
solution
inside a body cannot be made to match the external solution. In this
case,
if transition is performed to the reference system related to a
falling
test body, then it turns out to be that the test body will never
reach the
surface of the body, that is the source of the gravitational field.
Precisely this circumstance leads to the conclusion that the radius
of a
body cannot be inferior to the Schwarzschild radius. All this issue
will be
dealt with in detail in this section.}

We now write equations (\ref{76}), (\ref{77}) in the form
\ba
&&R^\mu_\nu -\f{1}{2}\delta^\mu_\nu R+\f{1}{2}
\left (
\f{mc}{\hbar}
\right )^2
\left (
\delta^\mu_\nu +g^{\mu\alpha}
\gamma_{\alpha\nu}-{}\vphantom{\frac12}\right.\nonumber \\
&&\left.-\f{1}{2}\delta^\mu_\nu
g^{\alpha\beta}\gamma_{\alpha\beta}
\right )
=\kappa T^\mu_\nu,  \label{276}
\ea
\be
D_\mu\tilde g^{\mu\nu}=0\;. \label{277}
\ee
Here $\tilde g^{\mu\nu}=\sqrt{-g} g^{\mu\nu},\;
g=\mbox{det}\;  g_{\mu\nu},\;R^\mu_\nu$ is the Ricci tensor,
$\kappa =\f{8\pi G}{c^2},\;\; G$ is the gravitational constant,
$D_\mu$ is the covariant derivative in Minkowski space,
$\gamma_{\mu\nu}(x)$ is the metric tensor of Minkowski space in
arbitrary
curvilinear coordinates.

Let us now determine the gravitational field created by a spherically
symmetric static source. The general form of an interval of effective
Riemannian space for such a source has the form
\be
ds^2=g_{00} dt^2+2g_{01}dt dr+
g_{11}dr^2+g_{22}d\Theta^2+g_{33}d\Phi^2, \label{278}
\ee
We introduce the notation
\ba
&&g_{00}(r)=U(r),\; g_{01}(r)=B(r),\;
\nonumber \\
&&g_{11}(r)=-\left [
V(r)-\f{B^2(r)}{U(r)}
\right ] \;\;\;, \label{279} \\
&&g_{22}(r)=-W^2(r),\;\;\;
g_{33}(r,\Theta)=-W^2(r)\sin^2\Theta.\nonumber
\ea
The components of the contravariant metric tensor are
\ba
&&g^{00}(r)=\frac{1}{U}
\left (
1-\f{B^2}{UV}
\right ),\;\;
g^{01}(r)=-\frac{B}{UV},\;\;
g^{11}(r)=-\frac{1}{V},\;\nonumber
\\*[-0.15cm] \label{280} \\*[-0.3cm]
&&g^{22}(r)=-\f{1}{W^2},\;\;
g^{33}(r,\Theta)=-\f{1}{W^2\sin^2\Theta}.\nonumber
\ea
The determinant of the metric tensor $g_{\mu\nu}$ is
\be
g=\mbox{det} g_{\mu\nu}=-UVW^4\sin^2\Theta\;. \label{281}
\ee
For a solution to have physical meaning the following condition must
be
satisfied:
\be
g<0\;. \label{282}
\ee

In the case of spherical coordinates $g$ can turn to zero only at the
point
$r=0$. On the basis of (\ref{280}) and (\ref{281}) we find the
density
components of the metric tensor
\be
\tilde g^{\mu\nu}=\sqrt{-g}g^{\mu\nu}. \label{283}
\ee
One have the form
\ba
&&\tilde g^{00}=\f{W^2}{\sqrt{UV}}
\left (
V-\f{B^2}{U}
\right )
\sin\Theta,\;
\tilde g^{01}=-\f{BW^2}{\sqrt{UV}} \sin\Theta,\; \nonumber\\
 &&\tilde g^{11}=-\sqrt{\f{U}{V}}W^2
\sin\Theta,\; \label{284} \\
&&\tilde g^{22}=-\sqrt{UV}\sin\Theta, \;\;
\tilde g^{22}=-\sqrt{UV}\sin\Theta,\;
\tilde g^{33}=-\f{\sqrt{UV}}{\sin\Theta}\;.\nonumber
\ea

We shall carry out all reasoning in an inertial reference system in
spherical coordinates. An interval in Minkowski space has the form
\be
d\sigma^2=dt^2-dr^2-r^2
(d\Theta^2+\sin^2\Theta d\Phi^2)\;. \label{285}
\ee
The Christoffel symbols in Minkowski space that differ from zero and
that
are determined by the formula
\be
\gamma^\lambda_{\mu\nu}=\f{1}{2}
\gamma^{\lambda\sigma}
(\pa_\mu\gamma_{\sigma\nu}
+\pa_\nu\gamma_{\sigma\mu}-\pa_\sigma\gamma_{\mu\nu})\; \label{286}
\ee
are equal to
\ba
&&\gamma^1_{22}=-r\;
\gamma^1_{33}=-r\sin^2\Theta,\;
\gamma^2_{12}=\gamma^3_{13}=\f{1}{r},\;\nonumber \\*
[-0.3cm] \label{287} \\[-0.3cm]
&&\gamma^2_{33}=-\sin\Theta\cos\Theta,\;
\gamma^3_{23}=\cot \Theta\;. \nonumber
\ea
We write equation (\ref{277}) in an expanded form,
\be
D_\mu\tilde g^{\mu\nu}=\pa_\mu \tilde g^{\mu\nu}
+\gamma^\nu_{\lambda\sigma}\tilde g^{\lambda\sigma}=0\;. \label{288}
\ee
In Galilean coordinates of Minkowski space they have the form
\be
\pa_\mu\tilde g^{\mu\nu}=0\;. \label{289}
\ee
In the case of a static gravitational field we have from (\ref{289})
\be
\pa_i\tilde g^{i\nu}=0,\;
i=1,2,3\;. \label{290}
\ee

Applying the tensor transformation law, it is possible to express
the components $\tilde g^{i0}$ in Cartesian coordinates in terms
of the components in spherical coordinates
\be
\tilde g^{i0}=-\f{BW^2}{\sqrt{UV}}\cdot
\f{x^i}{r^3}\;,
\sqrt{-g}=\sqrt{UV}W^2 r^{-2}. \label{291}
\ee
Here $x^i$ are spatial Cartesian coordinates. Assuming $\nu=0$ in
(\ref{290})
and integrating over the spherical volume upon application of the
Gauss-Ostrogradsky theorem, we obtain the following integral over the
spherical surface:
\be
\oint \tilde g^{i0}ds_i=
-\f{BW^2}{r^3\sqrt{UV}}\oint (\vec x  d \vec s)=0\;. \label{292}
\ee
Taking into account the equality
\be
\oint(\vec x d\vec s)=4\pi r^3, \label{293}
\ee
we obtain
\be
\f{BW^2}{\sqrt{UV}}=0\;. \label{294}
\ee

Since equation (\ref{289}) holds valid both within matter and outside
it,
(\ref{294}) should be valid for any value of $r$. But since, owing to
(\ref{282}), $U,V$ and $W$ cannot be equal to zero, then from
(\ref{294}) it
follows that
\be
B=0\;. \label{295}
\ee
The interval (\ref{278}) of effective Riemannian space assumes the form
\be
ds^2=Udt^2-Vdr^2-W^2(d\Theta^2+\sin^2\Theta d\Phi^2)\;. \label{296}
\ee

From (\ref{295}) it follows that no static solution exists of the
Hilbert--Einstein equations in harmonic coordinates, that contains in
the
interval a term of the form
\be
B(r)dtdr\;. \label{297}
\ee
The energy-momentum tensor of matter has the form
\be
T^\mu_\nu=\left (
\rho+\f{p}{c^2}\right )
v^\mu v_\nu-\delta^\mu_\nu\cdot \f{p}{c^2}\;. \label{298}
\ee
In expression (\ref{298}) $\rho$ is the mass density of matter, $p$
is the
isotropic pressure, and
\be
v^\mu=\f{dx^\mu}{ds} \label{299}
\ee
is the four-velocity satisfying the condition
\be
g_{\mu\nu}v^\mu v^\nu=1\;. \label{300}
\ee

From equations (\ref{276}) and (\ref{277}) follows
\be
\nabla_\mu T^\mu_\nu=0\;, \label{301}
\ee
where $\nabla_\mu$ is the covariant derivative in effective Riemannian
space
with the metric tensor $g_{\mu\nu}$. In the case of a static body
\be
v^i=0,\;\; i=1,2,3;\;\; v^0=\f{1}{\sqrt{U}}\;, \label{302}
\ee
and therefore
\ba
&&T^0_0=\rho (r),\;\; T^1_1=T^2_2=T^3_3=
-\f{p(r)}{c^2}\;,\;\;\nonumber  \\
[-0.2cm] \label{303} \\ [-0.2cm]
&&T^\mu_\nu=0,\;\; \mu\not =\nu\;. \nonumber
\ea

For the interval (\ref{296}) the Christoffel symbols, differing from
zero, are
\ba
&&\Gamma^0_{01}=\f{1}{2U}\f{dU}{dr}\;,\;
\Gamma^1_{00}=\f{1}{2V}\f{dU}{dr}\;,\;
\Gamma^1_{11}=\f{1}{2V}\f{dV}{dr}\;,\;
\;\nonumber \\
&&\Gamma^1_{22}=-\f{W}{V}\f{dW}{dr},
\Gamma^1_{33}=\sin^2\Theta\cdot \Gamma^1_{22}\;,
\label{304}
\\
&&\Gamma^2_{12}=\Gamma^3_{13}=\f{1}{W}\f{dW}{dr}\;,\;
\Gamma^2_{33}=-\sin\Theta\cos\Theta\;,\;\;\;
\Gamma^3_{23}=\cot\Theta.\nonumber
\ea
Applying the following expression for the Ricci tensor:
\ba
&&R_{\mu\nu}=\pa_\sigma\Gamma^\sigma_{\mu\nu}
-\pa_\nu \Gamma^\sigma_{\mu\sigma} + \nonumber \\
[-0.2cm] \nonumber \\ [-0.2cm]
&&+  \Gamma^\sigma_{\mu\nu}\Gamma^\lambda_{\sigma\lambda}
- \Gamma^\sigma_{\mu\lambda}\Gamma^\lambda_{\sigma\nu}\;,\;
R^\mu_\nu=g^{\mu\lambda}R_{\lambda\nu} \label{305}
\ea
and substituting into it expressions for the Christoffel symbols from
(\ref{304}), it is possible to reduce equations (\ref{276}) for
functions
$U,V$ and $W$ to the form
\ba
&&\f{1}{W^2}-\f{1}{VW^2}
\left (
\f{dW}{dr}\right )^2
-\f{2}{VW}\f{d^2W}{dr^2}
-\f{1}{W}\f{dW}{dr}\f{d}{dr}
\left (
\f{1}{V}\right )+\nonumber\\*[-0.3cm]
&&+\f{1}{2}
\left (
\f{mc}{\hbar}\right )^2
\left [
1+\f{1}{2}
\left (
\f{1}{U}-\f{1}{V}\right )
-\f{r^2}{W^2}\right ]
=\kappa \rho\;,\label{306}
\ea
\ba
&&\!\!\!\!\!\!\!\!\!\!\f{1}{W^2}-\f{1}{VW^2}
\left (
\f{dW}{dr}\right )^2
-\f{1}{UVW}\f{dW}{dr}\;\, \f{dU}{dr}+\nonumber
\\*
&&\!\!\!\!\!\!\!\!\!\!+\f{1}{2}
\left (
\f{mc}{\hbar}\right )^2
\left [
1-\f{1}{2}
\left (
\f{1}{U}-\f{1}{V}\right )
-\f{r^2}{W^2}\right ]
=-\kappa\f{p}{c^2}\;,\label{307}
\ea
\ba
&&-\f{1}{VW} W''-\f{1}{2UV}U''
+\f{1}{2WV^2}W'V'+\f{1}{4VU^2}
(U')^2+\nonumber  \\
&&+\f{1}{4UV^2}U'V'
-\f{1}{2UVW}W'U'+ \nonumber  \\[0.2cm]
&&+\f{1}{2}
\left (
\f{mc}{\hbar}\right )^2
\left [
1-\f{1}{2}
\left (
\f{1}{U}+\f{1}{V}\right )\right ]=-\kappa\f{p}{c^2}\;.\label{308}
\ea
Equation (\ref{288}), with account of (\ref{287}), (\ref{284}) and
(\ref{295}), can be reduced to the form
\be
\f{d}{dr}\left (
\sqrt{\f{U}{V}}W^2\right )=2r\sqrt{UV}\;. \label{309}
\ee

We note that by virtue of the Bianchi identity and of equation
(\ref{277})
one of the equations (\ref{306})~--~(\ref{308}) is a consequence of
the
other
ones. We shall further take equations (\ref{306}), (\ref{307}) and
(\ref{309}) to be independent.

We write equation (\ref{301}) in an expanded form
\be
\nabla_\mu T^\mu_\nu\equiv \pa_\mu T^\mu_\nu
+\Gamma^\mu_{\alpha\mu} T^\alpha_\nu
-\Gamma^\alpha_{\mu\nu}T^\mu_\alpha =0\;. \label{310}
\ee
Making use of expressions (\ref{303}) and (\ref{304}) we obtain
\be
\f{1}{c^2}\cdot\f{dp}{dr}=-\f{\rho+\f{p}{c^2}}{2U}
\cdot\f{dU}{dr}\;. \label{311}
\ee
Taking into account the identity
\ba
&&\f{1}{W^2\left (\f{dW}{dr}\right )}\cdot \f{d}{dr}
\; \left [
\f{W}{V} \left (
\f{dW}{dr}\right )^2\right ]=\f{1}{VW^2}
\left (
\f{dW}{dr}\right )^2+ \nonumber \\
&&+\f{2}{VW}\;
\f{d^2W}{dr^2}
+\f{1}{W}\f{dW}{dr}\f{d}{dr}
\left (\f{1}{V}\right )\;, \label{312}
\ea
equation (\ref{306}) can be written as
\ba
&&1-\f{d}{dW}
\left [
\f{W}{V\left (\f{dr}{dW}\right )^2}\right ]
+\f{1}{2}
\left (
\f{mc}{\hbar}\right )^2
\Biggl [
W^2-r^2+ \Biggr.\nonumber \\
&&\left.+\f{W^2}{2}
\left (
\f{1}{U}-\f{1}{V}\right )\right ]
=\kappa W^2\rho\;. \label{313}
\ea
In a similar manner we transform equation (\ref{307}):
\ba
&&1-\f{W}{V\left (\f{dr}{dW}\right )^2}
\; \f{d}{dW}\ln
(UW)+ \f{1}{2}\left (
\f{mc}{\hbar}\right )^2
\biggl [
W^2-r^2- \Biggr.
\nonumber \\
&&\biggl. -\f{1}{2}
\left (\f{1}{U}-\f{1}{V}\right )\biggr ]
=-\kappa\f{W^2p}{c^2}\;. \label{314}
\ea
We write equations (\ref{309}) and (\ref{311}) in the form
\be
\f{d}{dW}\left (W^2\sqrt{\f{U}{V}}\right )
=2r\sqrt{UV}\f{dr}{dW}\;. \label{315}
\ee
\be
\f{1}{c^2}\cdot \f{dp}{dW}=
-\left (\rho +\f{p}{c^2}\right )\f{1}{2U}
\cdot \f{dU}{dW}\;. \label{316}
\ee

In equations (\ref{313})~--~(\ref{316}) we pass to dimensionless
variables. Let
$l$ be the Schwarzschild radius of the source, the mass of which
equals $M$,
then
\be
l=\f{2GM}{c^2}\;. \label{317}
\ee
We introduce new variables $x$ and $z$, equal to
\be
W=lx,\;\; r=lz. \label{318}
\ee
Equations (\ref{313})~--~(\ref{316}) assume the form
$$
\!\!\!\!\!\!\!\!\!\!\!\! 1-\f{d}{dx}
\left (
\f{x}{V\left (\f{dz}{dx}\right )^2}
\right )
+\epsilon \biggl [
x^2-z^2+
$$
\begin{aequation}{313}{a}
\left.+ \f{1}{2}x^2
\left (
\f{1}{U}-\f{1}{V} \right )\right ]
=\tilde \kappa x^2\rho (x), \label{313a}
\end{aequation}
$$
\!\!1-\f{x}{V\left (\f{dz}{dx}\right )^2} \f{d}{dx}\ln (xU)+\epsilon
\Biggl [
x^2-z^2- \Biggr.
$$
\begin{aequation}{314}{a}
~~\left.-\f{x^2}{2}\left (
\f{1}{U}-\f{1}{V}\right )\right ]
=-\tilde \kappa \f{x^2p(x)}{c^2}\;, \label{314a}
\end{aequation}
\begin{aequation}{315}{a}
\f{d}{dx}
\left (
x^2\sqrt{\f{U}{V}}\right )
=2z\f{dz}{dx} \sqrt{UV}\;,\label{315a}
\end{aequation}
\begin{aequation}{316}{a}
\f{1}{c^2}\f{dp}{dx}=-
\left (
\rho+\f{p}{c^2}\right )\f{1}{2U}\f{dU}{dx}\;.\label{316a}
\end{aequation}
Here $\epsilon$ is a dimensionless constant equal to
\be
\epsilon =\f{1}{2}
\left (
\f{2GMm}{\hbar c}\right )^2,\;\; \tilde \kappa =\kappa l^2.
\label{319}
\ee

The sum of and the difference between equations (\ref{313a}) and
(\ref{314a}) are
\ba
&2&-\f{d}{dx}
\left [
\f{x}{V\left (\f{dz}{dx}\right )^2}
\right ]
- \f{x}{V\left (\f{dz}{dx}\right )^2}
\f{d}{dx}\ln (xU)+\nonumber \\
&+& 2\epsilon (x^2-z^2)
=\tilde \kappa x^2 \left (\rho-\f{p}{c^2}\right )\;, \label{320}
\ea
\ba
&&\f{d}{dx}
\left [
\f{x}{V\left (\f{dz}{dx}\right )^2}
\right ]
-\f{x}{V\left (\f{dz}{dx}\right )^2}
\f{d}{dx}\ln (xU) -\nonumber \\
&&-\;\epsilon x^2
\left (\f{1}{U}-\f{1}{V} \right )
=-\tilde \kappa x^2 \left (\rho +\f{p}{c^2}\right )\;. \label{321}
\ea
We introduce the new functions $A$ and $\eta$:
\be
U=\f{1}{x\eta A},\;\; V=\f{x}{A\left (\f{dz}{dx}\right )^2}\;.
\label{322}
\ee
In these new variables equation (\ref{320}) assumes the form
\be
A\f{d\ln \eta}{dx}+2+2\epsilon (x^2-z^2)
=\tilde\kappa x^2 \left (\rho-\f{p}{c^2}\right )\;. \label{323}
\ee
Equation (\ref{313a}) is written in the form
\be
\f{dA}{dx}=1+\epsilon (x^2-z^2)
+\epsilon \f{x^2}{2}
\left (
\f{1}{U} -\f{1}{V}\right ) -\tilde\kappa\cdot x^2\rho (x)\;.
\label{324}
\ee

In accordance with the causality condition (see~Addendum)
\be
\gamma_{\mu\nu}U^\mu U^\nu =0\;, \label{325}
\ee
\begin{aequation}{325}{a}
g_{\mu\nu}U^\mu U^\nu \leq 0\;,\label{325a}
\end{aequation}
it is easy to establish the inequality
\be
U\leq V\;. \label{326}
\ee
For our problem it suffices to consider only the values of $x$ and
$z$
from the interval
\be
0\leq x \ll \f{1}{\sqrt{2\epsilon}},\;\;
0\leq z \ll \f{1}{\sqrt{2\epsilon}}.\;\; \label{327}
\ee
These inequalities impose an upper limit on $r, W$:
\be
r,W \ll \f{\hbar}{mc}\;. \label{328}
\ee
In the case of such a restriction equation (\ref{324}) assumes the
form
\be
\f{dA}{dx}=1+\epsilon \f{x^2}{2}
\left (
\f{1}{U}-\f{1}{V}\right )-\tk x^2 \rho (x)\;. \label{329}
\ee
Outside matter we have
\be
\f{dA}{dx}=1+\epsilon\f{x^2}{2}
\left (
\f{1}{U}-\f{1}{V}\right )\;. \label{330}
\ee

By virtue of causality (\ref{326}) the following inequality holds
valid
beyond matter
\be
\f{dA}{dx} \geq 1. \label{331}
\ee
Integrating (\ref{329}) over the interval $(0,x)$ we obtain
\be
A(x)=x+\f{\epsilon}{2} \int\limits^{x}_{0}{x'}^2 \left (
\f{1}{U}-\f{1}{V}\right ) dx'-\tk \int\limits^{x}_{0}{x'}^2\rho
(x')dx'. \label{332} \ee $A(0)$ in (\ref{332}) is set equal to
zero, since if it were different from zero, function $V(x)$ would
turn to zero as $x$ tends toward zero, which is inadmissible from
a physical standpoint. On the basis of (\ref{331}) function $A(x)$
beyond matter grows in monotonic way with $x$, and it therefore
can only have the sole root
\be
A(x_1)=0,\;\; x_1>x_0. \label{333}
\ee
On the basis of (\ref{332}) we have
\be
x_1=1-\f{\epsilon}{2}
\int\limits^{x_1}_{0}{x'}^2
\left (
\f{1}{U}-\f{1}{V}\right )
dx'. \label{334}
\ee
We have here taken into account that when $l$ is chosen to be equal
to
(\ref{317})
$$
\tk
\int\limits^{x_0}_{0}{x'}^2
\rho (x')dx'=1.
$$

The matter is concentrated inside the sphere $0\leq x\leq x_0$. We
shall
further consider the case, when the radius of the body, $x_0$, is
less than $x_1$. Precisely in this case in vacuum, i.e. outside the body,
there
will exist a singularity which cannot be removed by a choice of
reference system.

Owing to the graviton mass the zero of function $A$ is shifted inward
the
Schwarzschild sphere. Since as $x$ tends toward $x_1,\;\; V(x)$ tends
toward infinity, owing to $A(x)$ tending toward zero, there will
exist
such a vicinity about the point $x_1$
\be
x_1(1-\lambda_1)\leq x\leq x_1
(1+\lambda_2),\;\; \lambda_1>0, \lambda_2>0\;, \label{335}
\ee
($\lambda_1$ and $\lambda_2$ assume small fixed values), inside which
the
following inequality holds valid:
\be
\f{1}{U}\gg \f{1}{V}\;. \label{336}
\ee
In this approximation we obtain
\be
A(x)=x-x_1+\f{\epsilon}{2}
\int\limits^{x}_{x_1}dx'{x'}^2\f{1}{U}\;. \label{337}
\ee
Substituting into this expression $U$ in the form (\ref{322}) we find
\be
A(x)=x-x_1+\f{\epsilon}{2}
\int\limits^{x}_{x_1}dx'{x'}^3\eta (x')A(x'). \label{338}
\ee

In the region of variation of $x$ one can substitute $x^3_1$ for
$x^3$
within the interval (\ref{335}):
\be
A(x)=x-x_1+\f{\epsilon}{2}x^3_1
\int\limits^{x}_{x_1}\eta (x')A(x')dx'. \label{339}
\ee
Hence, we obtain
\be
\f{dA}{dx}=1+\f{\epsilon}{2}x^3_1\eta (x)A(x). \label{340}
\ee
In the considered approximation (\ref{327}) equation (\ref{323})
assumes
the form
\be
A\f{d\ln\eta}{dx}+2=0\;. \label{341}
\ee
We now introduce a new function
\be
f(x)=\f{x^3_1}{2}\eta (x)A(x)\;. \label{342}
\ee

Equation (\ref{340}) assumes the form
\be
\f{dA}{dx}=1+\epsilon f(x)\;, \label{343}
\ee
and equation (\ref{341}) assumes the form
\be
\f{A}{f}\cdot \f{df}{dx}-\f{dA}{dx}=-2\;. \label{344}
\ee
From equations (\ref{343}) and (\ref{344}) we find
\be
A(x)=-\f{(1-\epsilon f)f}{\left (\f{df}{dx}\right )}\;. \label{345}
\ee
From expression (\ref{342}) we obtain
\be
\eta (x)=-\f{2\f{df}{dx}}{x^3_1(1-\epsilon f)}\;. \label{346}
\ee
Substituting (\ref{345}) and (\ref{346}) into (\ref{322}) we find
\be
U=\f{x^3_1}{2xf},\;\;
V=-\f{x\f{df}{dx}}{f(1-\epsilon f)\left (\f{dz}{dx}\right )^2}\;.
\label{347}
\ee
Making use of these expressions we can rewrite the determinant $g$ as
\be
g=
\f{x^3_1\f{df}{dx}x^4}
{2f^2\left (\f{dz}{dx}\right )^2(1-\epsilon f)}\sin^2\Theta<0\;.
\label{348}
\ee

For condition (\ref{282}) to be satisfied, it is necessary for
expressions
$\f{df}{dx}$ and $(1-\epsilon f)$ to have opposite signs. Substituting
(\ref{345}) into (\ref{343}) we obtain
\be
\f{d}{dx}\ln
\left |
\f{df}{dx}\right |
-\f{d}{dx}\ln
\left |f(1-\epsilon f)\right |=
\f{1+\epsilon f}{f(1-\epsilon f)}\cdot \f{df}{dx}\;. \label{349}
\ee
Hence, we find
\be
\f{d}{dx}\ln
\left |
\f{(1-\ep f)\f{df}{dx}}{f^2}\right |=0\;. \label{350}
\ee
Thus,
\be
\left |
\f{(1-\ep f)\f{df}{dx}}{f^2}\right |=C_0>0\;. \label{351}
\ee
Taking into account that the quantities $(1-\ep f)$ and $\f{df}{dx}$
must
have opposite signs we find
\be
\f{df}{dx}=-\f{C_0 f^2}{(1-\ep f)}\;. \label{352}
\ee
Substituting this expression into (\ref{345}) we find
\be
A(x)=\f{(1-\ep f)^2}{C_0 f},\;\;
A(x_1)=0\;\;\;\mbox{at}\;\;\; f=\f{1}{\ep}\;.  \label{353}
\ee
With account of (\ref{353}), expression (\ref{322}) for the function
$V$
assumes the form
\be
V=\f{C_0 xf}{(1-\ep f)^2\left (\f{dz}{dx}\right )^2}\;. \label{354}
\ee
Integrating (\ref{352}) and taking into account (\ref{353}) we obtain
\be
C_0\cdot (x-x_1)=\f{1}{f}+\ep\ln\ep |f|-\ep\;. \label{355}
\ee

Relation (\ref{355}) has been obtained for the domain of $x$ values
determined by equalities (\ref{335}), however, it is also valid in
the
region where the graviton mass can be neglected.

In accordance with (\ref{335}) the domain $C_0(x-x_1)$ is within the
limits
\be
-C_0 x_1\lambda_1\leq C_0 (x-x_1)\leq C_0 x_1\lambda_2,  \label{356}
\ee
when $f$ is positive, it satisfies the inequalities
\be
\tilde C\leq f\leq \f{1}{\ep}\;. \label{357}
\ee
Making use of (\ref{355}), in accordance with (\ref{356}), we have
$$
\f{1}{f}+\ep\ln\ep f-\ep\leq C_0x_1\lambda_2.
$$
Hence, we can find $\tilde C$:
\be
\f{1}{\tilde C}+\ep\ln\ep \tilde C-\ep=C_0x_1\lambda_2. \label{358}
\ee
From expression (\ref{358}) we find the approximate value for $\tilde
C$:
\be
\tilde C=\f{1}{C_0x_1\lambda_2}\;. \label{359}
\ee

For negative values of $f$, the value $|f|$, determined from the
following
equation, corresponds to the point $x=x_1$:
\be
-\f{1}{|f|}+\ep\ln\ep|f|-\ep=0\;. \label{360}
\ee
Hence, we find
\be
|f|=\f{a}{\ep},\;\;\ln a=\f{1+a}{a}\;. \label{361}
\ee
In accordance with (\ref{356}), the following inequality should be
satisfied:
\be
-C_0x_1\lambda_1\leq -\f{1}{|f|}
+\ep\ln\ep |f|-\ep\;. \label{362}
\ee
Hence it is possible to find the lower boundary for $|f|=D$:
\be
-C_0x_1\lambda_1=-\f{1}{D}+\ep\ln\ep D-\ep\;. \label{363}
\ee
From expression (\ref{363}) we find the approximate value for $D$:
\be
D=\f{1}{C_0x_1\lambda_1}\;. \label{364}
\ee
This means that the quantity $|f|$ satisfies the inequality
\begin{aequation}{364}{a}
|f|\geq D=\f{1}{C_0x_1\lambda_1}\;. \label{364a}
\end{aequation}

Let us now establish the dependence of variable $z$ upon $x$.
Substituting
(\ref{322}) into (\ref{315a}) and taking into account (\ref{323}) we
obtain
\ba
&&A\f{d}{dx}
\left (x\f{dz}{dx}\right )
=2z-x\f{dz}{dx}
\left [1+\ep (x^2-z^2)-\right.\nonumber \\
&&\left.-\f{1}{2} \tk x^2 \left (\rho-\f{p}{c^2}\right )\right ]\;.
\label{365}
\ea
In the approximation (\ref{327}), outside matter, equation
(\ref{365})
assumes the form
\be
A\f{d}{dx}
\left (
x\f{dz}{dx}\right )+x\f{dz}{dx}-2z=0\;. \label{366}
\ee

We have to find the regular solution $z(x)$ of equation (\ref{366}).
In
equation (\ref{366}) we pass from variable $x$ to $f$. Applying
relation (\ref{355})
\be
x=\f{1}{C_0f}
[C_0x_1 f+1-\ep f+\ep f\ln \ep |f|]\;, \label{367}
\ee
and taking into account (\ref{340}), (\ref{341}) and (\ref{358}),
equation
(\ref{366}) can be represented in the form
\be
\f{d^2z}{df^2}+\f{C_0xf+\ep f-1}{C_0f^2 x}
\cdot \f{dz}{df}-\f{2z}{C_0 f^3 x}=0\;. \label{368}
\ee

By direct substitution one can establish that the expression
\be
z=\f{x_1}{2}+\f{1}{C_0 f}
[1-\ep f+\ep f\ln\ep |f|] \label{369}
\ee
satisfies equation (\ref{368}) with an accuracy up to the quantity
\be
\ep\f{(1-\ep f+\ln\ep |f|)}{C_0^2 xf^3}\;, \label{370}
\ee
that is extremely small in the vicinity of the point $x_1$. From
expressions
(\ref{367}) and (\ref{369}) we find
\be
z=x-\f{x_1}{2}\;. \label{371}
\ee
Taking this relation, as well as (\ref{354}) and (\ref{347}), into
account
we obtain
\be
U=\f{x_1^3}{2xf},\;\;
V=\f{C_0xf}{(1-\ep f)^2}. \label{372}
\ee
For negative values of $f$ the causality condition (\ref{326})
assumes the
form
\be
|f|^2(2x^2C_0-\ep^2x_1^3)-2\ep x^3_1 |f|-x^3_1\leq 0. \label{373}
\ee

Inequality (\ref{373}) is not satisfied, since it does not comply
with
inequality (\ref{364a}). Thus, the causality principle is violated in
the
domain of negative values of $f$. This means that in the region
$x_1(1-\lambda_1)\leq x< x_1$ the solution has no physical sense. If
$x_0<x_1 (1-\lambda_1)$, the situation arises, when the physical
solution
inside the body, $0\leq x\leq x_0$, cannot be made to match the
physical
solution in the region $x>x_1$, since there exists an intermediate
region
$x_1(1-\lambda_1)\leq x < x_1$, within which the solution does not
satisfy
the causality principle. Hence it necessarily follows that $x_0\geq
x_1$.
From a physical point of view, it is necessary to exclude the
equality
$x_0=x_1$, since the solution inside the body should undergo smooth
transition to the external solution. Consequently, the variable $f$
only
assumes positive values. For values from the region $x\geq
x_1(1+\lambda_2)$
one may, in equations (\ref{313a}) and (\ref{314a}), drop the terms
containing the small parameter $\ep$. Thus, we arrive at
Schwarzschild's
external solution
\be
z_s=(x-\omega)
\left [
1+\f{b}{2\omega}
\ln\f{x-2\omega}{x}\right ]\;, \label{374}
\ee
\be
V_s=\f{x}{\left (\f{dz}{dx}\right)^2(x-2\omega)}\;,
U_s=\f{x-2\omega}{x}\;. \label{375}
\ee
Here, {\tt "}$\omega${\tt "} and {\tt "}$b${\tt "} are certain
constants
that are determined from the condition that solution (\ref{371}),
(\ref{372}) is made to match solution (\ref{374}), (\ref{375}). At
point
$x=x_1(1+\lambda_2)$ the function $z$ from (\ref{371}) is
\be
z=x_1\left (\f{1}{2}+\lambda_2\right )\;, \label{376}
\ee
At the same point $z_s$ equals
\be
z_s=[x_1(1+\lambda_2)-\omega]
\left [
1+\f{b}{2\omega}\ln
\f{x_1(1+\lambda_2)-2\omega}{x_1(1+\lambda_2)}\right ]. \label{377}
\ee
From the condition that (\ref{376}) and (\ref{377}) match we find
\be
\omega=\f{x_1}{2},\;\; b=0\;. \label{378}
\ee

At point $x=x_1(1+\lambda_2)$ the function $U$ from (\ref{372})
equals
\be
U=\f{x^3_1}{2x_1(1+\lambda_2)\tilde C}\;, \label{379}
\ee
since $\tilde C$, in accordance with (\ref{359}), is
\be
\tilde C=\f{1}{C_0x_1\lambda_2}\;. \label{380}
\ee
Substituting (\ref{380}) into (\ref{379}) we obtain
\be
U=\f{C_0x^3_1\lambda_2}{2(1+\lambda_2)}\;, \label{381}
\ee
At the same point, with account of (\ref{378}), $U_s$ is
\be
U_s=\f{\lambda_2}{1+\lambda_2}\;. \label{382}
\ee

From the condition that (\ref{381}) and (\ref{382}) match we find
\be
C_0=\f{2}{x^3_1}\;. \label{383}
\ee

At point $x=x_1(1+\lambda_2)$ the function $V$ from (\ref{372})
equals
\be
V=C_0x_1(1+\lambda_1)\tilde C\;. \label{384}
\ee
Substituting into (\ref{384}) the value $\tilde C$ from (\ref{380})
we
obtain
\be
V=\f{1+\lambda_2}{\lambda_2}\; \label{385}
\ee
at the same point, with account of (\ref{374}) and (\ref{378}),
$V_s$ equals
\be
V_s=\f{1+\lambda_2}{\lambda_2}\;, \label{386}
\ee
i.e. the solution for $V$ matches the solution for $V_s$.

Thus, if the radius of a body exceeds the Schwarzschild radius, then
the
graviton mass can be neglected, and the interval of effective Riemannian
space
in an inertial reference system in spherical coordinates outside the
body
in the region (\ref{328}) has the form:
\ba
&&ds^2=\f{r-GM}{r+GM}dt^2-\f{r+GM}{r-GM}dr^2-
\nonumber \\[2mm]
&&-(r+GM)^2[(d\Theta)^2+\sin^2\Theta (d\varphi)^2].
\nonumber
\ea

This expression is determined unambiguously from the complete set of
equations (\ref{276}) and (\ref{277}), and, here, there exists no
arbitrariness. When the solution inside the body is made to match the
solution outside the body it is also necessary, as first shown by
R.~Avakian, to take into account the logarithmic term (\ref{374})
which
arises when the solution of equations (\ref{277}) is sought. However,
since
the radius of the Sun exceeds the Schwarzschild radius significantly,
we can do not  take it into account in calculations of gravitational
effects
in the Solar system.

Now consider (\ref{367}) for values of $\ep f$ close to unity:
\be
f=\f{1}{\ep \left(1+\f{y}{\ep}\right) },\;\;
\f{y}{\ep}\ll 1. \label{387}
\ee
Substituting this expression into (\ref{367}) and expanding in
$\f{y}{\ep}$
we obtain
\be
y^2=2\ep C_0 (x-x_1). \label{388}
\ee
Inequality (\ref{387}) signifies, that the quantity
$(x-x_1)=\delta\!\ll\!\ep$,
i.e.
\be
\f{y}{\ep}=\sqrt{2C_0}\cdot \sqrt{\f{x-x_1}{\ep}}\ll 1\;. \label{389}
\ee
Substituting (\ref{388}) into (\ref{387}), and then $f$  into
(\ref{372}),
we obtain for $U$ and $V$ the following expressions:
\ba
U=\f{x^3_1[\ep +\sqrt{2\ep C_0(x-x_1)}]}{2x}\;,\;\; \nonumber \\[2mm]
\label{390} \\[-0.2cm]
V=\f{x [\ep +\sqrt{2\ep C_0(x-x_1)}]}{2\ep(x-x_1)}\;.
\nonumber
\ea
Hence, within the domain of variable $x$ satisfying inequality
(\ref{389}),
we have
\be
U=\f{\ep x^3_1}{2x},\;\;
V=\f{x}{2(x-x_1)}\;. \label{391}
\ee

We see that the presence of the graviton mass essentially alters the
character of the solution in the region close to the gravitational
radius.
At the point, where the function $V$, in accordance with (\ref{391}),
has
a pole, the function $U$ differs from zero, while in general
relativity
theory it equals zero. It is precisely owing to this circumstance,
that
an irreversible gravitational collapse arises in GRT, during which
there
appear ``black holes" (objects that have no material boundaries and
that
are ``cut off" from the external world). In RTG ``black holes" are
impossible.

If one takes into account (\ref{317}), (\ref{318}), (\ref{371}) and
neglects
the second term in (\ref{334}), then expressions
(\ref{391}) for $U$ and $V$ assume the form:
\be
U=\left (
\f{GMm}{\hbar c}\right )^2,\;\;
V=\f{1}{2}\cdot \f{r+\f{GM}{c^2}}{r-\f{GM}{c^2}}\;, \label{392}
\ee
which coincides with the formulae of~[2]. We note that
the
residue at the pole of function $V$ at $\ep\not=0$ equals
$\f{GM}{c^2}$,
while at $\ep=0$ it equals $\f{2GM}{c^2}$. This is so, because, when
$\ep=0$, the pole of function $V$ at point $x=x_1$ is due to function
$f$,
which at this point has a pole, while, if $\ep\not=0$, it is due to
function $(1-\ep f)$, which, in accordance with (\ref{367}), turns to
zero
at the point $x=x_1$.

We shall now compare the character of motion of test bodies in
effective
Riemannian space with the metric (\ref{392}) and with the Schwarzschild
metric.
We write the interval (\ref{296}) of Riemannian space in the form
\be
ds^2=Udt^2-\tilde VdW^2-W^2(d\Theta^2
+\sin^2\Theta d\Phi^2)\;. \label{393}
\ee
Here $\tilde V$ is
\be
\tilde V (W)=V\left (\f{dr}{dW}\right )^2. \label{394}
\ee
The motion of a test body proceeds along a geodesic line of Riemannian
space
\be
\f{dv^\mu}{ds}+\Gamma^\mu_{\alpha\beta}v^\alpha v^\beta=0\;,
\label{395}
\ee
where
\be
v^\mu=\f{dx^\mu}{ds}\;, \label{396}
\ee
the velocity four-vector $v^\mu$ satisfies the condition
\be
g_{\mu\nu}v^\mu v^\nu=1\;. \label{397}
\ee

Now consider radial motion, when
\be
v^\Theta =v^\Phi=0\;. \label{398}
\ee
Taking into account (\ref{304}), from equation (\ref{395}) we find
\be
\f{dv^0}{ds}+\f{1}{U}\cdot \f{dU}{dW} v^0v^1=0\;, \label{399}
\ee
where
\be
v^1=\f{dW}{ds}\;. \label{400}
\ee
From equation (\ref{399}) we find
\be
\f{d}{dW}\ln (v^0U)=0\;. \label{401}
\ee
Hence, we have
\be
v^0=\f{dx^0}{ds}=\f{U_0}{U}\;, \label{402}
\ee
where $U_0$ is the integration constant.

Taking into account (\ref{402}), condition (\ref{397}) for radial
motion
assumes the form
\be
\f{U^2_0}{U}-1=\tilde V\cdot \left (
\f{dW}{ds}\right )^2. \label{403}
\ee
If we assume the velocity of a falling test body to be zero at
infinity,
then we obtain $U_0=1$. From (\ref{403}) we find
\be
\f{dW}{ds}=-\sqrt{\f{1-U}{U\tilde V}}\;. \label{404}
\ee
Taking into account (\ref{354}), (\ref{371}), (\ref{372}) and
(\ref{383})
we have
$$
U=\f{x^3_1}{2xf},\;\;
\tilde V=\f{2xf}{x^3_1(1-\epsilon f)^2}\;.
$$
Substituting these expressions into (\ref{404}) we obtain
\be
\f{dW}{ds}=-\sqrt{1-U}(1-\ep f)\;. \label{405}
\ee
Applying (\ref{383}), (\ref{387}) and (\ref{388}), in the vicinity of
the
point $x_1$ we have
\be
\f{dW}{ds}=-\f{2}{x_1}\sqrt{\f{x-x_1}{\ep x_1}}. \label{406}
\ee
Passing from the variable $x$ to $W$, in accordance with (\ref{318})
and
taking into account (\ref{319}), we obtain
\be
\f{dW}{ds}=-\f{\hbar c^2}{mGM}
\sqrt{\f{W}{GM}\left (1-\f{2GM}{c^2W}\right )}\;. \label{407}
\ee

Hence there evidently arises a turning point. Dif\-fe\-ren\-ti\-a\-ting
(\ref{407})
with respect to $s$ we find
\be
\f{d^2W}{ds^2}=\f{1}{2GM}
\left (
\f{\hbar c^2}{mGM}\right )^2. \label{408}
\ee
At the turning point, the acceleration (\ref{408}) is very large, and
it is
positive, i.e. there occurs repulsion. Integrating (\ref{407}), we
obtain
\be
W=\f{2GM}{c^2}+
\left (
\f{\hbar c^2}{2mGM}\right )^2\cdot
\f{1}{GM}(s-s_0)^2. \label{409}
\ee

Formulae (\ref{407})~--~(\ref{409}) coincide with the formulae of
ref.~\cite{17}. The presence of the Planck constant in formula
(\ref{407})
is due to the wave nature of matter, in our case, of gravitons
exhibiting
rest mass. From formula (\ref{409}) it is evident that a test body
cannot
cross the Schwarzschild sphere. In GRT the situation is totally
different.
From the Schwarzschild solution and expression (\ref{404}) it follows
that a test body will cross the
Schwarzschild sphere, and that a ``black hole" will form. Test bodies
or
light can only cross the Schwarzschild sphere inwards, and then they
can
never leave the Schwarzschild sphere any more. The same result is
obtained,
if we pass to a synchronous set of freely falling test bodies with
the aid
of the transformations
\be
\tau = t+\int dW
\left [
\f{\tilde V (1-U)}{U}\right ]^{1/2}. \label{410}
\ee
\be
R = t+\int dW
\left [
\f{\tilde V}{U(1-U)}\right ]^{1/2}. \label{411}
\ee
In this case the interval (\ref{393}) assumes the form
\be
ds^2=d\tau^2-(1-U)dR^2-W^2
(d\Theta^2+\sin^2\Theta d\Phi^2)\;. \label{412}
\ee

In this form, the singularities of the metric coefficients disappear
both
for the Schwarzschild solution, when $\ep=0$, and for the solution in
our
case, when $\ep\not=0$. However, while in GRT the variable $W$ may
turn to
zero, in RTG, by virtue of expression (11.134), it is always larger,
than
the Schwarzschild radius.

Subtracting from expression (\ref{411}) expression (\ref{410}) we
obtain
\be
R-\tau =\int
dW\sqrt{\f{U\tilde V}{(1-U)}}\;. \label{413}
\ee
Differentiating equality (\ref{413}) with respect to $\tau$, we find
\be
\f{dW}{d\tau}=-\sqrt{\f{(1-U)}{U\tilde V}}\;. \label{414}
\ee
Thus, we arrive at the same initial equation (\ref{404}). Taking into
account that $r=W-\f{GM}{c^2}$, on the basis of expressions (11.117),
we
obtain from equation (11.139) the following:
$$
W=\f{2GM}{c^2}+\f{1}{4}
\left (
\f{\hbar c^2}{GMm}
\right )^2
\cdot
\f{\left ( R-c\tau\right )^2}{GM} \eqno{(11.134a)}
$$
Hence, it is also evident that, if $\epsilon\not=0$, then a falling
test
body can never cross the Schwarzschild sphere. In that case, when
$\ep=0$,
the Schwarzschild singularity in the metric does not influence the
motion of
the test body in a falling synchronous reference system. In GRT the
following expression will occur, instead of formula (11.134a):
$$
W=
\left [
\f{3}{2}
(R-c\tau)
\right ]^{2/3}
\left (
\f{2GM}{c^2}
\right )^{1/3},
$$
which testifies that a test body will reach the point $W=0$ in a
finite
interval of proper time. The falling particles, here, will only cross
the
Schwarzschild sphere in one direction, inward. We shall now calculate
the propagation time of a light signal from a certain point $W_0$ to
the
point $W_1=\f{2GM}{c^2}$, given by the clock of a distant observer.
From
the expression $ds^2=0$ we have the following for the Schwarzschild
solution:
\be
\f{dW}{dt}=-c
\left (
1-\f{2GM}{c^2W}\right )\;. \label{415}
\ee
Integrating this equation we obtain
\be
W_0-W+\f{2GM}{c^2}\ln
\f{W_0-\f{2GM}{c^2}}{W-\f{2GM}{c^2}}
=c(t-t_0)\;. \label{416}
\ee
Hence, it is obvious that an infinite time, by the clock of a distant
observer, is required in GRT in order to reach the gravitational
radius
$W_1=\f{2GM}{c^2}$. In RTG, as we established earlier, the
Schwarzschild
solution is valid up to the point $W=W_1(1+\lambda_2)$, so the time
required to reach this point is
\be
c(t-t_0)=W_0-W_1(1+\lambda_2)+
\f{2GM}{c^2}\ln
\f{W_0-\f{2GM}{c^2}}{\lambda_2\f{2GM}{c^2}}\;. \label{417}
\ee

The propagation time of a light ray from the point
$W=W_1(1+\lambda_2)$ to
the point $W_1$ can be calculated making use of formulae (\ref{372})
and
(\ref{383}). Within this interval we have
\be
\f{dW}{dt}=-c\f{x^3_1}{2xf}(1-\ep f). \label{418}
\ee
Hence, upon integration and a change of variable, we obtain
\be
\f{2MG}{c^2}\int\limits^{1/\epsilon}_{f}\f{xdf}{f}=c(t_1-t)\;.
\label{419}
\ee
In accordance with (\ref{359}) and (\ref{383}) the lower integration
limit
is
\be
f=\tilde C=\f{x^2_1}{2\lambda_2}\;. \label{420}
\ee
The integral (\ref{419}) is readily calculated and with a good
accuracy
leads to the following relation:
\be
c (t_1-t)=W_1\lambda_2+\f{2GM}{c^2}\ln \f{2\lambda_2}{\epsilon}\;.
\label{421}
\ee

On the basis of (\ref{417}) and (\ref{421}), the time required for a
light
signal to cover the distance between the points $W_0$ and
$W_1=\f{2GM}{c^2}$, is equal to the sum of expressions (\ref{417})
and
(\ref{421}),
\be
c(t_1-t_0)=W_0-W_1
+\f{2GM}{c^2}\ln
\f{W_0-\f{2GM}{c^2}}{\epsilon \f{GM}{c^2}}\;. \label{422}
\ee
Hence it is seen, that in RTG, unlike GRT, the propagation time of a
light
signal to the Schwarzschild sphere is finite, even if measured by the
clock
of a distant observer. From formula (\ref{422}) it is evident that
the
propagation time is not enhanced significantly by the influence of
the
gravitational field.

On the basis of the above presentation it is clear that, if the
graviton
mass exists, $\ep\not= 0$, then the solution in RTG differs
essentially from
the Schwarzschild solution owing to the presence on the Schwarzschild
sphere
of a singularity, that cannot be removed by a choice of reference
system.
Thus, in the case we have considered, when the radius of a body is
smaller
than the Schwarzschild radius, or to be more precise, when $x_0<x_1$,
a test
particle can never reach the surface of the body, by virtue of
(\ref{409}).
Owing to the presence of a singularity, the physical condition $g<0$
is
violated outside the body, and precisely for this reason a physical
solution
for a static spherically symmetric body is possible only in the case,
when
the point $x_1$ is inside the body. This conclusion is conserved,
also, for
a synchronous reference system, when the metric coefficients (see
(11.134a)) are functions of time.

Thus, in accordance with RTG, no Schwarzschild sin\-gu\-la\-ri\-ty for a body
of
arbitrary mass exists, since the radius of the body is greater than
the
Schwarzschild radius, and so the formation of ``black holes" (objects
without
material boundaries and ``cut off" from the external world) is
impossible.
This conclusion complies with the conclusion made by A.~Einstein in
1939,
most likely based on his physical intuition, than on GRT logic. He
wrote:
``{\it Schwarzschild's singularity does not exist, since matter cannot
be
concentrated in an arbitrary manner; otherwise clustering particles
would
achieve the velocity of light}"$\,$\footnote{Einstein A. Collection
of
scientific works, M.: Nauka, 1966, vol.2, art.119, p.531.}.
A.~Einstein, naturally, saw that the
existence of the Schwarzschild singularity violated his main
principle:
``{\it to acknowledge all conceivable} (we shall not, here, deal with
certain
restrictions, following from the requirement of uniqueness and
continuity)
{\it reference systems to be essentially equivalent for describing
nature}"$\,$\footnote{Einstein A. Collection of scientific works,
M.:~Nauka, 1965, vol.1, art.38, p.459.}.
Precisely for this reason, he considered, from a physical point of
view,
that no Schwarzschild singularity in the metric coefficients should
exist in
a reference system related to a distant observer, also. All this,
however,
is realized in RTG, but not in GRT.

In accordance with RTG, as a field theory of gravity, a body of
arbitrary
mass cannot undergo compression indefinitely, and therefore no
gravitational collapse involving the formation of a ``black hole" is
possible.
{\bf This means that a collapsing star cannot go beneath its
gravitational
radius.} Spherically symmetric accretion of matter onto such a body,
at
its final stage of evolution (when nuclear resources are exhausted),
will be
accompanied by a great release of energy, owing to matter falling
onto
the surface of the body. According to RTG, gravitational absorption
of light
is impossible. In GRT, when spherically symmetric accretion of matter
onto
a ``black hole" takes place, the energy release is quite low, since
the
falling matter brings energy into the ``black hole". Gravitational
absorption
of light takes place. Gravitational self-closure of the object
occurs.
Observational data on such objects could provide the answer, as to
what
happens with stars of large mass at their final stage of evolution,
when all
nuclear resources are exhausted.

\thispagestyle{empty}
\newpage
\section*{Addendum }\addcontentsline{toc}{section} {Addendum }
In spherical coordinates of Minkowski space the intervals of
Minkowski space
and of effective Riemannian space have the form
$$
d\sigma^2=dt^2-dr^2-r^2
(d\Theta^2+\sin^2\Theta d\Phi^2)\;, \eqno{(1)}
$$
$$
ds^2=U(r)dt^2-V(r)dr^2-W^2(r)
(d\Theta^2+\sin^2\Theta d\Phi^2)\;. \eqno{(2)}
$$
We now introduce the velocity vector
$$
v^i=\f{dx^i}{dt},\;\;
v^i=ve^i,\;\; (x^i=r,\Theta,\Phi)\;. \eqno{(3)}
$$
$e^i$ represents the unit vector with respect to the metric of the
spatial
part of Minkowski space
$$
\kappa_{ik}e^ie^k=1\;. \eqno{(4)}
$$
In the general case $\kappa_{ik}$ is
$$
\kappa_{ik}=-\gamma_{ik}+\f{\gamma_{0i}\gamma_{0k}}{\gamma_{00}}\;.
\eqno{(5)}
$$
In case (1)
$$
\kappa_{ik}=-\gamma_{ik}. \eqno{(6)}
$$
Condition (4) for metric (1) has the form
$$
(e^1)^2+r^2[(e^2)^2+\sin^2\Theta\cdot
(e^3)^2]=1\;. \eqno{(7)}
$$

We define the velocity four-vector by the equality
$$
v^\mu=(1, ve^i)\; \eqno{(8)}
$$
and require that it be isotropic in Minkowski space
$$
\gamma_{\mu\nu}v^\mu v^\nu=0\;. \eqno{(9)}
$$
Substituting (8) into (9) and taking into account (7) we find
$$
v=1\;. \eqno{(10)}
$$
Thus, the isotropic four-vector $v^\mu$ is equal to
$$
v^\mu=(1, e^i)\;. \eqno{(11)}
$$

Since, in accordance with special relativity theory, motion always
proceeds inside or on the boundary of the Minkowski causality cone,
then,
in the case of a gravitational field, the causality principle
$$
g_{\mu\nu}v^\mu v^\nu\leq 0\; \eqno{(12)}
$$
will be valid, i.e.
$$
U-V(e^1)^2-W^2[(e^2)^2+(e^3)^2\sin^2\Theta]\leq 0\;. \eqno{(13)}
$$
Taking into account (7), expression (13) may be written as
$$
U-\f{W^2}{r^2}-
\left (
V-\f{W^2}{r^2}\right )
(e^1)^2\leq 0\;. \eqno{(14)}
$$
Let
$$
V-\f{W^2}{r^2}\geq 0\;. \eqno{(15)}
$$
Owing to arbitrariness, $0\leq (e^1)^2\leq 1$, inequality (14) will
be
satisfied only if
$$
U-\f{W^2}{r^2}\leq 0\;. \eqno{(16)}
$$
From inequalities (15) and (16) follows
$$
U\leq V\;. \eqno{(17)}
$$
In the case, when
$$
V-\f{W^2}{r^2} < 0\;, \eqno{(18)}
$$
we write inequality (14) as
$$
U-V-\left (
\f{W^2}{r^2}-V\right )
(1-(e^1)^2)\leq 0\;. \eqno{(19)}
$$
Owing to the arbitrariness of $e^1$, expression (19) will hold valid
for
any values $0\leq (e^1)^2\leq 1$, only if
$$
U\leq V\;.\eqno{(20)}
$$

Thus, the causality principle in RTG always results in the inequality
$$
U(r)\leq V(r)\;. \eqno{(21)}
$$

\thispagestyle{empty}
\newpage
\section{Gravitational effects in the Solar system}
Before proceeding to examine such effects, we shall first dwell upon
certain general assertions of RTG and GRT, which are explicitly
manifested
in calculations of gravitational effects. The RTG equations
(\ref{76}) and
(\ref{77}) are universally covariant under arbitrary transformations
of
coordinates and form-invariant under the Lorentz transformations. In
other
words, the situation in RTG is the same as in electrodynamics. If in
two
inertial reference systems the respective distributions of matter in
Galilean coordinates, $T_{\mu\nu}[x, g_{\alpha\beta}(x)]$ and
$T_{\mu\nu}[x', g_{\alpha\beta}(x')]$, are identical, then by virtue
of
the form-invariance of the equations relative to the Lorentz
transformations
we obtain identical equations, which in identical conditions of the
problem
provide for the relativity principle to be satisfied. On the other
hand, if
in a certain inertial reference system and for a given distribution
of matter
$T_{\mu\nu}(x)$ the solution we have is $g_{\mu\nu}(x)$, then
applying the
Lorentz transformations to another inertial reference system we
obtain
the metric $g'_{\mu\nu}(x')$, but it corresponds to the distribution
of
matter $T'_{\mu\nu}(x')$. Owing to the equations  being form-invariant
under
the Lorentz transformations we can return to the initial variables
$x$
and obtain a new solution $g'_{\mu\nu}(x)$ corresponding to the
distribution
of matter $T'_{\mu\nu}(x)$. This means that a unique correspondence
exists
between the distribution of matter and the metric. When the
distribution of
matter changes, the metric changes also. An essential point in RTG is
the
presence of the metric of Minkowski space in the equations. Precisely
this
circumstance permits performing a comparison of the motion of matter
in a
gravitational field with the motion of matter in the absence of any
gravitational field.
{\sloppy

}
In GRT the situation is quite different. The GRT equations outside
matter
are form-invariant relative to arbitrary transformations of
coordinates, and
therefore, if for the distribution of matter $T_{\mu\nu}(x)$ the
solution we
have is $g_{\mu\nu}(x)$, then by transforming coordinates, so that
in the region of matter they coincide with the initial ones, and
outside
matter differ from them, our solution in the new coordinates will
have the
form $g'_{\mu\nu}(x')$. Owing to the equations being form-invariant
outside
matter, we can go back to the initial variables $x$, and,
consequently,
obtain a new solution $g'_{\mu\nu}(x)$ for the same distribution of
matter,
$T_{\mu\nu}(x)$. To these two metrics (any amount of metrics can be
constructed) there correspond differing intervals:
$$
ds^2_1=g_{\mu\nu}(x) dx^\mu dx^\nu
$$
$$
ds^2_2=g'_{\mu\nu}(x) dx^\mu dx^\nu.
$$

Which interval must be chosen? The point is that the geodesic lines
of these
intervals differ from each other. In this connection, attempts are
made to identify the gravitational field in GRT with the class of
equivalent
diffeomorphic metrics $g_{\mu\nu}(x), g'_{\mu\nu}(x)...$, obtained
with the
aid of transformations of coordinates. From the point of view of
mathematics
this is obvious, but what about the physical interpretation?

Thus, there exists a fundamental difference between the conclusions
in RTG
and GRT, and its essence consists in that the RTG equations are not
form-invariant with respect to arbitrary coordinate transformations,
while
the GRT equations outside matter are form-invariant relative to such
transformations. The RTG equations are only form-invariant relative to
such
transformations of coordinates that leave the Minkowski metric
$\gamma_{\mu\nu}(x)$ form-invariant. Hence, for example, follows the
form-invariance of equations with respect to the Lorentz
transformations.

The issue of the multiplicity of solutions
$g_{\mu\nu}(x), g'_{\mu\nu}(x)...$ worried A.~Einstein seriously, and
he
discussed this issue in detail in 1913--1914 in four
articles~\cite{20} and
arrived at the conclusion that the choice of coordinate reference
systems is
limited, since he considered that from the universal covariance for
one and
the same distribution of matter $T_{\mu\nu}(x)$ there arises a whole
set of
metrics, which is physically inadmissible. However, the main reason
for
ambiguity is not related to the general covariance, it is related to
the
form-invariance of GRT equations outside matter with respect to
arbitrary
transformations of coordinates. To remove this ambiguity, there is no
need
to renounce general covariance, since it is not the cause, but it is
necessary to restrict the form-invariance of equations in accordance
with
the relativity principle. Precisely this is done in RTG on the basis
of the
field approach. A simple example from electrodynamics can be
presented.
Assume that for a current $j_\mu (x)$ we have the solution $A_\mu
(x)$.
Upon performing transformations to the new variables $x'$ coinciding
with
the initial variables $x$ in the region of the distribution of
current
$j_\mu (x)$ and differing from them in the region outside the
current, our
solution assumes the form $A'_\mu (x')$. But it is absolutely obvious
that $A'_\mu (x)$ will not be a solution of the equations of
electrodynamics
in the coordinates $x$, since the equations of electrodynamics are
not
form-invariant with respect to arbitrary transformations of
coordinates.

This means that in electrodynamics for one and the same distribution
of
current $j_\mu(x)$ in identical conditions there exists only one
distribution of the electromagnetic field $\vec E,\vec H$. If in GRT
the
distribution of matter, determined by the tensors $T_{\mu\nu}(x)$ and
$T_{\mu\nu}(x')$, is the same in two arbitrary reference systems,
then
the GRT equations being form-invariant outside matter, we can,
in identical conditions, for example, have identical metric
coefficients
$g_{\mu\nu}(x)$ and  $g_{\mu\nu}(x')$. Precisely this circumstance
permitted
A.~Einstein to put forward the general relativity principle for all
physical processes. However, the requirement that the metric
coefficients
be identical results in a strong restriction being imposed on the
structure
of Riemannian space, it turns out to be a space with constant curvature.

Since Riemannian space in GRT does not have this property, in the general case, then the
general relativity principle, as a physical principle, is not
realized in
Nature. This also follows from the equations of electrodynamics, for
example, not being form-invariant with respect to {\bf arbitrary}
transformations of coordinates. The relativity principle, as a
physical
principle, is not related to universal covariance, but to the
form-invariance
of equations and of the  {\bf metric} relative to the transformations
of
coordinates. V.A.~Fock was right, when he wrote: {\it ``a general
relativity principle, as a physical principle, that could be valid
with
respect to arbitrary reference systems, is not possible"}~\cite{14}.
In GRT for
one
and the same distribution of matter $T_{\mu\nu}(x)$ there exists a
whole
range of solutions of the GRT equations for the metric coefficients
$g_{\mu\nu}(x), g'_{\mu\nu}(x),...$ Outside matter the geodesic
lines for
these solutions will be different.

The issue of the multiplicity of metrics in GRT in one coordinate
system was widely discussed in 1921-1922 by P.Painlev\'e,
M.Chazy,~ J.Becquerel,~ A.Gullstrand,~ E.Kretschmann. The essence
of the polemic actually reduced to the question: with which radial
variable in the GRT equations is it necessary to identify the
astronomically determined distance between the Sun and a planet?
It must be noted that this arbitrariness in the first order in the
gravitational constant does not influence certain gravitational
effects: the deflection of a light ray, the shift of the
perihelion of Mercury, the precession of a gyroscope. Ho\-we\-ver,
it does, already in the first order in $G$, influence the delay
effect of a radiosignal.

Thus, depending on the choice of
solutions in the Schwarz\-schild form or in harmonic coordinates we
will
obtain different values for the delay time. We will further see that
there
exists no such arbitrariness in RTG, and that effects are determined
unambiguously.

The reason for the multiplicity of metrics is not general
covariance, but the form-invariance of equations with respect to
arbitrary transformations of coordinates. There exists no such
ambiguity in RTG, since the metric $g_{\mu\nu}(x)$ is
unambiguously determined by the distribution of matter
$T_{\mu\nu}(x)$. In section~11 it was shown that since the radius
of a static spherically symmetric body exceeds the Schwarzschild
radius, then the external solution of RTG equations in an inertial
reference system in spherical coordinates in the region
(\ref{328}) has the form $$
ds^2=\f{r-MG}{r+MG}(dx^0)^2-\f{r+MG}{r-MG}(dr^2)- $$ $$
\!\!\!\!\!\!\!\!\!\!\!\!\!\!\!\!-(r+MG)^2
[(d\theta)^2+\sin^2\Theta (d\varphi)^2].\eqno{(\alpha)} $$
Precisely such a solution in the post-Newtonian approximation
yields expressions for the metric coefficients of effective
Riemannian space that coincide with the previously obtained
formulae (\ref{151a}) applied for explanation of gravitational
effects in the Solar system.

An essential point is that, when the gravitational field is switched
off
(for instance, the body is removed), we necessarily turn out to be in
Minkowski space in an inertial reference system with the metric
$$
d\sigma^2=(dx^0)^2-(dr)^2-r^2
[(d\theta)^2+\sin^2\Theta (d\varphi)^2].
$$

In calculating gravitational effects in the Solar system we have
to calculate the trajectory of motion in effective Riemannian
space, that is determined by the interval $ds$, and to compare it
with the corresponding trajectory determined by the interval
$d\sigma$. The metric of Minkowski space is present in the RTG
equations. This is precisely how the deflection angle of a light
ray and the delay time of a radiosignal, due to the influence of
the gravitational field of the Sun, are determined. As to
calculating the shift in the perihelion of a planet, here it is
necessary to compare the trajectory of motion of a test body
around the Sun calculated within RTG with the trajectory obtained
from Newton's theory of gravity. Precisely in these calculations
there exists a difference between the RTG and GRT conclusions,
since within GRT one cannot say in which reference system
(inertial or non-inertial) of Minkowski space one happens to be,
when the gravitational field is switched off. For calculating the
gravitational effect it is necessary to compare in one coordinate
system the motion along a geodesic line in Riemannian space with
motion along the geodesic line in Minkowski space with gravity
switched off. But to this end it is necessary to know exactly both
the metric $g_{\mu\nu}(x)$ and the metric $\gamma_{\mu\nu}(x)$.

{\bf However, in GRT, owing to the multiplicity of solutions both for
$g_{\mu\nu}(x)$ and for $\gamma_{\mu\nu}(x)$ we cannot with
definiteness
say which Riemannian metric $g_{\mu\nu}(x)$ it is necessary to take for
the
chosen metric $\gamma_{\mu\nu}(x)$ in order to find the geodesic
lines in
Riemannian space and in Minkowski space. This is actually the essence of
the
ambiguity in predictions of gravitational effects in GRT.} Sometimes
errors
are avoided in GRT by considering the initial reference system to be
an
inertial reference system in
Cartesian coordinates (but no such coordinates exist in GRT) and then
dealing with a weak gravitational field against this background. No
such
difficulty exists in RTG, since for the chosen metric
$\gamma_{\mu\nu}(x)$,
with the aid of equations (\ref{76}) and (\ref{77}), under
appropriate
conditions, the metric $g_{\mu\nu}(x)$ of effective Riemannian space is
determined unambiguously, which permits to determine unambiguously
the
gravitational effect.

In calculations of effects in the gravitational field of the Sun one
usually
takes as the idealized model of the Sun a static spherically
symmetric body
of radius $R_\odot$. The general form of the metric of Riemannian space
in an
inertial reference system in spherical coordinates is
\ba
&&ds^2=U(r)(dx^0)^2-V(r)(dr)^2-\nonumber \\
&&-W^2(r)
[(d\theta)^2+\sin^2\Theta (d\varphi)^2].\label{12.1}
\ea
In the absence of a gravitational field the metric has the form
\begin{aequation}{12.1}{a}
d\sigma^2=(dx^0)^2-(dr)^2
-r^2
[(d\theta)^2+\sin^2\Theta (d\varphi)^2].\label{12.1a}
\end{aequation}
Substituting (\ref{12.1}) and  (\ref{12.1a}) into equations
(\ref{76}) and
(\ref{77}) we precisely obtain the external solution for the Sun
($\alpha$).

In section 5 it was shown that from the RTG equations (\ref{76}) and
(\ref{77}) follow directly the equations of motion for matter,
\be
\nabla_\nu T^{\mu\nu}=0.\label{12.2}
\ee
Hence it is easy to obtain the equations of motion for a test body in
a
static gravitational field. The energy-momentum tensor for matter,
$T^{\mu\nu}$, in this case assumes the form
\be
T^{\mu\nu}=\rho U^\mu U^\nu,\;\;\; U^\mu=\f{dx^\mu}{ds}.\label{12.3}
\ee
Substituting (\ref{12.3}) into (\ref{12.2}) we obtain
\be
U^\mu\nabla_\nu (\rho U^\nu)+\rho U^\nu \nabla_\nu
U^\mu=0.\label{12.4}
\ee
Multiplying this equation by~ $U_\mu$~ and taking into account
$U_\mu U^\mu=1$ we obtain
\begin{aequation}{12.4}{a}
\nabla_\nu (\rho U^\nu)+\rho U^\nu U_\mu \nabla_\nu U^\mu
=0.\label{12.4a}
\end{aequation}
Since
$$
\nabla_\nu (U_\mu U^\mu)=2 U_\mu\nabla_\nu U^\mu=0.
$$
from equation (\ref{12.4a}) we have
\be
\nabla_\nu (\rho U^\nu)=0.\label{12.5}
\ee
Substituting (\ref{12.5}) into (\ref{12.4}) we find
\be
U^\nu\nabla_\nu U^\mu =0. \label{12.6}
\ee
Applying the definition of a covariant derivative, equations
(\ref{12.6})
may be written as
\be
\left [
\f{\pa U^\mu}{\pa x^\nu}+\Gamma^\mu_{\nu\sigma}
U^\sigma \right ]\f{dx^\nu}{ds}=0.\label{12.7}
\ee
Taking into account the definition of a total differential we have
\be
dU^\mu =\f{\pa U^\mu}{\pa x^\nu}dx^\nu.\label{12.8}
\ee
On the basis of (\ref{12.8}) equation (\ref{12.7}) assumes the form
\be
\f{dU^\mu}{ds}+\Gamma^\mu_{\nu\sigma}
U^\nu U^\sigma =0,\;\;\;
U^\mu =\f{dx^\mu}{ds}.\label{12.9}
\ee

The equation of motion of a test body, (\ref{12.9}), is an equation
of
geodesic lines in the space with the metric $g_{\mu\nu}$. The
Christoffel
symbols are determined by the formula
\be
\Gamma^\mu_{\nu\sigma}=\f{1}{2}g^{\mu\lambda}
(\pa_\nu g_{\sigma\lambda} +\pa_{\sigma}g_{\nu\lambda}
-\pa_\lambda g_{\nu\sigma}). \label{12.10}
\ee
On the basis of (\ref{12.1}) and (\ref{12.10}) it is easy to obtain
the
Christoffel symbols of interest to us:
\ba
&&\Gamma^2_{12}=\f{1}{W}\f{dW}{dr},\;\;\;
\Gamma^2_{33}=-\sin\Theta\cos\Theta,\;\;\;
\nonumber
\\*[-0.1cm] \label{12.11} \\*[-0.1cm]
&&\Gamma^3_{13}=\f{1}{W}\f{dW}{dr},\;\;\;
\Gamma^3_{23}=\cot\Theta,\;\;
\Gamma^0_{01}=\f{1}{2U}\f{dU}{dr}.\nonumber
\ea

Of the four equations (\ref{12.9}) only three are independent, since
the
following relation is valid:
\be
g_{\mu\nu}U^\mu U^\nu=1.\label{12.12}
\ee
We shall further make use of this circumstance by choosing the three
simplest equations from (\ref{12.9}). From equations (\ref{12.9}) we
shall
take equations
\ba
&&\f{d^2x^0}{ds^2}+\f{1}{U}
\f{dU}{dr}\f{dx^0}{ds}\f{dr}{ds}=0, \label{12.13}\\[0.2cm]
&&\f{d^2\Theta}{ds^2}+\f{2}{W}
\f{dW}{dr}\f{dr}{ds}\f{d\Theta}{ds}
-\sin\Theta\cos\Theta
\left (\f{d\varphi}{ds}\right )^2\!\!=0,\qquad
 \label{12.14}\\[0.2cm]
&&\f{d^2\varphi}{ds^2}+\f{2}{W}
\f{dW}{dr}\f{dr}{ds}\f{d\varphi}{ds}
+2\cot\Theta
\f{d\Theta}{ds}\f{d\varphi}{ds}=0 \label{12.15}
\ea
and supplement them with equation (\ref{12.12}) in the form
\ba
&&U \left (\f{dx^0}{ds}\right )^2
-V\left (\f{dr}{ds}\right )^2
-W^2\left (\f{d\Theta}{ds}\right )^2-\nonumber \\
&&-W^2\sin^2\Theta
\left (\f{d\varphi}{ds}\right )^2=1.\label{12.16}
\ea

Since the gravitational field is spherically symmetric, it is natural
to
choose the reference system so as to make the motion take place in
the
equatorial plane, i.e.
\be
\Theta=\f{\pi}{2}.
\ee
In the case of our choice equation (\ref{12.14}) is identically
satisfied.

Equations (\ref{12.13}) and (\ref{12.15}) can be respectively written
as
\be
\f{d}{ds}
\left [
\ln\f{dx^0}{ds}U\right ]=0,\label{12.18}
\ee
\be
\f{d}{ds}
\left [
\ln\f{d\varphi}{ds}W^2\right ]=0. \label{12.19}
\ee
Hence, we find the first integrals of motion $E,J$:
\be
\f{dx^0}{ds}U=\f{1}{\sqrt{E}},\;\;\;
\f{d\varphi}{ds}W^2=\f{J}{\sqrt{E}}.\label{12.20}
\ee
Substituting these expressions into (\ref{12.16}) we obtain
\be
\f{1}{EU}-V
\left (\f{dr}{ds}\right )^2-\f{J^2}{W^2E}=1.\label{12.21}
\ee
From the second relation (\ref{12.20}) we find
\be
ds=d\varphi \f{\sqrt{E}W^2}{J}.\label{12.22}
\ee
Passing in (\ref{12.21}), with the aid of (\ref{12.22}), to the
variable
$\varphi$ we obtain
\be
\f{V}{W^4}
\left (\f{dr}{d\varphi}\right )^2
+\f{1}{W^2}-\f{1}{J^2U}
+\f{E}{J^2}=0.\label{12.23}
\ee
From the first relation (\ref{12.20}) we find
\be
(ds)^2=EU^2(dx^0)^2. \label{12.24}
\ee
Hence it follows that $E>0$ for test bodies and $E=0$ for light.

Gravitational effects in the Solar system have been calculated within
GRT by various methods. Here, we shall follow S.~Weinberg's method of
calculation ~\cite{21}.

\subsection{Deflection of light rays by the Sun}
Consider a photon from a distant region passing near the Sun. What is
the trajectory of the light ray?  It is determined from equation
(\ref{12.23}), when $E=0$, and it has the form
\be
d\varphi=dr\sqrt{\f{UV}{W^2\left (\f{W^2}{J^2}-U\right )}}.
\label{12.25}
\ee
At the point of the light ray's trajectory (see the figure) closest
to the
Sun
\be
\f{dr}{d\varphi}\Bigl |_{r_0}=0.\label{12.26}
\Bigr.
\ee

\vspace*{11mm}
\begin{figure}[th]
\setlength{\unitlength}{1mm}
\begin{center}
\begin{picture}(100,50)
\put(0,0){\epsfxsize=10cm \epsfbox{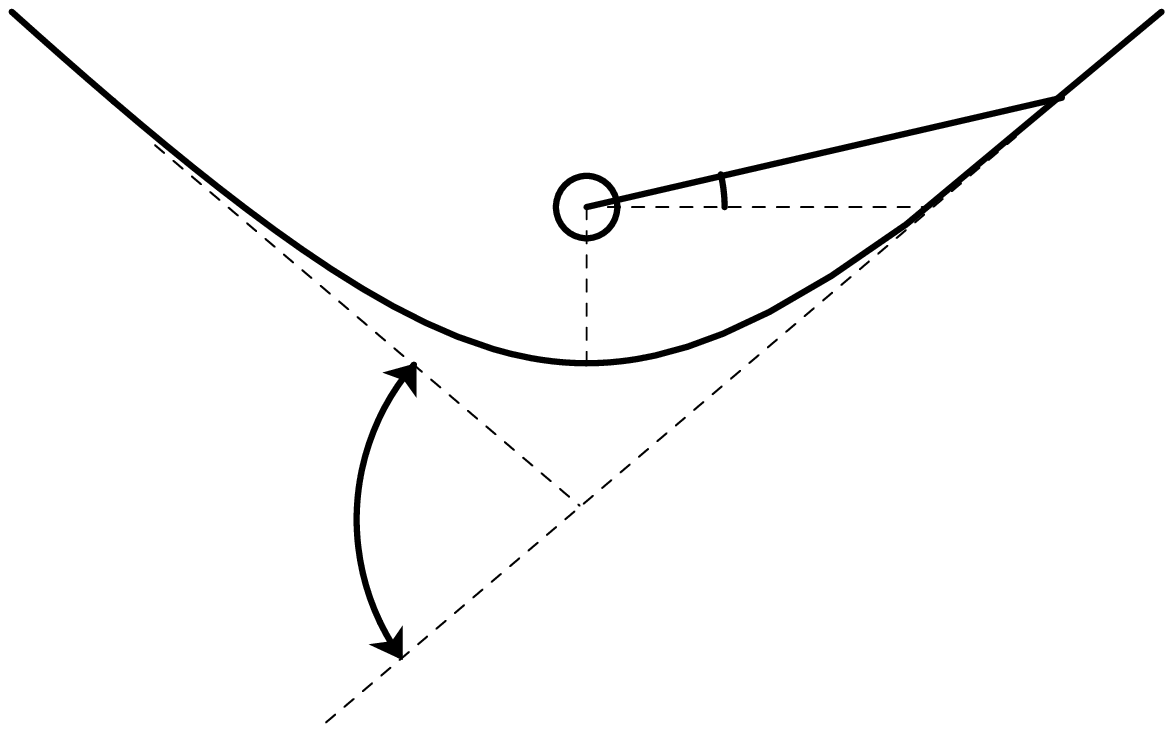}}
\put(20,22){$\Delta\varphi$}
\put(35,5){$l$}
\put(52,40){$r_0$}
\put(95,54){$l$}
\put(75,55){$r$}
\put(67,48.5){$\varphi(r)$}

\end{picture}
\end{center}
\vspace*{-4mm}
\centerline{\small\bf Deflection of the light ray}
\end{figure}

\vspace*{4mm}
The integral of motion $J$ is expressed via the metric parameters
$U_0$ and $W_0$:
\be
J^2=\f{W^2(r_0)}{U(r_0)}=\f{W^2_0}{U_0}.\label{12.27}
\ee
Integrating (\ref{12.25}) we obtain
\be
\varphi (r)=\varphi (\infty)+
\int\limits^{\infty}_{r}dr
\left [
\f{V}{W^2\left (
\f{W^2}{W^2_0}\f{U_0}{U}-1\right )}\right ]^{1/2}.\label{12.28}
\ee

The deflection angle of a light ray is
\be
\Delta\varphi =2|\varphi (r_0)-\varphi (\infty)|-\pi. \label{12.29}
\ee

Here, we have taken into account that in the absence of a
gravitational
field a light ray propagates along a straight line $l$, and precisely
for
this reason $\pi$ has appeared in (\ref{12.29}). For the Sun, from
the RTG
equations we have
\be
U(r)=\f{r-GM}{r+GM},\;\;
V=\f{r+GM}{r-GM},\;\;
W^2=(r+GM)^2. \label{12.30}
\ee
For calculations it is sometimes convenient to make use of an
independent
variable $W$:
\be
U(W)=1-\f{2GM}{W},\;\;
V(W)=\f{1}{1-\f{2GM}{W}}.\label{12.31}
\ee
In the first approximation in the gravitational constant $G$ the
metric
coefficients are
\ba
&&U(r)=1-\f{2GM}{r},\;\;
V(r)=1+\f{2GM}{r},\;\;\nonumber \\
[-0.2cm] \label{12.32} \\ [-0.2cm]
&&W^2=r^2\left (1+\f{2GM}{r}\right ). \nonumber
\ea
Substituting these expressions into the integral (12.28) we obtain
for it
the following expression:
\be
I=r_0
\int\limits^{\infty}_{r_0}\f{dr}{r\left [
r^2\left (1-\f{4MG}{r_0}\right )+4MGr-r^2_0\right ]^{1/2}}.
\label{12.33}
\ee
Performing a change of variables, $r=\f{1}{t}$, we obtain
\be
I=
\int\limits^{1/r_0}_{0}\f{dt}
{\sqrt{ r^{-2}_0\left (1-\f{2MG}{r_0}\right )^2-
\left (t-\f{2MG}{r_0^2}\right )^2}}. \label{12.34}
\ee
Making use of the tabular integral
$$
\int\f{dx}{\sqrt{m^2-\left  (x-\f{b}{2}\right )^2}}=
\arcsin \f{x-\f{b}{2}}{m}+c,
$$
we find
\be
I=\f{\pi}{2}+\f{2MG}{r_0}.\label{12.35}
\ee
On the basis of (\ref{12.29}) we obtain
\be
\Delta\varphi=\f{4M_\odot G}{c^2r_0}, \label{12.36}
\ee
taking into account
\be
\f{M_\odot G}{c^2}=1,475\cdot 10^5\mbox{cm},\;\;
R_\odot G=6,95\cdot 10^{10}\mbox{cm}, \label{12.37}
\ee
we find
\be
\Delta\varphi =\f{R_\odot}{r_0}\sum_\odot,\;\;
\sum_\odot=\f{4M_\odot G}{R_\odot c^2}=1,75''.\label{12.38}
\ee

Thus, the deflection of a light ray by the gravitational field of the
Sun
is equal to
\be
\Delta\varphi =1,75''\cdot \f{R_\odot}{r_0}.\label{12.39}
\ee
In calculating the deflection angle of a light ray we took into
account that
in the absence of a field in an inertial reference system, a light
ray
travels, by virtue of the metric (\ref{12.1a}), along a straight line
$l$.
Precisely the deviation from this straight line is the gravitational
effect.

\subsection{The delay of a radiosignal}
I.I.~Shapiro~\cite{22}  proposed and implemented an experiment for
measurement of
the time required for a radiosignal to reach the planet Mercury and,
upon
reflection, to return to the Earth. We shall calculate this time on
the
basis of RTG equations.

We shall pass from the independent variable $\varphi$ to the
independent
variable $x^0$. To this end, making use of (\ref{12.22}) and
(\ref{12.24}),
we shall obtain
\be
(d\varphi)^2=\f{J^2U^2}{W^4}(dx^0)^2.\label{12.40}
\ee
With the aid of (\ref{12.40}), equation (\ref{12.23}) assumes the
form
\be
ct(r,r_0)=
\int\limits^{r}_{r_0} dr
\left [
\f{V}
{\left (1-\f{W^2_0}
{W^2}\cdot\f{U}{U_0}\right )U}
\right ]^{1/2}.\label{12.41}
\ee
Substituting expressions (\ref{12.32}) into the integral
(\ref{12.41}),
we find
\be
ct(r,r_0)=
\int\limits^{r}_{r_0} \f{rdr}{\sqrt{r^2-r^2_0}}
\left [
1+\f{2MG}{r}+\f{2MG}{r}\f{r_0}{r+r_0}\right ]. \label{12.42}
\ee
Applying the tabular integrals
\ba
\int\limits^{r}_{r_0} \f{dr}{\sqrt{r^2-r^2_0}}=
\ln\f{r+\sqrt{r^2-r^2_0}}{r_0},\;\;\nonumber \\
[-0.2cm]  \label{12.43} \\ [-0.2cm]
\int\limits^{r}_{r_0} \f{dr}{\sqrt{r^2-r^2_0}}
\f{r_0}{r+r_0}=
\left [
\f{r-r_0}{r+r_0}
\right ]^{1/2},\nonumber
\ea
we obtain
\ba
&&ct(r,r_0)=
\sqrt{r^2-r^2_0} +2MG\ln
\f{r+\sqrt{r^2-r^2_0}}{r_0} + \nonumber \\
&&+2MG
\left [
\f{r-r_0}{r+r_0}
\right ]^{1/2}. \label{12.44}
\ea

Let $r_e, r_p$ be the heliocentric coordinates of the Earth and of
Mercury.
Since $r_e, r_p>> r_0$, then in the summands of expressions
(\ref{12.44}),
that contain the gravitational constant, the influence of $r_0$
present
under the square root sign can be neglected, which will result in
\ba
&&ct(r_p,r_e)=\sqrt{r^2_e-r^2_0}
+\sqrt{r^2_p-r^2_0}+\nonumber\\
&&+\f{2MG}{c^2}\ln
\f{4r_er_p}{r^2_0}
+\f{4MG}{c^2}. \label{12.45}
\ea
We shall drop a perpendicular $r_\bot$ from the center of the source
of
the gravitational field onto the straight line connecting points
$r_e$
and $r_p$. Then, according to Pythagoras' theorem, we have
\be
r^2_e=R^2_e+r^2_\bot,\;\;
r^2_p=R^2_p+r^2_\bot. \label{12.46}
\ee
In the first order in $G$
\be
r_0\simeq r_\bot +R_e\f{\Delta\varphi}{2},\;\;
r^2_0-r^2_\bot \simeq R_e r_0 \Delta\varphi, \label{12.47}
\ee
$\Delta\varphi$ is the deflection angle of a light ray due to the
influence
of a source of the gravitational field (see (\ref{12.36})).
\ba
&&\sqrt{r^2_e-r^2_0}=R_e\sqrt{1-
\f{r_0}{R_e}\Delta\varphi}\simeq\nonumber \\
&&\simeq R_e-r_0\f{\Delta\varphi}{2}
=R_e-\f{2MG}{c^2}, \label{12.48}
\ea
similarly
\be
\sqrt{r^2_p-r^2_0}\simeq R_p - r_0
\f{\Delta\varphi}{2}=R_p- \f{2MG}{c^2}.\label{12.49}
\ee
With account of (\ref{12.48}) and (\ref{12.49}) expression
(\ref{12.45})
assumes the form
\be
ct(r_p, r_e)-R=\f{2MG}{c^2}\ln
\f{4r_er_p}{r^2_0}, \label{12.50}
\ee
here $R=R_e+R_p$ is the distance between the planets.

{\baselineskip=1.18\normalbaselineskip
The delay time of a radiosignal propagating from the Earth to Mercury
and
back is
\ba
&&\Delta\tau=2[t(r_p, r_e)-R/c]=\f{4MG}{c^3}\ln
\f{4r_er_p}{r_0}, \nonumber \\
[-0.15cm] \label{12.51}\\ [-0.15cm]
&&r_e=r_\oplus=15\cdot 10^{12}\mbox{cm},\;\;
r_p=r_{\imercury} =5,8\cdot 10^{12}\mbox{cm},\nonumber
\ea
as $r_0$ it is possible to take the radius of the Sun, $R_\odot$:
$$
R_\odot =6,95\cdot 10^{10}\mbox{cm}.
$$
Substituting these values into (\ref{12.51}) and taking into account
that
$$
\f{4M_\odot G}{c^2}=5,9\cdot 10^5\mbox{cm},
$$
we obtain
\be
\Delta\tau=\f{4M_\odot G}{c^3}\ln
\f{4r_\oplus r_{\imercury}}{R^2_\odot}=219,9\; \mbox{$\mu$\,s}.
\label{12.52}
\ee

In calculating the delay effect of a radiosignal we have taken into
account
that in the absence of a gravitational field a light ray, by virtue
of
(\ref{12.1a}), travels from point $e$ to point $p$ along a straight
line in
an inertial reference system. Comparison with such motion is
precisely how
the gravitational effect is determined. It is precisely for this
reason,
that the summand $\f{2R}{c}$ has appeared to the left in
(\ref{12.51}).
In observations, the time $\f{2R}{c}$ is determined during a period,
when
the Sun moves away from the trajectory of the light ray, so its
influence
is significantly reduced.

In GRT, if the solution is sought of the Hilbert--Einstein equations
for
a static spherically symmetric body of mass $M$, then within one and
the
same coordinate system it is possible to obtain the external solution for
the
metric, that involves two arbitrary functions
$$
\!\!\!\!\!\!\!\!\!\!\!\!\!\!\!\!\!\!\!\!\!\!\!\!\!\!\!\!\!\!\!\!\!\!
ds^2=g_{00}dt^2+2g_{01}dt
dr +
$$
\begin{aequation}{12.31}{a}
+g_{11}dr^2
+g_{22}(d\Theta)^2+g_{33}(d\varphi)^2,  \label{12.31a}
\end{aequation}
where
\ba
&&g_{00}=1-\f{2GM}{W(r)},\;\;
g_{01}=-B(r),\;\;\nonumber \\
&&g_{11}=-\left (
1-\f{2GM}{W}\right )^{-1}
\cdot
\left [\left (\f{dW}{dr}\right )^2-B^2\right ],
\nonumber \\[1mm]
&&g_{22}=-W^2(r),\;\;
g_{33}=-W^2\sin\Theta.\nonumber
\ea

Thus, within one and the same coordinate system for a body of mass $M$
there
exists an infinite number of solutions. Here, functions $B(r)$ and
$W(r)$
are, generally speaking, arbitrary, they are not determined by GRT.
P.~Painlev\'e wrote about all this some 80~years ago and stressed
that
the
choice of initial formulae is purely arbitrary. Hence it is obvious
that
within GRT neither Newton's law nor the post-Newtonian approximation
(\ref{151a}), used in GRT for explaining gravitational effects in the
Solar
system, follow unambiguously from the exact external solution of
equations.
Hence, also, it follows, for example, that an infinitesimal period of
true physical
time in GRT,
$$
d\tau =\left [
dt\sqrt{1-\f{2GM}{W(r)}} -
\f{B(r)dr}{\sqrt{1-\f{2GM}{W(r)}}}
\right ]
$$
will differ depending on the choice of the arbitrary functions $W(r)$
and
$B(r)$. This means that for a static spherically symmetric body of mass
$M$
the course of physical time for one or another process is not
determined
unambiguously.

As we can see, the situation in GRT differs totally from the
situation in electrodynamics, where the Coulomb law unambiguously
follows from the equations of theory. We shall clarify the
situation making use of the example of Minkowski space. From the
point of view of geometry, Minkowski space both in an inertial and
in a noninertial reference system remains essentially the same,
since the tensor of Riemann curvature equals zero. By virtue of
the form-invariance of the Riemann tensor we will have in one and
the same coordinate system an infinite amount of metrics
$\gamma_{\mu\nu}(x), \;\;\gamma'_{\mu\nu}(x)...$ and so on, that
reduce the Riemann tensor to zero. But, depending on the choice of
metric we will have different geodesic lines in one and the same
coordinate system, i.e. the physics will vary. All this is obvious
and well known, since the dynamics in an inertial reference system
differs from the dynamics in a non-inertial reference system owing
to the appearance of forces of inertia. Precisely for this reason
the choice of the non-inertial reference system in
four-dimensional Minkowski space alters the physics. In Minkowski
space, however, there exist inertial reference systems, and
observational astronomical data are referred precisely to an
inertial reference system. The choice of reference system in
physical equations is based precisely on this circumstance. In GRT
Riemannian geometry there exist no such reference systems, so it
is absolutely unclear, which coordinates it is necessary to choose
so as to compare theoretical calculations and observational data.
The geometry does not depend on the choice of reference system
(or, in other words, in our particular case, on the choice of
functions $B(r)$ and $W(r)$)), it remains, like before, Riemannian
geometry, however, the physics changes. Naturally, it is possible,
in our case, to select arbitrary functions $B(r)$ and $W(r)$ so as
to provide for Newton's law of gravity to be satisfied, and for
the post-Newtonian approximation to have the form (\ref{151a}).
However, such a choice in GRT is, regrettably, arbitrary, since it
is not imposed by any physical conditions. It is not possible to
formulate physical requirements to be imposed on the behaviour of
Riemannian metric, if it is not of a field origin, because such
behaviour even depends on the choice of three-dimensional space
coordinates. V.A.~Fock resolved the issue of the choice of
coordinates for island systems, with the aid of the harmonicity
conditions. But why precisely they have to be chosen, instead of
some other ones, remained unclear.

Now, let us go back to the analysis of a concrete example
demonstrating the ambiguity of GRT in calculations of the
gravitational delay effect of a radiosignal traveling from the
Earth to Mercury and back. The predictions of theory depend on the
choice of solution. For simplicity we shall advantage of the
simplest partial case $$ B(r)=0,\;\; W(r)=r+(\lambda+1)M,\;\; r>
R_0, $$ $\lambda$ is an arbitrary parameter, $R_0$ is the radius
of the body dealt with. If an appropriate choice is made of
function $W(r)$ in the vicinity of the body, then this solution
can be made to match the solution inside the body. If the previous
calculations are repeated, then for this metric we find $$
ct(r,r_0)=\sqrt{r^2-r^2_0} +2MG\ln \f{r+\sqrt{r^2-r^2_0}}{r_0}+ $$
$$
\!\!\!\!\!\!\!\!\!\!\!\!\!\!\!\!\!\!\!\!\!\!\!\!\!\!\!\!\!\!\!\!\!\!
\! \!\!\!\! \!\!\!\!+2MG \left ( 1+\f{\lambda}{2}\right ) \left [
\f{r-r_0}{r+r_0}\right ]^{1/2}.\eqno{(A)} $$ Comparing this
expression with (\ref{12.44}) we see that already in the first
order in $G$ an ambiguity due to the influence of the source of
the gravitational field arises in GRT in the description of the
delay effect of a radiosignal. On the basis of (A) the delay time
of a radiosignal traveling from the Earth to Mercury and back, in
accordance with GRT, equals the following: $$ \Delta\tau =2\left [
t(r_e,r_p)-\f{R}{c}\right ] =\f{4MG}{c^3}\ln \f{4r_er_p}{r^2_0}
+\f{2MG}{c^3}\lambda. \eqno{(B)} $$

We note that in the first order in $G$ the physical distance $l$,
determined
by the expression
$$
l=\int\limits^{r}_{R_0}\sqrt{g_{11}} dr\simeq
r-R_0 +\f{GM}{c^2}\ln\f{r}{R_0}
$$
does not depend on the parameter $\lambda$, and precisely for this
reason
the first term in (B) will also not depend on $\lambda$, and,
consequently,
owing to the presence of the second term, we will have different
predictions
for the delay time of a radiosignal, depending on the choice of the
constant $\lambda$. Expression (B) differs essentially from the
result
(\ref{12.51}), which follows exactly from RTG and complies with
experimental
data~\cite{22}. For the Schwarzschild solution $\lambda=-1$ the
difference from (12.51) will be $\f{2GM}{c^3}$, which for the Sun
amounts
to about 10~microseconds. In calculations of the following
gravitational
effects:  of the deflection of a light ray by the Sun, of the shift
of a
planet's perihelion, of the precession of a gyroscope, of the shift
of
spectral
lines, an ambiguity also arises, but in the second order in the
gravitational constant $G$. All the above has been discussed in
detail
with prof.~Yu.M.~Loskutov (see, also, monograph~\cite{1_3}).

Thus, GRT cannot, in principle, provide definite predictions on
gravitational effects, which is still another of its fundamental
defects. Certain attempts to relate the gravitational field in GRT to
the
equivalence class of diffeomorphic metrics do not remove this
ambiguity,
since they do not discard the fundamental defect of GRT --- the
form-invariance of the Hilbert--Einstein equations outside matter with
respect to arbitrary transformations of coordinates$\,$\footnote{See,
for
instance: J.Stachel. Conference ``Jena-1980", 1981 (DDR).}.
Precisely this
circumstance results in the entire set of diffeomorphic metrics
arising
within one coordinate system for one and the same distribution of matter,
while
this, according to the Weyl--Lorentz--Petrov theorem (see end of
section~14), leads to different geodesic lines in identical
conditions
of the problem, which is physically inadmissible. {\bf The essence
of the
issue does not consist in the general covariance, which must always
exist,
but in whether the form-invariance of physical equations
relative to arbitrary transformations of coordinates is admissible?}

Since reference systems are not equivalent in the presence of
forces of inertia, no sense whatsoever can be attributed to the
form-invariance of physical equations with respect to arbitrary
transformations of coordinates. General covariance is a
mathematical requirement, while form-invariance has a profound
physical content. Actually, in all physical theories the
form-invariance of equations and of {\bf metrics} holds valid
relative to the Lorentz transformations --- precisely this is the
essence of the relativity principle.

The non-equivalence of various reference systems is especially
evident, if
one considers pseudo-Euclidean geometry, for which the curvature
tensor is
zero. From the equality of the curvature tensor to zero it is
possible, by
virtue of form-invariance, to obtain a set of solutions for the metric
tensor
in one and the same coordinate system. But it is quite obvious that they
are
physically not equivalent, since some of them are inertial reference
systems, while others are non-inertial reference systems. All the
above,
involving significant complications, also takes place in the case of
Riemannian
geometry.

{\bf It is important to stress once more that in GRT, owing to the
equations
being, outside the distribution of matter, form-invariant relative to
arbitrary transformations of coordinates, there arises a situation
when for
one and the same distribution of matter there exists, in one and the
same
coordinate system, an indefinite amount of metrics.} No such situation
exists in
any other physical theory, since form-invariance within them is
admissible
only with respect to transformations of coordinates leaving the
metric
$\gamma_{\mu\nu}(x)$ form-invariant. From this fact, for example,  the
form-invariance of equations relative to the Lorentz transformations follows.

\subsection{The shift of a planet's perihelion}

Consider the motion of a test body on a solar orbit. At the
perihelion the
heliocentric distance of the test body is at its minimum and is
$r_-$,
while at the aphelion it reaches its maximum and equals $r_+$. Since
at the
perihelion and at the aphelion $\f{dr}{d\varphi}=0$, from equation
(\ref{12.23}) we obtain
\ba
&&\f{1}{W(r_+)}-\f{1}{J^2U(r_+)}=-
\f{E}{J^2},\;\;\nonumber\\*
[-0.2cm]\label{12.53}\\*[-0.2cm]
&&\f{1}{W(r_-)}-\f{1}{J^2U(r_-)}=-\f{E}{J^2}. \nonumber
\ea
Hence, we find
\be
J^2=\f{\f{1}{U_+}-\f{1}{U_-}}
{\f{1}{W^2_+}-\f{1}{W^2_-}}.\label{12.54}
\ee
Now we write equations (\ref{12.53}) in another form:
\be
J^2=W^2_+\left (
\f{1}{U_+}-E\right ),\;\;
J^2=W^2_-
\left (\f{1}{U_-}-E\right ).\label{12.55}
\ee
Hence, we obtain
\be
E =\f{\f{W^2_+}{U_+}-\f{W^2_-}{U_-}}
{W^2_+-W^2_-}.\label{12.56}
\ee
By integration of equation (\ref{12.23}) we find
\be
\varphi (r)=\varphi (r_-)+
\int\limits^{r}_{r_-}\sqrt{V}
\left [
\f{1}{J^2U}-\f{1}{W^2}-\f{E}{J^2}\right
]^{-1/2}\f{dr}{W^2}\;.\label{12.57}
\ee
For convenience of calculations we introduce a new independent
variable
\be
W=r+GM.\label{12.58}
\ee
Substituting (\ref{12.54}) and  (\ref{12.56}) into (\ref{12.57}) and
passing
to the new independent variable $W$ we find
\ba
&&\varphi (W)=\varphi (W_-)+ \nonumber  \\[2mm]
&&+\int\limits^{W}_{W_-}
\left \{
\f{W^2_- [U^{-1}(W)-U^{-1}_-]-W^2_+[U^{-1}(W)-U^{-1}_+]}
{W^2_-W^2_+[U^{-1}_+-U^{-1}_-]}\right\}^{-1/2}\times
\nonumber \\[2mm]
&&\times
\f{\sqrt{V}dW}{W^2}.  \label{12.59}
\ea
On the basis of (\ref{12.31}) we have for function $U^{-1}(W)$ in the
second
order in the gravitational constant $G$ the following:
\be
U^{-1}(W)=1+\f{2GM}{W}+\f{(2GM)^2}{W^2}. \label{12.60}
\ee
We must take into account the second order in $U^{-1}(W)$, since, if
we
only consider the first order in $G$, then the expression under the
root
sign in figure brackets will be independent of the gravitational
constant.
But this means that in calculations this circumstance results in
our
losing
the term containing $G$ in the first order. We will have taken into
account
only the term of the first order in $G$ entering into function $V$.
For the
metric coefficient of $V$ it suffices to take into account only the
first
order in $G$,
\be
V(W)=1+\f{2GM}{W}. \label{12.61}
\ee

In the approximation (\ref{12.60}) the numerator of the expression
under
the root sign in the figure brackets of (12.59) is a quadratic
function of the variable
$\f{1}{W}$ of the following form:
\ba
&&2GM W_-W_+ (W_+-W_-)
\left [
\f{1}{W^2}-\f{1}{W}\left (
\f{1}{W_-}+\f{1}{W_+}\right )+ \right.\nonumber\\
&&\left. +\f{1}{W_-W_+}\right ]= 2GM W_-W_+(W_+- W_-)
\times\nonumber \\
&&\times\left (
\f{1}{W}-\f{1}{W_-}\right )
\left (
\f{1}{W}-\f{1}{W_+}\right ). \label{12.62}
\ea
The denominator of the expression under the root sign in
(\ref{12.59}) is
\ba
&&W^2_-W^2_+\ [U^{-1}_+ - U^{-1}_-]=2GMW_-W_+ (W_--W_+)\times
\nonumber \\[1mm]
&&\times\left [1+2GM\left (\f{1}{W_-}+\f{1}{W_+}\right )\right
].\label{12.63}
\ea

Taking (\ref{12.62}) and (\ref{12.63}) we find the expression under
the root
sign in the figure brackets (\ref{12.59}),
\be
\f{1}
{\left [1+2GM\left (\f{1}{W_-}+\f{1}{W_+}\right )\right ]}
\left (\f{1}{W_-}- \f{1}{W}\right )
\left (\f{1}{W}-\f{1}{W_+}\right ). \label{12.64}
\ee
Substituting (\ref{12.61}) and (\ref{12.64}) into (\ref{12.59}) and
considering only terms of the first order in the gravitational
constant $G$,
we obtain
\ba
&&\varphi (W)=\varphi (W_-)+
\left [1+\f{1}{2}
\left (\f{1}{W_-}+\f{1}{W_+}\right )2GM\right ]
\times\nonumber \\
&&\times \int\limits^{W}_{W_-}
\f{\left (1+\f{MG}{W}\right )dW}
{W^2\left [\left (\f{1}{W_-}-\f{1}{W}\right )
\left (\f{1}{W}-\f{1}{W_+}\right )\right ]^{1/2}}. \label{12.65}
\ea
For calculating the integral in (\ref{12.65}) we introduce a new
variable $\psi$
\be
\f{1}{W}=\f{1}{2}
\left (\f{1}{W_+} + \f{1}{W}\right )
+\f{1}{2}
\left (\f{1}{W_+}- \f{1}{W_-}\right )\sin \psi. \label{12.66}
\ee
Applying (\ref{12.66}) we obtain
\ba
&&\left (\f{1}{W_-} - \f{1}{W}\right )
\left (\f{1}{W} - \f{1}{W_+}\right )=\nonumber \\
&&=\f{1}{4}
\left (\f{1}{W_-}- \f{1}{W_+}\right )^2\cos^2 \psi. \label{12.67}
\ea
Upon substitution of (\ref{12.66}) and (\ref{12.67}) we find
\ba
&&I(W)=\int\limits^{W}_{W_-}
\f{\left (1+\f{MG}{W}\right )dW}
{W^2
\left [
\left
(\f{1}{W_-}-\f{1}{W}
\right )
\left (
\f{1}{W}-\f{1}{W_+}
\right )
\right ]^{1/2}}=
\nonumber \\
&&=\psi+GM
\left \{
\f{1}{2}
\left (
\f{1}{W_-}+\f{1}{W_+}
\right )\psi+
\right.
\nonumber \\
&&+\f{1}{2}
\left (
\f{1}{W_-}
-\f{1}{W_+}
\right )
\cos\psi
\left.\right \}
\left |^\psi_{-\pi/2}.\right. \nonumber
\ea
Hence we obtain
\be
I(W_+)=\pi +GM\f{1}{2}\left (
\f{1}{W_-}+\f{1}{W_+}\right )\pi. \label{12.68}
\ee
Making use of (\ref{12.68}), from (\ref{12.65}) we have
\be
\varphi (W_+)- \varphi (W_-)=\pi +\f{3}{2}\pi GM
\left (\f{1}{W_-}+\f{1}{W_+}\right ). \label{12.69}
\ee

Hence the change in angle $\varphi$ during one revolution is
\be
2|\varphi (W_+)-\varphi (W_-)|=2\pi +6\pi GM\f{1}{2} \left
(\f{1}{W_-}+\f{1}{W_+}\right ). \label{12.70} \ee The shift of the
perihelion during one revolution will amount to \ba
&&\delta\varphi=2|\varphi (W_+) - \varphi (W_-)|- 2\pi=\nonumber
\\ &&=6\pi GM\f{1}{2}\left (\f{1}{W_+}+\f{1}{W_-}\right
),\label{12.71} \ea or, going back to the variable determined from
(\ref{12.58}), we obtain the following in the same approximation
in $G$:
\be
\delta \varphi =6\pi GM \f{1}{2} \left (
\f{1}{r_+}+\f{1}{r_-}\right ).\label{12.72} \ee The quantities
$r_-$ and $r_+$ are expressed via the large semiaxis $a$ and the
eccentricity $e$ $e$
\be
r_{\pm}=(1\pm e)a.\label{12.73}
\ee
Usually a focal parameter is introduced:
\be
\f{1}{L}=\f{1}{2}
\left (
\f{1}{r_+}+\f{1}{r_-}\right ).\label{12.74}
\ee
Using (\ref{12.73}) we find
\be
L=(1-e^2)a.\label{12.75}
\ee
Substituting (\ref{12.75}) into (\ref{12.74}) we obtain
\be
\f{1}{2}
\left (
\f{1}{r_+}+\f{1}{r_-}\right )=\f{1}{(1-e^2)a}.\label{12.76}
\ee
Taking into account (\ref{12.76}) in (\ref{12.72}) we find
\be
\delta\varphi =\f{6\pi GM}{c^2(1-e^2)a}. \label{12.77}
\ee

In formula (\ref{12.77}) we have restored the dependence upon the
velocity
of light. For Mercury
\be
e=0,2056,\; a=57,91\cdot 10^{11}\mbox{cm}. \label{12.78}
\ee
Substituting these values into formula (12.77) we obtain the shift
of Mercury's perihelion during one revolution:
\be
\delta\varphi_{\imercury}=0.1037''.\label{12.79}
\ee
In one century Mercury undergoes 415 revolutions, therefore the shift
of
Mercury's perihelion in 100 years amounts to
\be
\Delta\varphi=43.03''.\label{12.80}
\ee

Modern data confirm this result with an accuracy up to  1\%.
Astronomers
have been studying the shift of Mercury's perihelion for several
centuries.
In 1882 S.~Newcomb established the difference between observations
and
theoretical calculations, that turned out to be $43''$ per century.
At
present optical observations, that have been under way for over
200~years,
yield an uncertainty in the determination of the precession velocity
of
approximately $0,4''$ per century.

To conclude this section we shall write equation (\ref{12.23}) in the
variables $u=\f{1}{W},\;\; W=r+GM$.
\be
\left (\f{du}{d\varphi}\right )^2+\f{u^2}{V}-\f{1}{J^2UV}+
\f{E}{J^2V}=0. \label{12.81}
\ee
For a static spherically symmetric field, by virtue of (\ref{12.31})
we
have
\be
U=V^{-1}=(1-2GMu).\label{12.82}
\ee
Differentiating (\ref{12.81}) with respect to  $\varphi$ and taking
into
account (\ref{12.82}) we obtain
\be
\f{d^2u}{d\varphi^2}+u=\f{EGM}{J^2c^2}+\f{3GM}{c^2}u^2.\label{12.83}
\ee
Here, we have restored the dependence upon  the velocity of light.
This
equation differs from the equation obtained on the basis of Newton's
theory
of gravity by an additional term $\f{3GM}{c^2}u^2$. As we see, this
term is
relativistic. Precisely this term leads to a shift in a planet's
perihelion.

Expressing the integrals of motion $E$ and $J^2$ in terms of the
eccentricity and the superior semiaxis in the nonrelativistic
approximation, we have
\be
\f{GME}{c^2J^2}=\f{1}{a(1-\epsilon^2)}. \label{12.84}
\ee
Thus, in the nonrelativistic approximation we obtain the equation of
Newton's theory of gravity
\be
\f{d^2\sigma}{d\varphi^2}+\sigma=\f{1}{a(1-\epsilon^2)},
\sigma=\f{1}{r}.
\label{12.85}
\ee
Precisely such an expression is found in classical mechanics, if the
initial
Newton equations are referred to an inertial re\-fe\-rence system. In our
calculation this is natural, since the initial RTG equations are also
written in an inertial reference system.

Comparing the motion complying with (\ref{12.83}) with the motion
(\ref{12.85}), we precisely determine the shift effect of the
perihelion
for a single revolution of the body about the Sun. In calculating the
shift
of Mercury's perihelion and the deflection of a light ray by the Sun,
A.~Einstein intuitively considered gravity to be a weak physical
field
against the background of Minkowski space. Precisely such an approach
brought him to the well-known formulae for these gravitational
effects.
Ho\-we\-ver, these formulae are not unambiguous consequences of the GRT
equations. In deriving them A.~Einstein rather followed his physical
intuition, than the logic of his theory.  However, upon finding these
effects
in 1915 he anyhow noted: ``{\it Consider a material point (the Sun) at
the
origin of the reference system. The gravitational field created by
this
material point can be calculated from equations by successive
approximations. However, it can be assumed that for a given mass of
the
Sun, $g_{\mu\nu}$ are not quite fully determined by equations (1) and
(3). } (Here the equations $R_{\mu\nu}=0$, given the restriction
$|g_{\mu\nu}|=-1$, are intended. -- A.L.) {\it This follows from
these
equations
being covariant with respect to any transformations with a
determinant~1.
Nevertheless, we, most likely, are justified in assuming that by such
transformations all these solutions can transform into each other and
that, consequently, (for given boundary conditions) they differ from
each
other only formally, but not physically. Following this conviction, I
shall
first restrict myself, here, to obtaining {\bf one} of the solutions,
without going into the issue of whether it is the {\bf sole} possible
solution"}$\,$\footnote{Einstein A. Collection of scientific works,
Moscow: Nauka, 1965, vol.1, art.36, p.440.}.

Later the issue of other possible external solutions arose in
the twenties, when the French mathematician P.~Painlev\'e criticized
A.~Einstein's results. Following P.~Painlev\'e, we shall consider
this
issue
from the point of view of the exact external solution (12.31a) of the
GRT
equations for a static spherically symmetric body.

In GRT, calculation of the shift of Mercury's perihelion on the basis
of the
exact external solution (12.31a), for a choice of the arbitrary
functions
$B(r)$ and $W(r)$ in the simplest form
$$
B(r)=0,\;\; W(r)=r+(\lambda+1)GM,
$$
would lead to the following result:
$$
\delta\varphi =\f{6\pi GM}{L}
\left [
-\f{(1+\lambda)GM}{L}
(1+e^2) )+ \right.
$$
\begin{aequation}{12.72}{a}
\!\!\!\!\!\!\!\!\!\!\!\!\!\!\!\!\!\!\!\!\!
\!\!\!\!\!\!\!\!\!\!\!\!\!\!\!\!\!\!\!\!\!\!\!\!\!\!\!\!\!\!\!\!
+\f{9GM}{2L}
\left (
1+\f{e^2}{18}
\right )\Biggl.\Biggr]. \label{12.72a}
\end{aequation}
This expression is presented in monograph~[10], therein, also,
references to original articles can be found. From formula
(\ref{12.72a}) it can be seen that in GRT, also, ambiguity exists
in predicting the shift effect of Mercury's perihelion, but it 
manifests itself in the second order in $G$, instead of the first, and
therefore is beyond the accuracy limits of modern observational
data, if one is restricted to small values of the arbitrary
parameter $\lambda$. However, from the point of view of principle
it is seen that the ambiguity is also present in the case of such
a choice of $\lambda$, when Newton's law of gravity holds valid.
But from GRT it does not follow that the parameter $\lambda$
should be small. Since in the solution (12.31a) the arbitrary
functions $B(r)$ and $W(r)$ are not determined in GRT, therefore
for a chosen partial case the parameter $\lambda$ can assume any
values. If it is chosen sufficiently large, so that the second
term in the brackets in expression (\ref{12.72a}) is of the order
of $10^{-1}$, we would arrive at a contradiction both with
observational data on the shift of Mercury's perihelion and with
Newton's universal law of gravity. But this means, owing to the
arbitrariness of $\lambda$, that Newton's law is not the only
possible consequence of GRT. If it were unknown to us, then from
GRT, as a theoretical scheme, we would never obtain neither it,
nor any corrections to it. The maximum, that we could establish,
would be the asymptotics at infinity. {\bf All this reveals that
although GRT happened to be an important landmark in gravity after
the works of I.~Newton, it nevertheless turned out to be an
incomplete scheme, from the point of view of both its physical
aspects and its main equations, applied for explaining and
predicting gravitational phenomena.}

After the
sharp~~ criticism~~ of GRT~~ (in the twenties)~~ by P.~Pain\-lev\'e and
A.~Gullstrand
concerning the ambiguity in determining gravitational effects,
V.A.~Fock
(in the thirties) clearly understood the essence of GRT and its not
complete
definiteness. While studying island systems in the distribution of
matter in
GRT, V.A.~Fock added to the Hilbert--Einstein equations harmonic
coordinate
conditions (actually, certain equations were taken in Galilean
coordinates
of Minkowski space, and thus departure beyond the limits of GRT was
performed) and obtained a complete set of gravitational equations. In
RTG,
in studying island systems of the distribution of matter precisely
such a
set of equations arises in an inertial reference system (in Galilean
coordinates) from the least action principle. Thus, it becomes clear,
why
the harmonic conditions in Galilean (Cartesian) coordinates are
universal
equations.

In studying island systems, A.~Ein\-stein and L.~Infeld applied
other coordinate conditions, however, in the post-New\-to\-ni\-an
approximation
they are close to the harmonic conditions, and therefore within this
approximation they yield the same result. Thus, V.A.~Fock's theory of
gravity permitted to unambiguously determine all the effects in the
Solar
system. But his approach was not consistent.

{\bf The RTG way consists in total renunciation of\break A.~Einstein's
ideas on
inertia and gravity and returning to the physical gravitational field
in
the spirit of Faraday--Maxwell, exact conservation of special
relativity
theory, proclaiming a universal conserved quantity --- the
energy-momentum
tensor of all matter, including the gravitational field, the
source of the gravitational field.} Precisely such an approach leads
to a
new set of equations of the theory of gravity, removes the
fundamental
difficulties of GRT, discards ambiguity in the determination of
gravitational effects, predicts an another (unlike GRT) development
of the
collapse and of the Universe, and at the same time retains what is
most
valuable in GRT: the tensor character of gravity and Riemannian space.
But
now it already stops being the starting point and fundamental, but
becomes
only effective, that arises because the energy-momentum tensor of all
matter, including the gravitational field, is the source of the
gravitational field.

All this is reflected in the complete set of gravitational equations
(5.19)
and (5.20), that differ from the GRT equations. The effective Riemannian
space,
that arises in RTG owing to the influence of the gravitational field,
only
has a simple topology. This means that, in principle, no ``miracles",
that
are possible in GRT owing to the complex topology of Riemannian space,
can take
place in RTG.

\subsection{The precession of a gyroscope}
In the works of Pugh ~\cite{23} and Schiff~ \cite{24} a proposal was
made to
put a gyroscope on an orbit around the Earth and to examine its
precession
for studying the Earth's gravitational field and for testing general
relativity theory. Precisely in this effect, was the existence to be
revealed of an inertial reference system connected with distant
stars. For
simplicity, we shall consider the gyroscope to be a pointlike test
body. The
equation for the angular momentum of the gyroscope, $S_\mu$, has the
following form:
\be
\f{dS_\mu}{ds}=\Gamma^\lambda_{\mu\nu} S_\lambda
\f{dx^\nu}{ds}.\label{12.86}
\ee
In a reference system connected with the gyroscope it undergoes no
precession, which is reflected in equation (\ref{12.86}). The angular
momentum of the gyroscope, $\vec J$, does not change in value, below
it
will be shown to be expressed via the angular momentum $\vec S$ and
the
velocity $\vec v$. In the rest frame of the test body $S_\mu=(0,\vec
S)$,
and therefore
\be
S_\mu U^\mu =0,\;\; U^\mu=\f{dx^\mu}{ds}.\label{12.87}
\ee
From equality (12.87) we obtain
\be
S_0=-\f{1}{c}S_iv^i,\;\; v^i=\f{dx^i}{dt}.\label{12.88}
\ee
Equation (\ref{12.86}) for $\mu=i$ assumes the form
\be
\f{dS_i}{dt}=c\Gamma^j_{i0}S_j-\Gamma^0_{i0}v^jS_j+
\Gamma^j_{ik}v^k S_j-
\Gamma^0_{ik}v^{j} S_j\f{1}{c}. \label{12.89}
\ee

For a static spherically symmetric source of the gravitational
field in the linear approximation in the gravitational constant we
have
\be
g_{00}=1+2\Phi,\;\; g_{11}=g_{22}=g_{33}=1-2\Phi,\;\;
\Phi=-\f{GM}{c^2r}.
\label{12.90}
\ee
Applying these expressions we calculate the Christoffel symbols
\ba
&&\Gamma^j_{i0}=0,\;\; \Gamma^0_{ik}=0,\;\;
\Gamma^0_{i0}=\f{\pa \Phi}{\pa x^i},\;\;\nonumber \\
[-0.2cm] \label{12.91} \\ [-0.2cm]
&&\Gamma^j_{ik}=\f{\pa \Phi}{\pa x^k}\delta_{ij}-
\f{\pa \Phi}{\pa x^i}\delta_{jk}+
\f{\pa \Phi}{\pa x^j}\delta_{ik}.\nonumber
\ea
Substituting these expressions into equation (\ref{12.89}) we obtain
\be
\f{d\vec S}{dt}=-2(\vec v\vec S)\nabla \Phi
-(\vec v\nabla\Phi) \vec S+(\vec S\nabla\Phi)\vec v.\label{12.92}
\ee
The following expression will be the integral of motion of this
equation:
\be
\vec J^2=\vec S^2+2\Phi\vec S^2-(\vec v\vec S)^2.\label{12.93}
\ee
This is readily verified by differentiating it with respect to time:
\ba
&&2\vec J\f{d\vec J}{dt}=2\vec S\f{d\vec S}{dt}+4\Phi
\vec S\f{d\vec S}{dt} +\nonumber \\
&&+2(\vec v\vec S)
\left\{(\vec S\nabla\Phi)
+\vec v\left (\f{d\vec S}{dt}\right)\right\}. \label{12.94}
\ea
Retaining the principal terms in this expression, we have
\be
2\vec J\f{d\vec J}{dt}=2\vec S\f{d\vec S}{dt}+2
(\vec v\vec S)(\vec S\nabla\Phi).\label{12.95}
\ee
Multiplying equation (\ref{12.92}) by $\vec S$ and retaining the
principal
terms, we obtain
\be
\vec S\f{d\vec S}{dt}=-(\vec v\vec S)(\vec S\nabla\Phi).\label{12.96}
\ee
Substituting this expression into (\ref{12.95}) we find
\be
\vec J\f{d\vec J}{dt}=0.\label{12.97}
\ee

Thus, we have established that expression (\ref{12.93}) is an
integral of
motion of equation (\ref{12.92}). On the basis of (\ref{12.93}) it is
possible to construct the vector $\vec J$. Within the limits of
accuracy
it has the form
\be
\vec J=(1+\Phi)\vec S-\frac{1}{2}\vec v(\vec v\vec S). \label{12.98}
\ee
Differentiating (\ref{12.98}) with respect to time, within the limits
of our accuracy, we obtain
\be
\f{d\vec J}{dt}=[\vec\Omega,\vec J],\;\;
\vec\Omega =-\f{3}{2}[\vec v,\nabla\Phi].\label{12.99}
\ee

The vector $\vec J$, while remaining the same in absolute value,
undergoes
precession with a velocity $|\vec\Omega|$ about the direction of
vector
$\vec\Omega$. At present such an experiment is at the stage of
preparation.
The precession of a gyroscope, determined by formula (\ref{12.99}),
shows
that a reference system, connected with a gyroscope undergoing free
motion, is not inertial. From the point of view of RTG this is
obvious,
since the motion of a gyroscope in the gravitational field represents
an
accelerated motion with respect to the inertial reference system
related to
distant stars. Precisely for this reason, the reference system
connected
with the gyroscope will be non-inertial, which is what causes
precession of
the gyroscope. In GRT a reference system connected with a gyroscope,
undergoing free motion, is considered inertial. But then it is
absolutely
unclear, why this inertial reference system rotates with an angular
velocity
of $|\vec\Omega|$ relative to distant stars.

\subsection{The gravitational shift\hfill\break of spectral lines}
Consider a stationary gravitational field, i.e. when the metric
coefficients
are independent of time. Let radiation be emitted from point $e$ by
the
source, and let it be received at point $p$ by a receiver. If the
source
emits radiation during a time interval $(dt)_e$, then the receiver
will
also perceive it during an identical time interval, since the
gravitational field is stationary.

The proper time at point $e$ is
\be
(d\tau)_e=(\sqrt{g_{00}}dt)_e,\label{12.100}
\ee
and at point $p$ it will be
\be
(d\tau)_p=(\sqrt{g_{00}}dt)_p.\label{12.101}
\ee
But, since the time $(dt)_e=(dt)_p$, from formulae (\ref{12.100}) and
(\ref{12.101}) we obtain
\be
\f{(d\tau)_e}{(d\tau)_p}=\sqrt{\f{(g_{00})_e}{(g_{00})_p}}.
\label{12.102}
\ee

Thus, the proper time interval, during which the source emits the
signal,
is not equal to the proper time interval, during which the signal is
received, since the gravitational field differs from point $e$ to
$p$.

If we pass to the light frequency $\omega$, then we obtain
\be
\f{\omega_e}{\omega_p}=\sqrt{\f{(g_{00})_p}{(g_{00})_e}}.
\label{12.103}
\ee
Here $\omega_e$ is the frequency of light measured at the source
point $e$,
and $\omega_p$ is the frequency of the light that arrives from point
$e$ and
is measured at the receiver point $p$. The change in frequency is
characterized by the quantity
\be
\f{\delta\omega}{\omega}=\f{\omega_e-\omega_p}{\omega_p}.
\label{12.104}
\ee
On the basis of (\ref{12.103}) and (\ref{12.104}) we find
\be
\f{\delta\omega}{\omega}=\sqrt{\f{(g_{00})_p}{(g_{00})_e}}-1.
\label{12.105}
\ee
For a weak gravitational field we have
\be
g_{00}=1-2U. \label{12.106}
\ee
Substituting this expression into (\ref{12.105}) we obtain
\be
\f{\delta\omega}{\omega}=U_e-U_p.\label{12.107}
\ee

If the source (for example, an atom) is in a strong gravitational
field,
and the receiver is in a weaker field, then a red shift is observed,
and
the quantity $\delta\omega/\omega$ will be positive.

\thispagestyle{empty}
\newpage
\section{Some other physical conclusions of RTG}

At large distances $r$ from a static spherically symmetric body the
metric
coefficients have the form
\ba
&&U(r)=1-\frac{2M}{r}e^{-mr},\,\,\,\,
V(r)=1+\frac{2M}{r}e^{-mr},\,\,\,\,\,\nonumber \\[0.1cm]
&&W=r\left (1+\frac{M}{r}e^{-mr}\right ).\nonumber
\ea
We shall now deal with the problem of radiation of weak gravitational
waves,
when the graviton has mass. It has long been well known that in
linear
tensor theory introduction of the graviton mass is always accompanied
by
``ghosts". However, in refs.~\cite{25, 25_2,25_1} it is shown that the
intensity of the gravitational radiation of massive gravitons in nonlinear
theory is
a positive definite quantity, equal to
\ba
&&\frac{dI}{d\Omega}=\frac{2}{\pi}
\int\limits^{\infty}_{\omega_{\min}}d\omega\omega^2q
\{|T^1_2|^2+\frac{1}{4}|T^1_1-^2_2|^2+ \nonumber \\
&&+\frac{m^2}{\omega^2}
(|T^1_3|^2+|T^2_3|^2)+\frac{3m^4}{4\omega^4}
|T^3_3|^2\}\;, \label{425}
\ea
here $q=\left(1-\frac{m^2}{\omega^2}\right)^{1/2}$.

In RTG, like in GRT, outside matter the density of the
energy-momentum
tensor of the gravitational field in Riemannian space equals zero:
\be
T^{\mu\nu}_{g}=-2\frac{\delta L_g}{\delta g_{\mu\nu}}=0\;.
\label{426}
\ee
However, from expression (13.2) no absence of the gravitational field
follows. It is precisely in this expression, that the difference
between
the gravitational field and other physical fields is especially
revealed.
But this means that the energy flux of the gravitational field in the
theory of gravity is not determined by the density components of the
tensor $T_g^{0i}$, calculated with the aid of the solutions of
equations
(\ref{426}), since they are equal to zero. The problem of determining
the
energy flux in the theory of gravity, unlike other theories, requires
a different approach.

Yu.M.~Loskutov~\cite{25,25_2,25_1} finds the solution of (\ref{426})
in the form
\be
\tilde\Phi^{\mu\nu}=\chi^{\mu\nu}+\psi^{\mu\nu}\;, \label{427}
\ee
where the quantities $\chi^{\mu\nu}$ and $\psi^{\mu\nu}$ are of the
same
order of smallness, and $\psi^{\mu\nu}$ describes waves diverging
from the
source, while $\chi^{\mu\nu}$ characterizes the background. Energy
transport
is only realized by divergent waves. In ref.~\cite{25} it is shown
that
the flux of gravitational energy is actually determined by the
quantity
$T^{0i}_{g}(\psi)$ calculated not on the solutions of
equations
(\ref{426}) themselves, but only on that part of solutions, that describes
divergent
waves $\psi^{\mu\nu}$. Here it is taken into account that gravitons
do
not travel in Minkowski space, like in linear theory, but in
effective
Riemannian space. Therefore, in the linear approximation the following
equality is satisfied:
\ba
&&7\gamma_{\mu\nu}
\frac{dx^\mu}{ds}\cdot
\frac{dx^\nu}{ds}-1=
\frac{d\sigma^2-ds^2}{ds^2}\simeq\nonumber \\
&&\simeq -\frac{1}{2}\gamma_{\mu\nu}\Phi^{\mu\nu}
+\Phi^{\mu\nu}
\frac{dx^\alpha}{d\sigma}\cdot
\frac{dx^\beta}{d\sigma}\gamma_{\mu\alpha}\gamma_{\nu\beta}\;.
\ea
Precisely taking this circumstance into account consistently in the
course
of finding the intensity has led the author of ref.~\cite{25} to the
positively definite energy flux, determined by formula (\ref{425}),
the
obtained result is of fundamental importance, since it alters
conventional
ideas and, therefore, it necessarily requires further analysis.

It must be noted that the set of gravitational equations (\ref{76})
and
(\ref{77}) is hyperbolic, and precisely the causality principle
provides
for existence, throughout the entire space, of a spacelike surface,
which is
crossed by each nonspacelike curve in Riemannian space only once, i.e.,
in other
words, there exists a global Cauchy surface, precisely on which the
initial
physical conditions are given for one or another problem. R.~Penrose
and
S.~Hawking~\cite{12_1} proved, for certain general conditions,
singularity
existence theorems in GRT. On the basis of equations (\ref{78a})
outside
matter, for isotropic vectors of Riemannian space, by virtue of the
causality
conditions (\ref{91a}), the following inequality holds valid:
\be
R_{\mu\nu}v^{\mu} v^{\nu}\leq 0\;. \label{428}
\ee
The conditions of the aforementioned theorems are contrary to
inequality
(\ref{428}), so they are not applicable in RTG.

In RTG spacelike events in the absence of a gravitational field can
never
become timelike under the influence of the gravitational field. On
the
basis of the causality principle effective Riemannian space in RTG will
exhibit isotropic and timelike geodesic completeness. In accordance
with
RTG, an inertial reference system is determined by the distribution
of
matter and of the gravitational field in the Universe (Mach's
principle).

In GRT the fields of inertia and of gravity are inseparable.
A.~Einstein
wrote about this: ``{\it ... there exists no real division into inertia and
gravity,
since the answer to the question, of whether a body at a certain
moment is
exclusively under the influence of inertia or under the combined
influence
of inertia and gravity, depends on the reference system, i.e. on the
method
of dealing with it"}$\,$\footnote{Einstein A. Collection of
scientific
works, M.: Nauka, 1965, vol.1, art.33, p.422.}. Fields of inertia
satisfy the Hilbert-Einstein
equations. In RTG the gravitational field and the fields of inertia,
determined by the metric tensor of Minkowski space, are separated,
they
have nothing in common. They are of different natures. The fields of
inertia
are not solutions of RTG equations (\ref{76}) and (\ref{77}). In RTG
the
fields of inertia are given by the metric tensor $\gamma_{\mu\nu}$,
while
the gravitational field $\tilde\Phi^{\mu\nu}$ is determined from the
equations of gravity (\ref{76}) and (\ref{77}).

In conclusion it must be noted that the idea that one can arbitrarily
choose
both the geometry (G) and the physics (P), since the sum (G+P)
apparently
seems to be the sole test object in the experiment, is not quite
correct.
The choice of pseudo-Euclidean geometry with the metric tensor
$\gamma_{\mu\nu}$ is dictated both by fundamental physical principles
---
the integral conservation laws of energy-mo\-me\-n\-tum and of
angular
momentum,
and by other physical phenomena. Thus, physics (at the present-day
stage)
unambiguously determines the structure of the space-time geometry,
within
which all physical fields, including the gravitational field,
develop. In
accordance with RTG, the universal gravitational field creates
effective
Riemannian space with a simple topology, and Minkowski space does not
vanish,
here, but is manifested in the equations of theory and reflects a
fundamental principle --- the relativity principle. Effective Riemannian
space
is of a field nature.

\bigskip
On the basis of RTG one can draw the following general conclusions:
\\

Representation of the gravitational field as a physical field
possessing
the energy-momentum tensor, has drastically altered the general
picture of
gravity, earlier worked out on the basis of GRT. First, the theory of
gravity now occupies its place in the same row as other physical
theories
based on the relativity principle, i.e. the primary space is
Minkowski space.
Hence, it immediately follows that for all natural phenomena,
including
gravitational phenomena, there exist fundamental physical laws of
energy-momentum and angular momentum conservation. Since a universal
quantity --- the conserved energy-momentum tensor of matter
(including the
gravitational field) is the source of the gravitational field, then
there
arises effective Riemannian space-time, which is of a field nature.
Since the
formation of effective Riemannian space-time is due to the influence of
the
gravitational field, it automatically has a simple topology and is
described
in a sole coordinate system. The forces of inertia, unlike GRT, have
nothing
in common with the forces of gravity, since they differ in nature,
the
first arise owing to the choice of reference system in Minkowski
space,
while the latter are due to the presence of matter. The theory of the
gravitation, like all other physical theories (unlike GRT) satisfy the
equivalence principle.

Second, the complete set of equations of the theory of gra\-vi\-ty
permits
to determine unambiguously gravitational effects in the Solar system
and
leads to other (differing from those of GRT) predictions both on the
evolution of objects of large mass and on the development of a
homogeneous
and isotropic Universe. Theory reveals the formation of ``black holes"
(objects without material boundaries that are ``cut off" from the
external
world) to be impossible and predicts the existence in the Universe of
a large
hidden mass of ``dark" matter. From the theory it follows that there was
no
Big Bang, while some time in the past (about ten-fifteen billion
years ago)
there existed a state of high density and temperature, and the
so-called
``expansion" of the Universe, observed by the red shift, is not
related to
the relative motion of matter, but to variation in time of the
gravitational
field. Matter is at rest in an inertial reference system. The
peculiar
velocities of galaxies relative to inertial reference systems are due
to
inhomogeneities in the density distribution of matter, which is
precisely
what led to the accumulation of matter during the period, when the
Universe
became transparent.

The universal integral conservation laws of energy-mo\-men\-tum and
such
universal properties of matter, as, for example, gravitational
interactions,
are reflected in the metric properties of space-time. While the first
are embodied in the pseudo-Euclidean geometry of space-time, the
latter are
reflected in effective Riemannian geometry of space-time, that arose
owing to
the presence of the gravitational field in Minkowski space.\break
Eve\-ry\-thing that has
a character common to all matter can be considered as a part of the
structure of the effective geometry. But, here, Minkowski space will
be
present for certain, which is precisely what leads to the integral
conservation laws of energy-momentum and angular momentum, and, also,
provides for the equivalence principle to be satisfied, when the
gravitational field, as well as other fields, are switched off.

\thispagestyle{empty}
\newpage
\renewcommand{\theequation}{${\mbox{A}}.$\arabic{equation}}
\begin{flushright}
{\bf Appendix A}
\end{flushright}
\addcontentsline{toc}{section}{\hskip 1.8em Appendix A}
\setcounter{equation}{0}
\noindent
Let us establish the relation
\be
\frac{\delta L}
{\delta\gamma_{\mu\nu}}=
\frac{\delta L}
{\delta g_{\alpha\beta}}\cdot
\frac{\partial g_{\alpha\beta}}
{\partial\gamma_{\mu\nu}}+
\frac{\delta^\star L}
{\delta\gamma_{\mu\nu}}\;,
\ee
here
\be
\frac{\delta L}{\delta\gamma_{\mu\nu}}=
\frac{\partial L}{\partial\gamma_{\mu\nu}}-
\partial_\sigma\left (\frac{\partial
L}{\partial\gamma_{\mu\nu,\sigma}}
\right )\;,
\ee
\be
\frac{\delta L}{\delta g_{\mu\nu}}=
\frac{\partial L}{\partial g_{\alpha\beta}}-
\partial_\sigma\left (\frac{\partial L}{\partial
g_{\alpha\beta,\sigma}}
\right ),
\ee
the asterisk in the upper formula~(A.1)  indicates the variational
derivative of
the density of the Lagrangian with respect to the metric
$\gamma_{\mu\nu}$
explicitly occurring in $L$. Upon differentiation we obtain
\be
\frac{\partial L}{\partial\gamma_{\mu\nu}}=
\frac{\partial^\star L}{\partial\gamma_{\mu\nu}}+
\frac{\partial L}{\partial g_{\alpha\beta,\sigma}}\cdot
\frac{\partial g_{\alpha\beta,\sigma}}{\partial\gamma_{\mu\nu}}+
\frac{\partial L}{\partial g_{\alpha\beta}}\cdot
\frac{\partial g_{\alpha\beta}}{\partial\gamma_{\mu\nu}}\;,
\ee
\be
\frac{\partial L}{\partial\gamma_{\mu\nu,\sigma}}=
\frac{\partial^\star L}{\partial\gamma_{\mu\nu,\sigma}}+
\frac{\partial L}{\partial g_{\alpha\beta,\tau}}\cdot
\frac{\partial
g_{\alpha\beta,\tau}}{\partial\gamma_{\mu\nu,\sigma}}\;.
\ee
We substitute these expressions into formula (A.2):
\ba
&&\frac{\partial L}{\partial\gamma_{\mu\nu}}-
\partial_\sigma\left (\frac{\partial
L}{\partial\gamma_{\mu\nu,\sigma}}
\right )=
\frac{\delta^\star L}{\delta\gamma_{\mu\nu}}+
\frac{\partial L}{\partial g_{\alpha\beta,\sigma}}\cdot
\frac{\partial g_{\alpha\beta,\sigma}}{\partial\gamma_{\mu\nu}}+
\nonumber \\
&&+\frac{\partial L}{\partial g_{\alpha\beta}}\cdot
\frac{\partial g_{\alpha\beta}}{\partial\gamma_{\mu\nu}}-
\partial_\sigma\left (
\frac{\partial L}{\partial g_{\alpha\beta,\tau}}\cdot
\frac{\partial g_{\alpha\beta,\tau}}{\partial\gamma_{\mu\nu,\sigma}}
\right )=
\frac{\delta^\star L}{\delta\gamma_{\mu\nu}}+
\nonumber \\
&&+\frac{\partial L}{\partial g_{\alpha\beta}}\cdot
\frac{\partial g_{\alpha\beta}}{\partial\gamma_{\mu\nu}}-
\partial_\sigma
\left (\frac{\partial L}{\partial g_{\alpha\beta,\tau}}\right )\cdot
\frac{\partial g_{\alpha\beta,\tau}}{\partial\gamma_{\mu\nu,\sigma}}+
\nonumber \\
&&+\frac{\partial L}{\partial g_{\alpha\beta,\sigma}}
\left [
\frac{\partial g_{\alpha\beta,\sigma}}{\partial\gamma_{\mu\nu}}-
\partial_\rho\left (
\frac{\partial g_{\alpha\beta,\sigma}}{\partial\gamma_{\mu\nu,\rho}}
\right )\right ].
\ea
Now, consider expression
\be
\frac{\partial g_{\alpha\beta,\sigma}}{\partial\gamma_{\mu\nu}}
-\partial_\rho
\left (
\frac{\partial g_{\alpha\beta,\sigma}}{\partial\gamma_{\mu\nu,\rho}}
\right ).
\ee
For this purpose we shall write the derivative
$g_{\alpha\beta,\sigma}$
in the form
\be
g_{\alpha\beta,\sigma}=
\frac{\partial g_{\alpha\beta}}{\partial\gamma_{\lambda\omega}}
\partial_\sigma\gamma_{\lambda\omega}+
\frac{\partial g_{\alpha\beta}}{\partial\Phi_{\lambda\omega}}
\partial_\sigma\Phi_{\lambda\omega}\;,
\ee
hence it is easy to find
\be
\frac{\partial g_{\alpha\beta,\sigma}}{\partial\gamma_{\mu\nu,\rho}}=
\frac{\partial g_{\alpha\beta}}{\partial\gamma_{\mu\nu}}
\cdot \delta^\rho_\sigma\;.
\ee
Upon differentiating this expression we have
\be
\partial_{\rho}\left (
\frac{\partial g_{\alpha\beta,\sigma}}
{\partial\gamma_{\mu\nu,\rho}}\right )=
\frac{\partial^2
g_{\alpha\beta}}{\partial\gamma_{\mu\nu}\partial\gamma_{\lambda\omega
}}
\partial_\sigma\gamma_{\lambda\omega}+
\frac{\partial^2 g_{\alpha\beta}}{\partial\gamma_{\mu\nu}
\partial\Phi_{\lambda\omega}}
\partial_\sigma\Phi_{\lambda\omega}\;.
\ee

On the other hand, differentiating (A.8) with respect to
$\gamma_{\mu\nu}$
we have
\be
\frac{\partial g_{\alpha\beta,\sigma}}
{\partial\gamma_{\mu\nu}}=
\frac{\partial^2 g_{\alpha\beta}}
{\partial\gamma_{\mu\nu}\partial\gamma_{\lambda\omega}}
\partial_\sigma\gamma_{\lambda\omega}+
\frac{\partial^2 g_{\alpha\beta}}
{\partial\gamma_{\mu\nu}\partial\Phi_{\lambda\omega}}
\partial_\sigma\Phi_{\lambda\omega}\;.
\ee
Comparing (A.10) and (A.11) we find
\be
\frac{\partial g_{\alpha\beta,\sigma}}
{\partial\gamma_{\mu\nu}}-\partial_\rho\left (
\frac{\partial g_{\alpha\beta,\sigma}}
{\partial\gamma_{\mu\nu,\rho}}\right )=0\;.
\ee
Taking this relation into account, we obtain in (A.6)
\be
\frac{\delta L}{\delta\gamma_{\mu\nu}}=
\frac{\delta^\star L}{\delta\gamma_{\mu\nu}}+
\frac{\partial L}{\partial g_{\alpha\beta}}\cdot
\frac{\partial g_{\alpha\beta}}{\partial\gamma_{\mu\nu}}-
\partial_\sigma
\left (\frac{\partial L}{\partial g_{\alpha\beta,\tau}}\right )\cdot
\frac{\partial g_{\alpha\beta,\tau}}
{\partial\gamma_{\mu\nu,\sigma}}\;.
\ee
Substituting (A.9) into (A.13) we find
\be
\frac{\delta L}{\delta\gamma_{\mu\nu}}=
\frac{\delta^\star L}{\delta\gamma_{\mu\nu}}+
\left [
\frac{\partial L}{\partial g_{\alpha\beta}}-
\partial_\sigma
\left (\frac{\partial L}{\partial g_{\alpha\beta,\sigma}}\right
)\right ]
\frac{\partial g_{\alpha\beta}}
{\partial\gamma_{\mu\nu}}\;,
\ee
i.e.,
\be
\frac{\delta L}{\delta\gamma_{\mu\nu}}=
\frac{\delta^\star L}{\delta\gamma_{\mu\nu}}+
\frac{\delta L}{\delta g_{\alpha\beta}}\cdot
\frac{\partial g_{\alpha\beta}}{\partial
\gamma_{\mu\nu}}.
\ee
The following is calculated in a similar manner:
\be
\frac{\delta L}{\delta g_{\alpha\beta}}=
\frac{\delta L}{\delta\tilde g^{\lambda\rho}}\cdot
\frac{\partial\tilde g^{\lambda\rho}}{\partial
g_{\alpha\beta}}\;.
\ee
Making use of (A.16), one can write expression (A.15) as follows:
\be
\frac{\delta L}{\delta\gamma_{\mu\nu}}=
\frac{\delta^\star L}{\delta\gamma_{\mu\nu}}+
\frac{\delta L}{\delta \tilde g^{\lambda\rho}}\cdot
\frac{\partial \tilde g^{\lambda\rho}}{\partial
\gamma_{\mu\nu}}\;.
\ee

\thispagestyle{empty}
\newpage
\renewcommand{\theequation}{${\mbox{B}}.$\arabic{equation}}
\begin{flushright}
{\bf Appendix B}
\end{flushright}
\addcontentsline{toc}{section}{\hskip 1.8em Appendix B}
\setcounter{equation}{0}
\noindent
The density of the Lagrangian of the gravitational field proper has
the form
\ba
&&L_g =L_{g0}+L_{gm}\;,\\
[-0,3cm]\nonumber \\
&&L_{g0}=-\frac{1}{16\pi}\tilde g^{\alpha\beta}
\left
(G^\tau_{\lambda\alpha}G^\lambda_{\tau\beta}-G^\tau_{\alpha\beta}
G^\lambda_{\tau\lambda} \right),
\\
[-0,3cm] \nonumber \\
&&L_{gm}=-\frac{m^2}{16\pi}
\left (\frac{1}{2}\gamma_{\alpha\beta}\tilde g^{\alpha\beta}
-\sqrt{-g}-\sqrt{-\gamma}\right ).
\ea
The third-rank tensor $G^\tau_{\alpha\beta}$ is
\be
G^\tau_{\alpha\beta}=\frac{1}{2}g^{\tau\lambda}
(D_\alpha g_{\beta\lambda}+D_\beta g_{\alpha\lambda}-D_\lambda
g_{\alpha\beta})\;,
\ee
it is expressed via the Christoffel symbols of Riemannian space and of
Minkowski space:
\be
G^\tau_{\alpha\beta}=\Gamma^\tau_{\alpha\beta}-
\gamma^\tau_{\alpha\beta}\;.
\ee

Let us calculate the variational derivative of $L_g$ with respect to
the explicitly present metric of Minkowski space, $\gamma_{\mu\nu}$:
\be
\frac{\delta^\star L_{g0}}{\delta\gamma_{\mu\nu}}=
\frac{\partial L_{g0}}{\partial\gamma_{\mu\nu}}-\partial_\sigma
\left (
\frac{\partial L_{g0}}{\partial\gamma_{\mu\nu,\sigma}}\right ).
\ee
For this purpose we perform certain preparatory calculations:
\ba
\!\!\!\!\!\!&&\frac{\partial
G^\lambda_{\alpha\beta}}{\partial\gamma_{\mu\nu}}=
-\frac{\partial
\gamma^\lambda_{\alpha\beta}}{\partial\gamma_{\mu\nu}}
=\frac{1}{2}(\gamma^{\lambda\mu}\gamma^{\nu}_{\alpha\beta}
+\gamma^{\lambda\nu}\gamma^\mu_{\alpha\beta})\;,\nonumber
\\[-2.5mm]
 \\ [-2.5mm]
\!\!\!\!\!\!&&\frac{\partial
G^\lambda_{\alpha\lambda}}{\partial\gamma_{\mu\nu}}=
-\frac{\partial
\gamma^\lambda_{\alpha\lambda}}{\partial\gamma_{\mu\nu}}
=\frac{1}{2}(\gamma^{\lambda\mu}\gamma^{\nu}_{\lambda\alpha}
+\gamma^{\lambda\nu}\gamma^\mu_{\alpha\lambda})\;,
\nonumber
\ea
\ba
~~~~&&\frac{\partial
G^\lambda_{\alpha\beta}}{\partial\gamma_{\mu\nu,\sigma}}=
-\frac{\partial
\gamma^\lambda_{\alpha\beta}}{\partial\gamma_{\mu\nu,\sigma}}=
-\frac{1}{4}
\left[\gamma^{\lambda\mu} (\delta^\nu_\alpha\delta^\sigma_\beta+
\delta^\sigma_\alpha\delta^\nu_\beta)+\right. \nonumber \\
~~~~&&\left.+\gamma^{\lambda\nu}
 (\delta^\mu_\alpha\delta^\sigma_\beta+
\delta^\sigma_\alpha\delta^\mu_\beta)
- \gamma^{\lambda\sigma} (\delta^\mu_\alpha\delta^\nu_\beta+
\delta^\nu_\alpha\delta^\mu_\beta)\right]\;, \\
~~~~&&\frac{\partial
G^\lambda_{\alpha\lambda}}{\partial\gamma_{\mu\nu,\sigma}}=
-\frac{\partial
\gamma^\lambda_{\alpha\lambda}}{\partial\gamma_{\mu\nu,\sigma}}=
-\frac{1}{2}\gamma^{\mu\nu}\delta^\sigma_\alpha\;.\nonumber
\ea
Differentiating (B.2) we obtain
\ba
&&\frac{\partial L_{g0}}{\partial\gamma_{\mu\nu}}=
-\frac{1}{16\pi}\tilde g^{\alpha\beta}
\left [\frac{\partial
G^\tau_{\alpha\lambda}}{\partial\gamma_{\mu\nu}}
G^\lambda_{\tau\beta}+G^\tau_{\lambda\alpha}
\frac{\partial G^\lambda_{\tau\beta}}{\partial\gamma_{\mu\nu}}\,-
\right.\nonumber \\
&&\left.-\frac{\partial
G^\tau_{\alpha\beta}}{\partial\gamma_{\mu\nu}}
G^\lambda_{\tau\lambda}-G^\tau_{\alpha\beta}
\frac{\partial
G^\lambda_{\tau\lambda}}{\partial\gamma_{\mu\nu}}\right ].
\nonumber
\ea
Substituting into this expression formulae (B.7) we find
\ba
&&\frac{\partial L_{g0}}{\partial\gamma_{\mu\nu}}
=-\frac{1}{16\pi}\tilde g^{\alpha\beta}
\biggl \{G^\tau_{\lambda\alpha}\gamma^{\lambda\mu}
\gamma^\nu_{\tau\beta}+G^\tau_{\lambda\alpha}
\gamma^{\lambda\nu}\gamma^{\mu}_{\tau\beta}-
\frac{1}{2}G^\lambda_{\tau\lambda}\gamma^{\tau\mu}
\gamma^\nu_{\alpha\beta}-\nonumber
\\
\!&&-\frac{1}{2}G^\lambda_{\tau\lambda}
\gamma^{\tau\nu}\gamma^\mu_{\alpha\beta}\!-\!
\frac{1}{2}G^\tau_{\alpha\beta}\gamma^{\lambda\mu}
\gamma^\nu_{\tau\lambda}
\!-\!\frac{1}{2}G^\tau_{\alpha\beta}\gamma^{\lambda\nu}
\gamma^\mu_{\tau\lambda}\biggr\}
\!=\!\frac{1}{32\pi}B^{\mu\nu}.\!\!
\ea
With the aid of the derivatives (B.8) we obtain
\be
\frac{\partial L_{g0}}{\partial\gamma_{\mu\nu,\sigma}}
=\frac{1}{32\pi}A^{\sigma\mu\nu},\,\,
\ee
where
$$
\vspace{1mm}
\!\!\!\!\!\!\!\!\!\!\!\!\!\!\!\!\!\!\!\!\!\!\!
A^{\sigma\mu\nu}=\gamma^{\tau\mu}
(G^\sigma_{\tau\beta}\tilde g^{\nu\beta}
+G^\nu_{\tau\beta}\tilde g^{\sigma\beta}-G^\lambda_{\tau\lambda}
\tilde g^{\sigma\nu})-
\vspace{1mm}
$$
$$
\vspace{1mm}
-\gamma^{\mu\nu}G^\sigma_{\alpha\beta}\tilde
g^{\alpha\beta}+
\gamma^{\tau\nu}(G^\sigma_{\tau\beta}\tilde g^{\mu\beta}
+G^\mu_{\tau\beta}\tilde g^{\sigma\beta}-
G^\lambda_{\tau\lambda}\tilde g^{\sigma\mu}) +
$$
$$
\vspace{1mm}
\!\!\!\!\!\!\!\!\!\!\!\!\!\!\!\!\!\!\!\!\!\!\!\!\!\!\!\!\!\!\!\!\!\!
\!
\!\!
+
\gamma^{\tau\sigma}
(G^\lambda_{\tau\lambda}\tilde g^{\mu\nu}
-G^\mu_{\tau\beta}\tilde g^{\nu\beta}-G^\nu_{\tau\beta}\tilde
g^{\mu\beta}).
\eqno{\mbox{(B.10$'$)}}
\vspace{1mm}
$$

The density of the tensor $A^{\sigma\mu\nu}$ is symmetric in indices
$\mu$ and $\nu$. The ordinary derivative of this density can be
represented
in the form
$$
\partial_\sigma A^{\sigma\mu\nu}=D_\sigma A^{\sigma\mu\nu}-
\gamma^\mu_{\sigma\rho}A^{\sigma\rho\nu}-\gamma^\nu_{\sigma\rho}
A^{\sigma\mu\rho}.
$$
Substituting into (B.6) expressions (B.9) and (B.10) we find
\ba
&&\frac{\delta^\star L_{g0}}{\delta\gamma_{\mu\nu}}=
\frac{1}{32\pi}B^{\mu\nu}-\frac{1}{32\pi}D_\sigma A^{\sigma\mu\nu}+
\nonumber
\\
&&+\frac{1}{32\pi}\gamma^\mu_{\sigma\rho}A^{\sigma\rho\nu}
+\frac{1}{32\pi}\gamma^\nu_{\sigma\rho}A^{\sigma\mu\rho}.
\ea
Now, we write the density of the tensor $A^{\sigma\rho\nu}$ in the
form
\ba
&&A^{\sigma\rho\nu}=(G^\sigma_{\tau\beta}\gamma^{\tau\rho}
\tilde g^{\nu\beta}-G^\rho_{\tau\beta}\gamma^{\tau\sigma}
\tilde g^{\nu\beta})+
(G^\nu_{\tau\beta}\gamma^{\tau\rho}
\tilde g^{\sigma\beta}\!-
G^\nu_{\tau\beta}\gamma^{\tau\sigma}
\tilde g^{\rho\beta})\! -\nonumber \\
&&\,-(G^\lambda_{\tau\lambda}\gamma^{\tau\rho}
\tilde g^{\sigma\nu}-G^\lambda_{\tau\lambda}\gamma^{\tau\sigma}
\tilde g^{\rho\nu})+G^\sigma_{\tau\beta}\gamma^{\tau\nu}
\tilde g^{\rho\beta}+G^\rho_{\tau\beta}\gamma^{\tau\nu}\tilde
g^{\sigma\beta} -
\nonumber \\
&&-G^\lambda_{\tau\lambda}\gamma^{\tau\nu}\tilde g^{\sigma\rho}
-G^\sigma_{\alpha\beta}\gamma^{\rho\nu}\tilde
g^{\alpha\beta},\nonumber
\ea
in the brackets, terms antisymmetric in indices $\sigma$ and $\rho$
have
been formed. Writing in such a way facilitates finding the expression
for
the quantity $\gamma^\mu_{\sigma\rho}A^{\sigma\rho\nu}$, since the
terms
antisymmetric in indices $\sigma$ and $\rho$ vanish automatically,
here.
\ba
&&\gamma^\mu_{\sigma\rho}A^{\sigma\rho\nu}=
2G^\sigma_{\tau\beta}\gamma^\mu_{\sigma\rho}\gamma^{\tau\nu}
\tilde g^{\rho\beta}-\nonumber \\
&&-
G^\lambda_{\tau\lambda}\gamma^\mu_{\sigma\rho}
\gamma^{\tau\nu}\tilde
g^{\sigma\rho}-G^\sigma_{\alpha\beta}\gamma^\mu_{\sigma\rho}
\gamma^{\nu\rho}\tilde g^{\alpha\beta}.
\ea

Representing in a similar manner $A^{\sigma\mu\rho}$ as
\ba
&&A^{\sigma\mu\rho}=(G^\sigma_{\tau\beta}\gamma^{\tau\rho}
\tilde g^{\mu\beta}\!-\!
G^\rho_{\tau\beta}\gamma^{\tau\sigma}
\tilde g^{\mu\beta})+
(G^\mu_{\tau\beta}\gamma^{\tau\rho}
\tilde g^{\sigma\beta}\!-\!
G^\mu_{\tau\beta}\gamma^{\tau\sigma}
\tilde g^{\rho\beta})+
\nonumber \\
&&
+(G^\lambda_{\tau\lambda}\gamma^{\tau\sigma}
\tilde g^{\mu\rho}-
G^\lambda_{\tau\lambda}\gamma^{\tau\rho}
\tilde g^{\sigma\mu})+
G^\sigma_{\tau\beta}\gamma^{\tau\mu}
\tilde g^{\rho\beta}+
G^\rho_{\tau\beta}\gamma^{\tau\mu}
\tilde g^{\sigma\beta}-
\nonumber \\
&&-G^\lambda_{\tau\lambda}\gamma^{\tau\mu}\tilde
g^{\sigma\rho}-G^\sigma_{\alpha\beta}
\gamma^{\mu\rho}\tilde g^{\alpha\beta},\nonumber
\ea
where again in the brackets terms antisymmetric in the indices
$\sigma$ and
$\rho$ are formed, we obtain
\ba
&&\gamma^\nu_{\sigma\rho}A^{\sigma\mu\rho}=2G^\sigma_{\tau\beta}
\gamma^\nu_{\sigma\rho}\gamma^{\tau\mu}
\tilde g^{\rho\beta}-\nonumber \\
&&-G^\lambda_{\tau\lambda}\gamma^\nu_{\sigma\rho}
\gamma^{\tau\mu}\tilde g^{\sigma\rho}-
G^\sigma_{\alpha\beta}\gamma^\nu_{\sigma\rho}
\gamma^{\mu\rho}\tilde g^{\alpha\beta}.
\ea
Summing (B.12) and (B.13) one readily verifies the following
equality:
\be
\gamma^\mu_{\sigma\rho}A^{\sigma\rho\nu}+
\gamma^\nu_{\sigma\rho}A^{\sigma\mu\rho}
=-B^{\mu\nu}. 
\ee
Taking this equality into account we write expression (B.11) in the
form
\be
\frac{\delta L_{g0}}{\delta \gamma_{\mu\nu}}=-\frac{1}{32\pi}
D_\sigma A^{\sigma\mu\nu}. 
\ee
Taking into account the equalities
$$
G^\lambda_{\tau\lambda}=\frac{1}{2}g^{\lambda\rho}D_\tau
g_{\lambda\rho},\,\,
D_\tau\sqrt{-g}=\sqrt{-g}G^\lambda_{\tau\lambda}\;,
$$
we find
\ba
&&G^\sigma_{\tau\beta}\tilde g^{\nu\beta}+G^\nu_{\tau\beta}
\tilde g^{\sigma\beta}-
G^\lambda_{\tau\lambda}\tilde g^{\sigma\nu}=-D_\tau
\tilde g^{\nu\sigma},\nonumber \\
&&G^\sigma_{\tau\beta}\tilde g^{\mu\beta}+G^\mu_{\tau\beta}
\tilde g^{\sigma\beta}-
G^\lambda_{\tau\lambda}\tilde g^{\sigma\mu}=-D_\tau
\tilde g^{\mu\sigma}
\\
&&G^\nu_{\tau\beta}\tilde g^{\mu\beta}+G^\mu_{\tau\beta}
\tilde g^{\nu\beta}-
G^\lambda_{\tau\lambda}\tilde g^{\mu\nu}=-D_\tau
\tilde g^{\mu\nu}.\nonumber
\ea
Substituting these expressions into (B.10$'$) we obtain
$$
A^{\sigma\mu\nu}=\gamma^{\tau\sigma}D_\tau
\tilde g^{\mu\nu}+\gamma^{\mu\nu}D_\tau\tilde g^{\tau\sigma}
-\gamma^{\tau\mu}D_\tau \tilde g^{\nu\sigma}
-\gamma^{\tau\nu}D_\tau\tilde g^{\mu\sigma}.
$$
Substituting this expression into (B.15) we find
\be
\frac{\delta^\star L_{g0}}{\delta\gamma_{\mu\nu}}=
\frac{1}{32\pi}J^{\mu\nu},\,\,
\ee
$$
\mbox{where}\;\; J^{\mu\nu}=-D_\sigma D_\tau
(\gamma^{\tau\sigma}\tilde g^{\mu\nu}
+\gamma^{\mu\nu}\tilde g^{\tau\sigma}-
\gamma^{\tau\mu}\tilde g^{\nu\sigma}-
\gamma^{\tau\nu}\tilde g^{\mu\sigma})\;.
$$
On the basis of (B.3) we have
$$
\frac{\delta^\star L_{gm}}{\delta\gamma_{\mu\nu}}=
-\frac{m^2}{32\pi}
(\tilde g^{\mu\nu}-\tilde\gamma^{\mu\nu})=-\frac{m^2}{32\pi}
\tilde\Phi^{\mu\nu}.\eqno{\mbox{(B.18)}}
$$

Thus, taking into account (B.1) and applying (B.17) and (B.18) we
find
$$
\frac{\delta^\star L_{g}}{\delta\gamma_{\mu\nu}}=
\frac{1}{32\pi}(J^{\mu\nu}
-m^2\tilde\Phi^{\mu\nu})\;,\eqno{\mbox{(B.19)}}
$$
and, consequently,
$$
-2\frac{\delta^\star L_{g}}{\delta\gamma_{\mu\nu}}=
\frac{1}{16\pi}(-J^{\mu\nu}+m^2\tilde\Phi^{\mu\nu})\;.
\eqno{\mbox{(B.20)}}
$$

\newpage
\begin{flushright}
{\bf Appendix B$^*$}
\end{flushright}
\addcontentsline{toc}{section}{\hskip 1.8em Appendix B$^*$}
\renewcommand{\theequation}{${\mbox{B}^*}\!.$\arabic{equation}}
\setcounter{equation}{0}
\noindent
In this Appendix we shall make use of expressions (B.2) and (B.3) for
the
density of the Lagrangian $L_{g0}$ and $L_{gm}$ in order to establish
the
following equalities:
\be
\f{\delta L_{g0}}{\delta\tilde g^{\alpha\beta}}=
-\f{1}{16\pi}R_{\alpha\beta},\;\;
\f{\delta L_{gm}}{\delta\tilde g^{\alpha\beta}}=
\f{m^2}{32\pi}(g_{\alpha\beta} -
\gamma_{\alpha\beta}),  \label{b1}
\ee
here, by definition, the tensors
$\f{\delta L_{g0}}{\delta\tilde g^{\alpha\beta}},\;\;
\f{\delta L_{gm}}{\delta\tilde g^{\alpha\beta}}$ are equal to
\be
\f{\delta L_{g0}}{\delta\tilde g^{\alpha\beta}}=
\f{\partial L_{g0}}{\partial\tilde g^{\alpha\beta}}
- \partial_\sigma
\f{\partial L_{g0}}{\partial \tilde g^{\alpha\beta}_{,\sigma}},\;\;
\f{\delta L_{gm}}{\delta\tilde g^{\alpha\beta}}
=\f{\partial L_{gm}}{\partial\tilde g^{\alpha\beta}}
-\partial_\sigma
\f{\partial L_{gm}}{\partial \tilde g^{\alpha\beta}_{,\sigma}}.
\label{b2}
\ee
The tensor relations (\ref{b1}) are most readily established in a
local
Riemann reference system, where the derivatives of the components of
the
metric tensor $g_{\mu\nu}$ with respect to the coordinates are zero
and,
consequently, the Christoffel symbols $\Gamma^\lambda_{\mu\nu}$ are
also zero.

On the basis of the formula
\be
\f{\partial \Gamma^\tau_{\lambda\alpha}}{\pa g_{\mu\nu}}
=-\f{1}{2} \left (
g^{\mu\tau}\Gamma^\nu_{\alpha\lambda}+g^{\nu\tau}
\Gamma^\mu_{\alpha\lambda}
\right ).
\label{b3}
\ee
it is easy to establish that in the indicated reference system the
following equality holds valid:
\ba
&&\f{\pa L_{g0}}{\pa g_{\mu\nu}}
=\f{\sqrt{-g}}{16\pi}
\left (
g^{\alpha\mu} g^{\beta\nu}
-\f{1}{2}
g^{\mu\nu}g^{\alpha\beta}
\right ))\times \nonumber \\
&&\times\left (
\gamma^\tau_{\lambda\alpha} \gamma^\lambda_{\tau\beta}
-\gamma^\tau_{\alpha\beta}\gamma^\lambda_{\tau\lambda}
\right ). \label{b4}
\ea
Here, $\gamma^\lambda_{\tau\beta}$ are the Christoffel symbols of
Minkowski space. Making use of
expression
\ba
&&\f{\partial\Gamma^\tau_{\alpha\lambda}}{\partial g_{\mu\nu,\sigma}}
=\f{1}{4}
\left \{
g^{\tau\mu}
(\delta^\nu_\alpha\delta^\sigma_\lambda
+\delta^\nu_\lambda \delta^\sigma_\alpha)
+g^{\tau\nu}
(\delta^\mu_\alpha\delta^\sigma_\lambda+
\right.\nonumber \\[1mm]
&&\left.
+\delta^\mu_\lambda \delta^\sigma_\alpha)
-g^{\tau\sigma}
(\delta^\mu_\alpha\delta^\nu_\lambda
+\delta^\mu_\lambda \delta^\nu_\alpha)
\right \}, \label{b5}
\ea
we obtain
\ba
&-&\f{\partial L_{g0}}{\partial g_{\mu\nu,\sigma}}=
\f{\sqrt{-g}}{16\pi}
\left  [
(g^{\alpha\mu} g^{\beta\nu} -\f{1}{2}
g^{\mu\nu} g^{\alpha\beta})
(\Gamma^\sigma_{\alpha\beta} -\gamma^\sigma_{\alpha\beta})-\right.
\nonumber\\[-0.2cm]
    \label{b6} \\ [-0.2cm]
&-& \left.(g^{\alpha\mu}g^{\sigma\nu}-\f{1}{2}
g^{\mu\nu}g^{\alpha\sigma})
(\Gamma^\lambda_{\alpha\lambda}-\gamma^\lambda_{\alpha\lambda})
\right].\nonumber
\ea
Hence in a local Riemann reference system we find
\ba
&-&\partial_\sigma\f{\partial L_{g0}}{\partial g_{\mu\nu,\sigma}}
=\f{\sqrt{-g}}{16\pi}
\left (
g^{\alpha\mu}g^{\beta\nu} -\f{1}{2}
g^{\mu\nu} g^{\alpha\beta}\right )\cdot \nonumber \\[-0.2cm]
 \label{b7} \\ [-0.2cm]
&&\times \left [
(\pa_\sigma\Gamma^\sigma_{\alpha\beta}
-\pa_\beta\Gamma^\lambda_{\alpha\lambda})
-(\pa_\sigma\gamma^\sigma_{\alpha\beta}
-\pa_\beta \gamma^\lambda_{\alpha\lambda})
\right ].  \nonumber
\ea
On the basis of (\ref{b4}) and (\ref{b7}) the tensor (\ref{b2}) in
the
local Riemann reference system is
\ba
&&\f{\delta L_{g0}}{\delta g_{\mu\nu}}
=\f{\pa L_{g0}}{\pa g_{\mu\nu}}
-\pa_\sigma \f{\pa L_{g0}}{\pa g_{\mu\nu,\sigma}}
=\f{\sqrt{-g}}{16\pi}
\left  (g^{\alpha\mu} g^{\beta\nu}-\right.\nonumber \\[-0.2cm]
 \label{b8} \\ [-0.2cm]
&&\left.-\f{1}{2}
g^{\mu\nu}g^{\alpha\beta} \right )
(\pa_\sigma\Gamma^\sigma_{\alpha\beta}
-\pa_\beta \Gamma^\lambda_{\alpha\lambda} - R_{\alpha\beta}
(\gamma)).\nonumber
\ea

In a local Riemann reference system the second-rank curvature
tensor of Riemann space, $R_{\alpha\beta}(g)$, has the form
\be
R_{\alpha\beta}(g)=\pa_\sigma \Gamma^\sigma_{\alpha\beta}
-\pa_\beta
\Gamma^\lambda_{\alpha\lambda}.\label{b9}
\ee
In expression (\ref{b8}), the second-rank tensor
$R_{\alpha\beta}(\gamma)$
is
$$
R_{\alpha\beta}(\gamma) =\pa_\sigma\gamma^\sigma_{\alpha\beta}
-\pa_\beta\gamma^\lambda_{\alpha\lambda} +\gamma^\tau_{\alpha\beta}
\gamma^\lambda_{\tau\lambda} -\gamma^\tau_{\lambda\alpha}
\gamma^\lambda_{\tau\beta}.
$$
In Minkowski space with the metric $\gamma_{\mu\nu}$ and Christoffel
symbols $\gamma^\sigma_{\mu\nu}$ the tensor $R_{\alpha\beta}(\gamma)$
equals
zero. Taking into account (\ref{b9}) and, also, that the tensor
$R_{\alpha\beta}(\gamma)$ equals zero, the tensor relation (\ref{b8})
assumes the form
\be
\f{\pa L_{g0}}{\pa g_{\mu\nu}}
=\f{\sqrt{-g}}{16\pi}
\left (
R^{\mu\nu}-\f{1}{2} g^{\mu\nu}R
\right ). \label{b10}
\ee
This equality has been established in a local Riemann reference
system,
but by virtue of its tensor character it holds valid in any reference
system.

Applying relation
\be
dg=-gg_{\alpha\beta} dg^{\alpha\beta}, \label{b11}
\ee
we find
\be
\f{\pa g}{\pa\tilde g^{\alpha\beta}}=g\tilde g_{\alpha\beta}.
\label{b12}
\ee
On the basis of equality
\be
\tilde g^{\mu\sigma}g_{\sigma\nu}=\delta^\mu_\nu\sqrt{-g} \label{b13}
\ee
it is easy to obtain the following relation:
\be
\f{\pa g_{\lambda\nu}}{\pa\tilde g^{\alpha\beta}}
=\f{1}{\sqrt{-g}}
\left \{
-\f{1}{2} (g_{\lambda\alpha} g_{\nu\beta}
+g_{\lambda\beta} g_{\nu\alpha})
+\f{1}{2} g_{\alpha\beta}g_{\nu\lambda}\right\}. \label{b14}
\ee
Since on the basis of Appendix A the following equality is valid:
\be
\f{\delta L_{g0}}{\delta \tilde g^{\alpha\beta}}
=\f{\delta L_{g0}}{\delta g_{\lambda\nu}}
\cdot \f{\pa g_{\lambda\nu}}{\pa\tilde g^{\alpha\beta}}, \label{b15}
\ee
then, upon applying expressions (\ref{b10}) and (\ref{b14}) we find
\be
\f{\delta L_{g0}}{\delta \tilde g^{\alpha\beta}}
=-\f{1}{16\pi} R_{\alpha\beta}. \label{b16}
\ee

In a similar manner we have
\be
\f{\delta L_{gm}}{\delta \tilde g^{\alpha\beta}}
=\f{\pa L_{gm}}{\pa \tilde g^{\alpha\beta}}
=\f{m^2}{32\pi} (g_{\alpha\beta}-\gamma_{\alpha\beta}). \label{b17}
\ee
Adding up expressions (\ref{b16}) and (\ref{b17}) we obtain
\be
\f{\delta L_g}{\delta \tilde g^{\alpha\beta}}
=-\f{1}{16\pi} R_{\alpha\beta}
+\f{m^2}{32\pi} (g_{\alpha\beta}-\gamma_{\alpha\beta}). \label{b18}
\ee
It is also easy to obtain the following relation:
\ba
&&\f{\delta L_{M}}{\delta \tilde g^{\alpha\beta}}
=\f{\delta L_{M}}{\delta  g_{\lambda\nu}}
\cdot \f{\pa g_{\lambda\nu}}{\pa\tilde g^{\alpha\beta}}
=-\f{1}{2} T^{\lambda\nu}
\f{\pa g_{\lambda\nu}}{\pa\tilde g^{\alpha\beta}}= \nonumber\\
&&=\f{1}{2\sqrt{-g}}
(T_{\alpha\beta} -\f{1}{2} g_{\alpha\beta}T).\label{b19}
\ea
Here $T^{\lambda\nu}=-2\f{\delta L_M}{\delta g_{\lambda\nu}}$ is the
density
of the energy-momentum tensor of matter in effective Riemannian space.
At  last we present the following relations:
\ba
&&D_\nu \sqrt{-g}=\pa_\nu\left (\sqrt{\f{g}{\gamma}}\right )
=\sqrt{-g} G^\lambda_{\nu\lambda},
\nonumber \\[-0.5cm]
 \label{b20} \\ [-0.2cm]
&&\pa_\nu \sqrt{-g}=\sqrt{-g}\Gamma^\lambda_{\nu\lambda},\;\;
\pa_\nu\sqrt{-\gamma}=\sqrt{-\gamma}\gamma^\lambda_{\nu\lambda},
\nonumber
\ea
that are applied in obtaining equality (\ref{72}).

\thispagestyle{empty}
\newpage
\renewcommand{\theequation}{${\mbox{C}}.$\arabic{equation}}
\begin{flushright}
{\bf Appendix C}
\end{flushright}
\addcontentsline{toc}{section}{ \hskip 1.8em Appendix C}
\setcounter{equation}{0}
\setcounter{equation}{0}
\noindent
For any given density of the Lagrangian $L$, in the case of an
infinitesimal
change in the coordinates, the variation of the action
$$
S =\int Ld^4x
$$
will be zero. We shall calculate the variation of the action of the
density
of the Lagrangian $L_M$
$$
S_M=\int L_M(\tilde g^{\mu\nu}, \Phi_A)d^4x
$$
of matter and establish a strong identity. In the case of the
following
transformation of the coordinates:
\be
x'^\mu=x^\mu+\xi^\mu(x)\;,
\ee
where $\xi^\mu(x)$ is an infinitesimal displacement four-vector, the
variation of the action under transformation of the coordinates
equals
\be
\delta_c S_M=\int d^4x
\left (\frac{\delta L_M}{\delta \tilde g^{\mu\nu}}
\delta_L\tilde g^{\mu\nu}+
\frac{\delta L_M}{\delta
\Phi_A}\delta_L\Phi_A+\mbox{div}\right)=0. 
\ee
In this expression div stands for the divergence terms, which are not
essential for our purposes.

The Euler variation is defined as usual:
$$
\frac{\delta L}{\delta\Phi}\equiv
\frac{\partial L}{\partial\Phi}-\partial_\mu
\frac{\partial L}{\partial(\partial_\mu\Phi)}+
\partial_\mu\partial_\nu
\frac{\partial L}{\partial(\partial_\mu\partial_\nu\Phi)}\;.
$$
The Lie variations $\delta_L\tilde g^{\mu\nu},\delta_L\Phi_A$ are
readily
calculated under changes of the coordinates, if the transformation
law of
the quantities $g^{\mu\nu},\Phi_A$ is applied:
\be
\begin{array}{lc}
\displaystyle
\delta_L\tilde g^{\mu\nu}=\tilde g^{\lambda\mu}
D_\lambda\xi^\nu +
\tilde g^{\lambda\nu}
D_\lambda\xi^\mu- D_\lambda
(\xi^\lambda\tilde g^{\mu\nu}), & \\
& \\
\displaystyle
\delta_L\Phi_A=
-\xi^\lambda D_\lambda\Phi_A+F^{B;\lambda}_{A;\sigma}
\Phi_BD_\lambda\xi^{\sigma}, & \\
\end{array}
\ee
$D_\lambda$ are covariant derivatives in Minkowski space.
Substituting these
expressions into (C.2) and integrating by parts we obtain
\ba
&&\delta S_M=
\int d^4x\left\{
-\xi^\lambda
\left [D_\alpha
\left (2\frac{\delta L_M}{\delta\tilde g^{\lambda\nu}}
\tilde g^{\alpha\nu}\right )
-D_\lambda
\left (\frac{\delta L_M}{\delta\tilde g^{\alpha\beta}}\right )
\tilde g^{\alpha\beta}+\right.\right. \nonumber \\ [1mm]
&&+\left.\left.
D_\sigma
\left (\frac{\delta
L_M}{\delta\Phi_A}F^{B;\sigma}_{A;\lambda}\Phi_B\right )+
\frac{\delta L_M}{\delta \Phi_A}D_\lambda\Phi_A\right
]+\mbox{div}\right\}=0\;.
\ea
Owing to the arbitrariness of vector $\xi^\lambda$, we derive from
this
equality a strong identity, which is valid independently of whether
the
equations of motion for the fields are satisfied or not. It has the
form
\ba
&&D_\alpha
\left (2\frac{\delta L_M}{\delta\tilde g^{\lambda\nu}}
\tilde g^{\alpha\nu}\right )-
D_\lambda
\left (\frac{\delta L_M}{\delta\tilde g^{\alpha\beta}}\right )
\tilde g^{\alpha\beta}=\nonumber \\
&&=-D_\sigma
\left (\frac{\delta L_M}{\delta\Phi_A}
F^{B;\sigma}_{A;\lambda}\Phi_B\right )-
\frac{\delta L_M}{\delta\Phi_A}D_\lambda\Phi_A.
\ea

We now introduce the notation
\ba
&&T_{\mu\nu}=2\frac{\delta L_M}{\delta g^{\mu\nu}},\,\,\,
T^{\mu\nu}=-2\frac{\delta L_M}{\delta g_{\mu\nu}}=
g^{\mu\alpha}g^{\nu\beta}T_{\alpha\beta},\,\,\,
\nonumber \\
&&T=T^{\mu\nu}g_{\mu\nu},\;\;
\tilde T_{\mu\nu}=2\frac{\delta L_M}{\delta\tilde g^{\mu\nu}},\\
&&\tilde T^{\mu\nu}=-2\frac{\delta L_M}{\delta \tilde g_{\mu\nu}}=
\tilde g^{\mu\alpha}
\tilde g^{\nu\beta}
\tilde T_{\alpha\beta},\,\,\,
\tilde T=\tilde T^{\alpha\beta}\tilde g_{\alpha\beta}.\nonumber
\ea
Taking into account this notation, one can write the left-hand side
of
identity (C.5) as
$$
D_\alpha (\tilde T_{\lambda\nu} \tilde g^{\alpha\nu})
-\frac{1}{2}\tilde g^{\alpha\beta}D_\lambda\tilde T_{\alpha\beta}=
\partial_\alpha (\tilde T_{\lambda\nu}\tilde g^{\alpha\nu})-
\frac{1}{2}\tilde g^{\alpha\beta}\partial_\lambda\tilde
T_{\alpha\beta}.
$$
The right-hand side of this equation is readily reduced to the form
\be
\partial_\alpha (\tilde T_{\lambda\nu}\tilde g^{\alpha\nu})
-\frac{1}{2}\tilde g^{\alpha\beta}\partial_\lambda\tilde
T_{\alpha\beta}=
\tilde g_{\lambda\nu}\nabla_\alpha
\left(\tilde T^{\alpha\nu}-\frac{1}{2}\tilde g^{\alpha\nu}\tilde
T\right)\;,
\ee
where $\nabla_\alpha$ is the covariant derivative in Riemannian space.
We
shall now represent the expression under the sign of the covariant
derivative $\nabla_\alpha$ in terms of the density of the
tensor $T^{\alpha\nu}$. To this end we take advantage of formula
(A.16):
\be
\frac{\delta L_M}{\delta g_{\mu\nu}}=
\frac{\delta L_M}{\delta \tilde g^{\alpha\beta}}\cdot
\frac{\partial \tilde g^{\alpha\beta}}{\partial
g_{\mu\nu}},
\ee
where
\be
\frac{\partial \tilde g^{\alpha\beta}}{\partial g_{\mu\nu}}=
\sqrt{-g}
\frac{\partial g^{\alpha\beta}}{\partial g_{\mu\nu}}-
\frac{1}{2\sqrt{-g}}\cdot
\frac{\partial g}{\partial
g_{\mu\nu}}g^{\alpha\beta}.
\ee
Using the relations
$$
g^{\alpha\beta}g_{\beta\sigma}=\delta^\alpha_\sigma,
$$
we find
\be
\frac{\partial g^{\alpha\beta}}{\partial g_{\mu\nu}}
=-\frac{1}{2}(g^{\alpha\mu}g^{\nu\beta}+g^{\alpha\nu}g^{\mu\beta})\;.
\ee
Applying the rule for differentiating determinants we find
\be
dg=gg^{\mu\nu}dg_{\mu\nu}, 
\ee
from which we find
\be
\frac{\partial g}{\partial g_{\mu\nu}}=gg^{\mu\nu}.
\ee
Substituting expressions (C.10) and (C.12) into (C.9) we obtain
\be
\frac{\partial \tilde g^{\alpha\beta}}{\partial g_{\mu\nu}}=
-\frac{1}{2}\sqrt{-g}
[g^{\alpha\mu}g^{\beta\nu}+g^{\alpha\nu}g^{\beta\mu}-
g^{\mu\nu}g^{\alpha\beta}]\;.
\ee
Using this relation in (C.8) we find
\be
\frac{\delta L_M}{\delta g_{\mu\nu}}=\sqrt{-g}
\left (\frac{\delta L_M}{\delta \tilde g^{\alpha\beta}}
g^{\alpha\mu}g^{\beta\nu}-\frac{1}{2}
\frac{\delta L_M}{\delta \tilde g^{\alpha\beta}}
g^{\alpha\beta}g^{\mu\nu}\right )\;.
\ee
With account of notation (C.6) this expression can be written in the
form
\be
\sqrt{-g}T^{\mu\nu}=\tilde T^{\mu\nu}-\frac{1}{2}\tilde
g^{\mu\nu}\tilde T.
\ee

On the basis of equality (C.15), the strong identity (C.5) assumes,
with
account of (C.7), the following form
\ba
\!\!\!\!\!\!\!\!\!\!&&g_{\lambda\nu}\nabla_\alpha
T^{\alpha\nu}=-D_\sigma
\left (\frac{\delta L_M}{\delta
\Phi_A}F^{B;\sigma}_{A;\lambda}\Phi_B\right )-
\frac{\delta L_M}{\delta \Phi_A}D_\lambda \Phi_A,
\;\;\mbox{or}\nonumber  \\[-0.2cm]
\!\!\!\!\!\!\!\!\!\!&&\phantom{aaa} \\ [-0.2cm]
\!\!\!\!\!\!\!\!\!\!&&\nabla_\alpha T^\alpha_\lambda=-D_\sigma
\left (\frac{\delta L_M}{\delta
\Phi_A}F^{B;\sigma}_{A;\lambda}\Phi_B\right )-
\frac{\delta L_M}{\delta \Phi_A}D_\lambda\Phi_A.\nonumber
\ea

\thispagestyle{empty}
\newpage
\renewcommand{\theequation}{${\mbox{D}}.$\arabic{equation}}
\setcounter{equation}{0}
\begin{flushright}
{\bf Appendix D}
\end{flushright}
\addcontentsline{toc}{section}{\hskip 1.8em Appendix D}

\noindent A second-rank curvature tensor $R_{\mu\nu}$ can be
written in the form \ba &&R_{\mu\nu}=\frac{1}{2} [\tilde
g^{\alpha\beta} (\tilde g_{\mu \kappa}\tilde g_{\nu \rho}
-\frac{1}{2}\tilde g_{\mu \nu}\tilde g_{\kappa\rho}) D_\alpha
D_\beta\tilde g^{\kappa\rho}- \nonumber \\ &&-\tilde
g_{\nu\rho}D_\kappa D_\mu \tilde g^{\kappa\rho} -\tilde
g_{\mu\kappa} D_\nu D_\rho \tilde g^{\kappa\rho}]
+\frac{1}{2}\tilde g_{\nu\omega} \tilde g_{\rho\tau} D_\mu \tilde
g^{\kappa\rho}D_\kappa \tilde g^{\omega\tau}+ \nonumber \\
&&+\frac{1}{2}\tilde g_{\mu\omega}\tilde g_{\rho\tau}D_\nu \tilde
g^{\kappa\rho}D_\kappa\tilde g^{\omega\tau} -\frac{1}{2}\tilde
g_{\mu\omega}\tilde g_{\nu\rho}D_\tau \tilde
g^{\omega\kappa}D_\kappa \tilde g^{\rho\tau}- \nonumber \\
&&-\frac{1}{4} (\tilde g_{\omega\rho}\tilde g_{\kappa\tau}
-\frac{1}{2}\tilde g_{\omega\tau}\tilde g_{\kappa\rho})D_\mu
\tilde g^{\kappa\rho}D_\nu \tilde g^{\omega\tau}- \nonumber \\
&&-\frac{1}{2}\tilde g^{\alpha\beta}\tilde g_{\rho\tau} (\tilde
g_{\mu\kappa}\tilde g_{\nu\omega}- \frac{1}{2}\tilde
g_{\mu\nu}\tilde g_{\kappa\omega}) D_\alpha \tilde
g^{\kappa\rho}D_\beta \tilde g^{\omega\tau}. \ea Raising the
indices by multiplying by $g^{\epsilon\mu}g^{\lambda\nu}$ and
taking into account the equation
\be
D_\mu \tilde g^{\mu\nu}=0\;,
\ee
we obtain
\ba
&&-gR^{\epsilon\lambda}=\frac{1}{2} \tilde g^{\alpha\beta}D_\alpha
D_\beta \tilde g^{\epsilon\lambda}-\frac{1}{4}
\tilde g^{\epsilon\lambda}
\tilde g_{\kappa\rho}\tilde g^{\alpha\beta}D_\alpha D_\beta
\tilde g^{\kappa\rho}+
\nonumber \\
&&+\frac{1}{2}\tilde g_{\rho\tau}\tilde g^{\epsilon\mu}D_\mu
\tilde g^{\kappa\rho}D_\kappa \tilde g^{\lambda\tau}+
\nonumber \\
&&+\frac{1}{2}\tilde g_{\rho\tau} \tilde g^{\lambda\nu}D_\nu
\tilde g^{\kappa\rho}D_\kappa \tilde g^{\epsilon\tau}
-\frac{1}{2}D_\tau \tilde g^{\epsilon\kappa}D_\kappa
\tilde g^{\lambda\tau}-\nonumber \\
 &&-\frac{1}{4}(\tilde g_{\omega\rho}\tilde
 g_{\kappa\tau}-
\frac{1}{2}\tilde g_{\omega\tau} \tilde g_{\kappa\rho})
\tilde g^{\epsilon\mu}\tilde g^{\lambda\nu}D_\mu
\tilde g^{\kappa\rho} D_\nu\tilde g^{\omega\tau}-
\nonumber \\
&&-\frac{1}{2}\tilde g_{\rho\tau}\tilde
 g^{\alpha\beta}D_\alpha
\tilde g^{\epsilon\rho}D_\beta \tilde g^{\lambda\tau}
\!\!+\!\frac{1}{4} \tilde g_{\rho\tau}\tilde g^{\epsilon\lambda}
\tilde g_{\kappa\omega}\tilde g^{\alpha\beta}D_\alpha
\tilde g^{\kappa\rho}D_\beta \tilde g^{\omega\tau}\!\!.
\ea
Hence, we find
\ba
&&-gR=\frac{1}{2} g_{\epsilon\lambda}
\tilde g^{\alpha\beta}D_\alpha D_\beta
\tilde g^{\epsilon\lambda}-g_{\kappa\rho}
\tilde g^{\alpha\beta}D_\alpha D_\beta
\tilde g^{\kappa\rho}+\frac{1}{2} g_{\rho\tau}
D_\mu \tilde g^{\kappa\rho} D_\kappa\tilde g^{\mu\tau} \!\!\!-
\nonumber \\
&&-\frac{1}{4}(\tilde g_{\omega\rho}\tilde g_{\kappa\tau}-
\frac{1}{2}\tilde g_{\omega\tau}\tilde g_{\kappa\rho})
\sqrt{-g} \tilde g^{\mu\nu} D_\mu \tilde g^{\kappa\rho}
D_\nu \tilde g^{\omega\tau}- 
\nonumber \\
&&-\frac{1}{2} \tilde g_{\rho\tau}\tilde g^{\alpha\beta}
g_{\epsilon\lambda} D_\alpha \tilde g^{\epsilon\rho}
D_\beta \tilde g^{\lambda\tau}+
\tilde g_{\rho\tau}g_{\kappa\omega}
\tilde g^{\alpha\beta}D_\alpha
\tilde g^{\kappa\rho} D_\beta \tilde g^{\omega\tau}.
\ea
With the aid of expressions (D.3) and (D.4) we find
\ba
&&-g(
\left(R^{\epsilon\lambda}-
\frac{1}{2}g^{\epsilon\lambda}R\right)
=\nonumber \\
&&=
-\frac{1}{2}\Biggl \{
\frac{1}{2}
\left  (\tilde g_{\nu\sigma} \tilde g_{\tau\kappa}
\frac{1}{2} \tilde g_{\nu\kappa}\tilde g_{\tau\sigma} \right )
\tilde g^{\epsilon\alpha}\tilde g^{\lambda\beta} D_\alpha
\tilde g^{\sigma\tau} D_\beta \tilde g^{\nu\kappa}-\Biggr.
\nonumber
\\
&&-\frac{1}{4} \tilde g^{\epsilon\lambda}
\tilde g^{\alpha\beta}
\left (\tilde g_{\nu\sigma} \tilde g_{\tau\kappa}-
\frac{1}{2} \tilde g_{\nu\kappa} \tilde g_{\tau\sigma}\right )
D_\alpha
\tilde g^{\tau\sigma}D_\beta \tilde g^{\nu\kappa}+\nonumber  \\
&&
+\tilde g^{\alpha\beta} \tilde g_{\sigma\tau}D_\alpha
\tilde g^{\epsilon\tau}D_\beta \tilde g^{\lambda\sigma}-
\tilde g^{\epsilon\beta}\tilde g_{\tau\sigma}D_\alpha
\tilde g^{\lambda\sigma}D_\beta \tilde g^{\alpha\tau}-
\nonumber \\
&&
-\tilde g^{\lambda\alpha} \tilde g_{\tau\sigma}D_\alpha
\tilde g^{\beta\sigma}D_\beta \tilde g^{\epsilon\tau}+
\frac{1}{2}\tilde g^{\epsilon\lambda}
\tilde g_{\tau\sigma}D_\alpha \tilde g^{\beta\sigma}D_\beta
\tilde g^{\alpha\tau}+\nonumber  \\
&&+D_\alpha \tilde g^{\epsilon\beta}D_\beta \tilde g^{\lambda\alpha}-
\tilde g^{\alpha\beta} D_\alpha D_\beta \tilde
g^{\epsilon\lambda}\Biggl. \Biggr\}\;.
\ea
It must  be especially stressed that in finding expression (D.5) we
made use
of equation (D.2). By substituting expression (D.5) into equation
(5.19) and
writing the thus obtained equation in the form (8.1) we find the
expression
for the quantity $-16\pi g\tau^{\epsilon\lambda}_{g}$:
\ba
&&-16\pi g\tau^{\epsilon\lambda}_{g}=\frac{1}{2}
(\tilde g^{\epsilon\alpha} \tilde g^{\lambda\beta}-\frac{1}{2}
\tilde g^{\epsilon\lambda} \tilde g^{\alpha\beta})
(\tilde g_{\nu\sigma}\tilde g_{\tau\mu}-\frac{1}{2}
\tilde g_{\tau\sigma}\tilde g_{\nu\mu})\times
\nonumber  \\
&&\times D_\alpha \tilde\Phi^{\tau\sigma}D_\beta
\tilde\Phi^{\mu\nu}+
+\tilde g^{\alpha\beta}\tilde g_{\tau\sigma} D_\alpha
\tilde\Phi^{\epsilon\tau} D_\beta \tilde\Phi^{\lambda\sigma}
-\tilde g^{\epsilon\beta} \tilde g_{\tau\sigma}D_\alpha
\tilde\Phi^{\lambda\sigma} D_\beta\tilde\Phi^{\alpha\tau}-\!\!
\nonumber  \\
&&
-\tilde g^{\lambda\alpha} \tilde g_{\tau\sigma}D_\alpha
\tilde\Phi^{\beta\sigma}D_\beta
\tilde\Phi^{\epsilon\tau}
\!+\!\frac{1}{2}\tilde g^{\epsilon\lambda}
\tilde g_{\tau\sigma} D_\alpha \tilde\Phi^{\sigma\beta}
D_\beta \tilde\Phi^{\alpha\tau}
\!+\!D_\alpha \tilde\Phi^{\epsilon\beta}
D_\beta \tilde\Phi^{\lambda\alpha}\!-\!
\nonumber  \\
&&
-\tilde\Phi^{\alpha\beta}D_\alpha D_\beta
\tilde\Phi^{\epsilon\lambda}- 
-m^2
\biggl  (
(\sqrt{-g}\tilde g^{\epsilon\lambda}
-\sqrt{-\gamma} \tilde\Phi^{\epsilon\lambda}
+\tilde g ^{\epsilon\alpha} \tilde g^{\lambda\beta}
\gamma_{\alpha\beta}-\biggr.\nonumber \\
&& -\frac{1}{2} \tilde g^{\epsilon\lambda}
\tilde g^{\alpha\beta}\gamma_{\alpha\beta})
\biggl. \biggr).
\ea

\thispagestyle{empty}
\newpage
\renewcommand{\theequation}{${\mbox{E}}.$\arabic{equation}}
\setcounter{equation}{0}
\begin{flushright}
{\bf Appendix E}
\end{flushright}
\addcontentsline{toc}{section}{\hskip 1.8em Appendix E}
\noindent
Let us write the RTG equation (5.20)
$$
D_\sigma \tilde g^{\sigma\nu}(y)
=\pa_\sigma \tilde g^{\sigma\nu}(y)
+\gamma^\nu_{\alpha\beta}(y)\tilde g^{\alpha\beta}(y)=0
\eqno{(\Sigma)}
$$
in a somewhat different form. For this purpose, making use of the
definition
of a Christoffel symbol,
\be
\Gamma_{\alpha\beta}^\nu (y)=\f{1}{2} g^{\nu\sigma}
(\pa_\alpha g_{\sigma\beta}
+ \pa_\beta g_{\sigma\alpha}
-\pa_\sigma g_{\alpha\beta}),
\ee
we find
\be
\Gamma^\nu_{\alpha\beta}\tilde g^{\alpha\beta}(y)
=\sqrt{-g}
\left (
g^{\nu\sigma} g^{\alpha\beta} \pa_\alpha
g_{\sigma\beta}-\f{1}{2} g^{\nu\sigma}
g^{\alpha\beta} \pa_\sigma g_{\alpha\beta}
\right ).
\ee
Taking into account the equalities
\ba
&&\Gamma^\lambda_{\sigma\lambda}=\f{1}{2}
g^{\alpha\beta} \pa_\sigma g_{\alpha\beta}
=\f{1}{\sqrt{-g}}
\pa_\sigma \sqrt{-g(y)},\;\; \nonumber\\[0.4mm]
&&\pa_\alpha g^{\alpha\nu}
=-g^{\nu\sigma} g^{\alpha\beta}
\pa_\alpha g_{\sigma\beta}
\ea
we rewrite (E.2) as
\be
\Gamma^\nu_{\alpha\beta}(y)\tilde g^{\alpha\beta}(y)
=-\sqrt{-g} \pa_\sigma g^{\sigma\nu}
-g^{\nu\sigma} \pa_\sigma\sqrt{-g}
=-\f{\pa\tilde g^{\sigma\nu}}{\pa y^\sigma}. 
\ee
With account of this equality the initial equation ($\Sigma$) assumes
the form
\be
(\Gamma^\nu_{\alpha\beta}(y)-\gamma^\nu_{\alpha\beta}(y))
g^{\alpha\beta}(y)=0. 
\ee

If we pass from coordinates ``$y$" to other curvilinear coordinates
``$z$",
then the Christoffel symbols assume the form
\ba
&&\Gamma^\lambda_{\mu\nu}(y)
=\f{\pa y^\lambda}{\pa z^\sigma}
\cdot \f{\pa z^\alpha}{\pa y^\mu}
\f{\pa z^\beta}{\pa y^\nu}
\Gamma^\sigma_{\alpha\beta} (z)+\nonumber \\
&&+\f{\pa^2z^\sigma}{\pa y^\mu\pa y^\nu}
\cdot \f{\pa y^\lambda}{\pa z^\sigma}. 
\ea
Applying this expression we find
\ba
&&\Gamma^\lambda_{\mu\nu}(y)g^{\mu\nu}(y)
=\f{\pa y^\lambda}{\pa z^\sigma}
\Biggl [
\Gamma^\sigma_{\alpha\beta} (z) g^{\alpha\beta} (z)
+\Biggr.\nonumber
\\
&&\left.+\f{\pa^2 z^\sigma}{\pa y^\mu \pa y^\nu}
\f{\pa y^\mu}{\pa z^\alpha}
\cdot \f{\pa y^\nu}{\pa z^\beta}
g^{\alpha\beta}(z).
\right ]
\ea
On the basis of (E.4) we write expression (E.7) in the form
\ba
&&\Gamma^\lambda_{\mu\nu}(y)g^{\mu\nu}(y)
=-\f{1}{\sqrt{-g}}\f{\pa}{\pa z^\mu}
\left (
\tilde g^{\mu\sigma} \f{\pa y^\lambda}{\pa z^\sigma}
\right )+\nonumber \\
&&+g^{\mu\sigma}
\f{\pa^2 y^\lambda}{\pa z^\mu \pa z^\sigma}
+\f{\pa y^\lambda}{\pa z^\sigma}\cdot
\f{\pa^2z^\sigma}{\pa y^\mu\pa y^\nu}
\f{\pa y^\mu}{\pa z^\alpha}
\f{\pa y^\nu}{\pa z^\beta}g^{\alpha\beta}(z).  
\ea
Upon differentiating equality
\be
\f{\pa z^\sigma}{\pa y^\mu}
\cdot \f{\pa y^\mu}{\pa z^\alpha}=\delta^\sigma_\alpha
\ee
with respect to the variable $z^\beta$ we obtain
\be
\f{\pa^2z^\sigma}{\pa y^\mu \pa y^\nu}
\f{\pa y^\mu}{\pa z^\alpha}\cdot
\f{\pa y^\nu}{\pa z^\beta}
=-\f{\pa z^\sigma}{\pa y^\mu}
\cdot \f{\pa^2 y^\mu}{\pa z^\alpha \pa z^\beta}.
\ee
Taking into account this equality, in the third term of (E.8) we find
\be
\Gamma^\lambda_{\mu\nu} (y)g^{\mu\nu}(y)
=-\f{1}{\sqrt{-g(z)}}\f{\pa}{\pa z^\nu}
\left (
\tilde g^{\nu\sigma}
\f{\pa y^\lambda}{\pa z^\sigma}
\right ).
\ee
Substituting this expression into (E.5) we obtain
\be
\dalam\; y^\lambda =-\gamma^\lambda_{\alpha\beta}(y)
g^{\alpha\beta}(y), 
\ee
where $\dalam$ denotes the operator
\be
\dalam=\f{1}{\sqrt{-g(z)}}\f{\pa}{\pa z^\nu}
\left (
\tilde g^{\nu\sigma}
\f{\pa}{\pa z^\sigma}
\right ).
\ee

\thispagestyle{empty}
\newpage
\renewcommand{\theequation}{\arabic{section}.\arabic{equation}}
\setcounter{equation}{0}

\section{Elements of tensor analysis and of Riemannian geometry}
Consider a certain coordinate system $x^\alpha, i=1,...n$ to be
defined in
$n$-dimensional space. Instead of this coordinate system one may also
choose another one defined by expression
\be
x'^\alpha =f(x^\alpha),\;\; \alpha=1,...n.\label{14.1}
\ee
These functions must be continuous and have continuous partial
derivatives
of order $N$. If the transformation Jacobian at each point,
\be
J=\det\left |\f{\pa f^\alpha}{\pa x^\beta}\right |, \label{14.2}
\ee
differs from zero, then in this condition the variables $x'^\alpha$
will be
independent, and, consequently, the initial variables $x^\alpha$ can
be
unambiguously expressed in terms of the new ones, $x'^\alpha$:
\be
x^\alpha =\varphi (x'^\alpha).\label{14.3}
\ee

Physical quantities must not depend on the choice of coordinate
system,
and therefore they must be expressed in terms of geometrical objects.
The simplest geometrical object is a scalar, that transforms in
transition
to the new coordinates as follows:
\be
\Phi'(x')=\Phi (x(x')). \label{14.4}
\ee
The gradient of a scalar function $\Phi (x)$ transforms in accordance
with the rule for the differentiation of composite functions,
\be
\f{\pa \Phi'(x')}{\pa x'^\alpha}=
\f{\pa \Phi}{\pa x^\beta}\cdot \f{\pa x^\beta}{\pa x'^\alpha}.
\label{14.5}
\ee
Here, summation is performed over identical indices $\beta$. The set
of
functions transforming under coordinate transformations by the rule
(\ref{14.5}) is termed the covariant vector
\be
A'_\alpha (x')=A_\beta (x)\f{\pa x^\beta}{\pa x'^\alpha}.
\label{14.6}
\ee
Correspondingly the quantity $B_{\mu\nu}$ is a covariant second-rank
tensor,
that transforms by the rule
\be
B'_{\mu\nu}(x')=B_{\alpha\beta}(x)\f{\pa x^\alpha}{\pa x'^\mu}
\cdot \f{\pa x^\beta}{\pa x'^\nu} \label{14.7}
\ee
and so on.

We shall now pass to another group of geometric objects. Consider
transformation of the differential of coordinates
\be
dx'^\mu =\f{\pa x'^\mu}{\pa x^\alpha} dx^\alpha. \label{14.8}
\ee
A set of functions transforming under coordinate transformations by
the
rule (\ref{14.8}) has been termed a contravariant vector,
\be
A'^\mu (x')=\f{\pa x'^\mu}{\pa x^\alpha}A^\alpha (x), \label{14.9}
\ee correspondingly, the quantity $B^{\mu\nu}$ a contravariant
second-rank tensor transforming by the rule
\be
B'^{\mu\nu}(x')=\f{\pa x'^\mu}{\pa x^\alpha}
\f{\pa x'^\nu}{\pa x^\beta}B^{\alpha\beta}(x) \label{14.10}
\ee
and so on. Expressions (\ref{14.6}), (\ref{14.7}), (\ref{14.9}) and
(\ref{14.10}) permit to write the transformation law of tensors of
any form.
For example,
\be
B'^\mu_\nu (x')=\f{\pa x'^\mu}{\pa x^\alpha}\cdot
\f{\pa x^\beta}{\pa x'^\nu}B^\alpha_\beta(x) \label{14.11}
\ee

From the transformational properties of a tensor it follows that, if
all
its components are equal to zero in one coordinate system, then they
equal
zero in another coordinate system, also. It is readily verified that
the
transformations of covariant and contravariant quantities exhibit the
group
property. For example:
\be
\begin{array}{lc}
\displaystyle
A'^\mu=\f{\pa x'^\mu}{\pa x^\alpha}A^\alpha(x),
A''^\nu(x'')=\f{\pa x''^\nu}{\pa x'^\mu}A'^\mu(x'), &
\\ &  \\ \label{14.12}
\displaystyle
A''^\nu(x'')=\f{\pa x''^\nu}{\pa x'^\mu}
\cdot \f{\pa x'^\mu}{\pa x^\alpha}
A^\alpha (x)=
\f{\pa x''^\nu}{\pa x^\alpha}
A^\alpha (x).
\end{array}
\ee
Now, let us pass to tensor algebra. Here, four operations are
possible:
addition, multiplication, convolution, and permutation of indices.

\noindent
\vspace*{0.2cm}
\underline{Addition and subtraction of tensors}

\vspace*{0.2cm}
\noindent
If we have tensors of identical structure, i.e. that have the same
number
of contravariant indices and the same number of covariant indices,
for
example,
$$
A^{\alpha\beta}_{\mu\nu\sigma},\;\;
B^{\alpha\beta}_{\mu\nu\sigma},
$$
then it is possible to form the tensor
\be
C^{\alpha\beta}_{\mu\nu\sigma}=
A^{\alpha\beta}_{\mu\nu\sigma}+
B^{\alpha\beta}_{\mu\nu\sigma}. \label{14.13}
\ee

\noindent
\vspace*{0.2cm}
\underline{Multiplication of tensors}

\vspace*{0.2cm}
\noindent
Tensors can be multiplied independently of their structure. For
example,
\be
C^{\alpha\beta\lambda}_{\mu\nu\sigma\rho}=
A^{\alpha\beta}_{\mu\nu\sigma}\cdot
B^{\lambda}_{\rho}. \label{14.14}
\ee
Here, both the order of multipliers and the order of indices must be
observed.

\noindent
\vspace*{0.2cm}
\underline{The convolution operation of tensors}

\vspace*{0.2cm}
\noindent
With the aid of the Kronecker symbol
\be
\delta^\mu_\nu=
\left\{
\begin{array}{lcl}
0 & \mbox{at} & \mu\not=\nu\\
1 & \mbox{at} & \mu=\nu,\\
\end{array}
\right.\label{14.15}
\ee
which is a tensor, it is possible to perform the convolution
operation of
indices, for example,
\be
A^{\alpha\beta}_{\mu\nu}\cdot
\delta^\nu_\sigma=A^{\alpha\beta}_{\mu\sigma}.\label{14.16}
\ee
Here, on the left, summation is performed over identical indices.

\noindent
\vspace*{0.2cm}
\underline{The permutation operation of indices}

\vspace*{0.2cm}
\noindent
By permutation of indices of the tensor we obtain another tensor, if
the
initial tensor was not symmetric over these indices, for example,
\be
B^{\mu\nu}_{\lambda\sigma}=A^{\mu\nu}_{\sigma\lambda}.\label{14.17}
\ee
With the aid of this operation, as well as addition, it is possible
to
construct a tensor that is symmetric over several indices. For
example,
\be
A_{(\mu\nu)}=\f{1}{2}
(A_{\mu\nu} + A_{\nu\mu}).\label{14.18}
\ee
It is also possible to construct a tensor, that is antisymmetric over
several indices. For example,
\be
A_{[\mu\nu]}=\f{1}{2}
(A_{\mu\nu}-A_{\nu\mu}).\label{14.19}
\ee
Such an operation is called antisymmetrization.

\vspace*{0.2cm}
\noindent
\underline{Riemannian geometry}

\vspace*{0.2cm}
\noindent
A Riemannian space $V_n$ is a real differentiable manifold, at each
point
of which there is given the field of a tensor
\be
g_{\mu\nu}(x)=g_{\mu\nu}(x^1,\ldots,x^n) \label{14.20}
\ee
twice covariant, symmetric and nondegenerate
\be
g_{\mu\nu}=g_{\nu\mu},\;\;
g=\det|g_{\mu\nu}|\not=0.\label{14.21}
\ee
The tensor $g_{\mu\nu}$ is called a metric tensor of Riemannian space.
The functions $g_{\mu\nu}$ are continuous and differentiable with
respect to all variables $x^1,\ldots,x^n$ up to the $n$-th order.

With the aid of the metric tensor in Riemannian space it is possible
to
introduce an invariant differential form termed an interval
\be
(ds)^2=g_{\mu\nu}(x)dx^\mu dx^\nu.\label{14.22}
\ee
With the aid of coordinate transformations this form at any fixed
point
can be reduced to a diagonal form. Here, in the general case, the
diagonal
components of the matrix $g_{\mu\nu}$ will not all be positive. But,
by
virtue of the law of inertia for quadratic forms the difference
between the
amounts of positive and of negative diagonal components will be
constant.
This difference is called the signature of a metric tensor. In an
arbitrary
Riemannian space $V_n$ the interval will exhibit alternating signs. We
shall
further call it timelike, if $ds^2>0$, spacelike, if $ds^2<0$,
isotropic,
if $ds^2=0$. These terms originated within special relativity theory,
where
space and time form a unique manifold, while the interval in
Cartesian
(Galilean) coordinates has the form
\be
d\sigma^2=(dx^0)^2-(dx^1)^2
-(dx^2)^2-(dx^3)^2.  \label{14.23}
\ee
In arbitrary coordinates it assumes the form
\be
d\sigma^2=\gamma_{\mu\nu}(x)dx^\mu dx^\nu. \label{14.24}
\ee
Since the determinant $|g_{\mu\nu}|\not= 0$, we can construct a
contravariant metric tensor with the aid of equations
\be
g_{\mu\sigma}g^{\sigma\nu}=\delta^\nu_\mu.\label{14.25}
\ee
With the aid of tensors $g_{\mu\nu}$ and $g^{\lambda\sigma}$ it is
possible
to raise and to lower indices
\be
A^\nu=g^{\nu\sigma}A_\sigma,\;\; A_\nu=g_{\nu\sigma}A^\sigma.
\label{14.26}
\ee

\vspace*{0.2cm}
\noindent
\underline{Geodesic lines in Riemannian space}

\vspace*{0.2cm}
\noindent
Geodesic lines in Riemannian space play the same role as straight lines
in
Euclidean space. They are called extremal lines. For defining an
extremal
we shall take advantage of variational calculus. The essence of
variational calculus consists in generalization of the concepts of
maximum and minimum. The issue is not finding the extremum of a
function,
but finding the extremum of a functional, i.e. finding such
functions, that
make it an extremum. The distance between close points in Riemannian
space is
determined by the interval $ds$. The quantity $ds$ is not a total
differential. The interval between points $a$ and $b$ is
\be
S=\int\limits^{b}_{a}ds=
\int\limits^{b}_{a}\sqrt{g_{\mu\nu}(x)dx^\mu dx^\nu}.\label{14.27}
\ee
The extremum is determined by the relation
\be
\delta \int\limits^{b}_{a}ds=
\int\limits^{b}_{a}\delta (ds)=0.\label{14.28}
\ee
Thus, such functions $g_{\mu\nu}(x)$ are sought, that provide for the
functional (integral) achieving its extremum:
\ba
&&\delta (ds^2)=2ds\delta (ds)=\delta (g_{\mu\nu}(x)dx^\mu
dx^\nu)=\nonumber \\
 && =\f{\pa g_{\mu\nu}}{\partial x^\sigma}\delta x^\sigma
dx^\mu
dx^\nu
 +2g_{\mu\nu} (x) dx^\mu \delta (dx^\nu). \label{14.29}
\ea

We note that
\be
\delta (dx^\nu)=d(\delta x^\nu).\label{14.30}
\ee
On the basis of (\ref{14.29}) and (\ref{14.30}) we have
\be
\delta (ds)=\f{1}{2}
\f{\pa g_{\mu\nu}}{\pa x^\sigma} U^\mu dx^\nu
\delta x^\sigma +g_{\mu\nu} U^\mu d
(\delta x^\nu),\;\;
U^\mu =\f{dx^\mu}{ds}.\label{14.31}
\ee
Substituting (\ref{14.31}) into (\ref{14.28}) we obtain
\be
\delta S=\int\limits^{b}_{a}
\left [
\f{1}{2}\f{\pa g_{\mu\nu}}{\pa x^\sigma} U^\mu U^\nu
\delta x^\sigma
+g_{\mu\nu} U^\mu \f{d (\delta x^\nu)}{ds}\right ]ds=0.\label{14.32}
\ee
Since
\be
g_{\mu\nu} U^\mu
\f{d (\delta x^\nu)}{ds}=\f{d}{ds}
(g_{\mu\nu}U^\mu \delta x^\nu)-\delta x^\nu
\f{d}{ds}(g_{\mu\nu}U^\mu ), \label{14.33}
\ee
and at the integration limits $\delta x^\nu=0$, from (\ref{14.32})
we obtain
\ba
&&\delta S=\int\limits^{b}_{a}
\left [
\f{1}{2}
\f{\pa g_{\mu\nu}}{\partial x^\sigma}U^\mu U^\nu -
g_{\mu\sigma}\f{dU^\mu}{ds}-\right.
\nonumber \\
&&-
\f{\pa g_{\mu\sigma}}
{\partial x^\lambda}
U^\mu U^\lambda
\left.\right ]ds\delta
x^\sigma=0.
\label{14.34}
\ea
We now represent the last term in (\ref{14.34}) as
\be
U^\mu U^\lambda \f{\pa g_{\mu\sigma}}{dx^\lambda}
=\f{1}{2}
\left (
\f{\pa g_{\mu\sigma}}{\pa x^\lambda}+
\f{\pa g_{\lambda\sigma}}{\pa x^\mu}\right )U^\mu
U^\lambda.\label{14.35}
\ee
Substituting (\ref{14.35}) into (\ref{14.34}) we find
\ba
&&\delta S=\int\limits^{b}_{a}
\left [ U^\mu U^\lambda\f{1}{2}
\left (
\f{\pa g_{\mu\sigma}}{\pa x^\lambda}+
\f{\pa g_{\lambda\sigma}}{\pa x^\mu}-
\f{\pa g_{\mu\lambda}}{\pa x^\sigma}\right ) +\right.\nonumber
\\
&&\left.+ g_{\mu\sigma} \f{dU^\mu}{ds}\right ] ds\delta
x^\sigma=0.\label{14.36}
\ea
Since the variation $\delta x^\sigma$ is arbitrary, the integral
(\ref{14.36}) turns to zero, only if
\be
g_{\mu\sigma}\f{dU^\mu}{ds}+\f{1}{2}
\left (
\f{\pa g_{\mu\sigma}}{\pa x^\lambda}+
\f{\pa g_{\lambda\sigma}}{\pa x^\mu}
-\f{\pa g_{\mu\lambda}}{\pa x^\sigma}
\right )U^\mu U^\lambda=0.\label{14.37}
\ee
Multiplying (\ref{14.37}) by $g^{\sigma\alpha}$ we obtain
\be
\f{dU^\alpha}{ds}+\Gamma^\alpha_{\mu\lambda}U^\mu
U^\lambda=0,\label{14.38}
\ee
where the Christoffel symbols $\Gamma^\alpha_{\mu\nu}$ are
\be
\Gamma^\alpha_{\mu\lambda}=\f{1}{2}g^{\alpha\sigma}
(\pa_\lambda g_{\mu\sigma}+\pa_\mu g_{\lambda\sigma}-\pa_\sigma
g_{\mu\lambda}).\label{14.39}
\ee

The Christoffel symbols are not tensor quantities. Precisely
equations
(\ref{14.38}) are the equations for a geodesic line. There are four
of them,
but not all are independent, since the following condition takes
place:
\be
g_{\mu\nu} (x) U^\mu U^\nu =1. \label{14.40}
\ee
By transformations of coordinates $x^\mu$ it is possible to equate
the
Christoffel symbols to zero along any not self-intersecting chosen
line~\cite{26}.

\vspace*{0.2cm}
\noindent
\underline{Covariant differentiation}

\vspace*{0.2cm}
\noindent
We now take an arbitrary covariant vector $A_\lambda$ and form its
convolution with the vector $U^\lambda$, and thus obtain the scalar
\be
A_\lambda U^\lambda, \label{14.41}
\ee
upon differentiating it with respect to $ds$ we also have a scalar:
\ba
&&\f{d}{ds} (A_\lambda U^\lambda)=\f{dA_\lambda}{ds}U^\lambda
+A_\nu \f{dU^\nu}{ds}=\nonumber \\
&&=\f{\pa A_\lambda}{\pa x^\sigma}U^\sigma U^\lambda
+A_\nu \f{dU^\nu}{ds}. \label{14.42}
\ea
Substituting into the right-hand side expression (\ref{14.38}) we
obtain
\be
\f{d}{ds}(A_\lambda U^\lambda) =\left [
\f{\pa A_\lambda}{\pa x^\sigma}-\Gamma^\nu_{\sigma\lambda}A_\nu\right
]
U^\sigma U^\lambda.\label{14.43}
\ee
Since (\ref{14.43}) is a scalar, and $U^\sigma$ is a vector, we hence
have
a second-rank tensor
\be
A_{\lambda;\sigma}=\f{DA_\lambda}{dx^\sigma}=\f{\pa A_\lambda}{\pa
x^\sigma}-\Gamma^\nu_{\sigma\lambda}A_\nu. \label{14.44}
\ee

Here and further a semicolon denotes covariant differentiation. Thus,
we
have defined the covariant derivative of the covariant vector
$A_\lambda$.
We shall now define the covariant derivative of the contravariant
vector $A^\lambda$.

To this end we write the same scalar in the form
\ba
&&\f{d}{ds} (A^\mu U^\nu g_{\mu\nu})=
\f{\pa A^\mu}{\pa x^\sigma} U^\sigma U^\nu
g_{\mu\nu} +\nonumber \\
&&+A^\mu g_{\mu\lambda}
\f{dU^\lambda}{ds} +A^\mu U^\nu U^\sigma \pa_\sigma
g_{\mu\nu}.\label{14.45}
\ea
Substituting into the right-hand side expression (\ref{14.38}) we
obtain
\ba
&&\f{d}{ds}  (A^\mu U^\nu g_{\mu\nu})=U^\nu U^\sigma
\left [
g_{\mu\nu}\f{\pa A^\mu}
{\pa x^\sigma}
-\right. \nonumber
\\
&&- A^\mu
g_{\mu\lambda} \Gamma^\lambda_{\sigma\nu}+A^\mu
\pa_\sigma g_{\mu\nu}\Biggl.\Biggr].\label{14.46}
\ea
Taking into account expression (\ref{14.39}) we have
\ba
&&\f{d}{ds}  (A^\mu U^\nu g_{\mu\nu})=\left [
g_{\mu\nu}\f{\pa A^\mu}{\pa x^\sigma}+\right. \nonumber \\
&&\Biggl.+\f{1}{2}
(\pa_\sigma g_{\mu\nu} +\pa_\mu g_{\sigma\nu}
-\pa_\nu g_{\sigma\mu})A^\mu\Biggr]U^\nu U^\sigma. \label{14.47}
\ea

Representing $U^\nu$ in the form
\be
U^\nu=U_\lambda g^{\lambda\nu} \label{14.48}
\ee
and substituting it into relation (\ref{14.47}) we obtain
\be
\f{d}{ds}  (A^\mu U^\nu g_{\mu\nu})=\left [
\f{\pa A^\lambda}{\pa x^\sigma}+\Gamma^\lambda_{\sigma\mu}
A^\mu\right ] U^\sigma U_\lambda. \label{14.49}
\ee
Since this expression is a scalar, hence it follows that the
contravariant derivative is a tensor,
\be
A^\lambda_{;\sigma}=\f{DA^\lambda}{dx^\sigma}
=\f{\pa A^\lambda}{\pa
x^\sigma}+\Gamma^\lambda_{\sigma\mu}A^\mu.\label{14.50}
\ee
Thus, we have defined the covariant derivative of the contravariant
vector $A^\lambda$.

Applying formulae (\ref{14.44}) and (\ref{14.50}), it is also
possible to
obtain covariant derivatives of a second-rank tensor:
\ba
&&A_{\mu\nu;\sigma}=\f{\pa A_{\mu\nu}}{\pa x^\sigma}
-\Gamma^\lambda_{\sigma\mu}A_{\lambda\nu}-\Gamma^\lambda_{\sigma\nu}
A_{\lambda\mu}, \label{14.51}
\\[2mm]
&&A^{\mu\nu}_{;\sigma}=\f{\pa A^{\mu\nu}}{\pa x^\sigma}
+\Gamma^\mu_{\sigma\lambda}A^{\nu\lambda}+\Gamma^\nu_{\sigma\lambda}
A^{\mu\lambda}. \label{14.52}
\\[2mm]
&&A^\nu_{\rho;\sigma}=\f{\partial A^\nu_\rho}{\partial x^\sigma}
-\Gamma^\lambda_{\rho\sigma}A^\nu_\lambda
+\Gamma^\nu_{\sigma_\lambda}
A^\lambda_\rho. \label{14.53}
\ea
Making use of expression (\ref{14.51}) it is easy to show, that
$$
g_{\mu\nu;\sigma}\equiv 0,
$$
i.e. the covariant derivative of a metric tensor is equal to zero.

\vspace*{0.2cm}
\noindent
\underline{The Riemann-Christoffel curvature tensor}

\vspace*{0.2cm}
\noindent
In Riemannian space the operation of covariant differentiation is
noncommutative. Covariant differentiation of vector $A_\lambda$,
first,
with respect to the variable $x^\mu$ and, then, with respect to
$x^\nu$
leads to the following expression:
\be
A_{\lambda;\mu\nu}=\f{\pa A_{\lambda;\mu}}{\pa x^\nu}
-\Gamma^\tau_{\nu\lambda}
A_{\tau;\mu}-\Gamma^\tau_{\mu\nu}A_{\lambda;\tau},\label{14.54}
\ee
but since
\ba
&&A_{\lambda;\mu}=\f{\pa A_\lambda}{\pa
x^\mu}-\Gamma^\tau_{\lambda\mu}A_\tau,\;\;
A_{\tau;\mu}=\f{\pa A_\tau}{\pa x^\mu}
-\Gamma^\sigma_{\mu\tau}A_\sigma,\;\; \nonumber \\
[-0.3cm] \label{14.55}\\ [-0.3cm]
&&A_{\lambda;\tau}=\f{\pa A_\lambda}{\pa x^\tau}
-\Gamma^\sigma_{\lambda\tau}A_\sigma, \nonumber
\ea
upon substitution of these expressions into (\ref{14.54}) we have
\ba
&& A_{\lambda;\mu\nu}=\f{\pa^2 A_\lambda}{\pa
x^\mu\pa x^\nu}-\Gamma^\tau_{\lambda\mu}
\f{\pa A_\tau}{\pa x^\nu} -
\Gamma^\tau_{\nu\lambda}\f{\pa A_\tau}{\pa x^\mu}-
-\Gamma^\tau_{\mu\nu}
\f{\pa A_\lambda}{\pa x^\tau} -
\nonumber \\
&&
-A_\tau \f{\pa \Gamma^\tau_{\lambda\mu}}{\pa x^\nu}
+\Gamma^\tau_{\nu\lambda}
\Gamma^\sigma_{\mu\tau}
A_\sigma+
\Gamma^\tau_{\mu\nu}\Gamma^\sigma_{\lambda\tau}
A_\sigma.\label{14.56}
\ea

We shall now calculate the quantity $A_{\lambda;\nu\mu}$:
\be
A_{\lambda;\nu\mu}=\f{\pa A_{\lambda;\nu}}{\pa x^\mu}
-\Gamma^\tau_{\mu\lambda}A_{\tau;\nu}
-\Gamma^\tau_{\mu\nu}A_{\lambda;\tau},\label{14.57}
\ee
with account of the expression
\ba
&&A_{\lambda;\nu}=\f{\pa A_\lambda}{\pa x^\nu}
-\Gamma^\tau_{\lambda\nu}A_{\tau},\;\;
A_{\tau;\nu}=\f{\pa A_{\tau}}{\pa x^\nu}
-\Gamma^\sigma_{\tau\nu}A_{\sigma},\;\;\nonumber \\
[-0.2cm] \label{14.58}\\ [-0.2cm]
&&A_{\lambda;\tau}=\f{\pa A_{\lambda}}{\pa x^\tau}
-\Gamma^\sigma_{\lambda\tau}A_{\sigma},\nonumber
\ea
relation (\ref{14.57}) assumes the form
\ba
&&A_{\lambda;\nu\mu}=\f{\pa^2 A_\lambda}{\pa x^\mu\pa x^\nu}-
\Gamma^\tau_{\lambda\nu}\f{\pa A_\tau}{\pa x^\mu} -
\Gamma^\tau_{\mu\lambda}\f{\pa A_\tau}{\pa x^\nu} -
\Gamma^\tau_{\mu\nu}\f{\pa A_\lambda}{\pa x^\tau} -
\nonumber \\
&& -
A_\tau \f{\pa \Gamma^\tau_{\lambda\nu}}{\pa x^\mu} +
\Gamma^\tau_{\mu\lambda} \Gamma^\sigma_{\nu\tau}A_{\sigma}+
\Gamma^\tau_{\mu\nu}
\Gamma^\sigma_{\lambda\tau}A_{\sigma}.\label{14.59}
\ea
On the basis of (\ref{14.56}) and (\ref{14.59}), only the following
terms
are retained in the difference:
\ba
&&A_{\lambda;\mu\nu}-A_{\lambda;\nu\mu}=A_\sigma
\left [
\f{\pa \Gamma^\sigma_{\lambda\nu}}{\pa x^\mu} -
\f{\pa \Gamma^\sigma_{\lambda\mu}}{\pa x^\nu} +
\right.\nonumber \\
&&\Biggl.+\Gamma^\tau_{\nu\lambda} \Gamma^\sigma_{\mu\tau}
-\Gamma^\tau_{\mu\lambda}\Gamma^\sigma_{\nu\tau}
\Biggr]. \label{14.60}
\ea

The quantity $R^\sigma_{\lambda\mu\nu}$ is termed the Riemann
curvature
tensor
\be
R^\sigma_{\lambda\mu\nu}=
\f{\pa \Gamma^\sigma_{\lambda\nu}}{\pa x^\mu} -
\f{\pa \Gamma^\sigma_{\lambda\mu}}{\pa x^\nu} +
\Gamma^\tau_{\nu\lambda} \Gamma^\sigma_{\mu\tau} -
\Gamma^\tau_{\mu\lambda}\Gamma^\sigma_{\nu\tau}.  \label{14.61}
\ee
From this tensor it is possible, by convolution, to obtain a
second-rank
tensor, the Ricci tensor:
\be
R_{\lambda\nu}=
R^\sigma_{\lambda\sigma\nu}=
\f{\pa \Gamma^\sigma_{\lambda\nu}}{\pa x^\sigma} -
\f{\pa \Gamma^\sigma_{\lambda\sigma}}{\pa x^\nu} +
\Gamma^\tau_{\nu\lambda} \Gamma^\sigma_{\sigma\tau} -
\Gamma^\tau_{\sigma\lambda}\Gamma^\sigma_{\nu\tau}.  \label{14.62}
\ee
We note that for an interval of the form (\ref{14.23}) or
(\ref{14.24}) the
curvature tensor equals zero.

From expression (\ref{14.61}) it is obvious that the curvature tensor
is
antisymmetric with respect to the two last indices $\mu,\nu$:
$$
R^\sigma_{\lambda\mu\nu} =-R^\sigma_{\lambda\nu\mu}
$$
It is possible to construct a curvature tensor with lower indices:
$$
R_{\rho\lambda\mu\nu}=g_{\rho\sigma}R^\sigma_{\lambda\mu\nu}.
$$
It possesses the following symmetry properties:
$$
R_{\rho\lambda\mu\nu} =- R_{\lambda\rho\mu\nu}
=-R_{\rho\lambda\nu\mu},\;\;
R_{\rho\lambda\mu\nu} = R_{\mu\nu\rho\lambda}.
$$
We see that the curvature tensor is antisymmetric both with respect
to the
first pair of indices and with respect to the second. It is also
symmetric
with respect to permutation of index pairs, without any change of
their
order.

In Riemannian space there exists a local coordinate system, within which
the
first derivatives of the components of the metric tensor $g_{\mu\nu}$
are
equal to zero. Here, the Christoffel symbols are, naturally, also
equal to
zero. Such coordinates are called Riemann coordinates. They are
convenient
for finding tensor identities, since if it has been established, that
in
this coordinate system a certain tensor is zero, then, by virtue of
tensor
transformations, it will also be zero in any coordinate system.

The curvature tensor in a Riemann coordinate system is
\be
R^\sigma_{\lambda\mu\nu} =\partial_\mu \Gamma^\sigma_{\lambda\nu}
-\pa_\nu\Gamma^\sigma_{\lambda\mu}. \label{14.63}
\ee
The covariant derivative of the curvature tensor has the form
\be
R^\sigma_{\lambda\mu\nu;\rho} =\pa_\rho \partial_\mu
\Gamma^\sigma_{\lambda\nu}
-\pa_\rho\pa_\nu \Gamma^\sigma_{\lambda\mu}. \label{14.64}
\ee
Cyclically transposing indices $\mu,\nu,\rho$ and adding up the
obtained
expressions we obtain the Bianchi identity
\be
R^\sigma_{\lambda\mu\nu;\rho} + R^\sigma_{\lambda\rho\mu;\nu}
+R^\sigma_{\lambda\nu \rho;\mu}\equiv 0. \label{14.65}
\ee
Performing convolution of indices $\sigma$ and $\nu$ we obtain
\be
-R_{\lambda\mu;\rho} + R^\sigma_{\lambda\rho\mu;\sigma}
+R_{\lambda \rho;\mu}= 0. \label{14.66}
\ee
We multiply this expression by $g^{\lambda\alpha}$:
$$
-R^\alpha_{\mu;\rho} +
(g^{\lambda\alpha}R^\sigma_{\lambda\rho\mu})_{;\sigma}
+R^\alpha_{\rho;\mu}=0.
$$
We have, here, taken into account the previously established property
of
metric coefficients consisting in that they can, in case of covariant
differentiation, be freely brought or taken out from under the
derivative
sign.

Performing convolution of indices $\rho$ and $\alpha$ we obtain
\be
-R^\rho_{\mu;\rho} +
(g^{\lambda\rho}R^\sigma_{\lambda\rho\mu})_{;\sigma}
+\pa_\mu R\equiv 0, \label{14.67}
\ee
where
$$
R=R^\rho_\rho=R_{\mu\nu}g^{\mu\nu}
$$
is the scalar curvature.

Let us consider under the derivative sign the second term in identity
(\ref{14.67}):
$$
g^{\lambda\rho}R^\sigma_{\lambda\rho\mu}=
g^{\lambda\rho}g^{\nu\sigma}R_{\nu\lambda\rho\mu}
=g^{\nu\sigma}g^{\lambda\rho}R_{\lambda\nu\mu\rho}
=g^{\nu\sigma}R^\rho_{\nu\mu\rho}
=-R^\sigma_\mu.
$$
We have, here, applied the symmetry properties of the curvature
tensor
and the definition of the tensor $R_{\mu\nu}$. Substituting this
expression
into (\ref{14.67}) we obtain
\be
(R^\rho_\mu-\f{1}{2}\delta^\rho_\mu R)_{;\rho}=
\nabla_\rho
(R^\rho_\mu-\f{1}{2}\delta^\rho_\mu R)\equiv 0. \label{14.68}
\ee

We now introduce the notation
\be
G^\rho_\mu=R^\rho_\mu -\f{1}{2}
\delta^\rho_\mu R. \label{14.69}
\ee
On the basis of (\ref{14.53}), identity (\ref{14.68}) can be written
in
the expanded form
\be
\nabla_\nu G^\nu_\rho=G^\nu_{\rho;\nu}
=\f{\pa G^\nu_\rho}{\pa x^\nu} -
\Gamma^\lambda_{\rho\nu} G^\nu_\lambda +
\Gamma^\nu_{\nu\lambda}G^\lambda_\rho \equiv 0, \label{14.70}
\ee
taking into account that
\be
\Gamma^\nu_{\nu\lambda}=\f{1}{2}g^{\mu\nu}
\f{\pa g_{\mu\nu}}{\pa x^\lambda} \label{14.71}
\ee
and differentiating the determinant $g$,
\be
\f{\pa g}{\pa x^\lambda}=gg^{\mu\nu}\f{\pa g_{\mu\nu}}{\pa
x^\lambda},\label{14.72}
\ee
we find, by comparison of (\ref{14.71}) and (\ref{14.72}), the
following:
\be
\Gamma^\nu_{\nu\lambda}=\f{1}{2}\cdot \f{1}{g}
\f{\pa g}{\pa x^\lambda}=\f{1}{\sqrt{-g}}\pa_\lambda
(\sqrt{-g}).\label{14.73}
\ee
Substituting this expression into (\ref{14.70}) we obtain
\be
\nabla_\nu (\sqrt{-g}G^\nu_\rho)
=\pa_\nu (\sqrt{-g}G^\nu_\rho)
-\sqrt{-g} \Gamma^\lambda_{\rho\nu}G^\nu_\lambda\equiv 0.
\label{14.74}
\ee
Making use of expression (\ref{14.39}), we find for the Christoffel
symbol
\be
\pa_\nu (\sqrt{-g}G^\nu_\rho)+\f{1}{2}\cdot
\f{\pa g^{\lambda\sigma}}{\pa x^\rho}
\sqrt{-g}G_{\lambda\sigma}\equiv 0. \label{14.75}
\ee
Such an identity was first obtained by D.~Hilbert. It was necessary
for
constructing the equations of general relativity theory.

In conclusion we shall show that the quantity determining volume
\be
v'=\int\sqrt{-g'} dx^{0\;\prime}dx^{1\;\prime}
dx^{2\;\prime}dx^{3\;\prime} \label{14.76}
\ee
is an invariant under arbitrary transformations of coordinates. Under
coordinate transformations we have
$$
g'_{\mu\nu}(x')=g_{\lambda\sigma}(x)
\f{\pa x^\lambda}{\pa x'^\mu}\cdot
\f{\pa x^\sigma}{\pa x'^\nu}.
$$
We write this expression in the form
$$
g'_{\mu\nu}(x')=
\f{\pa x^\lambda}{\pa x'^\mu}\cdot g_{\lambda\sigma}
\f{\pa x^\sigma}{\pa x'^\nu}.
$$
We shall, now, calculate the determinant $g'=\det g'_{\mu\nu}$
\ba
&&g'=\det \left (
g_{\sigma\lambda}
\f{\pa x^\lambda}{\pa x'^\mu}\right )
\det
\left (
\f{\pa x^\sigma}{\pa x'^\nu}\right )
=\nonumber \\
&&=\det (g_{\lambda\sigma})
\det
\left (
\f{\pa x^\lambda}{\pa x'^\mu}\right )
\det
\left (
\f{\pa x^\sigma}{\pa x'^\nu}\right ).\nonumber
\ea
Hence, we have
\be
g'=gJ^2.\label{14.77}
\ee
Here $J$ is the transformation Jacobian,
\be
J=\f{\pa(x^0. x^1, x^2, x^3)}{\pa(x^{0\prime}, x^{1\prime},
x^{2\prime},
x^{3\prime})}. \label{14.78}
\ee
Thus,
\be
\sqrt{-g'}=\sqrt{-g}J. \label{14.79} \ee Substituting this
expression into (\ref{14.76}) we obtain \ba
&&\!\!\!\!v'=\!\int\sqrt{-g} \f{\pa(x^0. x^1, x^2,
x^3)}{\pa(x^{0\prime}, x^{1\prime}, x^{2\prime}, x^{3\prime})}
dx^{0\;\prime}dx^{1\;\prime} dx^{2\;\prime}dx^{3\;\prime}
=\nonumber \\
=
&&\int\sqrt{-g}
dx^0dx^1dx^2dx^3.
\label{14.80}
\ea
But the right-hand side represents volume
\be
v=\int\sqrt{-g}dx^0dx^1dx^2dx^3. \label{14.81}
\ee
Thus, we have established the equality
\be
v'=v. \label{14.82}
\ee
Hence it follows, that the quantity
\be
\sqrt{-g}d^4x \label{14.83}
\ee
is also an invariant relative to arbitrary coordinate
transformations.

Certain special features of Riemannian geometry should be noted. In the
general case, Riemannian space cannot be described in a sole coordinate
system.
For its description an atlas of maps is necessary. Precisely for this
reason, the topology of Riemannian space differs essentially from the
topology
of Euclidean space. In the general case, no group of
motion
exists in Riemannian space. In pseudo-Euclidean space, described by the interval
(\ref{14.23}) or
(\ref{14.24}), there exists a ten-parameter group of space motions.

The main characteristic of Riemannian geometry --- the curvature tensor
$R^\sigma_{\lambda\mu\nu}$ --- is a form-invariant quantity relative
to
coordinate transformations. The tensor $R_{\lambda\nu}$ is also a
form-invariant quantity. Here, form-invariance is not understood as one
and
the same functional dependence of the curvature tensor upon the
choice of
coordinate system, but identity in constructing the curvature tensor
for
a given expression $g_{\mu\nu}(x)$, similarly to how expression
$$
\dalam\; A^\nu(x)
$$
is written in the same way in Galilean coordinates in differing
inertial
reference systems for a given expression $A^\nu$. There exists an
essential
difference between invariance and form-invariance. for example, the
operator
$\gamma^{\mu\nu}(x)D_\mu D_\nu$ (where $\gamma^{\mu\nu}(x)$ is the
metric
tensor of Minkowski space) for arbitrary coordinate transformations
is
an invariant, i.e. a scalar, but it is not form-invariant. It will be
form-invariant only in the case of such coordinate transformations,
under
which the tensor $\gamma^{\mu\nu}(x)$ remains form-invariant, i.e.
$$
\delta\gamma^{\mu\nu}(x)=0.
$$

The curvature tensor varies under gauge transformations (3.16)
according
to the following rule:
$$
\begin{array}{l}
\displaystyle \delta_\epsilon R_{\mu\nu\alpha\beta}
=- R_{\sigma\nu\alpha\beta} D_\mu\epsilon^\sigma
-R_{\mu\sigma\alpha\beta} D_\nu\epsilon^\sigma- \\[-0.2cm]
\\[-0.2cm]
-R_{\mu\nu\sigma\beta} D_\alpha\epsilon^\sigma
-R_{\mu\nu\alpha\sigma} D_\beta\epsilon^\sigma
-\epsilon^\sigma D_\sigma R_{\mu\nu\alpha\beta}.
\end{array}
$$
This variation is due to arbitrary coordinate systems not being
physically
equivalent.

Within the text of this book there are encountered, together with
covariant
derivatives in Riemannian space, $\nabla_\lambda$, covariant derivatives
in
Minkowski space, $D_\lambda$. The difference consists in that in
constructing the covariant derivatives $D_\lambda$ it is necessary to
substitute into formulae (\ref{14.50}~-~\ref{14.53}) the Christoffel
symbols
of Minkowski space, $\gamma^\nu_{\alpha\beta}$, instead of the
Christoffel
symbols of Riemannian space, $\Gamma^\nu_{\alpha\beta}$,

In conclusion, we present the Weyl--Lorentz--Petrov
theorem~\cite{27}.
Coincidence of the respective equations of isotropic and timelike
geodesic lines for two Riemannian spaces with the metrics
$g_{\mu\nu}(x)$ and
$g'_{\mu\nu}(x)$ and with the same signature --2 leads to their metric
tensors only differing by a constant factor. From this theorem it
follows
that, if in one and the same coordinate system $x$ we have different
metric
tensors $g_{\mu\nu}(x)$ and $g'_{\mu\nu}(x)$, then, in identical
conditions,
different geodesic lines, and, consequently, different physics, will
correspond to them. Precisely for this reason, the situation, that
arises in GRT with the appearance of a multiplicity of metrics within
one coordinate system, leads to ambiguity in the description of
gravitational
effects.

\thispagestyle{empty}
\newpage
\section*{ADDENDUM}\addcontentsline{toc}{section}{ADDENDUM}
\subsection*{On the gravitational
force}\addcontentsline{toc}{section}
{\hskip 1.8em On the gravitational force}

The expression for the gravitational force is presented at page
66.
We shall now derive this expression from the equation of a geodesic
in
effective Riemannian space. The equation for the geodesic line has the
form
$$
\f{dp^\nu}{ds} + \Gamma^\nu_{\alpha\beta}p^\alpha p^\beta=0,\;\;
p^\nu=\f{dx^\nu}{ds},\;\;
ds^2=g_{\mu\nu} dx^\mu dx^\nu >0. \label{n1} \eqno{(1)}
$$
In accordance with the definition of a covariant derivative in
Minkowski
space we have
$$
\f{Dp^\nu}{ds}=
\f{dp^\nu}{ds}+\gamma^\nu_{\alpha\beta}
p^\alpha p^\beta. \label{n2}\eqno{(2)}
$$
Applying (1) and (2) we obtain
$$
\f{Dp^\nu}{ds}=-G^\nu_{\alpha\beta}
p^\alpha p^\beta. \label{n3} \eqno{(3)}
$$
Here
$$
G^\nu_{\alpha\beta}
=\Gamma^\nu_{\alpha\beta} -\gamma^\nu_{\alpha\beta}
\label{n4}\eqno{(4)}
$$
We shall write the left-hand side of relation (3) as
$$
\f{Dp^\nu}{ds} =
\left (
\f{d\sigma}{ds}
\right )^2
\left [
\f{DV^\nu }{d\sigma}
+V^\nu
\f{\f{d^2\sigma}{ds^2}}
{\left ( \f{d\sigma}{ds}\right )^2}
\right ],\;\;
V^\nu=\f{dx^\nu}{d\sigma}. \label{n5}\eqno{(5)}
$$
Here $V^\nu$ is the timelike velocity four-vector in Minkowski space,
that satisfies the condition
$$
\gamma_{\mu\nu} V^\mu V^\nu=1,\;\;
d\sigma^2>0. \label{n6} \eqno{(6)}
$$
Substituting (5) into (3) we obtain
$$
\f{DV^\nu }{d\sigma}
=-G^\nu_{\alpha\beta}
V^\alpha V^\beta
-V^\nu
\f{\f{d^2\sigma}{ds^2}}
{\left ( \f{d\sigma}{ds}\right )^2} \label{n7} \eqno{(7)}
$$
From (6) we have
$$
\left ( \f{d\sigma}{ds}\right )^2=\gamma_{\alpha\beta} p^\alpha
p^\beta.
\label{n8} \eqno{(8)}
$$
Differentiating this expression with respect to $ds$ we obtain
$$
\f{\f{d^2\sigma}{ds^2}}
{\left ( \f{d\sigma}{ds}\right )^2}=-\gamma_{\lambda\mu}
G^\mu_{\alpha\beta} V^\lambda V^\alpha V^\beta.
 \label{n9} \eqno{(9)}
$$
Substituting this expression into (7) we find~[4]
$$
\f{DV^\nu}{d\sigma}=
-G^\mu_{\alpha\beta} V^\alpha V^\beta
(\delta^\nu_\mu - V^\nu V_\mu).
 \label{n10} \eqno{(10)}
$$
Hence it is evident that the motion of a test body in Minkowski space
is due to the action of the force four-vector $F^\nu$:
$$
F^\nu=
-G^\mu_{\alpha\beta} V^\alpha V^\beta
(\delta^\nu_\mu - V^\nu V_\mu), \;\;
V_\mu=\gamma_{\mu\sigma} V^\sigma. \label{n11}\eqno{(11)}
$$
One can readily verify that
$$
F^\nu V_\nu =0. \label{n12} \eqno{(12)}
$$
By definition, the left-hand side of equation (10) is
$$
\f{DV^\nu}{d\sigma}=
\f{dV^\nu}{d\sigma}
+\gamma^{\nu}_{\alpha\beta}
V^\alpha V^\beta. \label{n13} \eqno{(13)}
$$
It must be especially noted that the motion of a test body along a
geodesic of effective Riemannian space can be represented as
motion in Minkowski space due to the action of the force $F^\nu$,
only if the causality principle is satisfied. The force of gravity
and the Riemann curvature tensor, arising from the gravitational
equations~(5.19) and (5.20), are correlated. Thus, if the
curvature tensor is zero, then, by virtue of equations~(5.19) and
(5.20), the gravitational force will also be equal to zero. When
the curvature tensor differs from zero and $R_{\mu\nu}\not= 0$,
the force of gravity will also not be zero. And, on the contrary,
if the force of gravity $F^\nu$, arising from equations~(5.19) and
(5.20), differs from zero, then the Riemann curvature is also not
zero. Equating the gravitational force $F^\nu$ to zero results in
the Riemann curvature tensor being equal to zero.

\subsection*{Is the metric field of a non-inertial reference system a
special case of the physical gravitational field?}

\addcontentsline{toc}{section}
{\hskip 1.8em Is the metric field of a non-inertial reference system
a
special case of the gravitational physical field?}

From the causality conditions~(6.10) and (6.11) it follows that, if
the
vector $L^\nu$ satisfies the condition
$$
\gamma_{\mu\nu} L^\mu L^\nu <0,\label{nn1} \eqno{(1)}
$$
then the inequality
$$
g_{\mu\nu} L^\mu L^\nu <0.\label{nn2} \eqno{(2)}
$$
should also be fulfilled. We now form the convolution of equation
(10.1)
with the aid of the vector $L^\nu$ defined by inequality (1),
$$
m^2 \gamma_{\mu\nu} L^\mu L^\nu =16\pi
\left (
T_{\mu\nu} -\f{1}{2}g_{\mu\nu} T
\right )-\nonumber \\
$$
$$
\hspace*{-17mm} -2R_{\mu\nu}
L^\mu L^\nu
+m^2 g_{\mu\nu}
L^\mu L^\nu. \label{nn3} \eqno{(3)}
$$
Since we are only considering metric fields  of
Minkowski
space, equation (3) is simplified:
$$
m^2\gamma_{\mu\nu}
L^\mu L^\nu =16\pi
\left (
T_{\mu\nu}  -\f{1}{2}
g_{\mu\nu} T
\right )
+m^2 g_{\mu\nu} L^\mu L^\nu. \label{nn4} \eqno{(4)}
$$
In the case of an ideal fluid, the energy-momentum tensor of matter
has
the form
\newpage
$$
\hspace*{-12.5mm} T_{\mu\nu} = (\rho +p) U_\mu U_\nu -pg_{\mu\nu},
$$
\vspace*{-8mm}
$$
\phantom{a} \eqno{(5)}
$$
\vspace*{-8mm}
$$
T=T_{\mu\nu} g^{\mu\nu}
=\rho -3p,\;\;
U^\nu =\f{dx^\mu}{ds}. \nonumber
$$
Substituting (5) into (4) we obtain
$$
m^2\gamma_{\mu\nu}
L^\mu L^\nu
=16\pi
(\rho+p)
(U_\mu L^\mu)^2-
$$
$$
\hspace{-11mm} -8\pi g_{\mu\nu}
L^\mu L^\nu
\left (
\rho - p - \f{m^2}{8\pi}
\right ). \label{nn6} \eqno{(6)}
$$
From conditions (1) and (2) it follows that the right-hand side of
equation
(6) is strictly positive, since
$$
\rho > p+\f{m^2}{8\pi},\label{nn7} \eqno{(7)}
$$
while the left-hand side of equation (6) is strictly negative. Hence
it
follows that in the presence of matter no metric field of Minkowski
space
satisfies the gravitational equations, and therefore the metric
fields
arising in non-inertial reference systems of Minkowski space cannot
be
considered gravitational fields. In the absence of matter,
$\rho=p=0$,
equation (6) has the sole solution
$$
g_{\mu\nu} (x)=\gamma_{\mu\nu} (x). \label{nn8} \eqno{(8)}
$$
\subsection*{On the covariant conservation law}
\addcontentsline{toc}{section}
{\hskip 1.8em On the covariant conservation law}

The covariant conservation law of  matter energy-momentum tensor 
density $T_{\mu}^{\nu}$ in General Relativity (GRT) takes in 
Riemannian space  the following form
$$
\nabla_{\nu}T_{\mu}^{\nu}=\partial_{\nu}T_{\mu}^{\nu}-
\frac{1}{2}T^{\sigma\lambda} \partial_{\mu} g_{\sigma\lambda}\,,\qquad
T^{\sigma\lambda}=-2\frac{\delta L_M}{\delta g_{\sigma\lambda}}\,.
\label{nn1}   \eqno{(1)}
$$
This equation is a straightforward consequence of Gilbert-Einstein equations. 
Though the equation has a covariant form, nevertheless the energy-momentum 
conservation law of matter and gravitational field taken together has in 
GRT a noncovariant appearance
$$
\partial_{\nu}(T_{\mu}^{\nu}+\tau_{\mu}^{\nu})=0. \label{nn2} \eqno{(2)}
$$
Just by this way the gravitational field pseudotensor $\tau_{\mu}^{\nu}$, 
which is not a covariant quantity, arises in GRT. It is impossible in 
principle to write conservation equations of the energy-momentum of 
matter and gravitational field in the generally covariant form. 
The idea that the gravitational energy cannot be localized in GRT 
has arisen from this fact.

If we will not use Eq. (8.2) in the derivation of Eqs. (8.1),
then gravitational equations will take the following form

$$
\sqrt{-\gamma}(-J^{\varepsilon\lambda}+m^2\tilde\phi^{\varepsilon\lambda})
=16\pi\sqrt{-g}(T^{\varepsilon\lambda}+t_g^{\varepsilon\lambda})\,.
\label{nn3} \eqno{(3)}
$$
Here $t_g^{\varepsilon\lambda}$ is the energy-momentum tensor density 
for the gravitational field.

$$
\begin{array}{l}
16\pi\sqrt{-g}t^{\varepsilon\lambda}=
-D_{\mu}D_{\sigma}(\tilde\phi^{\varepsilon\lambda}\tilde\phi^{\mu\sigma}
-\tilde\phi^{\varepsilon\mu}\tilde\phi^{\lambda\sigma})+\\
+D_{\sigma}\tilde\phi^{\varepsilon\lambda}D_{\mu}\tilde\phi^{\mu\sigma}
-D_{\mu}\tilde\phi^{\varepsilon\mu}D_{\sigma}\tilde\phi^{\lambda\sigma}+\\
+\frac{1}{2}g^{\varepsilon\lambda}g_{\rho\tau}
D_{\mu}\tilde\phi^{\alpha\rho}D_{\alpha}\tilde\phi^{\mu\tau}
-g_{\rho\tau}g^{\varepsilon\mu}
D_{\mu}\tilde\phi^{\alpha\rho}D_{\alpha}\tilde\phi^{\lambda\tau}-\\
-g_{\rho\tau}g^{\lambda\nu}
D_{\nu}\tilde\phi^{\alpha\rho}D_{\alpha}\tilde\phi^{\varepsilon\tau}
+g_{\rho\tau}g^{\alpha\beta}
D_{\alpha}\tilde\phi^{\varepsilon\rho}D_{\beta}\tilde\phi^{\lambda\tau}+\\
+\frac{1}{2}\left(g_{\beta\rho}g_{\alpha\tau}-
\frac{1}{2}g_{\beta\tau}g_{\alpha\rho}\right)\times\\
\times\left(g^{\varepsilon\mu}g^{\lambda\nu} 
-\frac{1}{2}g^{\varepsilon\lambda}g^{\mu\nu}\right)
D_{\mu}\tilde\phi^{\lambda\rho}D_{\nu}\tilde\phi^{\beta\tau}-\\
-m^2\left (\sqrt{-g}\tilde g^{\varepsilon\lambda}
-\sqrt{-\gamma}\tilde\phi^{\varepsilon\lambda}
+\tilde g^{\varepsilon\alpha}\tilde g^{\lambda\beta}\gamma_{\alpha\beta}
-\frac{1}{2}\tilde g^{\varepsilon\lambda}\tilde g^{\alpha\beta}\gamma_{\alpha\beta}\right).
\end{array} \label{nn4} \eqno{(4)}
$$
$$
J^{\varepsilon\lambda}=-D_{\mu}D_{\nu}
(\gamma^{\mu\nu}\tilde g^{\varepsilon\lambda}
+\gamma^{\varepsilon\lambda}\tilde g^{\mu\nu}
-\gamma^{\varepsilon\nu}\tilde g^{\mu\lambda}
-\gamma^{\varepsilon\mu}\tilde g^{\lambda\nu}).\label{nn5} \eqno{(5)}
$$
Let us mention that expression
$$
D_{\sigma}(\tilde\phi^{\varepsilon\lambda}\tilde\phi^{\mu\sigma}
-\tilde\phi^{\varepsilon\mu}\tilde\phi^{\lambda\sigma})
$$
is antisymmetric under permutation of indicies $\lambda$ ¨ $\mu$,
and so the following identity takes place:
$$
D_{\lambda}D_{\mu}D_{\sigma}(\tilde\phi^{\varepsilon\lambda}\tilde\phi^{\mu\sigma}
-\tilde\phi^{\varepsilon\mu}\tilde\phi^{\lambda\sigma})=0\,.
$$
It is also easy to get convinced, that the following equation is valid
$$
D_{\lambda}J^{\sigma\lambda}=0\,.\label{nn6} \eqno{(6)}
$$
Gravitational field equation (3) can be presented also in other form (5.19)
$$
\begin{array}{l}
\sqrt{-g}\left(R^{\mu\nu}-\frac{1}{2}g^{\mu\nu}R\right)+
\\
+\frac{m^2}{2}\left [\tilde g^{\mu\nu}
+\left (\tilde g^{\mu\alpha}g^{\nu\beta}
-\frac{1}{2}\tilde g^{\mu\nu}g^{\alpha\beta}\right)
\gamma_{\alpha\beta}\right]
=8\pi T^{\mu\nu}\,. 
\end{array} \label{nn7} \eqno{(7)}
$$
From Eqs. (7) it follows
$$
m^2\sqrt{-g}\left (g^{\mu\alpha}g^{\nu\beta}
-\frac{1}{2}g^{\mu\nu}g^{\alpha\beta}\right )
\nabla_{\mu}\gamma_{\alpha\beta}
=16\pi\nabla_{\mu}T^{\mu\nu}\,. \label{nn8} \eqno{(8)}
$$
By taking into account relation
$$
\nabla_{\mu}\gamma_{\alpha\beta}
=-G_{\mu\alpha}^{\sigma}\gamma_{\sigma\beta}
-G_{\mu\beta}^{\sigma}\gamma_{\sigma\alpha}\,, \label{nn9} \eqno{(9)}
$$
we find
$$
\begin{array}{l}
\left(g^{\mu\alpha}g^{\nu\beta}-
\frac{1}{2}g^{\mu\nu}g^{\alpha\beta}\right)\nabla_{\mu}\gamma_{\alpha\beta}=
\\
=-g^{\mu\alpha}g^{\nu\beta}G_{\mu\alpha}^{\sigma}\gamma_{\sigma\beta}-
g^{\mu\alpha}g^{\nu\beta}G_{\mu\beta}^{\sigma}\gamma_{\sigma\alpha}+
g^{\mu\nu}g^{\alpha\beta}G_{\mu\alpha}^{\sigma}\gamma_{\sigma\beta}\,.
\end{array} \label{nn10} \eqno{(10)}
$$
It is easy to see that the following identity takes place:
$$
-g^{\mu\alpha}g^{\nu\beta}G_{\mu\beta}^{\sigma}\gamma_{\sigma\alpha}
+g^{\mu\nu}g^{\alpha\beta}G_{\mu\alpha}^{\sigma}\gamma_{\sigma\beta}\,.
\label{nn11} \eqno{(11)}
$$
Therefore we have
$$
\left (g^{\mu\alpha}g^{\nu\beta}-
\frac{1}{2}g^{\mu\nu}g^{\alpha\beta}\right )\nabla_{\mu}\gamma_{\alpha\beta}
=-g^{\mu\alpha}g^{\nu\beta}G_{\mu\alpha}^{\sigma}\gamma_{\sigma\beta}\,.
\label{nn12} \eqno{(12)}
$$
By substituting expression  (4.5) instead of  $G_{\mu\alpha}^{\sigma}$ we get
$$
\begin{array}{l}
\left(g^{\mu\alpha}g^{\nu\beta}-
\frac{1}{2}g^{\mu\nu}g^{\alpha\beta}\right)\nabla_{\mu}\gamma_{\alpha\beta}=
\\
=\gamma_{\mu\lambda}g^{\mu\nu}
\left(D_{\sigma}g^{\sigma\lambda}
+G_{\alpha\sigma}^{\sigma}g^{\alpha\lambda}\right)\,.
\end{array} \label{nn13} \eqno{(13)}
$$
After taking into account Eq. (13), Eq. (8) takes the following form
$$
16\pi\nabla_{\mu}T^{\mu\nu}
=m^2\sqrt{-g}\,\gamma_{\mu\lambda}\,g^{\mu\nu}
\left (D_{\sigma}g^{\sigma\lambda}
+G_{\alpha\sigma}^{\sigma}g^{\alpha\lambda} \right )\,.
\label{nn14} \eqno{(14)}
$$
Applying equation
$$
\sqrt{-g}\left (D_{\sigma}g^{\sigma\lambda}
+G_{\alpha\sigma}^{\sigma}g^{\alpha\lambda} \right )
=D_{\sigma}\tilde \phi^{\sigma\lambda}\,,
\label{nn15} \eqno{(15)}
$$
we find
$$
m^2\gamma_{\nu\lambda}D_{\sigma}\tilde\phi^{\sigma\lambda}
=16\pi\nabla_{\mu}T_{\nu}^{\mu}\,.
\label{nn16} \eqno{(16)}
$$
According to Eq. (3) we have
$$
m^2\tilde\phi^{\sigma\lambda}
=J^{\sigma\lambda}+16\pi\sqrt{\frac{g}{\gamma}}
\left (T^{\sigma\lambda}+t_g^{\sigma\lambda} \right )\,.
\label{nn17} \eqno{(17)}
$$
Substituting Eq. (17) into Eq. (16) we get
$$
D_{\sigma}\left [\sqrt{\frac{g}{\gamma}}
\left (T^{\sigma\lambda}+t_g^{\sigma\lambda} \right )\right ]
=\gamma^{\nu\lambda}\nabla_{\mu}T_{\nu}^{\mu}\,.
\label{nn18} \eqno{(18)}
$$
When matter equations of motion are valid we have
$$
\frac{\delta L_M}{\delta \phi_A}=0\,.
\label{nn19} \eqno{(19)}
$$
According to strong identity ('.16) the following equality is valid
$$
\nabla_{\mu}T_{\nu}^{\mu}=0\,,
\label{nn20} \eqno{(20)}
$$
and therefore, according to Eq. (18), the covariant conservation
law for energy-momentum of matter and gravitational field  
taken together is as follows
$$
D_{\sigma}\left [\sqrt{-g}
\left (T^{\sigma\lambda}+t_g^{\sigma\lambda}\right )\right ]=0\,.
\label{nn21} \eqno{(21)}
$$

So, if in GRT covariant law (1) leads to a noncovariant conservation 
law for energy-momentum of matter and gravitational field (2),
and also to arising of a noncovariant quantity -- pseudotensor 
$\tau_{\mu}^{\nu}$ of the gravitational field, then in RTG
covariant law (1) together with gravitational equations
written in a form (3) or (7) exactly leads to the covariant
conservation law for energy-momentum of matter and gravitational
field written in a form (21). In Eq. (21) the gravitational 
component $\sqrt{-g}\,t_g^{\sigma\lambda}$ enters in additive 
form under the Minkowski space covariant derivative symbol, 
whereas  the gravitational component disappears from Eq. (20), 
it is used for generating the effective Riemannian space and 
so only the energy-momentum tensor density of matter in Riemannian
space stays under the covariant derivative symbol.
The presence of covariant conservation laws  (21) is just 
the point to see that the gravitational energy as also all 
other forms of energy is localizable.


\subsection*{H.Poincare On the dynamics of the electron}
(5 June 1905)\footnote{Poincare H. Sur la dynamique de 
l'electron // Comptes rendus hebdomadaires des seances de 
l'Akademie des sciences. -- Paris, 1905. -- V.140. -- P.1504--1508.}.
\addcontentsline{toc}{section}
{\hskip 1.8em H.Poincare On the dynamics of the electron}
\noindent
(The comments are italicized and indicated by an asterisk).\\

\noindent
It seems at first sight that the aberration of light and the 
related optical and electrical phenomena would provide us with
a means of determining the absolute motion of the Earth, or rather 
its motion not with respect to the other stars, but with respect to
the ether. Actually, this is not so: experiments in which only terms
of the first order in the aberration were taken into account first 
yielded negative results, which was soon given an explanation; 
but Michelson also, who proposed an experiment in which
terms depending on the square aberration were noticeable, 
also met with no luck. The impossibility to disclose 
experimentally the absolute motion of the Earth seems
to be a general law of Nature.

\hspace{-15pt}*\quad{\it 
``Experiment has provided numerous
facts justifying the following generalization: absolute motion
of matter, or, to be more precise, the relative motion of weighable 
matter and ether cannot be disclosed. All that can be done 
is to reveal the motion of weighable matter with respect to
weighable matter'' \char245\footnote{Poincare H. On Larmor's theory~// 
The relativity principle: Collection of works on special relativity 
theory.~--- Moscow, 1973.~--- P.7.}.

  The above words written by Poincare ten years earlier quite 
  clearly demonstrate that his vision of a general law determining 
  the impossibility of absolute motion of matter had been maturing since long ago.

In development of his idea on ¥the total impossibility of defining
absolute motionõ in relation to the new hypothesis, put forward by 
Lorentz and according to which all bodies should experience a
decrease in length by $1/2\cdot 10^{-9}$ in the direction of
motion of the Earth, Poincare wrote:

 ``Such a strange property seems to be a real
 {\bf coup de pouce} presented by Nature itself, 
 for avoiding the disclosure of absolute motion with the
 aid of optical phenomena. I can't be satisfied and I must
 here voice my opinion: I consider quite probable that optical 
 phenomena depend only on the relative motion of the material 
 bodies present, of the sources of light or optical instruments,
 and this dependence is not accurate up to orders of magnitude 
 of the square or cubic aberration, but rigorous. This principle
 will be confirmed with increasing precision, as measurements 
 become more and more accurate.

\addtocounter{footnote}{+2}
Will a new {\bf coup de pouce} or a new hypothesis 
be necessary for each approximation? Clearly this is not so:
a well formulated theory should permit proving a principle at
once with all rigour. The theory of Lorentz does not permit this yet. 
But, of all theories proposed it is the one nearest to achieving this goal''
\char245\footnote{Poincare H. Electricite et optique: 
La lumiere et les theories electrodynamiques. --- 2 ed., rev. 
et complettee par Jules Blondin, Eugene Neculcea. Paris: Gauthier--Villars, 1901.}.

In a report to the Congress of art and science held in Saint
Louis in 1904 Poincare among the main principles of theoretical
physics formulates the relativity principle, in accordance with which, 
in the words of Poincare, 
``the laws governing physical
phenomena should be the same for a motionless observer and
for an observer experiencing uniform motion, so there is no
way and cannot be any way of determining whether one experiences 
such motion or not''\char245 \footnote{Poincare H. L'etat et l'avenir 
de la Physique mathematique // Bulletin des Sciences Mathematiques.
--- Janvier 1904 --- V.28. Ser.2 --- P.302--324; The Monist. --- 1905. V.XV, N1.}.
}

An explanation has been proposed by Lorentz, who has introduced 
the hypothesis of a contraction experienced by all bodies
in the direction of the motion of the Earth; this contraction
should account for Michelson's experiment and for all other
relevant experiments performed to date. It would, however, 
leave place for other even more subtle experiments, more
simple to be contemplated than to be implemented, aimed at
revealing absolute motion of the Earth. But considering the
impossibility of such a claim to be highly probable one may 
foresee that these experiments, if ever they will be performed, 
to once again provide a negative result. Lorentz has attempted 
to complement and alter the hypothesis so as to establish a
correspondence between it and the postulate of total impossibility
of determining absolute motion. He has succeeded in doing so in
his article entitled \char165Electromagnetic phenomena in
a system moving with any velocity smaller than that of light
\char245\  (Proceedings de l'Academie d'Amsterdam, 27 May 1904).

The importance of this issue has induced me to consider it once again;
the results I have obtained are in agreeement with those obtained
by Lorentz in what concerns all the main points; I have only attempted 
to modify them somewhat and to complement them with some details.

The essential idea of Lorentz consists in that the equations of 
the electromagnetic field will not be altered by a certain transformation
(which I shall further term the Lorentz transformation) of the following form:
$$
x' = \gamma l (x-\beta t), \quad y' = ly, \, z' = lz, \quad t' = \gamma l (t - \beta x),
\eqno(1)
$$
where $x, y, z$ are the coordinates and $t$ is the time before 
the transformation; and $x', y', z'$ and $t'$ are the same after
the transformation. The quantity $\beta$ is a constant determined 
by the transformation
$$
\gamma = \frac{1}{\sqrt{1-\beta^2}},
$$
while $l$ is a certain function of $\beta$:

\hspace{-15pt}*\quad{\it {Poincare writes: 
``The idea of Lorentz'',
 but Lorentz never wrote such words
before Poincare. Here, Poincare formulated his own fundamental idea, 
but attributed it fully to Lorentz. Probably more than any other person, 
did he always highly esteem and note each person, who gave his thought 
an impetus and presented him with the happiness of creativity.
He was totally alien to issues of his own personal priority.

From formulae (1) it is immediately seen that the condition $x=\beta t$
corresponds to the origin $(x' = y' = z' = 0)$ of the new reference system. 
In other words, the new origin is shifted in the reference system $x, y, z$ with 
a speed $\beta$ along the $x$-axis.

Thus, the Lorentz tranformations relate the variables $(x, y, z, t)$ 
referred to one reference system and the variables $(x', y', z', t')$ 
in another system moving uniformly in a straight line along the x-axis
with the velocity $\beta$ relative to the first system.
{\sloppy

}
Proof of the statement that the equations of the electromagnetic field
do not alter under the Lorentz transformations signifies that 
electromagnetic phenomena are described in both reference systems 
by identical equations and that, consequently, no electromagnetic 
processes can be utilized to distinguish between the $(x, y, z, t)$
reference system and the $(x', y', z', t')$ reference system moving
uniformly in a straight line with respect to the first system.

We see that invariance of the equations of the electromagnetic
field under transformations of the Lorentz group results in 
the relativity principle being fulfilled in electromagnetic 
phenomena. In other words, the relativity principle for electromagnetic
phenomena follows from the Maxwell--Lorentz
equations in the form of a rigorous mathematical truth.}}

In this transformation the $x$-axis plays a particular role,
but it is clearly possible to construct such a transformation, 
in which this role will be assumed by a certain straight line 
passing through the origin. The set of all such transformations
together with all spatial rotations should form a group; but for
this to take place it is necessary that $l = 1$; hence one is
led to assume $l = 1$, which is precisely the consequence 
obtained by Lorentz in another way.

\hspace{-15pt}*\quad{\it{It must be underlined that, 
by having established the group nature of the set of 
all purely spatial transformations together with the 
Lorentz transformations, that leave the equations of 
electrodynamics invariant, Poincare thus discovered 
the existence in physics of an essentially new type 
of symmetry related to the group of linear space-time 
transformations, which he called the Lorentz group.

Supplemented with transformations of space coordinate and 
time translations, the Lorentz group forms a maximum group
of space-time transformations, under which all equations 
of motion for particles and fields remain invariant and 
which now is called the Poincare group, the name given
to it subsequently by E.Wigner. Richard Feynman wrote 
about this fact as follows: 
``Precisely Poincare 
proposed to find out what one can do with equations without 
altering their form. He was the person who had the idea 
to examine the symmetry properties of physical laws''.}}

Let $\rho $ be the charge density of the electron, and $v_x$, $v_y$, and $v_z$ 
the components of the electron velocity before the transformation; then,
after applying the transformation one has for these same quantities
$\rho'$, $v{_x}'$, $v{_y}'$, and $v{_z}'$ the following:
$$
\begin{array}{l}
\rho' = \gamma l ^{-3} \rho (1-\beta v_x), \qquad
\rho' v{_x}'= \gamma l ^{-3} \rho (v_x - \beta), \\
\rho' v{_y}' = l^{-3} \rho v_y,\qquad \rho' v{_z}' = l^{-3} \rho v_z.
\end{array} \label{nn2}\eqno(2)
$$

These formulas differ somewhat from the ones found by Lorentz.

Now, let $\vec {f}$ and $\vec {f'}$ be the three-force components 
before and after application of the transformation (the force 
is referred to unit volume); then
$$
f'_x = \gamma l ^{-5} (f_x - \beta \vec {f} \vec {v}); f'_y =
l^{-5} f_y; f'_z = l^{-5}f_z.
\eqno(3)
$$

These formulas also differ somewhat from the ones proposed by Lorentz; 
the additional term in $\vec {f} \cdot \vec {v}$ reminds the result 
earlier obtained by Lienard.

If we now denote by $\vec {F}$ and $\vec {F'}$ the force components 
referred to the electron mass unit, instead of unit volume, we obtain
$$
F'_x = \gamma l ^{-5} \frac{\rho}{\rho'}
(F_x - \beta \vec {F}\cdot \vec {v}); F'_y =
\frac{\rho}{\rho'}l^{-5} F_y; F'_y = \frac{\rho}{\rho'}l^{-5} F_x.
\eqno(4)
$$

\hspace{-15pt}*\quad {\it {Formulae (2), (3), and (4) comprise 
the relativistic transformation laws, first established by Poincare, 
for the charge density and velocity of motion of an electron, 
referred both to unit charge and unit volume.

Thus, Poincare, as compared with Lorentz, made the decisive step in 
this work and laid down the foundations of relativity theory.

It is notable that, while developing totally new ideas in articles on
the dynamics of the electron and correcting and complementing Lorentz, 
Poincare is careful to paid maximum tribute to Lorentz as the discoverer
and leaves it to others to judge about his own personal contribution 
to the creation of relativity theory.

The testimony of Lorentz himself is extremely important in this respect,
since it permits filling in the gap, sometimes left by certain authors 
writing about the history of the creation of relativity theory, and to
do justice to Poincare as the creator of relativistic mechanics 
and special relativity.

\addtocounter{footnote}{+4}
Thus, in discussing the relativistic transformation formulae for 
the velocities, charge densities, and current of the electron, 
Lorentz wrote:\footnote{Lorentz H.A. Two articles by Henri Poincare 
on mathematical physics // The relativity principle: Collection 
of works on special relativity theory. -- Moscow, 1973. -- P.189--196.\\
Formulae (4) and (7) dealt with by Lorentz are the transformation 
formulae for the electron velocities and charge densities, respectively.} 
``Formulae (4) and (7) are absent in my article published in 1904, 
since I didn't even think of a direct path leading to them, because 
I thought an essential difference existed between the systems $x, y, z, t$ 
and $x', y', z', t'$. In one of them -- such was my reasoning --
coordinate axes were used that had a fixed position in ether and
what could be called \char165true\char245\  time; in the other system,
on the contrary, one dealt with simply auxiliary quantities introduced 
only with the aid of a mathematical trick. Thus, for instance, the variable 
$t'$ could not be considered to be \char165time\char245\  in the same 
sense as the variable~$t$.

Given such reasoning, I had no intention of describing phenomena 
in the system $x', y', z', t'$ in precisely the same manner as in 
the system $x, y, z, t$...''
And further, in the same work:
``... I was unable to achieve total invariance of the equations;
my formulae remained cumbersome owing to additional terms, that should
have disappeared. These terms were too small to exert noticeable 
influence on the phenomena, and this supplied me with an explanation 
for their being independent of the Earth's motion revealed by obs
and universal truth.

Contrariwise, Poincare achieved total invariance of the equations 
of electrodynamics and formulated the
<<relativity postulate>>
--- a term introduced by him. Indeed, adopting the point of view, that
I had failed to take into account, he derived formulae (4) and (7). 
We should add, that in correcting the defects of my work he never 
reproached me for them''.}}

Lorentz also arrived at the necessity of assuming a moving electron
to have the shape of a compressed ellipsoid; the same hypothesis 
was made by Langevin, but while Lorentz assumed the two axes of 
the ellipsoid to be constant, in agreement with his hypothesis 
that $l = 1$, Langevin assumed, contrariwise, the volume of the 
ellipsoid to be constant. Both authors showed the two hypotheses
to be in the same good agreement with the experiments performed 
by Kaufmann, as the initial hypothesis of Abraham (the spherical electron). 
The advantage of Langevin's hypothesis consists in its being sufficient,
i.e. it suffices to consider the electron to be deformable and 
incompressible for explaining why it assumes an ellipsoidal
shape in motion. But I can show, without contradicting Lorentz, 
that this hypothesis cannot be consistent with the impossibility 
of revealing absolute motion. As I have already said, this occurs
because $l = 1$ is the only hypothesis for which the Lorentz 
transformations form a group.

But in the Lorentz hypothesis, also, the agreement between
the formulas does not occur just by itself; it is obtained
together with a possible explanation of the compression of
the electron under the assumption that {\bf {the deformed 
and compressed electron is subject to constant external pressure,
the work done by which is proportional to the variation of volume
of this electron.}}

Applying the principle of least action, I can demonstrate the 
compensation under these conditions to be complete, if inertia 
is assumed to be of a totally electromagnetic origin, as generally
acknowledged after Kaufmann's experiments, and if all forces are 
of an electromagnetic origin, with the exception of the constant 
pressure of which I just spoke and which acts on the electron. 
Thus, it is possible to explain the impossibility of revealing 
the absolute motion of the Earth and the contraction of all
bodies in the direction of the Earth's motion.

But this is not all. In the quoted work Lorentz considers
it necessary to complement his hypothesis with the assumption
that in the case of uniform motion all forces, of whatever origin, 
behave exactly like electromagnetic forces, and that, consequently,
the influence of the Lorentz transformation on the force components 
is determined by equations (4).

\hspace{-15pt}*\quad {\it {Here, Poincare in development of the
assumption expressed by Lorentz extends the Lorentz transformations 
to all forces, including, for instance, gravitational forces.

He was the first to point out that the relativity postulate requires
such a modification of the laws of gravity, according to which the 
propagation of forces of gravity is not instantaneous, but proceeds
with the speed of light.}}

It has turned out necessary to consider more carefully this
hypothesis and, in particular, to clarify which changes it
compels us to introduce into the laws of gravity. This is 
just what I attempted to determine: I was first induced to 
assume the propagation of gravity forces to proceed with the
speed of light, and not instantaneously. This seems to 
contradict the result obtained by Laplace who claims that
although this propagation may not be instantaneous, it is at 
least more rapid than the propagation of light. However,
the issue actually raised by Laplace differs significantly 
from the issue dealt with here by us. According to Laplace,
a finite propagation velocity was {\bf {the sole}} alteration, 
introduced by him to Newton's law. Here, also, a similar change
is accompanied by many others; hence, partial compensation 
between them is possible, and it actually does take place.

Consequently, if we speak about the position or velocity of 
a body exerting attraction, we shall bear in mind its position 
or velocity at the moment, when {\bf {the gravitational wave}}
departs from this body; if we speak about the position or 
velocity of a body being attracted, we shall intend its position 
or velocity at the moment, when this body being attracted
is overcome by the gravitational wave emitted by another body: 
the first moment clearly precedes the second.

\looseness=1
Hence, if $x, y, z$ are the projections onto three axes of 
the vector $\vec {r}$ connecting the two positions and if
$\vec {v} = (v_x, v_y, v_z)$ are the velocity components 
of the body attracted and $\vec {v_{1}} = (v_{1x}, v_{1y}, v_{1z})$ 
are the velocity components of the
attracting body, then the $1z$ three components of 
the attraction (which I may also call $\vec {F}$) will 
be functions of $\vec {r}, \vec {v}, \vec {v_1}$. 
The question is whether these functions can be defined 
in such a way that they behave under the Lorentz transformation
in accordance with equations (4) and that the conventional law of 
gravity be valid in all cases of the velocities $\vec {v}, \vec {v_1}$ 
being sufficiently small to allow neglecting their square values as
compared with the square speed of light?

The answer to this question must be affirmative. It has been revealed 
that the attraction, taking into account the correction, consists of
two forces, one of which is parallel to the components of the vector 
$\vec {r}$, and the other to the components of the velocity $\vec {v_1}$.

The disagreement with the conventional law of gravity, as I just pointed out,
is of the order of $v^2$; if, on the other hand, one assumes, as Laplace did, 
the propagation velocity to be equal to the speed of light, 
this divergence will be of the order of $v$, i.e. 10000 times greater.
Consequently, 
at first sight, it does not seem absurd to assume astronomical 
observations to be insufficiently precise for revealing the 
smallest imaginable divergence. Only a profound investigation 
can resolve this issue.

\hspace{-15pt}*\quad{\it{Poincare thus introduces the physical 
concept of gravitational waves, the exchange of which generates 
gravitational forces, and supplies an estimation of the contribution 
of relativistic corrections to Newton's law of gravity.

For example, he shows that the terms of first order in $v/c$ 
cancel out exactly and so the relativistic corrections to Newton's 
law are quantities of the order of ($v/c$)$^2$.

These results remove the difficulty noted previously by Laplace and 
permit making the conclusion that the hypothesis equating the speeds 
of light and of gravitational influence is not in contradiction with
observational data.\\
\indent Thus, in this first work Poincare already gave a general and 
precise formulation of the main points of relativity theory.
It is here that such concepts as the following first appeared: 
the Lorentz group, invariance of the equations of the electromagnetic 
field with respect to the Lorentz transformations, the transformation 
laws for charge and current, the addition formulae of velocities, 
the transformation laws of force. Here, also, Poincare extends 
the transformation laws to all the forces of Nature, whatever their 
origin might be.}}

\thispagestyle{empty}

\thispagestyle{empty}
\newpage
\pagestyle{empty}
\small
{\baselineskip=1.1\normalbaselineskip
\tableofcontents}

\end{document}